%% file: Shimajiri2015b.tex
\documentclass[11pt,preprint]{aastex}

\usepackage{multirow}

\newcommand{\ambienttracers}{CN, C$^{17}$O, C$_2$H, HNC, HN$^{13}$C, H$^{13}$CO$^+$, N$_2$H$^+$, c-C$_3$H$_2$, and CH$_3$CN }
\newcommand{\possibleambienttracers}{$^{13}$CS, H$^{15}$NC, HC$^{18}$O$^+$, NH$_2$D, CH$_3$CCH }
\newcommand{\outflowtracers}{CO, CS, HCN, and HCO$^{+}$ }
\newcommand{\shocktracers}{C$^{34}$S, SO, SiO, H$^{13}$CN, HC$^{15}$N, H$_2^{13}$CO, H$_2$CS, HC$_3$N, and CH$_3$OH }

\slugcomment{\today}

\shorttitle{Spectral-Line Survey toward OMC 2-FIR 4}
\shortauthors{Shimajiri et al.}

\begin{document}


\title{Spectral-Line Survey at Millimeter and Submillimeter Wavelengths \\ toward an Outflow-Shocked Region, OMC 2-FIR 4}


\author{Yoshito Shimajiri\altaffilmark{1,2,3}, Takeshi Sakai\altaffilmark{4}, Yoshimi Kitamura\altaffilmark{5},  Takashi Tsukagoshi\altaffilmark{6}, 
Masao Saito\altaffilmark{2,3,7}, Fumitaka Nakamura\altaffilmark{2},  Munetake Momose\altaffilmark{6}, Shigehisa Takakuwa\altaffilmark{8}, Takahiro Yamaguchi\altaffilmark{9},  
Nami Sakai\altaffilmark{9},
Satoshi Yamamoto\altaffilmark{9}, 
and Ryohei Kawabe\altaffilmark{2,7,10}}

\email{Yoshito.Shimajiri@cea.fr}

\altaffiltext{1}{Laboratoire AIM, CEA/DSM-CNRS-Universit$\acute{\rm e}$ Paris Diderot, IRFU/Service d'Astrophysique, CEA Saclay, F-91191 Gif-sur-Yvette, France}
\altaffiltext{2}{National Astronomical Observatory of Japan, 2-21-1 Osawa, Mitaka, Tokyo 181-8588, Japan}
\altaffiltext{3}{Nobeyama Radio Observatory, 462-2 Nobeyama, Minamimaki, Minamisaku, Nagano 384-1305, Japan}
\altaffiltext{4}{Graduate School of Informatics and Engineering, The University of Electro-Communications, Chofu, Tokyo 182-8585, Japan}
\altaffiltext{5}{Institute of Space and Astronoutical Science, Japan Aerospace Exploration Agency, 3-1-1 Yoshinodai, Chuo-ku, Sagamihara 252-5210, Japan}
\altaffiltext{6}{College of Science, Ibaraki University, 2-1-1 Bunkyo, Mito, Ibaraki 310-8512, Japan}
\altaffiltext{7}{SOKENDAI (The Graduate University for Advanced Studies), 2-21-1 Osawa, Mitaka, Tokyo 181-8588, Japan}
\altaffiltext{8}{Academia Sinica Institute of Astronomy and Astrophysics, P.O. Box 23-141, Taipei 10617, Taiwan}
\altaffiltext{9}{Department of Physics, Graduate School of Science, The University of Tokyo, 7-3-1 Hongo, Bunkyo, Tokyo 113-0033, Japan}
\altaffiltext{10}{Department of Astronomy, School of Science, University of Tokyo, Bunkyo, Tokyo 113-0033, Japan}


\begin{abstract}
We performed the first spectral-line survey at 82--106 GHz and 335--355 GHz toward the outflow-shocked region, OMC 2-FIR 4, the outflow driving source, FIR 3, and the northern outflow lobe, FIR 3N. We detected 120 lines of 20 molecular species. The line profiles are found to be classifiable into two types:  one is a single Gaussian component with a narrow ($<$ 3 km s$^{-1}$) width and another is  two Gaussian components with narrow and wide ($>$ 3km s$^{-1}$) widths. The narrow components for the most of the lines are detected at all positions, suggesting that they trace the ambient dense gas. For \outflowtracers, the wide components are detected at all positions, suggesting the outflow origin. The wide components of \shocktracers are detected only at FIR 4, suggesting the outflow-shocked gas origin. The rotation diagram analysis revealed that the narrow components of C$_2$H and H$^{13}$CO$^+$ show low temperatures of 12.5$\pm$1.4 K, while the wide components show high temperatures of 20--70 K. This supports our interpretation that the wide components trace the outflow and/or outflow-shocked gas. We compared observed molecular abundances relative to H$^{13}$CO$^+$ with those of the outflow-shocked region, L1157 B1, and the hot corino, IRAS 16293-2422. Although we cannot exclude a possibility that the chemical enrichment in FIR 4 is caused by the hot core chemistry, the chemical compositions in FIR 4 are more similar to those in L1157 B1 than those in IRAS 16293-2422.

\end{abstract}





\section{Introduction} \label{intro}
Enormous progress has been achieved in the past few decades in studying the chemical composition of dense molecular gas in star-forming regions. The chemical composition and evolution in the dense interstellar medium (ISM) themselves are of great interest. In addition, they are very useful for diagnostics of protostar or protoplanetary disk evolution, and also of shocks and energy sources of extragalactic nuclei. Shock chemistry is a key to understand the chemical composition of ISM, because shock waves are ubiquitous in astrophysical phenomena; evidence for shock is usually found in outflows and jets associated with star formation. 
One of the best-studied shocked regions is L 1157 B1. Recently, spectral line surveys were performed at the wavelengths of 3 mm and 500 $\mu$m, which revealed the chemical composition of the outflow-shocked region \citep{Codella01, Sugimura11,Yamaguchi12, Benedettini12}.  
However, observational studies of shocked regions focused on molecular compositions have been limited to a few sources, 
although theoretical studies on shock physics have been extensively made since the 1980s  \citep{McKee80, Neufeld95}. 
Physical and chemical properties of shocked gas depend on physical conditions of ambient medium and shock velocity. 
To unveil  the nature of the shocked gas, we need to observe various outflow-shocked regions. 

FIR 4 is located in OMC2, which is the northern part of the Orion A Molecular cloud \citep[$d$ = 400 pc,][]{Menten07,Sandstrom07,Hirota08}. 
This region is known as an active cluster-forming region where three 3.6-cm free-free emission sources and nine MIR sources are embedded\citep{Reipurth99,Nielbock03}.
\citet{Williams03} and  \citet{Takahashi08b} found a molecular outflow driven by FIR 3 in the $^{12}$CO (1--0, 3--2) lines. \citet{Shimajiri08} suggested that the outflow driven by FIR 3 interacts with the 0.07 pc-scale dense gas associated with FIR 4 (FIR 4 clump), from morphological, kinematical, chemical, and physical evidence. 
Morphologically, the length ($\sim$0.2 pc) of the northeast (NE) CO (1--0, 3--2) outflow lobe from FIR 3 is smaller than that of the  southwest (SW) lobe ($\sim$0.1 pc). Moreover, the FIR 4 clump is located at the tip of the SW lobe. 
These features strongly suggest that the SW lobe has been stemmed by the FIR 4 clump. 
In addition, the shock tracers of SiO ($v$=0, 2--1) and CH$_3$OH ($J_K$=7$_K$--6$_K$ ; $K$=-1, 0, 2) are detected at the interface between the SW lobe and the FIR 4 clump.
Moreover, the abrupt increase of the velocity width of CO, H$^{13}$CO$^{+}$, CH$_3$OH, and SiO at the interface between the SW lobe and the FIR 4 clump suggests the presence of the outflow interaction.
Consequently, it is likely that the high-velocity outflow collides with the quiescent dense clump. Hence, FIR 4 is a good target to investigate the chemistry of the outflow-shocked region.

This paper is organized as follows: 
In Sect. 2, we describe our targets, FIR 3N, FIR 3, and FIR 4. 
In Sect. 3, we describe the Nobeyama 45 m and Atacama submillimeter telescope experiment (ASTE) observations and data reduction procedures using the common astronomy software application (CASA). 
In Sect. 4, we present results of our line survey observations in the 82--106 GHz and 335--355 GHz toward FIR 3N, FIR 3, and FIR 4.
In Sect. 5, we discuss the nature of the dense gas in FIR 4 on the basis of the velocity widths, rotation temperatures, and fractional abundances of the detected molecules.
We compare rotation temperatures and column densities among molecules detected in the Nobeyama 45 m and ASTE spectral line surveys.
At the end of this section, we show the fractional abundances relative to H$^{13}$CO$^+$ in FIR 4. Then, we compare them with the corresponding abundances in the L 1157 B1 and IRAS 16293-2422 regions. 
In Sect. 6, we summarize this paper.

\section{Our Targets}
Our target sources are FIR 3N, FIR 3, and FIR 4 in OMC 2, as indicated in Fig. \ref{obspoint}.
Table  \ref{obstable} shows the positions of the three sources, the CO (1--0) intensities, and the dust continuum flux densities at 1.1 mm with the AzTEC  \citep[also see Sect. \ref{fractional_abundance}]{Shimajiri11}. Here, we describe the detailed features of the sources. 

\subsection{FIR 3N}
FIR 3N is a CO peak position in the northeastern lobe of the blue-shifted outflow driven by FIR 3 \citep{Takahashi08b}.
In this area, it is likely that no prominent shock occurs for the following reasons. No dense condensations have been found by the H$^{13}$CO$^{+}$ (1--0), N$_2$H$^+$ (1--0), and 1.1 mm dust continuum observations \citep{Ikeda07, Tatematsu08, Shimajiri15}.
Although \citet{Yu97} and \citet{Stanke02} found several H$_2$ knots, the shock tracers such as the SiO ($v$=0, 2--1) and CH$_3$OH ($J_K$=7$_K$--6$_K$;$K$=-1,0,1) lines have not been detected toward the outflow lobe \citep{Shimajiri08}.

\subsection{FIR 3}
FIR 3 is known as a Class 0/I object, which ejects the molecular outflow along the northeast-southwest direction with a dynamical time scale of 1.4 $\times$ 10$^4$ yr \citep{Chini97, Takahashi08b}. 
The momentum flux, $F_{\rm CO}$, of the northeastern blue-shifted lobe is 2.3 $\times$10$^{-3}$ $M_{\odot}$ km s$^{-1}$ yr$^{-1}$,  the highest value in the OMC 2/3 region.  
The 1.3 mm and 850 $\mu$m dust continuum emission and two MIR sources (MIR 21 and 22) are associated with the source FIR 3 \citep{Chini97, Johnstone99, Nielbock03}. 
FIR 3 has been identified as SOF 2N in the 4.5 $\mu$m and 37.1 $\mu$m emission by the SOFIA/FORCAST and its total luminosity is estimated to be 300 $L_{\odot}$ \citep{Adams12}.

\subsection{FIR 4}
FIR 4 is the strongest dust-continuum source in the OMC 2 region \citep{Chini97,Shimajiri15}. 
The Nobeyama Millimeter Array (NMA) observations at an angular resolution of $\sim$3$\arcsec$ have revealed that FIR 4 consists of eleven compact dust clumps \citep{Shimajiri08}. 
The source has been identified as SOF 3 in the 4.5 $\mu$m and 37.1 $\mu$m emission by the SOFIA/FORCAST, and its total luminosity is estimated to be 50 $L_{\odot}$. However, the clumps detected in the 3.3 mm dust continuum emission by \citet{Shimajiri08} do not coincide with the 8--160 $\mu$m emission sources \citep{Adams12}.
\citet{Shimajiri08} concluded that the molecular outflow driven from FIR 3 interacts with the dense gas associated with FIR 4 (see Sect. \ref{intro}). 
Thus, FIR 4 is a good target to investigate the chemistry of an outflow-shocked region.

\section{Observations and Data Reduction}


\subsection{Nobeyama 45m Observations}
The Nobeyama 45 m observations were conducted in 2011 January toward the FIR 4 region in addition to the driving source of the molecular outflow, FIR 3, and the northeast outflow lobe, FIR 3N, as shown in Fig. \ref{obspoint}.  In Table \ref{obstable}, the positions of FIR 4, FIR 3, and FIR 3N, are summarized. 
The position of FIR 3N corresponds to the CO peak position of the northeast lobe. Around FIR 3N, there is no sign of the interaction between the outflow and the ambient quiescent gas \citep{Shimajiri08}.
The position of FIR 4 corresponds to the CO peak position of the southeast lobe.

The data were taken in the position-switching mode. The T100 receiver, which is a dual polarization sideband-separating SIS receiver, was used in combination with the Fast Fourier Transform Spectrometer (SAM45) providing a total bandwidth of 16 GHz (=((4 GHz for LSB) + (4 GHz for USB)) $\times$ 2 polarization) and a frequency resolution of 488.24 kHz, corresponding to $\sim$1.5 km s$^{-1}$ at 100 GHz \citep{Nakajima08}. 
To cover the frequency range between 82 and 106 GHz, we used three different frequency settings (82--86 \& 94--98 GHz, 86--90 \& 98--102 GHz, and 90-94 \& 102--106 GHz).
Pointing was checked by observing the Ori-KL SiO maser emission every hour, and was shown to be accurate within a few arc-seconds. 
The parameters for the observations are summarized in Table \ref{obs}. 
The main-beam efficiency was 40\%. 
The system noise temperature was around 150 K, resulting in the rms noise level of 10--40 mK in $T_{\rm A}^*$, as summarized in Table \ref{rms_list}.

\subsection{ASTE Observations}
The ASTE observations at 335--355 GHz were conducted toward the same sources of FIR 4, FIR 3, and FIR 3N.
The data were taken in 2012 December in the position-switching mode. The CATS345 receiver, which is a side-band separating (2SB) mixer receiver, was used in combination with the MAC correlator providing a bandwidth of  512 MHz in each of 4 arrays and a frequency resolution of 0.5 MHz, corresponding to $\sim$0.5 km s$^{-1}$ at 345 GHz \citep{Sorai00,Inoue08}. 
The adjust frequency coverage overlaps by 112 MHz. Thus, one frequency setting can cover the frequency range of $\sim$1.6 GHz. We used 13 different frequency settings to cover  the frequency range of 20 GHz ($\sim$1.6 GHz $\times$ 13 arrays).
The main-beam efficiency of the telescope was $\sim$50\%. 
The system noise temperature was 200--500 K, and the rms noise level was 18--42 mK in $T_{\rm A}^*$, as summarized in Table \ref{rms_list}.
Pointing of the telescope was checked by observing the CO (3--2) line emission from O-Cet every two hours, and was shown to be accurate within 2$\arcsec$. 
The observational parameters are summarized in Table \ref{obs}.

\subsection{Data Reduction}
We used the 3.2.1 version of the Common Astronomy Software Application (CASA) package \citep{McMullin07} to reduce the data obtained with the Nobeyama 45 m and ASTE telescopes. The CASA package is developed for the Atacama Large Millimeter/submillimeter Array (ALMA), and is also available for reducing data obtained by a single-dish telescope such as the Nobeyama 45 m telescope and the ASTE telescope. Here, we describe detailed steps of the data reduction process.  
In the first step, edge channels of each correlator band were flagged out using the task $sdflag$, since the sensitivity of the edge channels drops.  
The SAM45 spectrometer of the Nobeyama 45 m telescope provides 4096 frequency channels, 
so that the data for the channel numbers from 1 to 300 and from 3796 to 4096 were flagged out.
The data of the MAC spectrometer of the ASTE telescope provides 1024 frequency channels. 
Hence, the data for the channel numbers from 1 to 100 and from 924 to 1024 were flagged out.
In the second step, 
we determined the baseline of each spectrum using the task  $sdbaseline$. 
The "auto" baseline mode automatically detects line emission regions and decides the baseline from the emission-free regions.
We adopted a signal-to-noise ratio (S/N) of 5 as the threshold level for detecting the signal regions.  
In the third step,  we used the task $sdaverage$\footnote{The task $sdaverage$ was integrated into the task "sdcal" in the 3.4.0 version of the CASA.} to average all the spectra in one observing sequence at each observed position with a weight of 1/$T_{\rm sys}^2$, where $T_{\rm sys}$ denotes the system noise temperature.
In the fourth step, we used the task $sdflag$ again to flag out bad channels in the averaged spectra. 
In the fifth step, we merged all the averaged spectra into one file using the task $sdcoadd$.
In the final step, we averaged all the averaged spectra with  a weight of 1/$T_{\rm sys}^2$ using the task $sdaverage$.




\section{Results and Analysis of Molecular Lines}
\subsection{Spectrum and Line Identification}\label{lineid}
Figures \ref{45m_all_spec} and \ref{ASTE_all_spec} show the compressed spectra from 82 to 106 GHz and from 335 to 355 GHz, respectively. 
The rms noise ranges from 10 mK to 41 mK and from 19 mK to 43 mK in $T_{\rm A}^*$, respectively. 
We adopted a criterion that a signal-to-noise ratio should be higher than 3 with the spectral line catalog by \citet{Lovas04} to identify molecular lines in our detected spectra.

As a result, our 3-mm and 850-$\mu$m line surveys have identified 120 lines of 20 molecular species including CH$_3$CCH, CH$_3$CHO, and C$_3$H$_2$ as well as S-bearing molecules such as H$_2$CS, SO, and HCS$^{+}$. 
In addition, 40 lines and 12 rare isotopic species ($^{13}$C, $^{15}$N, $^{17}$O, $^{18}$O, $^{34}$S, and D) have been detected. 
See Appendix for the descriptions of the detected species and their line profiles. 
The numbers of the detected lines and species at each position are listed in Table \ref{number_detected_list} and the detected molecules are summarized in Table \ref{detected_list}. 

\subsection{Line Parameters with Narrow and Wide Velocity Components}\label{lineid}
We found that all the identified molecular lines can be classified into the following two types according to their line profiles. 
One is a single narrow line whose velocity width is 3 km s$^{-1}$ or less, 
and the other is a line accompanying a wide component with the line width of 3 km s$^{-1}$ or larger in addition to the narrow component.
Figure \ref{example} shows two representative spectra of the H$^{13}$CO$^{+}$ (1--0) and HCO$^{+}$ (1--0) emission at FIR 4. 
The H$^{13}$CO$^{+}$ (1--0) line profile can well be fitted by a single Gaussian component with a narrow velocity width (1.85 km s$^{-1}$), while the HCO$^{+}$ (1--0) line profile consists of narrow (1.85 km s$^{-1}$) and wide (8.12 km s$^{-1}$) components.
We performed the Gaussian fitting for all the lines by the following steps.
First, we applied one Gaussian component with a narrow width to the observed spectrum.
When the signal-to-noise ratio (S/N) of the peak residual intensity is less than 2, we regard the spectral line as a single narrow line. 
If not, we applied two Gaussian components to the observed spectrum.
Hereafter, we simply refer to the Gaussian component with the narrow velocity width as $Narrow$, and the Gaussian component with the wide velocity width as $Wide$.
In the Gaussian fitting, we also obtained the line parameters such as peak intensity, systemic velocity, and velocity width, as shown in Tables  \ref{FIR3N_table}--\ref{FIR4_table}.
In this survey, the lines with hyperfine structure (HFS) such as the HCN, H$^{13}$CN, N$_2$H$^+$, and NH$_2$D lines are detected.  
For the N$_2$H$^+$ and NH$_2$D spectra having only the narrow components, we also applied the HFS fitting (See Sect. \ref{Appendix_HFS}). 
We assumed that all the HFS components of each line have the same LSR velocity and the same velocity width.
We used the frequency differences among the HFS components listed in the spectral line catalog by \citet{Lovas04}.
The obtained line parameters of $V_{\rm sys}$ and d$V_{\rm FWHM}$ are comparable to those obtained by the Gaussian fitting. However, the intensities obtained by the HFS fitting are different from those obtained by the Gaussian fitting, suggesting that the results of the Gaussian fitting are affected by the blending of the HFS components.

\subsubsection{Spectra Having One Narrow Component}
The \ambienttracers  lines are found to have one narrow component and are detected at the three positions (see Table \ref{relation_between} and Figs. \ref{cn}--\ref{ch3cn}). 
The velocity widths of the lines range from 0.6 to 4.8 km s$^{-1}$ (mean: 2.1$\pm$0.7 km s$^{-1}$). 
Only one channel in the c-C$_3$H$_2$ and CH$_3$CN spectra at FIR 3N has emission more than 3$\sigma$ due to poor velocity resolution. Thus, the fitting points (velocity channels) are not sufficient to make a reasonable fit by the single Gaussian component. Since these spectra at FIR 4 have only narrow components, these spectra at FIR 3N are also likely to have only narrow components.
The $^{13}$CS, HDCO, HNCO, CH$_3$CCH, and CH$_3$CHO lines also have one narrow component with velocity widths from 1.3 to 3.8 km s$^{-1}$ (mean: 2.6$\pm$0.7 km s$^{-1}$), although they are not detected at FIR 3N (see Table \ref{relation_between} and Figs. \ref{13cs}--\ref{ch3cho}). 
The HCS$^+$, NH$_2$D, and  H$_2$CO lines have one narrow component and detected only at FIR 4 (see Table \ref{relation_between} and Figs. \ref{hcsp}--\ref{h2co}).  The velocity widths of the lines range from 1.5 to 3.8 km s$^{-1}$ (mean: 2.1$\pm$1.0 km s$^{-1}$). 
The H$^{15}$NC line is detected at FIR 3N and FIR 4 and has one narrow component. The velocity width at FIR 4 is 1.82 km s$^{-1}$ (see Table \ref{relation_between} and Fig. \ref{h15nc}). The spectrum of H$^{15}$NC at FIR 3N is not fitted well by the single Gaussian component due to insufficient resolution. Since the H$^{15}$NC spectrum at FIR 4 has an only narrow component, the H$^{15}$NC one at FIR 3N is also likely to have an only narrow component. 
The HC$^{18}$O$^+$ line is detected only at FIR 3 and has one narrow component with a velocity width of 2.2 km s$^{-1}$ (see Table \ref{relation_between} and Fig. \ref{hc18op}).

\subsubsection{Spectra Having Narrow and Wide Components}
The \outflowtracers lines are found to have narrow and wide components, and they are detected at the three positions (see Table \ref{relation_between} and Figs. \ref{co}--\ref{hcop}). The narrow and wide velocity widths of the lines range from 0.9 to 5.6 km s$^{-1}$ (mean: 2.9$\pm$1.0 km s$^{-1}$) and from 3.3 to 20.2 km s$^{-1}$ (mean: 11.7$\pm$3.5 km s$^{-1}$), respectively. 
The \shocktracers lines also have narrow and wide components, but the wide components are detected only at FIR 4 (see Table \ref{relation_between} and Figs. \ref{c34s}--\ref{ch3oh}). The ranges of the narrow and wide velocity widths are 0.6--3.3 km s$^{-1}$ (mean: 1.9$\pm$0.7 km s$^{-1}$) and 4.2--17.3 km s$^{-1}$ (mean: 7.1$\pm$2.5 km s$^{-1}$), respectively. Figure \ref{FWHM_hist} shows histograms of the velocity widths of the fitted Gaussian components for the detected lines.

\subsection{Optical Depth}
Since the same rotational transitions of the normal species and their rare isotropic species are detected for HCO$^{+}$, CS, HCN, and HNC, we estimated optical depths of the transitions.
Assuming the same excitation temperature for the two isotopic species and the isotopic ratios, $R_{\rm i}$, of 22 for $^{32}$S/$^{34}$S and 62 for $^{12}$C/$^{13}$C \citep{Langer93}, we evaluated the optical depth of the normal species line by the following equation:

\begin{equation}
\frac{T_A^*({\rm i)}}{T_A^*({\rm n})}=\frac{1-e^{-\tau(\rm {n)/R_{\rm i}}}}{1-e^{-\tau(\rm{n})}}.
\end{equation}

\noindent Here, $T_{\rm A}^*({\rm n}$) and $\tau(\rm{n})$ are the observed antenna temperature and the optical depth of the normal species, whereas 
$T_{\rm A}^*({\rm i}$) is the observed antenna temperature of the rare isotopic species.
As a result, the normal species lines of the above molecules are found to be optically thick, as shown in Table \ref{optical}. In particular, the optical depth of the wide component of the HCN (1--0) line exceeds 10.
The large optical depth might be, however, artificial, because we assumed the same intensity ratios of the narrow to wide components for the hyperfine structure of HCN; The wide components of the HFS lines heavily overlap with one another, and cannot be well separated (see Fig. \ref{hcn} and \ref{h13cn} and  Appendix \ref{HCN_description}).

\subsection{Rotation Diagram} \label{rotation_section_result}\label{estimated_Trot_N}
To estimate the rotation temperature and column density of the detected molecules, we made rotation diagrams by using the data taken with the Nobeyama 45 m and ASTE telescopes. 
This method is based on the relationship among the beam-averaged column density, $N_{\rm mol}$, the rotation temperature, $T_{\rm rot}$,  and the line intensity, $\int$($T_{\rm R}^*$)$dv$, under the assumption of the LTE and optically thin conditions \citep{Turner91}:

\begin{equation}
\ln L= \ln (\frac{N_{\rm mol}}{Q(T_{\rm rot})}) - \frac{E_{\rm u}}{k} \frac{1}{T_{\rm rot} },
\label{rotation_eq}
\end{equation}

\noindent and

\begin{equation}
L= \frac{3k \int T_{\rm R}^*dv}{8\pi^3 \nu S \mu^2 g_I g_K}.
\end{equation}

\noindent Here, $E_{\rm u}$, $S$, and $\mu$ are the upper state energy, the intrinsic line strength and the relevant dipole moment. 
The values of $E_{\rm u}$ and $S \mu^2$ are obtained from Splatalogue\footnote{http://www.cv.nrao.edu/php/splat/}, except for CH$_3$CN and CH$_3$CCH. For CH$_3$CN and CH$_3$CCH, these values are from the Cologne Database for Molecular Spectroscopy (CDMS).
The factors of $g_{\rm I}$ and $g_{\rm K}$ are the reduced nuclear spin degeneracy and the $K$-level degeneracy, respectively. 
For linear molecules, $g_{\rm I}$=$g_{\rm K}$=1 for all energy levels.
We used the partition functions, $Q(T_{\rm rot})$=$\sum_{i=1}  g_{i} {\rm exp}(-E_{i}/kT_{\rm rot})$, where $g_{i}$ and $E_{i}$ are the degeneracy and energy of the $i$-th state. Rotational energies of each molecule are taken from CDMS.

For the line intensity, we adopted $T_{\rm R}^*$=$T_{\rm A}^*$/$\eta_{\rm MB}$/$f$, where $\eta_{\rm MB}$ and $f$ are the main-beam efficiency of the telescope and the beam filling factor of the source, respectively. 
The factor $f$ is given by $f=\theta_{\rm s}^2/(\theta_{\rm s}^2 + \theta_{\rm b}^2$), where $\theta_{\rm s}$ and $\theta_{\rm b}$ are the source size and the telescope beam size, respectively \citep{Kim00}. 
Here, we assumed that $\theta_{\rm s}$ is 17$\arcsec$ for FIR 3 and 19$\arcsec$ for FIR 4 which are the sizes of the dust condensations of FIR 3 and FIR 4 measured in the 1.1 mm dust continuum emission \citep{Chini97}. For FIR 3N, we assumed $f$=1 since the emission at FIR 3N seems to trace the extended structures of the outflow and ambient gas (see Sect. \ref{What_detected}).
Finally, we obtained the integrated intensity of each line by $\frac{1.06 \times T_{\rm A}^*  \times dV}{f \eta_{\rm MB}}$, where a correction factor of 1.06 is applied for the Gaussian profile and $dV$ is the FWHM of the velocity profile. 
According to Eq. (\ref{rotation_eq}), $N_{\rm mol}$ and $T_{\rm rot}$ can be determined 
by a best-fit straight line in a plot of log $L$ as a function of $E_{\rm u}/k$.
As a result, we obtained the rotation temperatures and the column densities for 12 molecules, as shown in Table \ref{rotation_results}.
The rotation diagrams of individual molecules are shown in Figs. \ref{cch_rotation} to \ref{ch3cch_rotation}.

Figure \ref{hist_temp} shows comparisons of the rotation temperatures between the narrow and wide components at FIR 3N, FIR 3, and FIR 4.
The rotation temperatures of the narrow components of C$_2$H and H$^{13}$CO$^+$ are estimated to be 12.5$\pm$1.4 K, which are similar to the typical gas kinetic temperature of cores in Orion A estimated from N$_2$H$^+$ \citep{Tatematsu08}. 
On the other hand, the rotation temperatures of the wide components are estimated to be 20--70 K, which is higher by a factor two or more than those of the narrow components.
The kinetic temperature at FIR 4 was estimated to be 23 K from the NH$_3$ data \citep{Li13}. The NH$_3$ emission is likely to trace the same region as in the ambient dense-gas tracers of C$_2$H, H$^{13}$CO$^+$, and HN$^{13}$C. The rotational temperatures at FIR 4 are estimated to be 12.3 K for C$_2$H, 13.0 K for H$^{13}$CO$^+$, 9.8 K for HN$^{13}$C. The rotational temperatures are lower than the kinetic temperature, suggesting that the gas is sub-thermally excited.
If we assume  $T$ = 23 K, the H$^{13}$CO$^+$ column density is estimated to be 8.2$\times$10$^{12}$ cm$^{-2}$. Hence, the H$^{13}$CO$^+$ column densities derived from the rotation diagram may be underestimated by a factor of two. 

Furthermore, we derived the column densities of CN, C$^{17}$O, $^{13}$CS, HCS$^+$, N$_2$H$^+$, HC$^{18}$O$^+$, HNC, H$^{15}$NC, H$_2$CO, H$_2^{13}$CO, HDCO, HNCO, NH$_2$D, c-C$_3$H$_2$, HC$_3$N, and CH$_3$CHO
on the assumption that the temperature is 10, 20, and 40 K and the source size is 16$\arcsec$, because the derived rotation temperatures for most of the molecules detected in several transitions are $\sim$10-40 K.
These results are summarized in Table \ref{LTE_results}.

For several shock tracers (see Sect. \ref{What_detected}), the rotation temperatures and column densities require the correction for the beam dilution.
The previous observations of FIR 4 in the SiO (2--1) line, which is a representative shock tracer, 
with an angular resolution of $\sim$5$\arcsec$ could not resolve the emission region \citep{Shimajiri08}. This means that the emission of the shock tracers should have a compact structure with a size of  $<$ 5$\arcsec$ in the FIR 4 region. 
Therefore, the line intensities of the shock tracers observed in this study are likely to be affected by the beam dilution effect. 
To correct for this effect, we assumed that the source sizes of the narrow and wide components are 3$\arcsec$. 
Table \ref{rotation_results_beam} shows the corrected rotation temperatures and column densities.

Figure \ref{abundance_hist} shows comparisons of the column densities among the three positions. 
In the comparison between FIR 3N and FIR 4 shown in Fig. \ref{abundance_hist} (a), 
the fractional column density ratios between molecules, $\frac{N_{\rm mol}({\rm FIR 3N})}{N_{\rm mol}({\rm FIR 4})}\frac{N_{\rm H^{13}CO^+}({\rm FIR 4})}{N_{\rm H^{13}CO^+}({\rm FIR 3N})}$, detected in FIR 3N and FIR 4 are similar within one order of magnitude (1.2$\pm$1.3 with a range of 0.1--5.3). 
The $^{13}$CS, C$^{34}$S, SiO, HC$^{15}$N, HC$^{18}$O$^+$, HCS$^+$, H$_2$CO, HDCO, HNCO, NH$_2$D, CH$_3$CCH, and CH$_3$CHO emission lines are not detected in FIR 3N. 
The 3$\sigma$ upper limits of C$^{34}$S, SiO, HC$^{15}$N, H$_2$CO, HDCO, NH$_2$D, and CH$_3$CCH are plotted below the line of $N_{\rm mol}$(FIR 3N)=$\frac{N_{\rm H^{13}CO^+}({\rm FIR\ 3N})}{N_{\rm H^{13}CO^+}({\rm FIR\ 4})}$ $N_{\rm mol}$(FIR 4) in Fig. \ref{abundance_hist} (a). 
Thus, the sensitivities in detecting these molecules seem sufficient even in FIR 3N, and the fractional abundances of the molecules in FIR 4 are likely higher than those in FIR 3N. Note that the fractional abundance ratio is equivalent to the column density ratio.
In contrast, the 3$\sigma$ upper limits of $^{13}$CS, HC$^{18}$O$^+$, HCS$^+$, HNCO, CH$_3$CCH, and CH$_3$CHO are plotted above the line of $N_{\rm mol}$(FIR 3N)=$\frac{N_{\rm H^{13}CO^+}({\rm FIR\ 3N})}{N_{\rm H^{13}CO^+}({\rm FIR\ 4})}$$N_{\rm mol}$(FIR 4) in Fig. \ref{abundance_hist} (a), 
suggesting that the sensitivities for the six molecules were poor in FIR 3N. Therefore, we cannot definitely conclude that the fractional abundance of the molecules in FIR 4 are higher than those in FIR 3N.
To access this issue, the improvement of the sensitivity is required.
In the comparison between FIR 3 and FIR 4 as shown in Fig. \ref{abundance_hist} (b), the fractional column density ratios between molecules, $\frac{N_{\rm mol}({\rm FIR 3})}{N_{\rm mol}({\rm FIR 4})}\frac{N_{\rm H^{13}CO^+}({\rm FIR 4})}{N_{\rm H^{13}CO^+}({\rm FIR 3})}$, detected in FIR 3 and FIR 4 are similar within one order of magnitude (0.7$\pm$0.3 with a range of 0.2--1.4), except for SO (0.03).
The SiO, HCS$^+$, H$_2$CO, and NH$_2$D emission lines were not detected at FIR 3. 
The 3$\sigma$ upper limits of these molecules are plotted below the line of $N_{\rm mol}$ (FIR 3)=$\frac{N_{\rm H^{13}CO^+} ({\rm FIR\ 3})}{N_{\rm H^{13}CO^+} ({\rm FIR\ 4})}$ $N_{\rm mol}$ (FIR 4) in Fig. \ref{abundance_hist} (b), indicating that the fractional abundances of these molecules in FIR 4 are higher than those in FIR 3.
The HC$^{18}$O$^+$ emission is detected only at FIR 3 and has the only narrow component. As shown in Fig. \ref{hc18op}, the emission-like feature with an S/N of 2.4 at FIR 4 can be marginally seen at the same LSR velocity as in FIR 3. Thus, the non-detection of the HC$^{18}$O$^+$ at FIR 4 is considered to be due to poor sensitivity. In Fig. \ref{abundance_hist}(a), the 3$\sigma$ upper limit of HC$^{18}$O$^+$ at FIR 3N is  plotted above the line of $N_{\rm mol}$ (FIR 3N)=$\frac{N_{\rm H^{13}CO^+}({\rm FIR\ 3N})}{N_{\rm H^{13}CO^+}({\rm FIR\ 4})}$$N_{\rm mol}$(FIR 4), suggesting that the sensitivity for HC$^{18}$O$^+$ is poor. 

\subsection{Fractional Abundances of Molecules}\label{fractional_abundance}

We estimated the fractional abundances relative to H$_2$ for the detected molecules.
For this purpose, the column densities of molecular hydrogen at FIR 3N, FIR 3, and FIR 4 were determined from the AzTEC 1.1-mm dust-continuum data \citep{Shimajiri11}. 
Assuming that the 1.1-mm dust-continuum emission is optically thin, the mean column density of H$_2$ within the AzTEC beam of 40$\arcsec$ in HPBW can be derived from the 1.1 mm flux density, S$_{\rm 1.1 mm}$ as: 

\begin{equation}
N(\rm H_2)=\frac{S_{\rm 1.1 mm}}{\Omega_{\rm bm} \mu m_{\rm H} \kappa_{\rm 1.1 mm} \it{B}_{\rm 1.1 mm}(\it{T}_{\rm d}),}
\end{equation}

\noindent where $\Omega_{\rm bm}$ is the beam solid angle in str and $B_{\rm 1.1 mm}(T_{\rm d}$) is the Planck function at 1.1 mm with a dust temperature $T_{\rm d}$.
We assumed that $T_{\rm d}$ is equal to the H$^{13}$CO$^+$ rotation temperature estimated in Sect. \ref{rotation_section_result}: the H$^{13}$CO$^+$ rotation temperatures are 10.2, 13.8, and 13.0 K at FIR 3N, 3, and 4, respectively (see Table \ref{obspoint}).
As we described in Sect. \ref{estimated_Trot_N}, the temperatures might be underestimated. As an upper limit of the dust temperature, $T_{\rm d}$, we also adopted the CO (1--0) peak intensity. 

We calculated the dust mass opacity coefficient at 1.1 mm, $\kappa_{\rm 1.1mm}$ by using the relation that $\kappa_{\nu}$=0.1(250$\mu$m/$\lambda$)$^{\beta}$ cm$^2$g$^{-1}$ \citep{Hildebrand83}.
\citet{Chini97} estimated the value of $\beta$ to be 2 by the spectral energy distributions toward FIR 1 and FIR 2 in the OMC-2 region. 
Since FIR 4 is located in the same molecular cloud filament of OMC-2, we used the $\beta$ value of 2.  The H$_2$ column densities at FIR 3N, FIR 3, and FIR 4, respectively, are estimated to be  1$\times$10$^{23}$, 3$\times$10$^{23}$, and 5$\times$10$^{23}$ cm$^{-2}$ for $T_{\rm d}$=$T_{\rm rot}$(H$^{13}$CO$^+$) and {0.2}$\times$10$^{23}$, 0.7$\times$10$^{24}$, and 1$\times$10$^{23}$ cm$^{-2}$ for $T_{\rm d}$=$T_{\rm co\ peak}$ (see Table \ref{obstable}). The uncertainty in $T_{\rm d}$ causes the uncertainty in column density of a  factor of 4.1 --5.8.
Using these H$_2$ column densities and the molecular column densities estimated in Sect. \ref{rotation_section_result}, 
we derived the fractional abundance $X_{\rm mol}$ as: $X_{\rm mol}$ = $N_{\rm mol}$/$N_{\rm H_2}$.
We note that the estimated H$_2$ column densities are lower limits and the fractional abundances are upper limits, since the beam size of the 1.1 mm data (40$\arcsec$) is larger than the beam size (16$\arcsec$) assumed to estimate the column densities of each molecule.
The results are summarized in Table \ref{rotation_results}.

\section{Discussion}
\subsection{What physical and chemical conditions do the detected molecular emission lines trace?} \label{What_detected}

By comparing the spatial distributions and velocity widths of the detected emission among the northern outflow lobe, FIR 3N, the driving source of the molecular outflow, FIR 3, and the outflow-shocked region, FIR 4, we found that the physical and chemical conditions the detected molecular emission traces can be categorized into the following five types:

\begin{enumerate}
\item[]{[Ambient dense-gas tracers]} The emission is detected at the three positions and has the only narrow components. These emission lines are considered to trace the ambient dense gas. 
\item[]{[Possible ambient dense-gas tracers]} The emission is detected at one or two positions and has the only narrow components. The velocity widths of these emission lines are similar to those of the ambient dense-gas tracers. The non-detection of these emission lines at one or two positions is likely due to poor sensitivity or low abundance. Thus, these emission lines likely trace the ambient dense gas.
\item[]{[Outflow tracers]} The emission is detected at the three positions and has the narrow and wide components. These emission lines are considered to trace the molecular outflow.
\item[]{[Shock tracers]} The wide components are detected only at the outflow-shocked region, FIR 4, indicating the outflow shock tracers. Although the narrow component is also associated, the wide component only at FIR 4 shows enhancement of the molecule in the shock.
\item[]{[Possible shock tracers]} The velocity widths of the narrow component at FIR 4 are larger than those at FIR 3N and FIR 3. Thus, these emission lines might have wide components at FIR 4 and might trace the outflow shock. If the wide component toward FIR 4 is not well separated owing to poor sensitivity, the single Gaussian fitting gives a broader line width. 
\end{enumerate}

Table \ref{phy_che_mol} and Figure \ref{Schematic} show a relationship between the physical environment and the chemical composition and its schematic illustration. We describe the details of each type in the following subsubsections.


\subsubsection{Ambient dense-gas tracers}
The CN, C$^{17}$O, C$_2$H, HNC, HN$^{13}$C, H$^{13}$CO$^+$, N$_2$H$^+$, c-C$_3$H$_2$, and CH$_3$CN molecules belong to this category.
These emission lines have critical densities of $>$ 10$^4$ cm$^{-3}$ and the rotation temperatures of the components of C$_2$H and H$^{13}$CO$^+$ for which several transitions are detected at the three positions, are indeed estimated to be 12.5$\pm$1.4 K, a typical temperature of ambient clouds.

\subsubsection{Possible ambient dense-gas tracers}

The $^{13}$CS and CH$_3$CCH emission is detected at FIR 3 and FIR 4 and have the only narrow component. As shown in Figs. \ref{13cs}, and \ref{ch3cch}, the emission-like feature with an S/N of 2.1 for $^{13}$CS and 2.2 for CH$_3$CCH at FIR 3N can be marginally seen at the same LSR velocity as in FIR 3 and FIR 4. Thus, the non-detection of $^{13}$CS, CH$_3$CCH at FIR 3N is considered to be due to  poor sensitivity. The velocity widths at FIR 3 and FIR 4 are 2.59 and 1.89 km s$^{-1}$, respectively, for $^{13}$CS and 2.02--3.14 and 1.98--2.91 km s$^{-1}$ for CH$_3$CCH, which are consistent with those of the ambient dense-gas tracers (2.1$\pm$0.7 km s$^{-1}$).
Thus, the $^{13}$CS and CH$_3$CCH emission possibly traces the ambient dense gas.

The narrow components of the NH$_2$D emission are detected only at FIR 4.  
Since the velocity width and antenna temperature of the narrow components are 1.54 km s$^{-1}$ and 0.4 K in $T_{\rm A}^*$, respectively, the NH$_2$D emission seems to trace the quiescent dense gas associated with FIR 4. 
As described in Sect. \ref{fractional_abundance} and Appendix \ref{nh2d_appendix}, the sensitivities in NH$_2$D seem sufficient in FIR 3N and FIR 3. Thus, the fractional abundance of NH$_2$D 
in FIR 4 is likely higher than those in FIR 3N and FIR 3 as shown in Fig. \ref{abundance_hist} and described in Sect. \ref{rotation_section_result}.

The narrow components of the H$^{15}$NC emission are detected at FIR 3N and FIR 4. 
The velocity width of the H$^{15}$NC at FIR 4 is 1.82 km s$^{-1}$, which agrees with those of the ambient dense-gas tracers. Thus, the H$^{15}$NC emission is likely trace the ambient dense gas. We note that the spectrum at FIR 3N cannot be fitted well by the single Gaussian component due to  insufficient velocity resolution, although the signal-to-noise ratio of the peak temperature is 3. 
As described in Appendix \ref{HCN_description}, it is possible that the fractional abundance of H$^{15}$NC in FIR 3 is lower than that in FIR 4.

The HC$^{18}$O$^+$ emission is detected only at FIR 3 and have only a narrow component. As shown in Fig. \ref{hc18op}, the emission-like feature with an S/N of 2.4 at FIR 4 can be marginally seen at the same LSR velocity as in FIR 3. Thus, the non-detection of the HC$^{18}$O$^+$ at FIR 4 is considered to be due to poor sensitivity. 
As described in Sect \ref{fractional_abundance} and Appendix \ref{HCOp_appendix} the sensitivity for HC$^{18}$O$^+$ is poor at FIR 3N. The velocity width of HC$^{18}$O$^+$ at FIR 3 is 2.2 km s$^{-1}$, which agrees with those of the ambient dense-gas tracers. Thus, the HC$^{18}$O$^+$ emission likely traces the ambient dense gas.

\subsubsection{Outflow tracers}\label{outflow_tracers}
The wide components of the CO (3--2), CS (2--1, 7--6), HCN (4--3), and HCO$^{+}$ (1--0) emission are detected at the three positions including the northern lobe of the FIR 3 outflow, FIR 3N, suggesting that these components trace a dense part of the molecular outflow driven from FIR 3.
In fact, the CO (3--2) and HCO$^+$ (1--0) molecular outflows driven by FIR 3 are detected in previous studies \citep{Aso00, Takahashi08b}. 
The NMA observations with a high angular resolution of $\sim$5$\arcsec$ have revealed that the distribution of the CS (2--1) emission is similar to that of the CO (1--0) outflow emission \citep{Shimajiri08}. 
Recently,  \citet{Takahashi12a} detected the HCN (4--3) molecular outflow in another region, MMS 6 in OMC-3, implying the HCN (4--3) emission indeed traces an outflow gas.

\subsubsection{Shock tracers}
The wide components of the C$^{34}$S, SO, SiO, H$^{13}$CN, HC$^{15}$N, H$_2^{13}$CO, H$_2$CS, HC$_3$N, and CH$_3$OH emission are detected only toward FIR 4, and their rotation temperatures are as high as 20--70 K.
In fact, mapping observations in the SiO and CH$_3$OH emission have revealed that the emission is distributed at the interface between the FIR 3 outflow and the FIR 4 dense gas \citep{Shimajiri08}. 
The peak velocities of the wide components are blueshifted with respect to those of the narrow components, except for C$^{34}$S (7--6) and H$_2^{13}$CO. 
This trend is consistent with the result of the interferometer NMA observations in SiO ($v$=0, 2--1) that the SiO emission distributed between the outflow and the FIR 4 clump is blueshifted \citep{Shimajiri08}. 
These results suggest that the components trace the outflow shock associated with FIR 4.

However, the wide components of the SO emission with higher upper energy levels are also seen at FIR 3 (see Fig. \ref{so}). Thus, there are two possibilities. First is that the SO emission traces the molecular outflow and the non-detection at FIR 3N of the wide components is due to poor sensitivity. Second is that the SO emission traces the outflow shock and the wide components at FIR 3 trace the outflow shock near the driving source where the outflow gas is being launched. The possible reason for the wide components of only SO being detected at FIR 3 is that the SO emission has higher upper energy levels ($E_{\rm u}$ $>$80 K) and traces the warm region. However, the H$_2$CS and CH$_3$OH emission lines with a higher upper energy level more than 80 K have only narrow components at FIR 3. The velocity widths of these narrow components (2.9 km s$^{-1}$) are 1.3 times larger than those at FIR 4 (2.3 km s$^{-1}$). Thus, it might be true that the wide components could not be detected at FIR 3 due to poor sensitivity.

While the wide components of the normal species, CS and HCN, are detected at the three positions and trace the molecular outflow as described in Sect. \ref{outflow_tracers}, the wide components of  their rare isotopic species, C$^{34}$S and H$^{13}$CN are detected only at FIR 4. To investigate whether the absence of the C$^{34}$S and H$^{13}$CN wide components at FIR 3N and FIR 3 is due to poor sensitivity or not, we compared the 3$\sigma$ upper limits of the column densities at FIR 3N and FIR 3 with the column density at FIR 4. If the C$^{34}$S and H$^{13}$CN emission trace the outflow, the wide components should be detected at the three positions. 
Here, to estimate the 3$\sigma$ upper limits of the column densities, we assume that the velocity widths of the wide components at FIR 3N and FIR 3 are the same as those at FIR 4. 
  The 3$\sigma$ upper limits of the C$^{34}$S column density with an assumption of $T$ = 20K are estimated to be 7.6$\times$10$^{12}$ cm$^{-2}$ at FIR 3N and 7.4$\times$10$^{12}$ cm$^{-2}$ at FIR 3, while the C$^{34}$S column density of the wide component at FIR 4 is 1.1$\times$10$^{13}$ cm$^{-2}$. 
 The 3$\sigma$ upper limits of the H$^{13}$CN column density with an assumption of $T$ = 20K are estimated to be 3.2$\times$10$^{13}$ cm$^{-2}$ at FIR 3N and 2.7$\times$10$^{13}$ cm$^{-2}$ at FIR 3, while the H$^{13}$CN column density of the wide component at FIR 4 is 6.4$\times$10$^{13}$ cm$^{-2}$. 
The 3$\sigma$ upper limits of C$^{34}$S and H$^{13}$CN at FIR 3N are plotted above the line of $N_{\rm mol}$(FIR 3N)=$\frac{N_{\rm H^{13}CO^+}({\rm FIR\ 3N})}{N_{\rm H^{13}CO^+}({\rm FIR\ 4})}$$N_{\rm mol}$(FIR 4), indicating that the non-detection of the C$^{34}$S and H$^{13}$CN wide components at FIR 3N might be due to the poor sensitivity. 
However, the 3$\sigma$ upper limits of C$^{34}$S and H$^{13}$CN at FIR 3 are plotted below the line of $N_{\rm mol}$(FIR 3)=$\frac{N_{\rm H^{13}CO^+}({\rm FIR\ 3})}{N_{\rm H^{13}CO^+}({\rm FIR\ 4})}$$N_{\rm mol}$(FIR 4), indicating the non-detection of C$^{34}$S and H$^{13}$CN wide components at FIR 3 is not due to the poor sensitivity. 
It is quite unlikely that the C$^{34}$S and H$^{13}$CN wide components trace the outflow because these components were not detected at FIR 3 in spite of the sufficient sensitivity we take into account for the sensitivity.  Thus, the C$^{34}$S and H$^{13}$CN wide components at FIR 4 are likely to trace the outflow shock.

\subsubsection{Possible shock tracers}
The velocity widths of the narrow components of the HCS$^+$, H$_2$CO, HNCO, and CH$_3$CHO emission toward FIR 4 are 1.5 times or more as larger as those at FIR 3N and FIR 3 and those of the ambient dense-gas tracers. The velocity widths of HCS$^+$, H$_2$CO, HNCO, and CH$_3$CHO emission lines in FIR 4 have 3.01, 3.84, 3.15, and 2.34--3.81 km s$^{-1}$, respectively, while the mean velocity width of the ambient dense-gas tracers is 2.1$\pm$0.7 km s$^{-1}$. 
Thus, these emission lines might have a wide component at FIR 4 and might trace the outflow shock. 
If the wide component is not seen due to poor sensitivity, the single Gaussian fitting gives a broader line width. In fact, the signal-to-noise ratios of the emission are too low to separate the two components.  
Although the narrow components of  the HDCO (5$_{1,4}$--4$_{1,3}$) emission are detected at FIR 3 and FIR 4, we consider HDCO as a possible shock tracer for the following reasons. 
These intensities and velocity widths are 0.100 K in $T_{\rm A}^*$ (S/N=3.0)  and 2.93 km s$^{-1}$, respectively, at FIR 3: 0.099 K  in $T_{\rm A}^*$ (S/N=4.1) and 3.45 km s$^{-1}$, respectively, at FIR 4.
The velocity widths at FIR 3 and FIR 4 are 1.5 times or more as broad as those of the ambient dense-gas tracers. 
As we described in Sect. \ref{fractional_abundance}, the non-detection of HDCO at FIR 3N is not due to the poor sensitivity. 
This emission might trace the outflow shock at the launching point of the FIR 3 outflow and at the interface between the FIR 3 outflow and the dense gas associated with FIR 4.

\subsection{Shock vs hot core chemistry}
On the basis of the molecular abundances, we discuss two possible origins for the chemical enrichment at FIR 4. One is the shock chemistry caused by the collision between the outflow from FIR 3 and the dense clump at FIR 4, and the other is the hot-core chemistry as suggested by \citet{Ceccarelli10}, \citet{Kama10}, \citet{Kama14}, and \citet{Lopes13}.

First, we compare the molecular abundances in FIR 4 with those in L 1157 B1 \citep[$d$ = 440 pc,][]{Viotti69} which is a well-studied outflow-shocked region \citep{Avery96, Bachiller01, Sugimura11}.  In L 1157 B1, the molecular outflow driven from IRAS 20386+6751 interacts with the ambient cloud. 
The total luminosities of the outflow driving sources, OMC 2-FIR 3 and L 1157, are estimated to be 300 and 4 $L_{\odot}$, respectively \citep{Adams12, Mundy86}.
The momentum ($P$) of both the blue and red lobes of FIR 3 (2.4 $M_{\odot}$ km s$^{-1}$), however, is comparable to that of L 1157 (4.7 $M_{\odot}$ km s$^{-1}$), 
where no corrections are made for projection angles of the outflows \citep{Aso00,Bachiller01}.

Figure \ref{comp_abundance} (Left) shows a comparison of the molecular fractional abundances relative to H$^{13}$CO$^+$ between FIR 4 and L 1157 B1.  Here, we adopted the abundances for L 1157 B1 from \citet{Bachiller97}. 
The abundances for FIR 4 are the sum of the abundances for the narrow and wide components. We estimate the H$^{13}$CN and HN$^{13}$C abundances from the HCN and HNC abundances on the assumption that [HCN]/[H$^{13}$CN]=[$^{12}$CO]/[$^{13}$CO], [HNC]/[HN$^{13}$C]=[$^{12}$CO]/[$^{13}$CO], and [$^{12}$CO]/[$^{13}$CO]=90.9 \citep{Bachiller97}.
The abundances of C$_2$H, H$^{13}$CN, HN$^{13}$C, and CH$_3$OH in FIR 4 are similar to those in L 1157 B1. On the other hand, the abundances of C$^{34}$S, SiO, SO, and H$_2$CS in FIR 4 are one order of magnitude lower than those in L 1157 B1.
These significant differences are, however, likely due to the beam dilution effect, as described in Sect. \ref{rotation_section_result}.
The corrected abundances in FIR 4 come to agree with the abundances in L 1157 B1, as shown in Fig. \ref{comp_abundance} (Left).
Hence, the chemical composition of FIR 4 could be caused by the outflow shock.

Next, we compare the molecular abundances in FIR 4 with those in IRAS 16293-2422 \citep[$d$ = 160 pc:][]{Whittet74} which is a well-studied low-mass hot corino. 
The total luminosity of IRAS 16293-2422 is comparable to that of FIR 4 (50 $L_{\odot}$ for FIR 4 by Adams et al. 2012 and 27 $L_{\odot}$ for IRAS 16293-2422 by Mundy et al. 1986).
Here, we adopted the molecular abundances in IRAS 16293-2422 from \citet{Schoier02}. 
Figure \ref{comp_abundance} (right) shows a comparison of the molecular fractional abundances relative to H$^{13}$CO$^+$ between FIR 4 and IRAS 16293-2422.  
In the comparison, the fractional abundances of C$^{34}$S, SO and H$_2$CS are similar to each other in the two regions within one order of magnitude. On the other hand, the abundances of C$_2$H, HN$^{13}$C, CH$_3$CCH, SiO, H$^{13}$CN, HC$^{15}$N, and CH$_3$OH in FIR 4 are one order of magnitude higher than those in IRAS 16293-2422. 
Here, we note that the abundances of the shock tracer molecules in \citet{Schoier02} seem to be affected by the beam dilution effect, because \citet{Chandler2005} revealed that shock tracer molecules have compact structures in IRAS 16293-2422.

\begin{figure*}
\begin{center}
\includegraphics[angle=0,scale=.45]{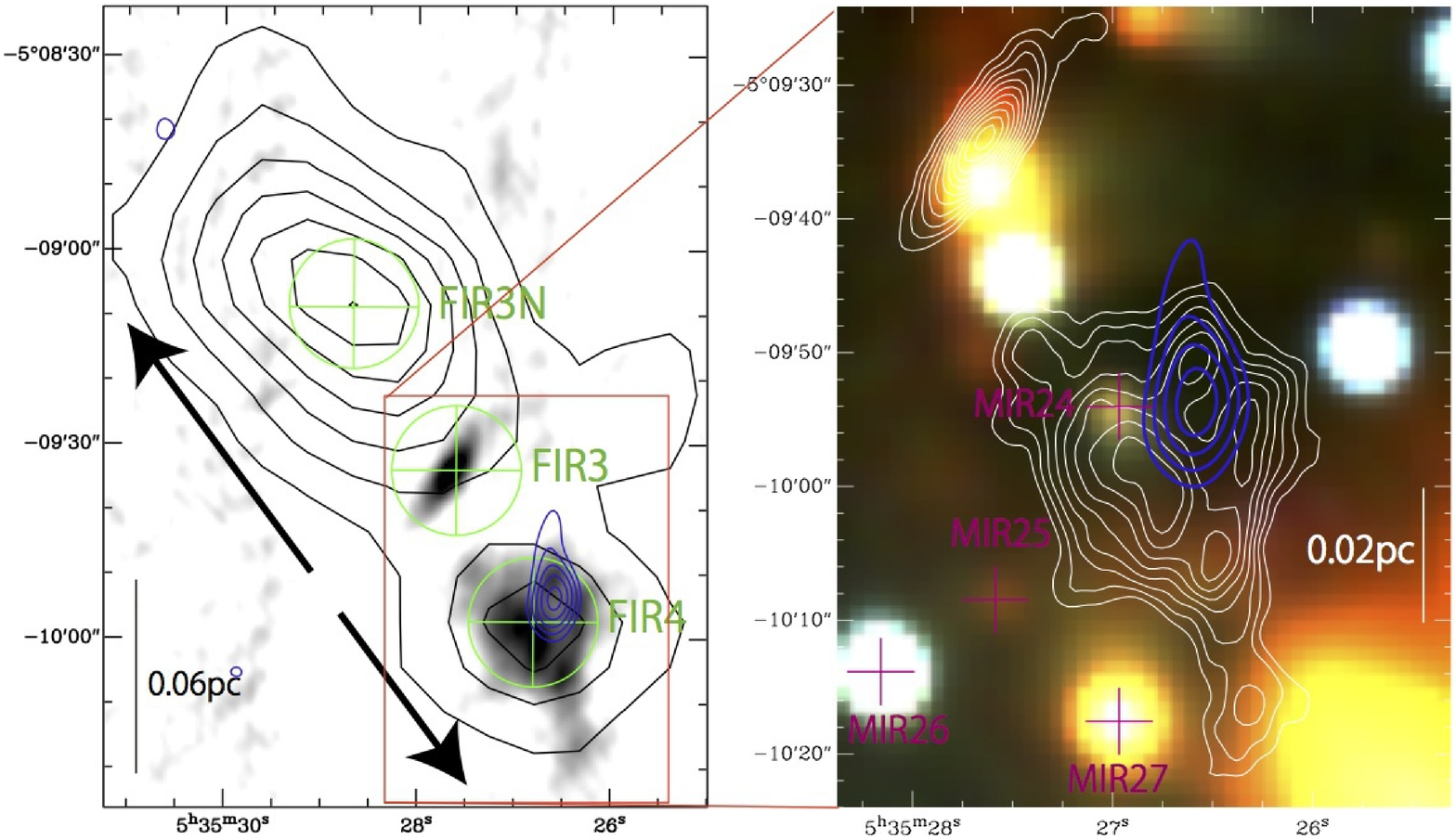}
\end{center}
\caption{Left: three observed positions (green crosses) for FIR 3N, 3, and 4 in the line-survey observations. The circles mean our beam size in HPBW. The black contours show the distribution of the $^{12}$CO (3--2) outflow gas from FIR 3 and the outflow direction is indicated by the arrows. The blue contours and the gray-scale map show the SiO (2--1) line and 3.3 mm dust continuum emission, respectively. Right: Magnification of the FIR 3 and 4 regions. The distributions of the SiO (2--1) line (blue contours) and 3.3 mm dust continuum (white contours) emission are superposed on the 2MASS color map in $J$, $H$, and $K$ bands. The crosses show the positions of the mid-infrared sources, MIR 24 to 27. }
\label{obspoint}
\end{figure*}

\begin{figure*}
\begin{center}
\includegraphics[angle=270,scale=.50]{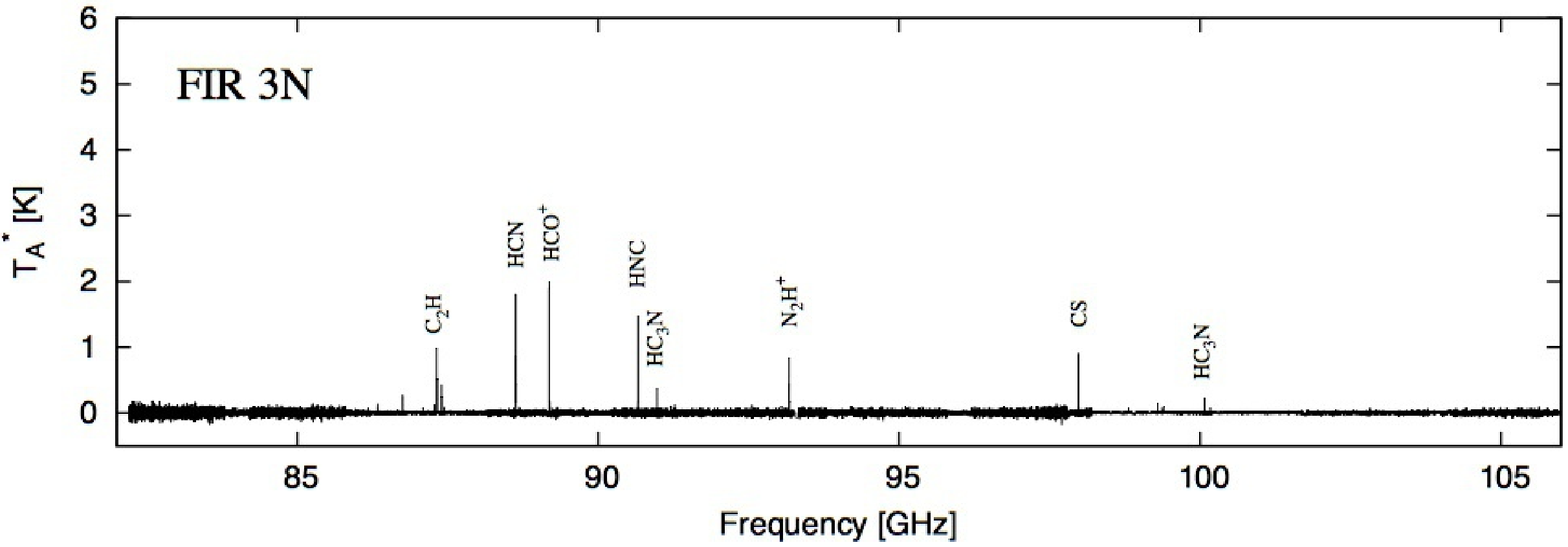}
\includegraphics[angle=270,scale=.50]{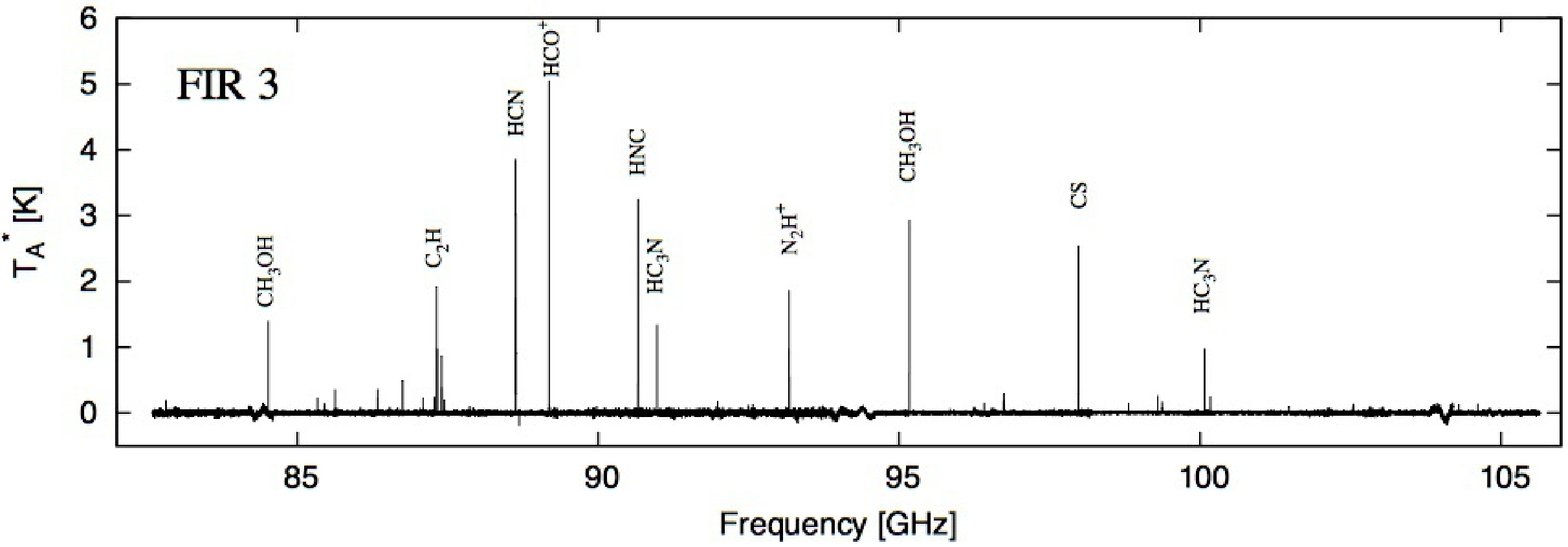}
\includegraphics[angle=270,scale=.50]{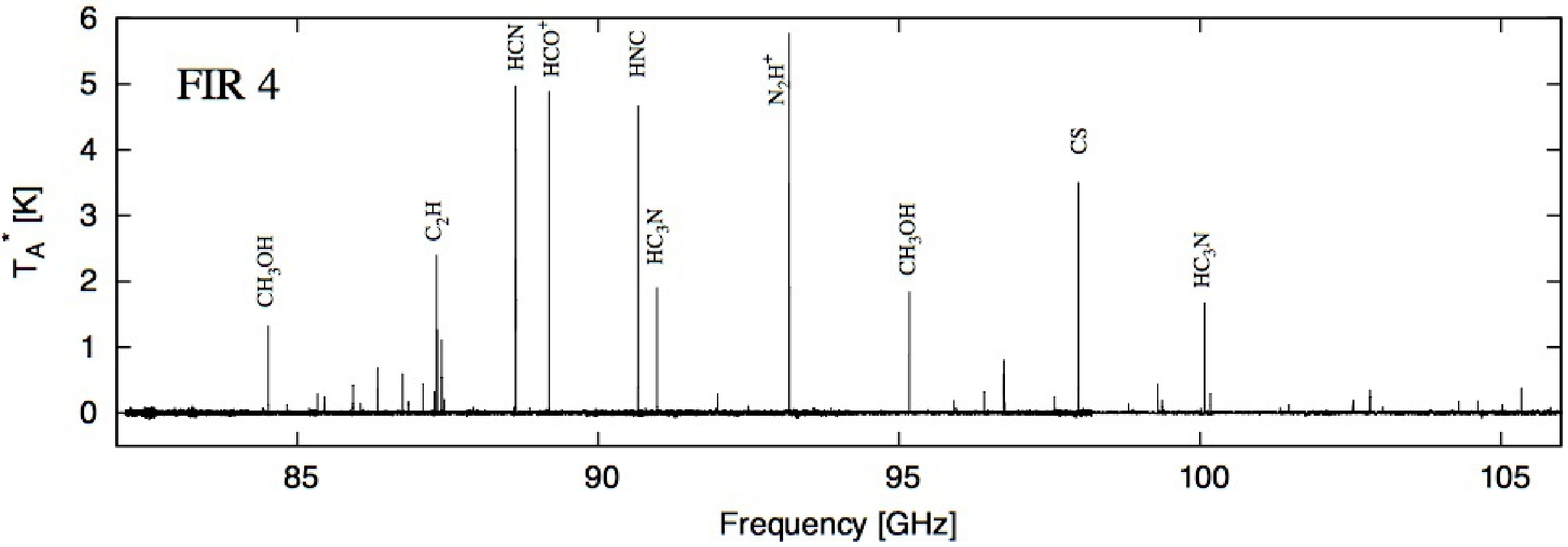}
\end{center}
\caption{Compressed spectra of FIR 3N, FIR 3, and FIR 4 observed in 82--106 GHz.}
\label{45m_all_spec}
\end{figure*}

\begin{figure*}
\begin{center}
\includegraphics[angle=270,scale=.50]{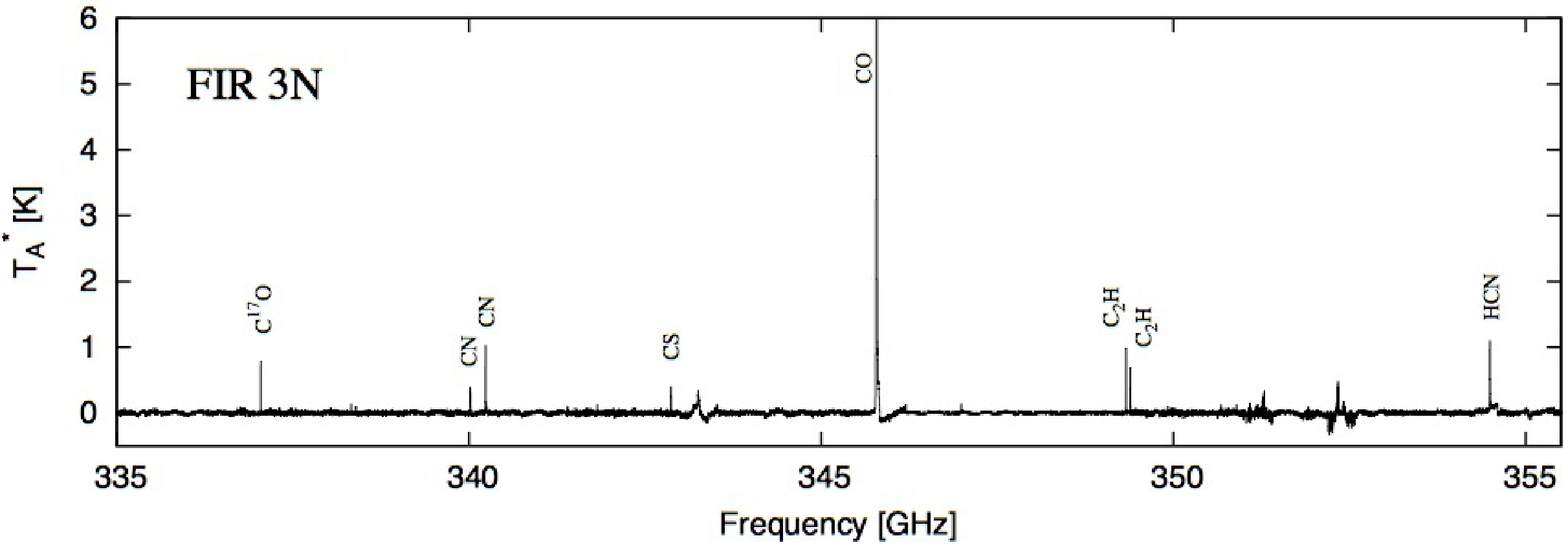}
\includegraphics[angle=270,scale=.50]{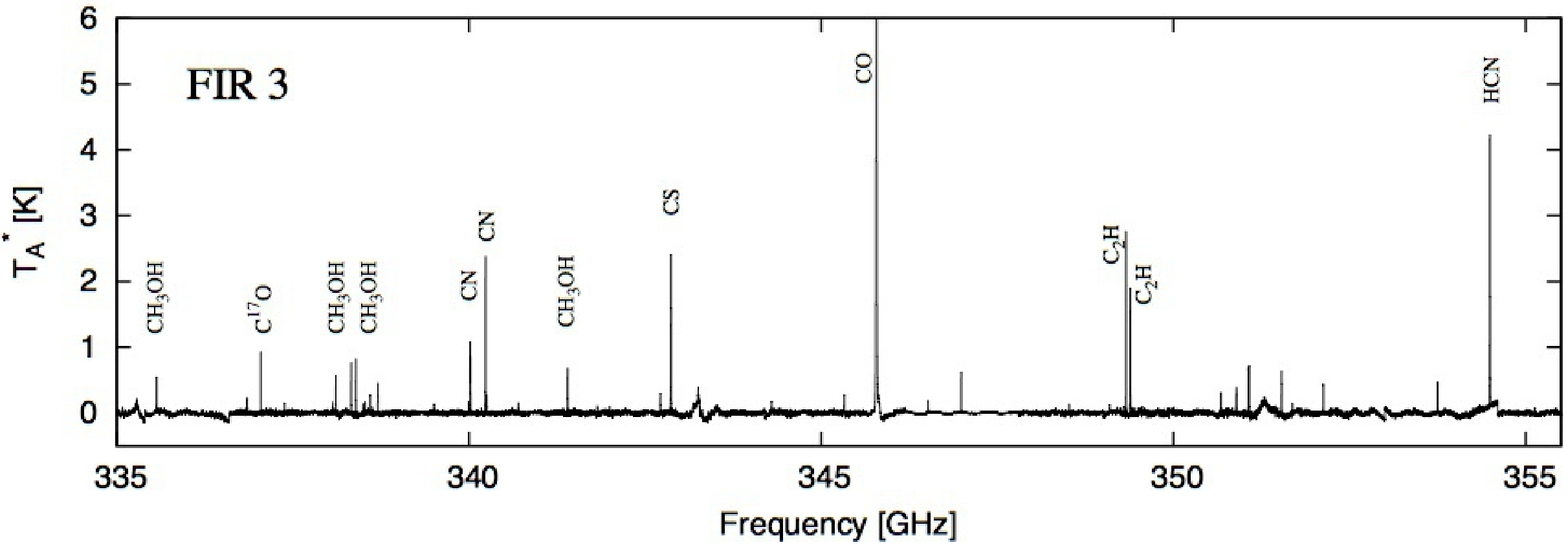}
\includegraphics[angle=270,scale=.50]{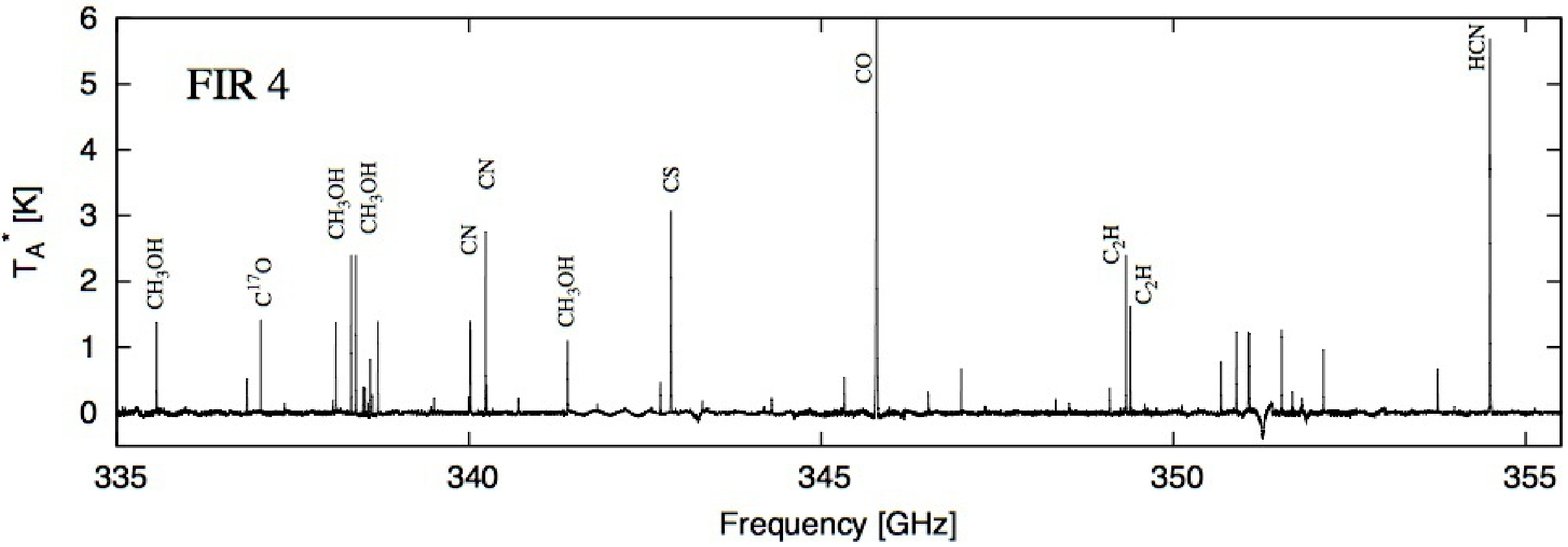}
\end{center}
\caption{Compressed spectra of FIR 3N, FIR 3, and FIR 4 observed in 335--355 GHz.}
\label{ASTE_all_spec}
\end{figure*}

\begin{figure*}
\begin{center}
\includegraphics[angle=0,scale=.6]{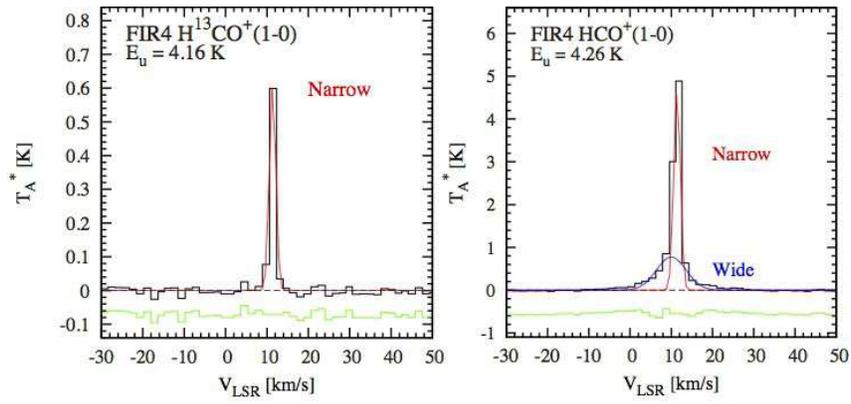}
\end{center}
\caption{Two representative examples of the Gaussian fitting. 
Left: one narrow component (red line) can be well fitted to the H$^{13}$CO$^+$ (1--0) line profile (black lines) at FIR 4. The green lines show the residual spectrum. 
Right: both narrow (red) and wide (blue) components are required for the HCO$^+$ (1--0) line.
}
\label{example}
\end{figure*}

\begin{figure*}
\begin{center}
\includegraphics[angle=0,scale=.3]{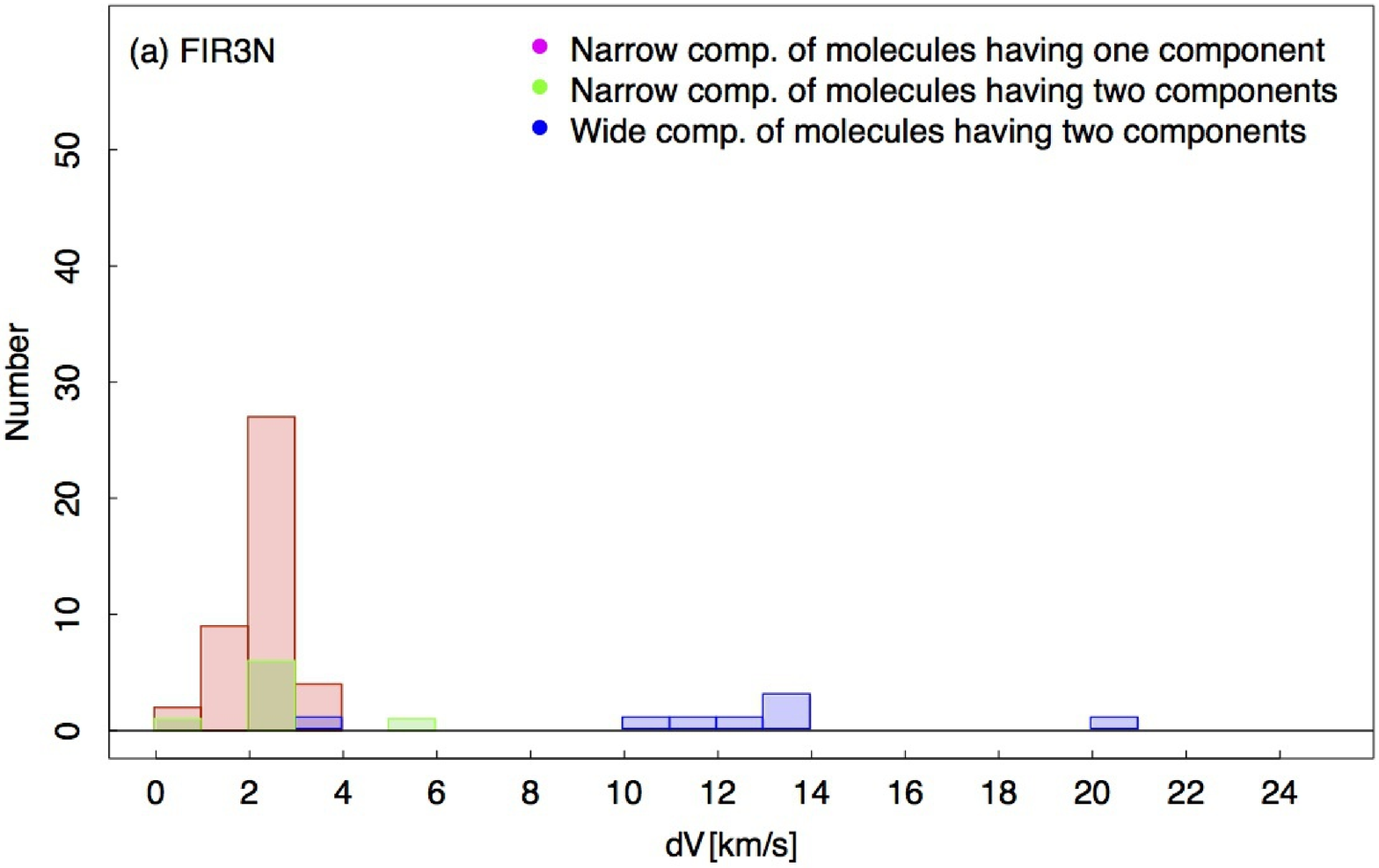}
\includegraphics[angle=0,scale=.3]{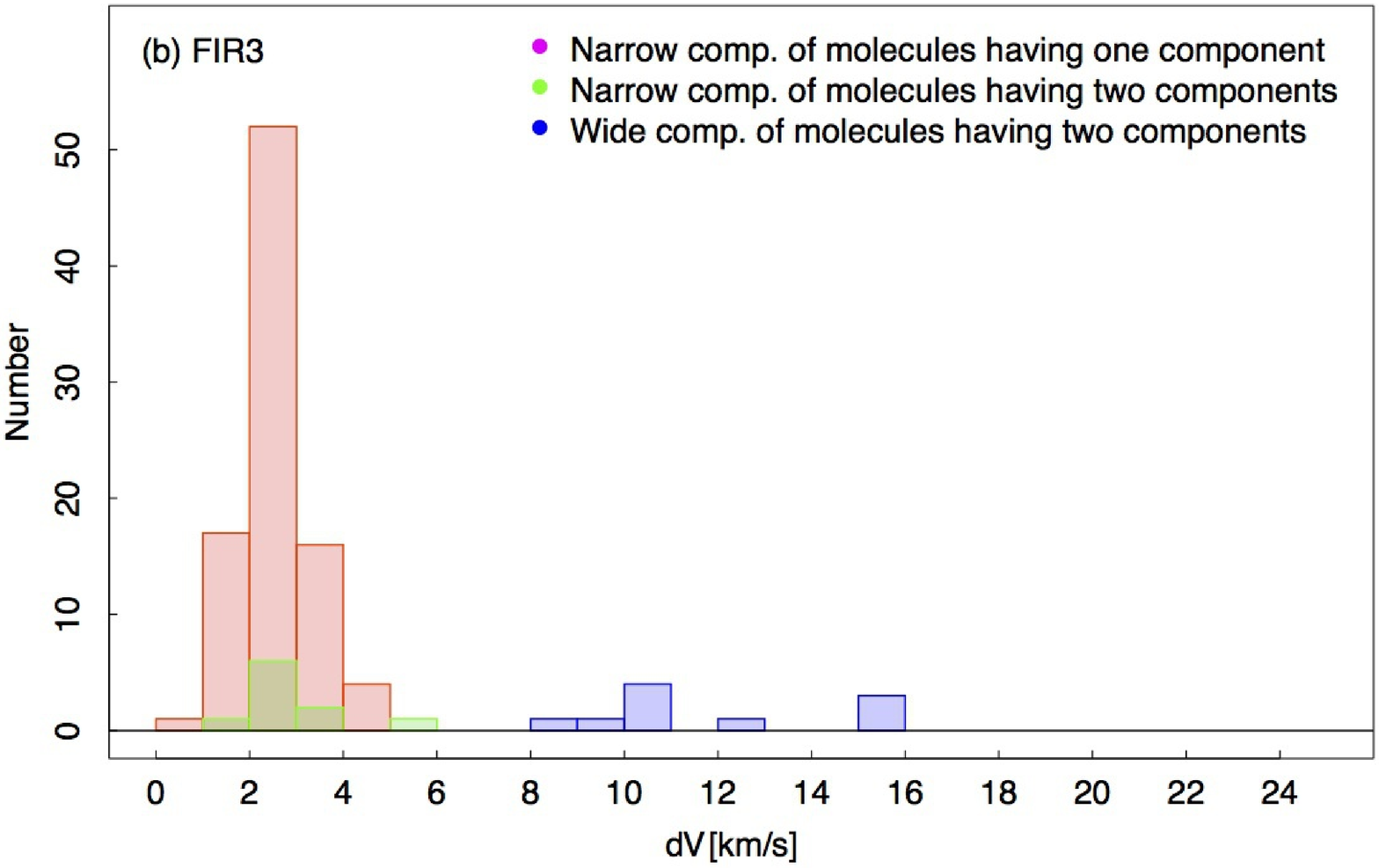}
\includegraphics[angle=0,scale=.3]{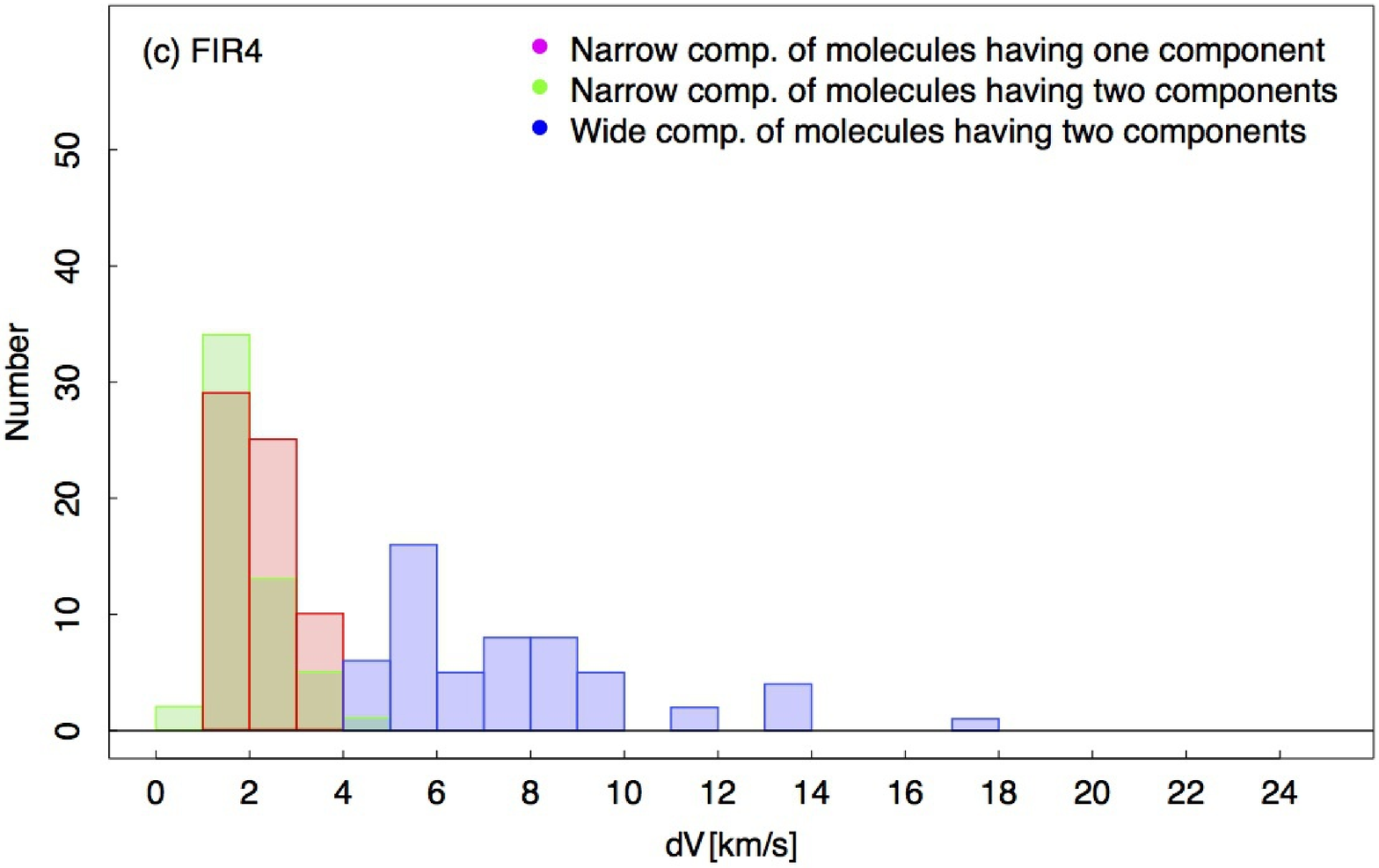}
\end{center}
\caption{Histograms of the velocity widths in FWHM for the detected lines toward (a) FIR 3N, (b) FIR 3, and (c) FIR 4.  The red histograms correspond to the narrow components of the molecules having an only one component, which are determined by Gaussian fitting, as shown in Fig. \ref{example}. The green and blue histograms correspond to the narrow and wide components of the molecules having two components.}
\label{FWHM_hist}
\end{figure*}

\begin{figure*}
\begin{center}
\includegraphics[angle=0,scale=.27]{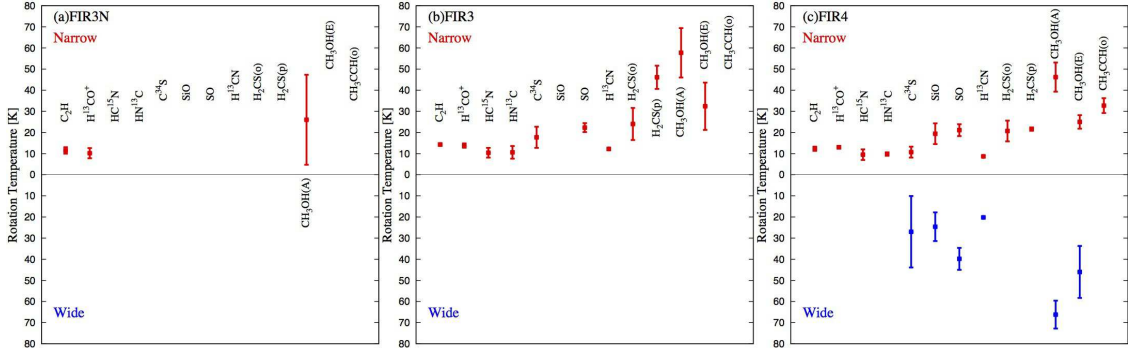}
\end{center}
\caption{Comparison of the rotation temperatures between the narrow and wide components for the 13 molecular lines at (a) FIR 3N, (b) FIR 3, and (c) FIR 4. We excluded the plots for CH$_3$CN and CH$_3$CCH (para), because the accuracies of the Gaussian fitting and the fitting in the rotation diagram are poor (see Table \ref{rotation_results}).}  
\label{hist_temp}
\end{figure*}

Although we cannot exclude a possibility that the chemical enrichment in FIR 4 is caused by the hot core chemistry, the chemical compositions in FIR 4 are more similar to those in L 1157 B1 than those in IRAS 16293-2422.
Although we have detected 120 molecular lines, it is still controversial whether the chemical enrichment in FIR 4 is caused by the outflow shock or by the hot core chemistry from the viewpoint of the molecular abundances. 
However, results of previous observations with angular resolutions of 3--5$\arcsec$ support the scenario that the outflow shock is the origin of the chemical enrichment in FIR 4 for the following reasons.
First, the SiO emission is distributed at the interface between the outflow driven from FIR 3 and the dense gas associated with FIR 4 (see Fig. \ref{obspoint}). 
Second, in the position-velocity diagram, the SiO emission is located at the tip of the CO outflow emission, suggesting that the SiO emission is kinematically related with the FIR 3 outflow \citep[see Fig. 8 in][]{Shimajiri08}.
Third, the distribution of the SiO emission does not coincide with that of the dusty cores traced in the 3.3-mm dust continuum emission.  
Furthermore, any infrared sources, i.e., the driving sources of the hot core chemistry, are not associated with the dusty cores \citep{Adams12}.

To firmly unveil the origin of the chemical enrichment in FIR 4, observations with an angular resolution of $<$ $\sim$3$\arcsec$ are urgent. 
If the molecules interpreted as shock tracer are found to be distributed at the interface between the FIR 3 outflow and the dense clump in FIR 4, the chemical enrichment in FIR 4 comes to be caused by the outflow shock. If the molecules are found to be concentrated at the hot core candidate, the hot core chemistry is dominated.

\section{Summary}
Using the Nobeyama 45 m and ASTE telescopes, we have conducted first line-survey observations at 82--106 GHz and 335--355 GHz toward an outflow-shocked region, FIR 4.  
To deeply characterize the chemistry in FIR 4, we have observed the two additional sources of FIR 3, an outflow driving source and FIR 3N, an outflow lobe without shock. The main results are summarized as follows:

The main results are summarized as follows.

\begin{enumerate}

\item Our 3-mm and 850-$\mu$m line surveys have identified 120 lines and 20 species including CH$_3$CCH, CH$_3$CHO, and C$_3$H$_2$ as well as S-bearing molecules such as H$_2$CS, SO, and HCS$^{+}$. 
In addition, 11 rare isotopic species including $^{13}$C-bearing, $^{15}$N-bearing, $^{17}$O-bearing, $^{18}$O-bearing, $^{34}$S-bearing, and deuterated species are detected. 
We have found that the line profiles of the molecules can be classified into the two types: one Gaussian component with a narrow ($<$ 3 km s$^{-1}$) velocity width and two Gaussian components with narrow and wide ($>$ 3 km s$^{-1}$) velocity widths. 

\item We have estimated the rotation temperatures of the detected molecules with multiple transitions by the rotation diagram method. 
The rotation temperatures of C$_2$H and H$^{13}$CO$^+$ having only one narrow component are found to be 12.5$\pm$1.4 K. 
On the other hand, the rotation temperatures of the wide components are estimated to be 20--70 K, suggesting that the components trace the outflow and/or the outflow shock.

\item On the basis of the spatial distributions, velocity widths, and rotation temperatures of the detected emission lines, we discussed what physical and chemical conditions the detected molecular emission lines trace. 
The narrow components of the \ambienttracers emission are detected at the three positions and the rotation temperatures of C$_2$H and H$^{13}$CO$^+$ are estimated to be 12.5$\pm$1.4 K. 
Thus, these are considered to trace the ambient dense gas. 
The narrow components of the \possibleambienttracers   emission are detected at one or two positions. Those velocity widths are similar to those of the ambient dense-gas tracers. Thus, the emission is likely to trace the ambient dense gas.
The wide components of the CO, CS, HCN, and HCO$^+$ emission are detected at the three positions and likely trace the molecular outflow. 
The wide components of the C$^{34}$S, SO, SiO, H$^{13}$CN, HC$^{15}$N, H$_2^{13}$CO, H$_2$CS, HC$_3$N, and CH$_3$OH emission are detected only at FIR 4 with the high rotation temperatures of 20--70 K and likely trace the outflow shock. 
Although the HCS$^+$, H$_2$CO, HDCO, HNCO, and CH$_3$CHO line profiles consist of one narrow Gaussian component, these velocity widths are twice or more broader than those of the ambient dense-gas tracers. 
Thus,  the emission lines may have a wide component tracing the outflow shock.

\item We have compared the molecular fractional abundances relative to H$^{13}$CO$^{+}$ in FIR 4 with those in the outflow-shocked region, L 1157 B1, and the hot core, IRAS 16293-2422. 
Although we cannot exclude a possibility that the chemical enrichment in FIR 4 is caused by the hot core chemistry, the chemical compositions in FIR 4 are more similar to those in L 1157 B1 than those in IRAS 16293-2422. 
It is still controversial whether the chemical enrichment in FIR 4 is caused by the outflow shock or not.  

\end{enumerate}

FIR 4 ($d$ = 400 pc) is one of the nearest outflow-shocked regions and we have revealed that FIR 4 is a chemically rich source. 
To reveal the nature and origin of the chemical enrichment and the relationship between the chemical and physical conditions, mapping observations with a higher angular resolution of $<<$ 3\arcsec are crucial. If the molecules of the possible-shock tracers are enhanced by the outflow shock, the emission is expected to be distributed at the interface between the FIR 3 outflow and the dense gas associated with FIR 4.

\acknowledgments
We would like to thank the anonymous referee for constructive comments that helped us to polish the manuscript significantly. 
We are grateful to the staff members at the Nobeyama Radio Observatory (NRO) for both operating the 45 m and helping us with the data reduction; NRO is a branch of the National Astronomical Observatory, National Institutes of Natural Sciences, Japan. We also acknowledge the ASTE staff members for both operating ASTE and helping us with the data reduction. Observations with ASTE were (in part) carried out remotely from Japan by using NTT's GEMnet2 and its partner R\&E (Research and Education) networks, which are based on the AccessNova collaboration of University of Chile, NTT Laboratories, and National Astronomical Observatory of Japan. We thank Kanako Sugimoto, Takeshi Nakazato, Jun Maekawa, George Kosugi, Daisuke Iono, Wataru Kawasaki, Shinnosuke Kawakami, Koji Nakamura,  and ALMA-J computing team for helping the data reduction using CASA. Y. S.'s work was supported by JSPS KAKENHI Grant Number 90610551. Part of this work was supported by the ANR-11-BS56-010 project "STARFICH".

{\it Facilities:} \facility{Nobeyama 45m}, \facility{ASTE}.

\clearpage





\clearpage

\begin{figure*}
\begin{center}
\includegraphics[angle=0,scale=.5]{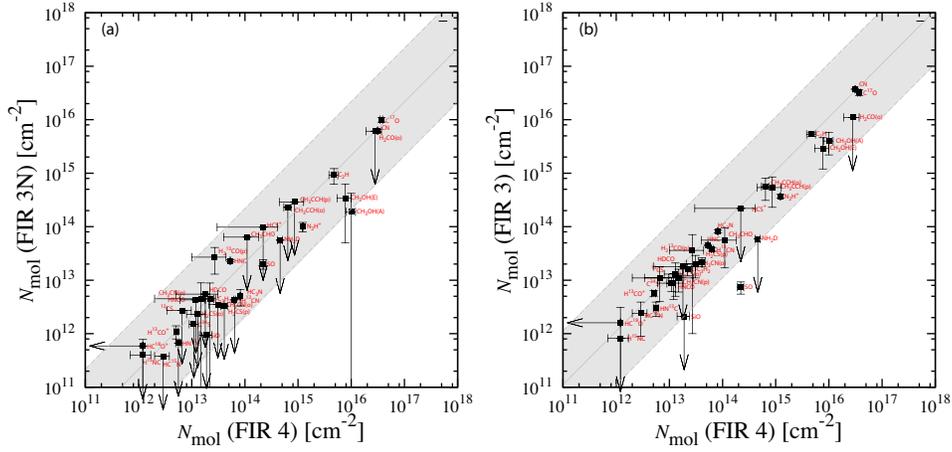}
\end{center}
\caption{
Comparisons of the column densities for the 33 molecules (a) between FIR 3N and FIR 4 and (b) between FIR 3 and FIR 4. 
The gray lines indicate $N_{\rm mol}$(FIR3N)=$N_{\rm H^{13}CO^+}$(FIR3N)/$N_{\rm H^{13}CO^+}$(FIR4)$\times$$N_{\rm mol}$(FIR4)  in panel (a) and  $N_{\rm mol}$(FIR3)=$N_{\rm H^{13}CO^+}$(FIR3)/$N_{\rm H^{13}CO^+}$(FIR4)$\times$$N_{\rm mol}$(FIR4)  in panel (b), and the gray areas show deviations from the lines by a factor 10. }
\label{abundance_hist}
\end{figure*}

\begin{figure*}
\begin{center}
\includegraphics[angle=0,scale=.5]{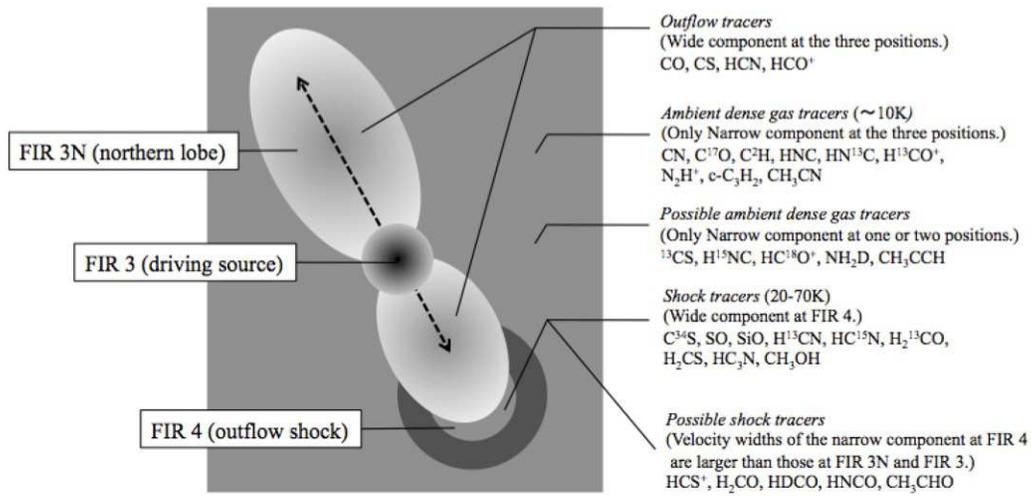}
\end{center}
\caption{Schematic illustration of the relationship between the physical environment  and the chemical composition.}
\label{Schematic}
\end{figure*}

\begin{figure*}
\begin{center}
\includegraphics[angle=0,scale=.6]{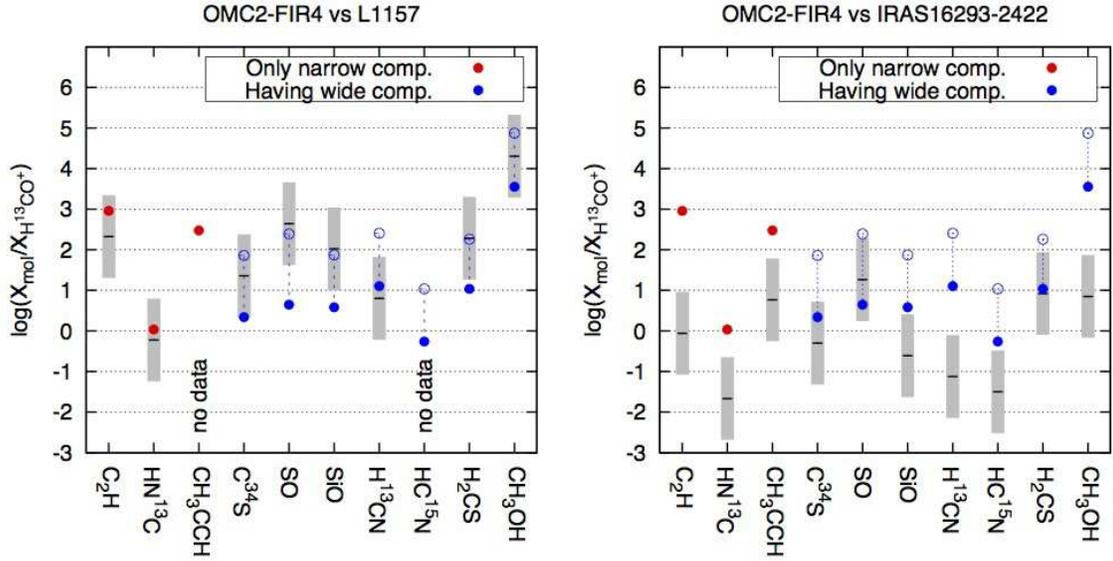}
\end{center}
\caption{Comparison of the fractional abundances relative to that of H$^{13}$CO$^{+}$ for the 10 molecules among OMC 2-FIR 4, L 1157 B1, and IRAS 16293-2422. 
The horizontal black bars show the molecular abundances of the molecule at L 1157 B1 and IRAS 16293-2422. The gray boxes show the uncertainties of one order of magnitude in the fractional abundances estimated for L 1157 B1 and IRAS 16293-2422 \citep{Bachiller97,Schoier02}.
The red circle shows the abundance of the molecule whose spectrum has a single narrow component at FIR 4. 
The blue circle shows the abundance of the molecule whose spectrum has narrow and wide components at FIR 4. 
The lower limit values (blue filled circles) for CH$_3$OH, SO, H$_2$CS, SiO, C$^{34}$S, HC$^{15}$N, and H$^{13}$CN in FIR 4 are the abundances derived on the assumption that the source size is  19$\arcsec$ for all the detected transitions of each molecule. 
The upper limit values (blue open circles) are the abundances corrected for the beam dilution assuming that the source size is 3$\arcsec$. 
 }
\label{comp_abundance}
\end{figure*}

\clearpage

\clearpage
\input{Table1.tex}
\input{Table2.tex}
\input{Table3.tex}
\input{Table4.tex}
\input{Table5.tex}
\input{Table6.tex}
\input{Table7.tex}
\input{Table8.tex}
\input{Table9.tex}
\input{Table10.tex}

\input{Table11.tex}

\clearpage
\appendix
\renewcommand{\thefigure}{A\arabic{figure}}
\setcounter{figure}{0}
\renewcommand{\thetable}{A\arabic{table}}
\setcounter{table}{0}

\section{Appendix}
\subsection{Hyper fine structure fitting}\label{Appendix_HFS}

As seen in Figs. \ref{n2hp} and \ref{nh2d}, the HFS components of N$_2$H$^+$ and NH$_2$D are blended due to the poor velocity resolution. 
To more accurately obtain the line parameters, 
we applied the HFS fitting to the N$_2$H$^+$ and NH$_2$D spectra according to the document for Class/Gildas\footnote{https://www.iram.fr/IRAMFR/GILDAS/doc/pdf/class.pdf} on the assumption that the all HFS components for each molecule have the same excitation temperature and velocity width, Gaussian line profiles as a function of velocity, and the multiple components do not overlap. 

The opacity of the $i$th component, $\tau_i$($v$), is expressed as

\begin{equation}
\tau_i (v)=  \tau_i^{\rm center} {\rm exp} \Biggl(-4{\rm ln}2 \biggl(\frac{v-(V_{\rm sys}+\delta v_i)}{dV_{\rm FWHM}}\biggr)^2 \Biggr),
\end{equation}

\noindent where $\tau_i^{\rm center}$, $dV_{\rm FWHM}$, $\delta v_i$, and $V_{\rm sys}$ are the opacity at the line center of the $i$th component, the velocity width in FWHM, the velocity offset of the component $i$ with respect to the main component, and the systemic velocity of the target.  
The opacity of the all components, $\tau$ can be written as

\begin{equation}
\tau (v)= \tau_{\rm main} \Sigma^N_{i=1} r_i {\rm exp}  \Biggl(-4{\rm ln}2 \biggl(\frac{v- \delta v_i - V_{\rm sys}}{dV_{\rm FWHM}}\biggr)^2 \Biggr),
\end{equation}

\noindent where $r_i$ is the relative strength of each component and $\Sigma^N_{i=1} r_i$=1
Thus, the antenna and excitation temperatures are related as

\begin{equation}
T_{\rm ant} (v)= \eta_{\rm mb} [T_{\rm ex} - T_{\rm bg}] (1-e^{-\tau(v)})
\end{equation}

and the excitation temperature $T_{\rm ex}$  is derived as 
\begin{equation}
T_{\rm ex} =  T_{\rm bg} + \frac{T_{\rm ant} \tau_{\rm total}}{\tau_{\rm main} \eta_{\rm mb}},
\end{equation}

\noindent where $\tau_{\rm total}$ is $\tau_{\rm main} \Sigma^N_{i=1} r_i$.

Table \ref{HFS_result} shows the HFS fitting results.
The obtained line parameters of $V_{\rm sys}$ and $dV_{\rm FWHM}$ are consistent with those by the Gaussian fitting (see Sect \ref{lineid}) within the uncertainties. However, as shown in Figs. \ref{n2hp} and \ref{nh2d}, there exist differences between the best-fit intensities in the Gaussian and HFS fitting. 
This is probably because the Gaussian fitting is affected by the blending of the HFS components.

\input{TableA1.tex}

\subsection{Description of Each Line} \label{appendixA}

\subsubsection{CN}
\begin{figure}
\begin{center}
\includegraphics[angle=0,scale=.5]{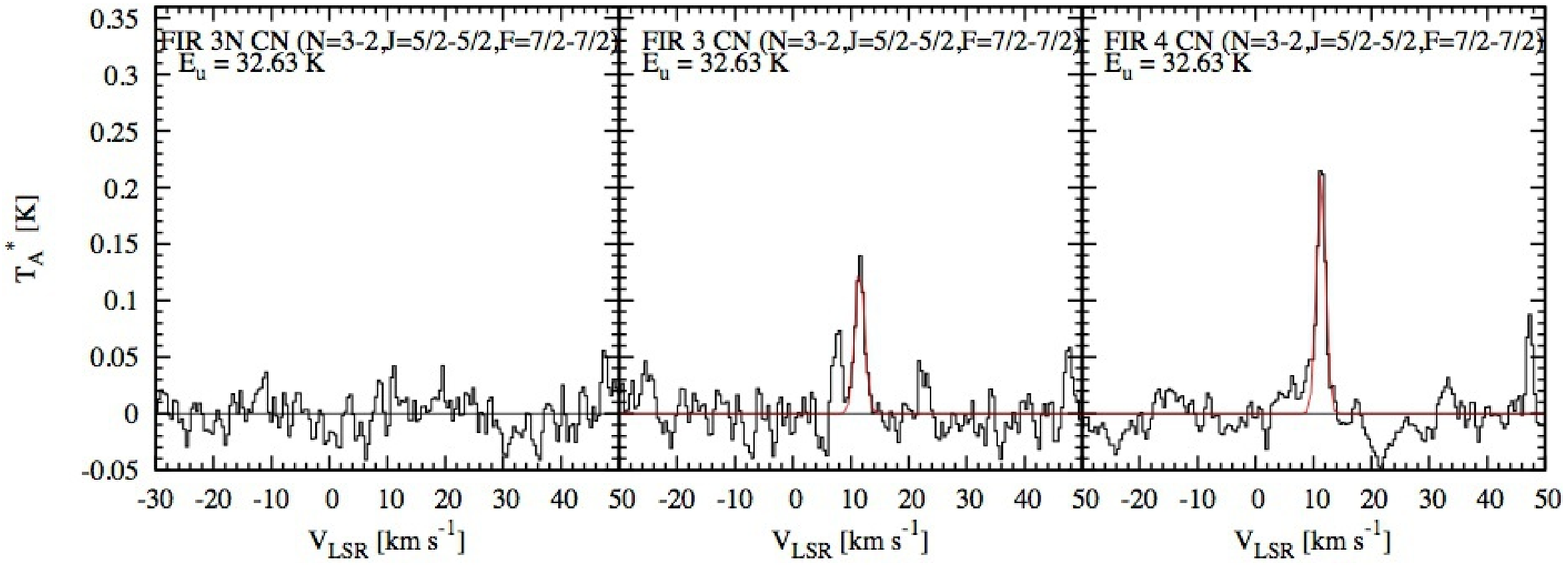}
\includegraphics[angle=0,scale=.5]{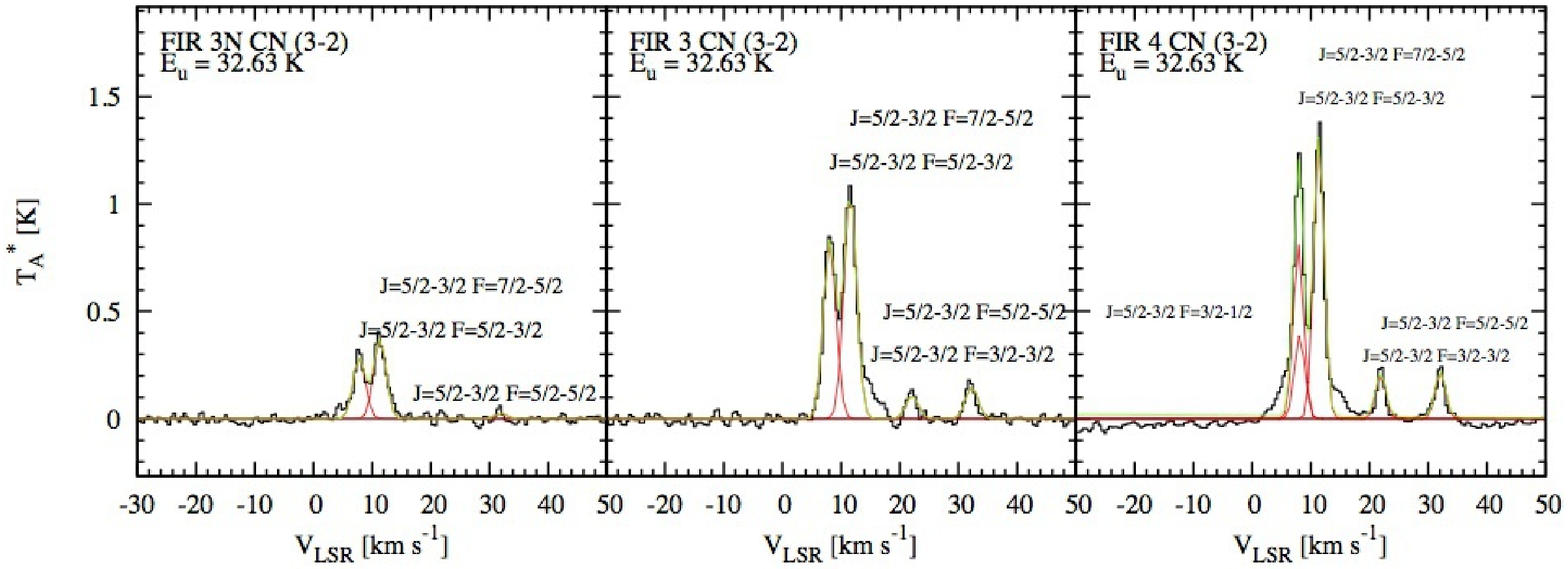}
\includegraphics[angle=0,scale=.5]{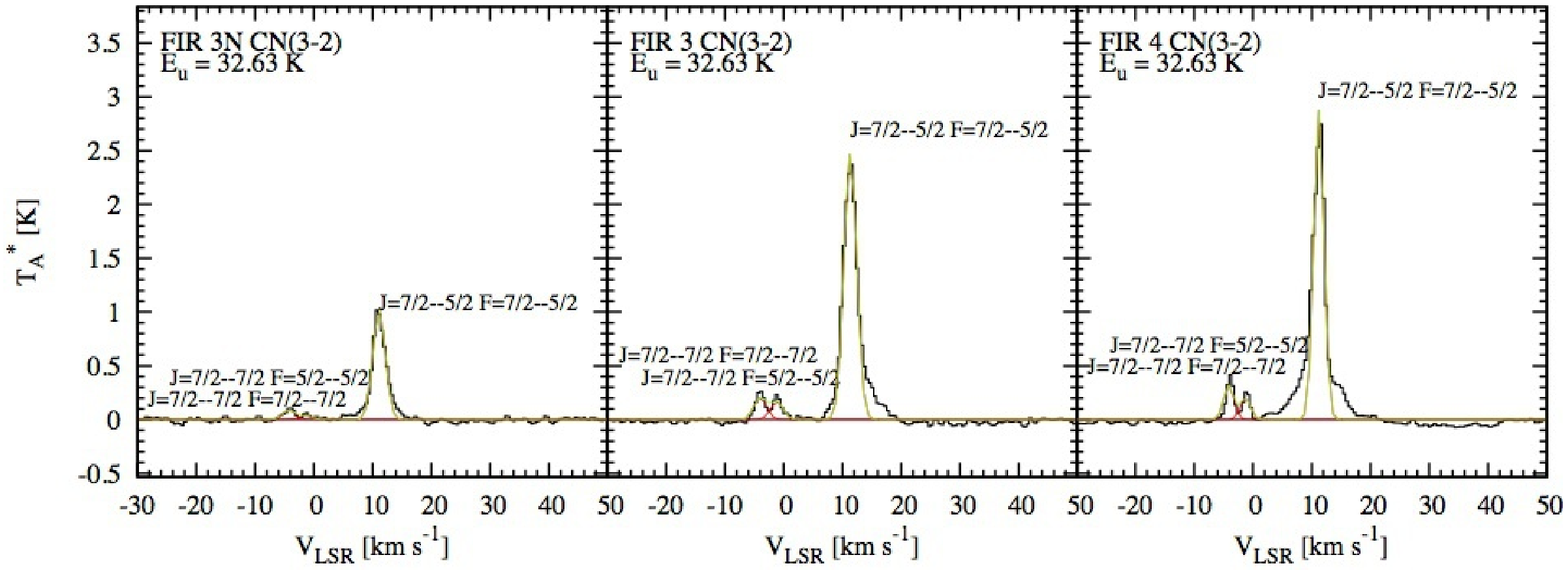}
\end{center}
\caption{CN spectra  with the upper energy levels in K. The best-fit narrow component of each transition is shown by the red line, and the green line shows the sum of all the components.}
\label{cn}
\end{figure}

The CN emission is detected at the three positions of FIR 3N, 3, and 4, and their profiles are shown in Fig. \ref{cn}. The CN emission is though to trace the ambient dense gas toward the regions.

\subsubsection{CO and C$^{17}$O}
\begin{figure}
\begin{center}
\includegraphics[angle=0,scale=.5]{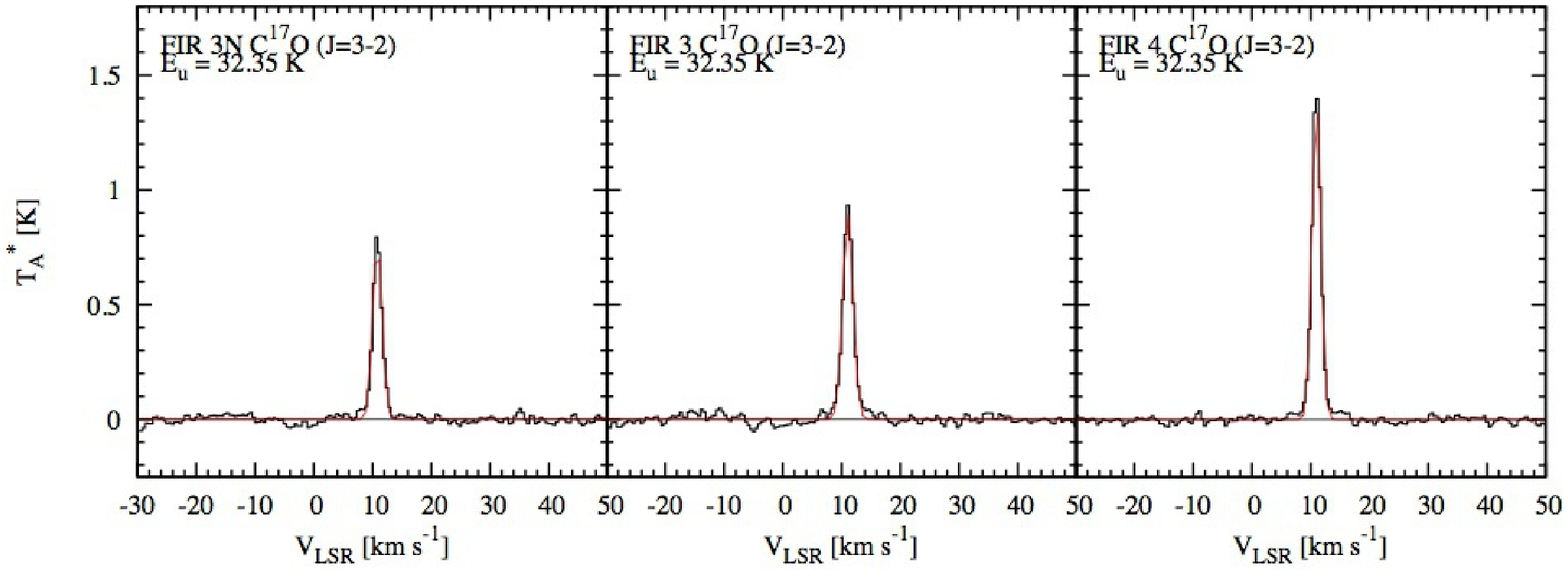}
\end{center}
\caption{C$^{17}$O spectra}
\label{c17o}
\end{figure}

Figure \ref{c17o} shows a comparison among the C$^{17}$O spectra at FIR 3N, 3, and 4. We note that the C$^{17}$O (3--2) emission line has HFS, which is not resolved in our observations. Figure \ref{co} shows the CO profiles at the three positions.
The CO (3--2) spectra at the three positions consist of the two Gaussian components with the narrow and wide velocity widths. It is likely that the wide components of the CO (3--2) emission trace the molecular outflow. In fact, a previous study has detected the molecular outflow driven from FIR 3 in the CO (3--2) emission \citep{Takahashi08b}. 
On the other hand, the C$^{17}$O spectra at the three positions have the single narrow (1.7--2.0 km s$^{-1}$) components, and are though to trace the ambient dense gas.

\subsubsection{CS, C$^{34}$S, and $^{13}$CS}
Figures \ref{13cs}, \ref{cs}, and \ref{c34s}, show the $^{13}$CS, CS, and C$^{34}$S spectra at the three positions, respectively.
The CS (2--1, 7--6) emission is detected at the three positions and have the narrow and wide components. Thus, we consider that the wide components trace the molecular outflow.
In fact, NMA observations with a low-velocity resolution of 50 km s$^{-1}$ have revealed that  the distribution of the CS (2--1) emission is similar to that of the CO (1--0, 3--2) molecular outflow \citep{Shimajiri08}. 
The C$^{34}$S emission is detected at FIR 3 and 4, but the wide components are found only at FIR 4.
This result suggests that the wide components trace the outflow shock. 
The $^{13}$CS emission is detected at FIR 3 and FIR 4 and has the only narrow component. As shown in Fig. \ref{13cs}, the emission-like feature with an S/N of 2.1 at FIR 3N can be marginally seen at the same LSR velocity as in FIR 3 and FIR 4. Thus, the non-detection of $^{13}$CS at FIR 3N is considered to be due to poor sensitivity. The velocity widths of the $^{13}$CS at FIR 3N and FIR 4 are 2.59 and 1.89 km s$^{-1}$, which are similar to those of the ambient dense-gas tracers having the only narrow component. Thus, the $^{13}$CS components likely trace the ambient dense gas.

Table \ref{optical} shows our estimated optical depths of the CS (2--1, 7--6) emission, assuming the same excitation temperature for the CS and C$^{34}$S lines and the [C$^{32}$S]/[C$^{34}$S] ratio of 22 \citep{Wilson94}.

\subsubsection{C$_2$H}
Figure \ref{c2h} shows a comparison of the C$_2$H lines among FIR 3N, FIR 3, and FIR 4. 
The C$_2$H emission is detected at the three positions and has the only narrow component. The velocity widths of C$_2$H (1--0 3/2--1/2 $F$=2--1) are 2.04, 2.17, and 1.84 km s$^{-1}$, 
which are similar to those in the typical dense-gas tracers such as the H$^{13}$CO$^{+}$ (1--0) line.
These results suggest that the C$_2$H emission traces the ambient dense gas.

\subsubsection{HCO$^{+}$, H$^{13}$CO$^{+}$, and HC$^{18}$O$^{+}$} \label{HCOp_appendix}
Figures \ref{h13cop}, \ref{hc18op}, and \ref{hcop},  show the H$^{13}$CO$^+$, HC$^{18}$O$^+$, and HCO$^{+}$ lines, respectively. 
The HCO$^{+}$ (1-0) emission is detected at FIR 3N, 3, and 4, and has the narrow and wide velocity components, suggesting that the emission traces the molecular outflow. 
Actually, previous mapping observations in the HCO$^+$ (1--0) line have detected the molecular outflow driven by FIR 3 \citep{Aso00}. 
The H$^{13}$CO$^{+}$ (1--0, 4--3) emission was also detected at the three positions. 
However, the H$^{13}$CO$^{+}$ emission has the single narrow velocity components with velocity widths of 1.6--1.8 km s$^{-1}$, suggesting that the emission traces the ambient dense gas. 
The HC$^{18}$O$^+$ emission is detected only at FIR 3 and has the only narrow component. As shown in Fig. \ref{hc18op}, the emission-like feature with an S/N of 2.4 at FIR 4 can be marginally seen at the same LSR velocity as in FIR 3. Thus, the non-detection of the HC$^{18}$O$^+$ at FIR 4 is considered to be due to poor sensitivity. In Fig. \ref{abundance_hist}(a), the 3$\sigma$ upper limit of HC$^{18}$O$^+$ at FIR 3N is  plotted above the line of $N_{\rm mol}$ (FIR 3N)=$\frac{N_{\rm H^{13}CO^+}({\rm FIR\ 3N})}{N_{\rm H^{13}CO^+}({\rm FIR\ 4})}$$N_{\rm mol}$(FIR 4), suggesting the sensitivity for HC$^{18}$O$^+$ is poor. The velocity width of HC$^{18}$O$^+$ at FIR 3 is 1.02 km s$^{-1}$, which is similar to those of molecules having the only narrow component. Thus, the HC$^{18}$O$^+$ component likely traces the ambient dense gas.

We estimated the optical depth of the HCO$^{+}$ (1--0) line, assuming the same excitation temperature for the HCO$^{+}$ (1--0) and H$^{13}$CO$^+$ (1--0) lines and the [$^{12}$C/$^{13}$C] ratio of 62 \citep{Langer93}.
Table \ref{optical} shows that  the optical depth of the HCO$^+$ (1--0) line toward FIR 3N, FIR 3, and FIR 4 is 9.7, 7.1, and 8.6, respectively, i.e., optically thick.
Note that we could not cover the frequency of the HCO$^{+}$ (4-3) line.

\subsubsection{HNC, HN$^{13}$C, H$^{15}$NC, HCN, H$^{13}$CN, and HC$^{15}$N } \label{HCN_description}
Figures \ref{hnc}, \ref{hn13c}, \ref{h15nc}, \ref{hcn}, \ref{h13cn}, and \ref{hc15n} show the HNC, HN$^{13}$C, H$^{15}$NC, HCN, H$^{13}$CN, and HC$^{15}$N spectra, respectively.
The emission of HNC and HN$^{13}$C are detected at the three positions. Their spectra consist of only narrow components with the velocity widths of $\sim$1 km s$^{-1}$, which are similar to those of the typical dense-gas tracers such as the H$^{13}$CO$^{+}$ (1--0) line.
These results suggest that the emission of HNC and HN$^{13}$C traces the ambient dense gas.
In contrast, the spectra of HCN at the three positions have the narrow and wide velocity components, suggesting that the emission traces the molecular outflow.
To analyze the hyper fine structure (HFS) of HCN, we assumed that the intensity ratio of the narrow to wide velocity components is the same for all the HFS components.
The wide components of the H$^{13}$CN (1--0, 4--3) and HC$^{15}$N emission are detected only at FIR 4, suggesting that the wide components trace the outflow shock.

The H$^{15}$NC emission is detected  at FIR 3N and FIR 4 and have the only narrow component.
The 3$\sigma$ upper limit of  H$^{15}$NC at FIR 3 is plotted below the line of  $N_{\rm mol}$(FIR 3)=$\frac{N_{\rm H^{13}CO^+}({\rm FIR\ 3})}{N_{\rm H^{13}CO^+}({\rm FIR\ 4})}$$N_{\rm mol}$(FIR 4) in Fig. \ref{abundance_hist}(b). 
Thus, we cannot conclude that the non-detection of H$^{15}$NC at FIR 3 is due to poor sensitivity. The velocity width of H$^{15}$NC at FIR 4 is 1.82 km s$^{-1}$, which is similar to those of molecules having the only narrow component. Thus, the H$^{15}$NC component likely traces the ambient dense gas.

We estimated the optical depths of the HNC emission, assuming the same excitation temperature for the HNC and HN$^{13}$C lines and the [$^{12}$C]/[$^{13}$C] ratio of 62 \citep{Langer93}.
The optical depth of the HNC (1--0) line with a transition of 1--0 is estimated to be 3.1, 4.5, and 6.2 toward FIR 3N, FIR 3, and FIR 4, respectively, i.e., optically thick.
Similarly, we estimated the optical depths of the HCN (1--0, 4--3) emission, assuming the same excitation temperature for the HCN and H$^{13}$CN lines. 
The HCN emission is shown to be optically thick. 
Particularly, the optical depths of the wide components of the HCN (1--0) lines are too high ($>$ 10) compared with those of the narrow components. This is probably due to our assumption that the intensity ratio of the narrow to wide velocity components is the same for all the HFS transitions.

\subsubsection{N$_2$H$^{+}$}
Figure \ref{n2hp} shows a comparison of the N$_2$H$^{+}$ spectra among FIR 3N, FIR 3, and FIR 4. 
The N$_2$H$^{+}$ emission is detected at the three positions and has the only narrow component, suggesting that the emission traces the ambient dense gas.

\subsubsection{c-C$_3$H$_2$}
Figure \ref{c3h2} shows a comparison of the c-C$_3$H$_2$ lines among FIR 3N, FIR 3, and FIR 4. The c-C$_3$H$_2$ (2--1) emission with the single narrow component is detected at the three positions. 
The velocity width and total integrated intensity at FIR 3 are similar to those at FIR 4.
These results suggest that the c-C$_3$H$_2$ emission traces the ambient dense gas.

\subsubsection{NH$_{2}$D} \label{nh2d_appendix}
Figure \ref{nh2d} shows the NH$_2$D spectra at FIR 3N, FIR 3, and FIR 4. 
The NH$_2$D emission is detected only at FIR 4. The velocity width is 1.5 km s$^{-1}$ which is similar to those of the typical dense-gas tracers. Thus, the NH$_2$D emission traces the dense gas. 
In Fig. \ref{abundance_hist}, the 3$\sigma$ upper limits of NH$_2$D at FIR 3N and FIR 3 are plotted below the lines of $N_{\rm mol}$(FIR 3N)=$\frac{N_{\rm H^{13}CO^+}({\rm FIR\ 3N})}{N_{\rm H^{13}CO^+}({\rm FIR\ 4})}$ and  $N_{\rm mol}$(FIR 4) and $N_{\rm mol}$(FIR 3)=$\frac{N_{\rm H^{13}CO^+} ({\rm FIR\ 3})}{N_{\rm H^{13}CO^+}({\rm FIR\ 4})}$$N_{\rm mol}$(FIR 4), suggesting that the sensitivities in NH$_2$D seem sufficient in FIR 3N and FIR 3, and the fractional abundance of NH$_2$D in FIR 4 is likely higher than those in FIR 3N and FIR 3.

\subsubsection{SiO}
Figure \ref{sio} shows the SiO (2--1, 8--7) spectra at FIR 3N, 3, and 4.
The SiO emission is detected only at FIR 4 and has the narrow and wide components.
The velocity widths of the wide components are 8--10 km s$^{-1}$, which are extremely larger than those of the H$^{13}$CO$^{+}$ narrow components (1.6--2.1 km s$^{-1}$).
The peak velocities of the narrow and wide components are 11 and 5 km s$^{-1}$, which is consistent with the NMA observations \citep[see Fig 8(c) in ][]{Shimajiri08}. The NMA observations in SiO revealed that the SiO emission is distributed at the interface between the outflow driven from FIR 3 and the FIR 4 clump. These facts suggest that the SiO emission traces the outflow shock.

\subsubsection{SO}
Figure \ref{so} shows a comparison of the SO lines among FIR 3N, FIR 3, and FIR 4.
We have succeeded in detecting the four transitions of SO.
At FIR 4, all the transitions have the wide velocity components, and the peak velocities of the wide components are blueshifted with respect to those of the narrow components.
The difference in peak velocity between the narrow and wide components is quite similar to that of the SiO emission at FIR 4, suggesting that the wide components of the SO emission trace the outflow shock. However, the wide components of the SO emission with higher upper energy levels are also seen at FIR 3. 
Thus, there are two possibilities.  First is that the SO emission traces the molecular outflow and the non-detection at FIR 3N and FIR 3 of the wide components is due to poor sensitivity. Second is that the SO emission traces the outflow shock and the wide components at FIR 3 trace the outflow shock near the driving source where the outflow gas is being launched. 
The possible reason for the wide components of only SO being detected at FIR 3 is that the SO emission have higher upper energy levels ($E_{\rm u}$ $>$ 80 K) and traces the warm region.

\subsubsection{H$_2$CS}
Figure \ref{h2cs} shows a comparison of the H$_2$CS lines among FIR 3N, FIR 3, and FIR4. 
The H$_2$CS emission is detected at the three positions. 
The H$_2$CS emission with an upper-state energy of $>$ 90 K has the narrow and wide components. 
According to our criteria of the line identification (see Sect. \ref{lineid}), the H$_2$CS (3$_{0,3}$--2$_{0,2}$, 3$_{1,3}$--2$_{1,2}$, 3$_{1,2}$--2$_{1,1}$) lines with a upper-state energy of $<$ 90 K are identified as spectra having a  single Gaussian component. 
The wide components of the H$_2$CS emission are detected only at FIR 4, suggesting that these components trace the outflow shock.

\subsubsection{HC$_3$N}
Figure \ref{hc3n} shows a comparison of the HC$_3$N (10--9, 11--10) lines among FIR 3N, FIR 3, and FIR 4. 
The two transitions are detected at the three positions. Only at FIR 4, the HC$_3$N emission has the wide velocity component, suggesting that the wide components trace
 the outflow shock. 
Owing to a slight difference between the two upper energy levels of the transitions, we could not derive the rotation temperature and column density of HC$_3$N from the rotation diagram.

\subsubsection{CH$_3$OH}
Figure \ref{ch3oh} shows a comparison among the spectra of the CH$_3$OH lines at FIR 3N, FIR 3, and FIR 4. 
The wide components of the CH$_3$OH lines are detected only at FIR 4, suggesting that the wide components trace the outflow shock.
In addition, the intensity ratio of the wide to narrow components of the CH$_3$OH line tends to increase with increasing upper state energy level in the transition.
This result suggests that the wide component of the CH$_3$OH line traces the high temperature part of the outflow-shocked region at FIR 4.
In fact, the rotation diagram analysis has revealed that the rotation temperatures of the CH$_3$OH wide components are as high as 46.4$\pm$6.8 K for CH$_3$OH A and 25.4$\pm$3.3 K for CH$_3$OH E.

\subsubsection{HCS$^{+}$}
Figure \ref{hcsp} shows the HCS$^+$ spectra at FIR 3N, FIR 3, and FIR 4. 
The HCS$^{+}$ emission is detected only at FIR 4. 
The intensity and velocity width are 0.07 K in $T_{\rm A}^*$ (corresponding to a signal-to-noise ratio of 3.5) and 3.0 km s$^{-1}$. 
The velocity width is twice as large more than twice as large as those of the ambient dense-gas tracers. 
It is possible that the HCS$^+$ emission has a wide component only at FIR 4 and traces the outflow shock. The reason for the non-detection of the wide component is considered to be due to poor sensitivity. In fact, the signal-to-noise ratio of the emission at FIR 4 is too low to separate the two components.  We cannot, however, exclude another possibility that the non-detection of the emission at FIR 3 and FIR 3N is due to poor sensitivity (see the center panel of Fig. \ref{hcsp}) and that the emission tracers the outflow. To reveal whether the emission has the wide component or not, higher sensitivity observations are required.

\subsubsection{H$_2$CO, H$_2^{13}$CO, and HDCO}\label{appendix_hdco}
Figures \ref{h2co}, \ref{h213co}, and \ref{hdco} show the H$_2$CO, H$_2^{13}$CO, and HDCO spectra, respectively.
The H$_2$CO emission is detected only at FIR 4. The velocity width of the H$_2$CO emission is 3.8 km s$^{-1}$ which is twice as large more than twice as large as the typical width of the ambient dense-gas tracers. 
Considering the weak intensity of the emission, there remains a possibility that the wide component of H$_2$CO is not detected due to poor sensitivity. Thus, we classify this molecule as a possible shock tracer.

The H$_2^{13}$CO emission is detected at the three positions. The wide component is, however, seen only at FIR 4. 
This result suggests that the wide component of the H$_2^{13}$CO emission traces the outflow shock. 

The HDCO emission is detected at FIR 3 and FIR 4. 
The velocity widths at FIR 3 and FIR 4 are 2.9 km s$^{-1}$ and 3.5 km s$^{-1}$, respectively. 
These values are twice as large more than twice as large as the typical velocity width of the ambient dense-gas tracers, and the HDCO emission possibly traces the outflow or the outflow shock.

\subsubsection{HNCO}
Figure \ref{hnco} shows the HNCO spectra at FIR 3N, FIR 3, and FIR 4. 
The HNCO emission is detected at FIR 3 and FIR 4. At FIR 3N, a emission-like feature with an S/N of 2.0 is seen, and thus, the non-detection at FIR 3N might be due to poor sensitivity.
The intensity and velocity width at FIR 3 are 0.114 K in $T_{\rm A}^*$ (S/N=6.1) and 1.48 km s$^{-1}$, respectively, and 0.083 K (S/N=6.1) and 3.15 km s$^{-1}$ at FIR 4.
The velocity width at FIR 4 is twice as large more than twice as large as that at FIR 3 and those of the ambient dense-gas tracers. 
We speculate that the emission has a wide component only at FIR 4 and traces the outflow shock. The reason for the non-detection of the wide component would be due to poor sensitivity. In fact, the signal-to-noise ratio of the emission is too low to separate the two components.  We cannot, however, exclude another possibility that the non-detection of the wide component at FIR 3 and FIR 3N is due to poor sensitivity. To reveal whether the emission has the wide component or not, higher sensitivity observations are required.

\subsubsection{CH$_3$CN}
Figure \ref{ch3cn} shows a comparison of the CH$_3$CN lines among FIR 3N, FIR 3, and FIR 4.
The CH$_3$CN emission are detected at the three positions and have the single narrow components. 
These results suggest that the CH$_3$CN emission traces the ambient dense gas.

\subsubsection{CH$_3$CCH}
Figure \ref{ch3cch} shows a comparison of the CH$_3$CCH lines among FIR 3N, FIR 3, and FIR 4.
The CH$_3$CCH emission is detected at FIR 3 and FIR 4 and has the single narrow components.
At FIR 3N, emission-like features of CH$_3$CCH (6$_1$-5$_1$) with an S/N of 2.2 are marginally seen at the same velocities as in FIR 3 and FIR 4. Thus, the non-detection of CH$_3$CCH at FIR 3N is considered to be due to poor sensitivity. 
These results suggest that the CH$_3$CCH emission traces the ambient dense gas.
However, the rotation temperatures of the CH$_3$CCH transitions at FIR 4 are estimated to be 32.5$\pm$3.8 K for ortho and 53.2$\pm$36.4 K for para, which are twice as large more than twice as large as the mean temperature of the dense-gas tracers of C$_2$H and H$^{13}$CO$^+$ (12.5$\pm$1.4 K) . We cannot exclude a possibility that the CH$_3$CCH traces the outflow and/or outflow shock.

\subsubsection{CH$_3$CHO}
Figure \ref{ch3cho} shows a comparison of the CH$_3$CHO lines among FIR 3N, FIR 3, and FIR 4.
The CH$_3$CHO lines are detected only at FIR 4 except the CH$_3$CHO (5$_{1,5}$--4$_{1,4}$A) transition. 
The velocity widths of the CH$_3$CHO lines at FIR 4 are 2--4 km s$^{-1}$, which twice as large more than twice as large as the mean velocity width of the ambient dense-gas tracers and the velocity width at FIR 3 (2.38 km s$^{-1}$). 
Thus, we consider the CH$_3$CHO emission as possible shock tracers.

\clearpage

\begin{figure}
\begin{center}
\includegraphics[angle=0,scale=.5]{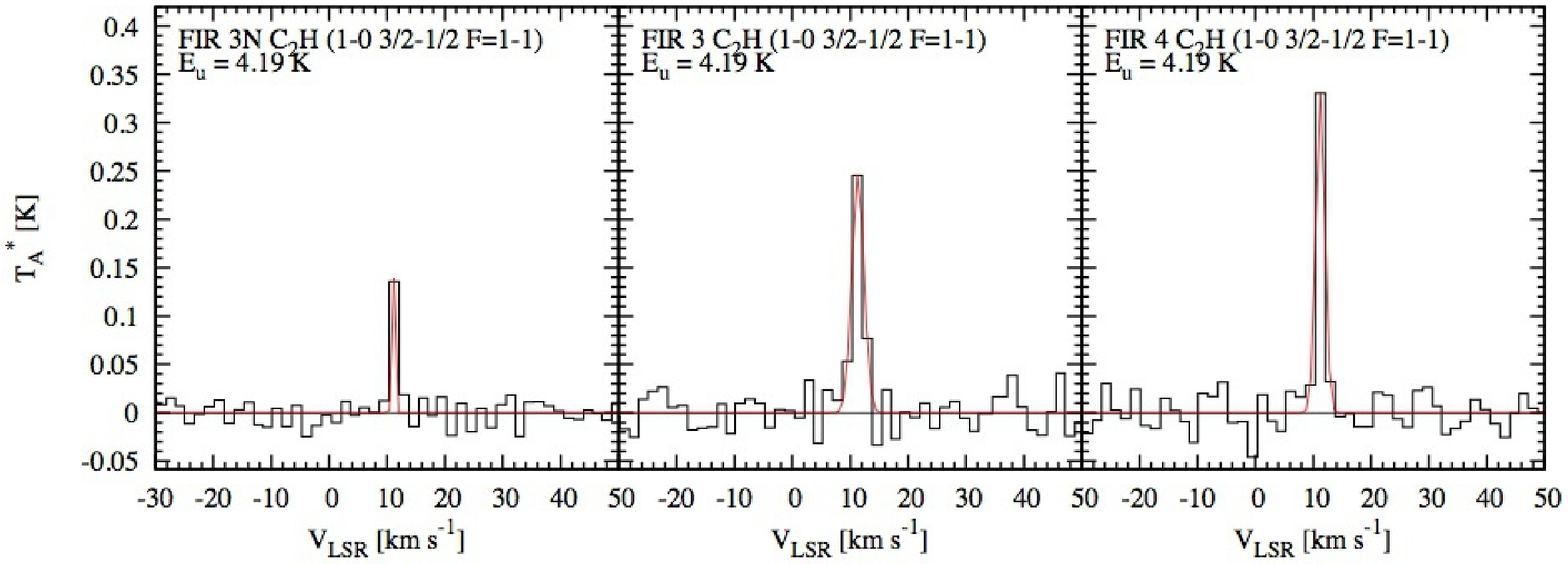}
\includegraphics[angle=0,scale=.5]{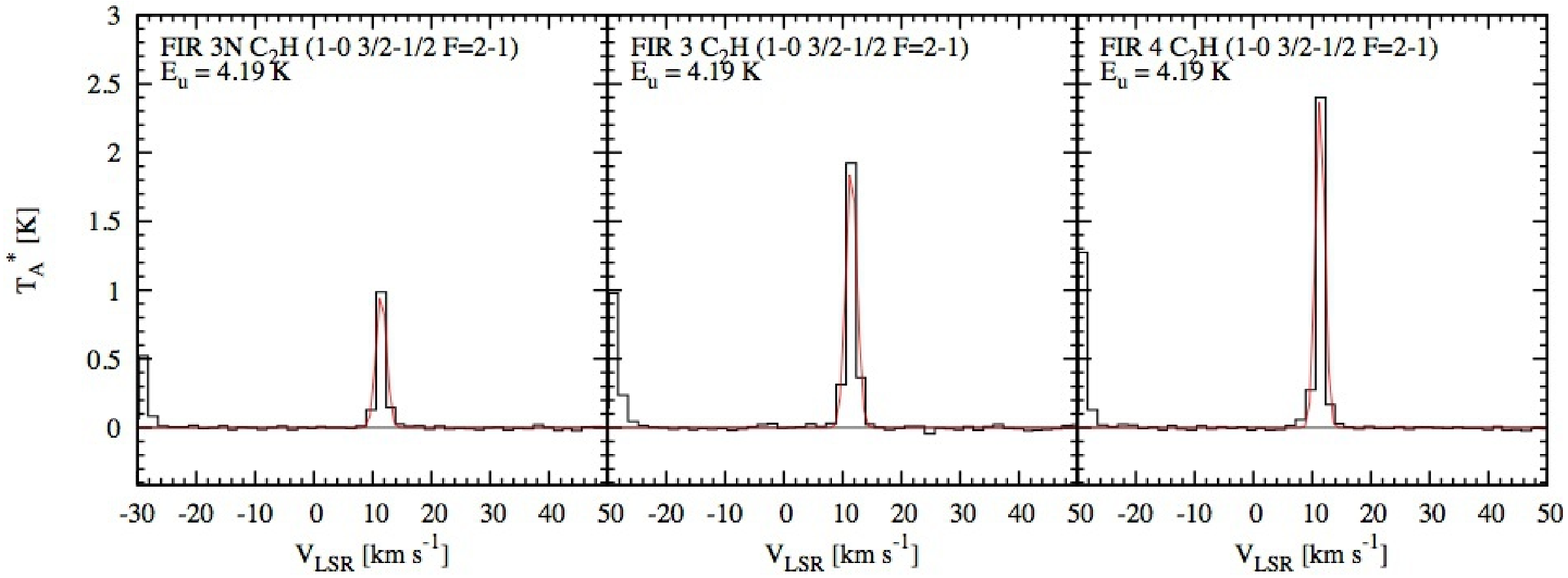}
\includegraphics[angle=0,scale=.5]{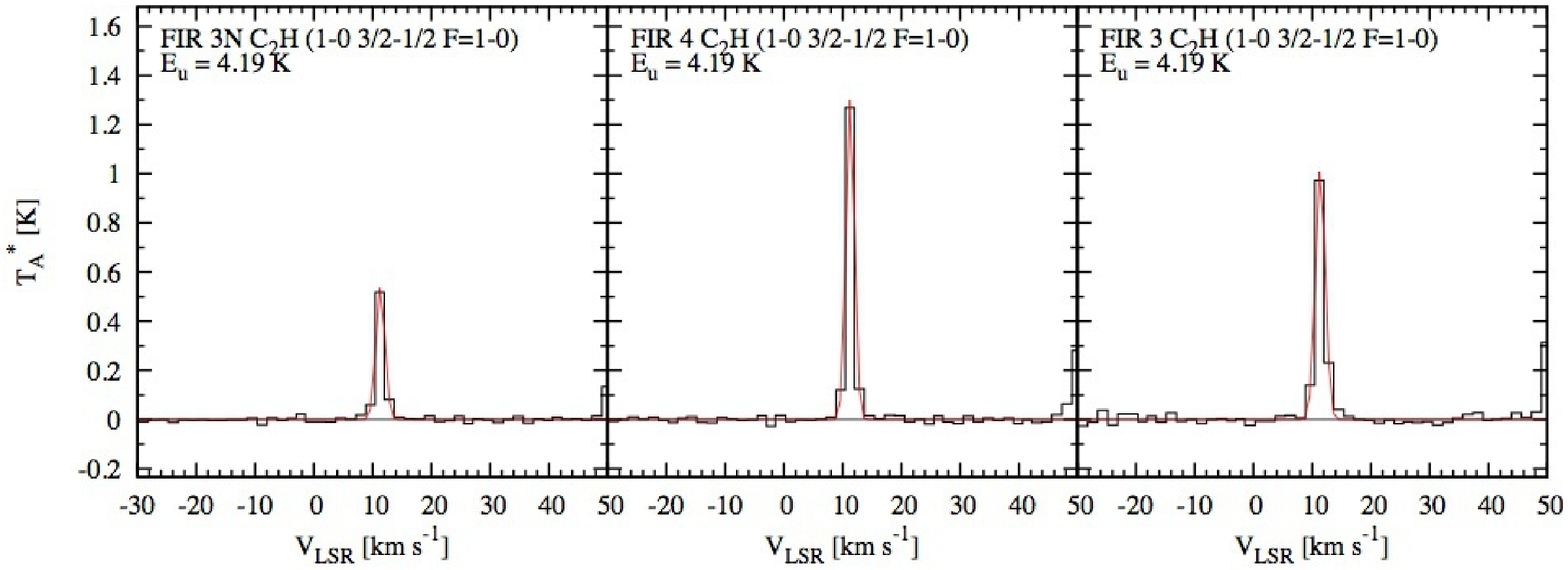}
\end{center}
\caption{C$_2$H spectra}
\label{c2h}
\end{figure}

\clearpage

\addtocounter{figure}{-1}
\begin{figure}
\begin{center}
\includegraphics[angle=0,scale=.5]{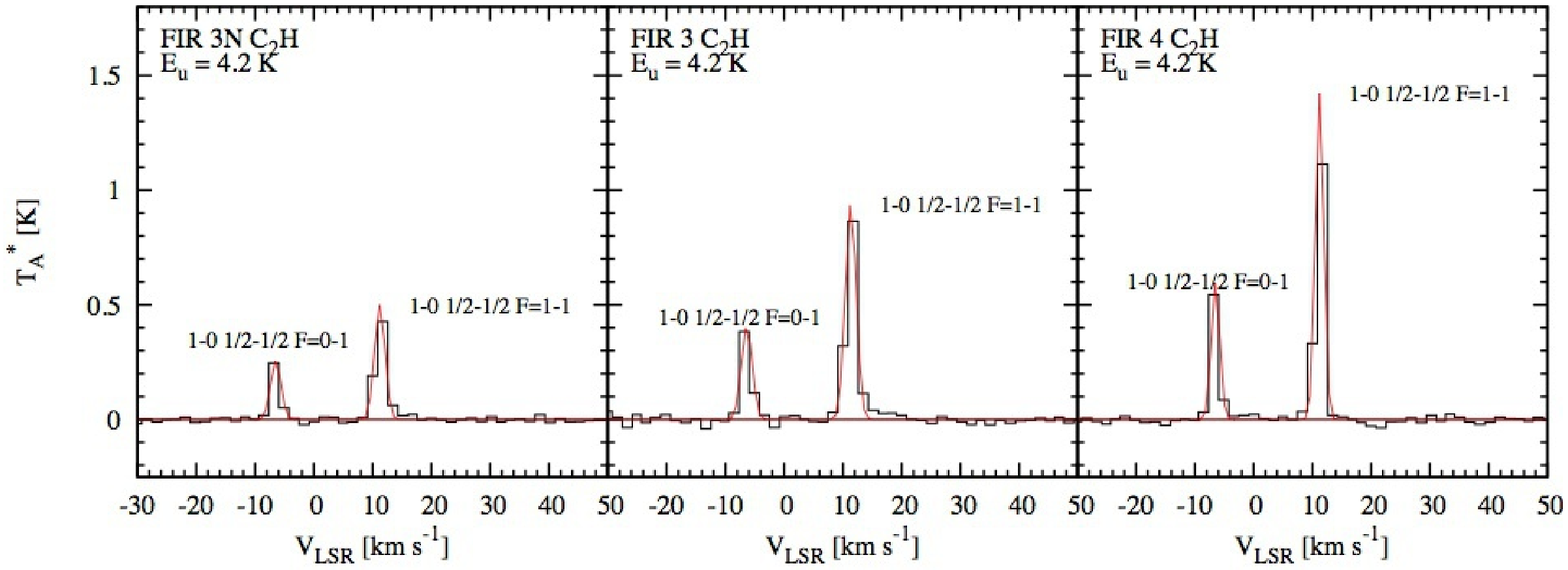}
\includegraphics[angle=0,scale=.5]{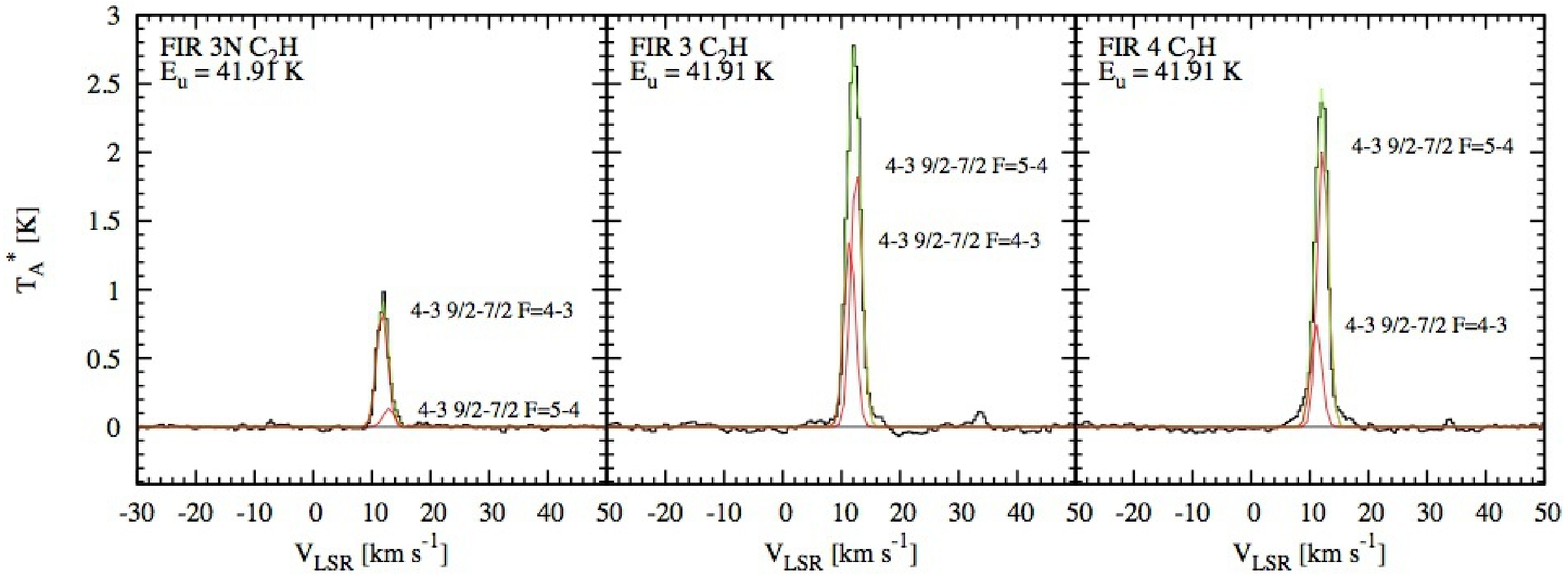}
\includegraphics[angle=0,scale=.5]{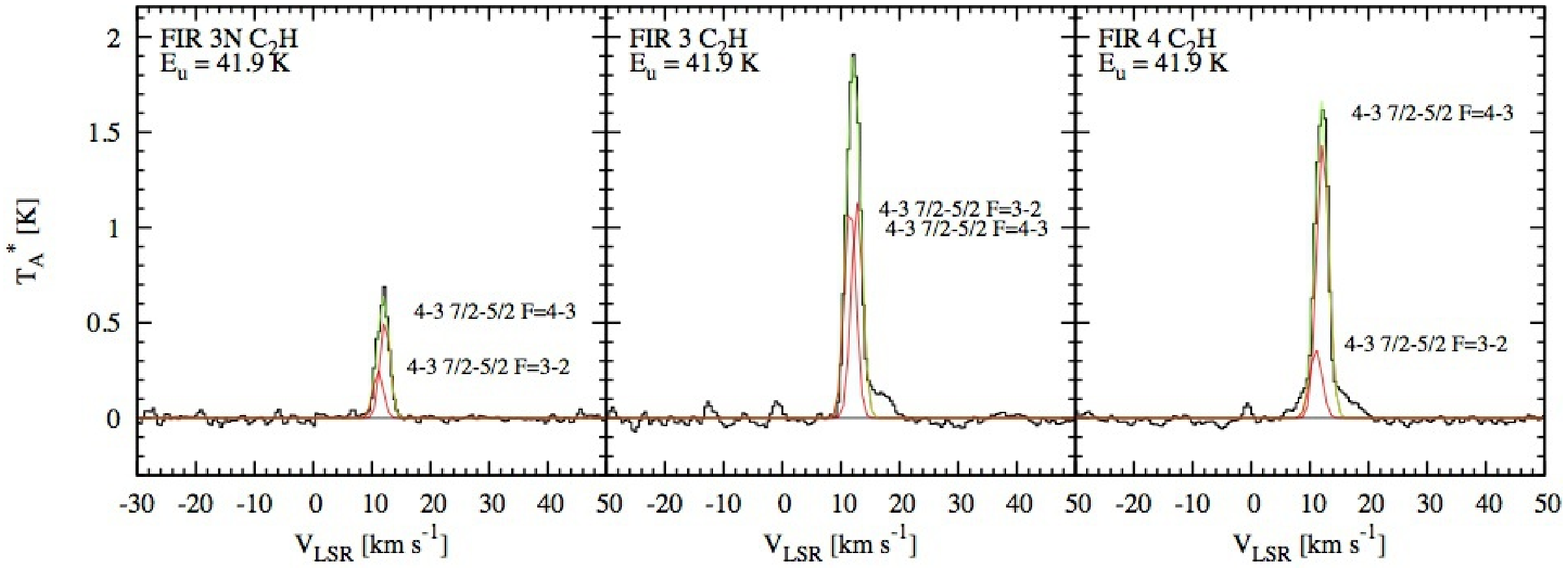}
\end{center}
\caption{\it Continued.}
\label{c2h}
\end{figure}

\begin{figure}
\begin{center}
\includegraphics[angle=0,scale=.5]{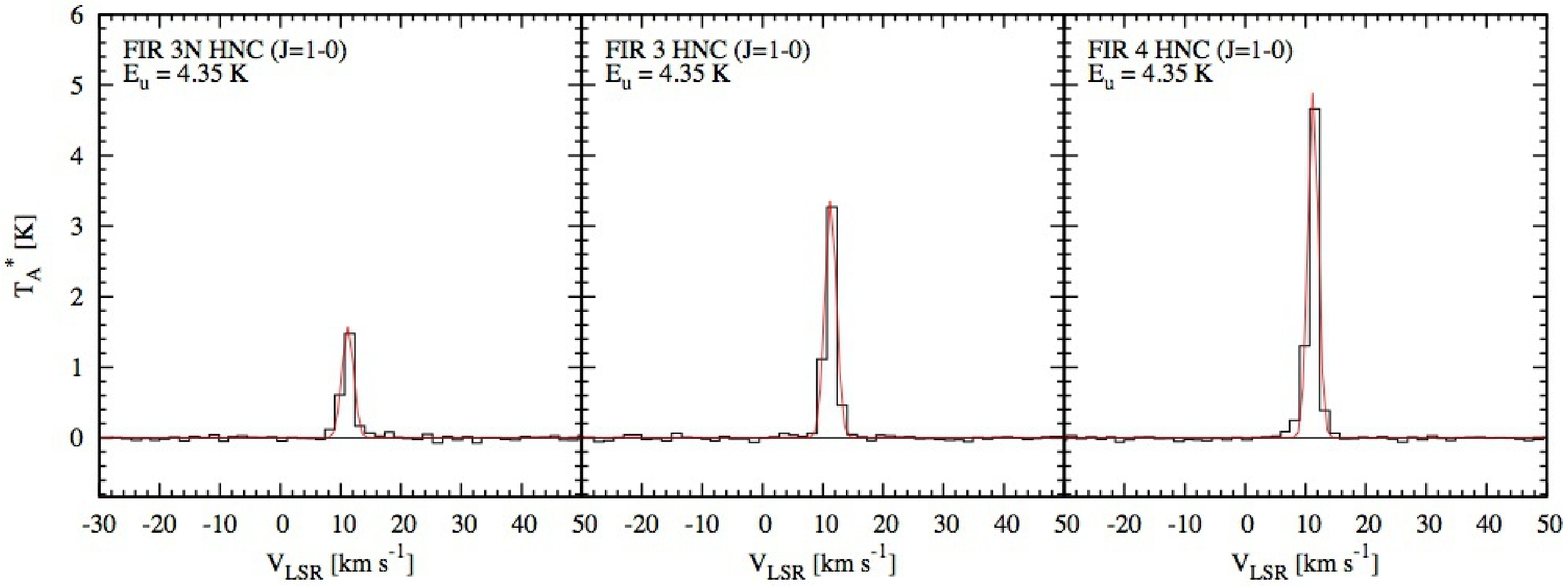}
\end{center}
\caption{HNC spectra}
\label{hnc}
\end{figure}

\begin{figure}
\begin{center}
\includegraphics[angle=0,scale=.5]{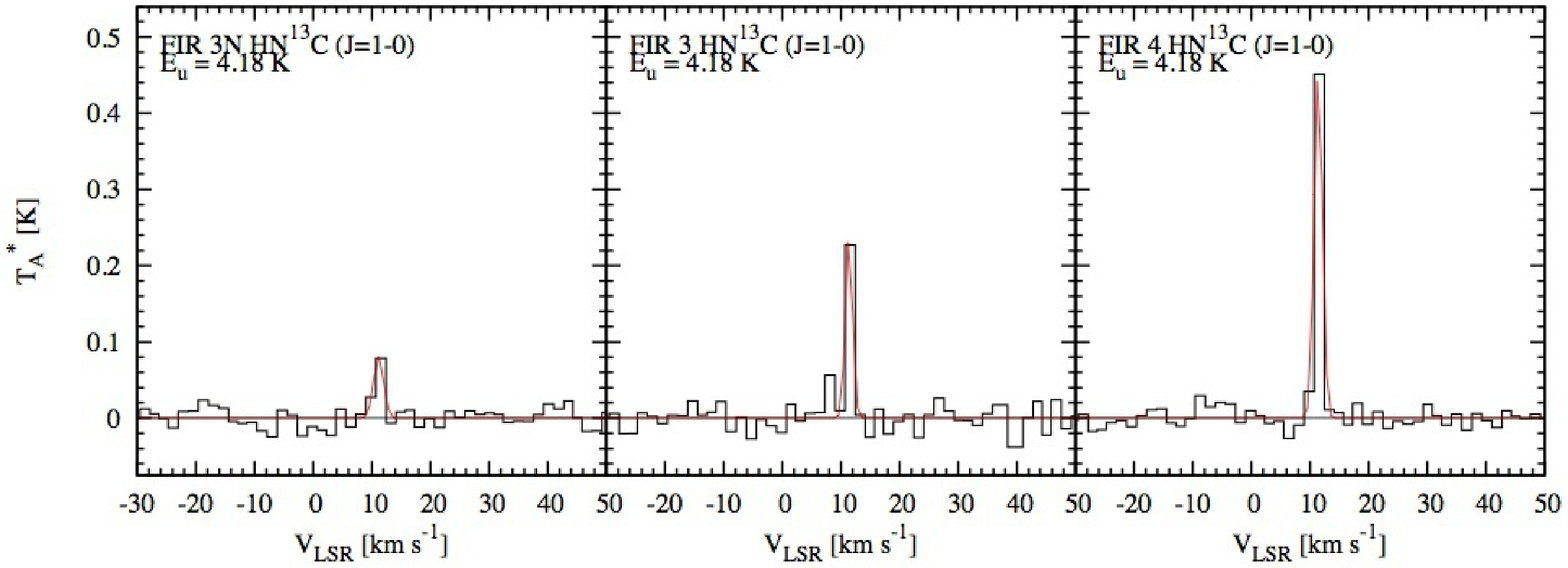}
\includegraphics[angle=0,scale=.5]{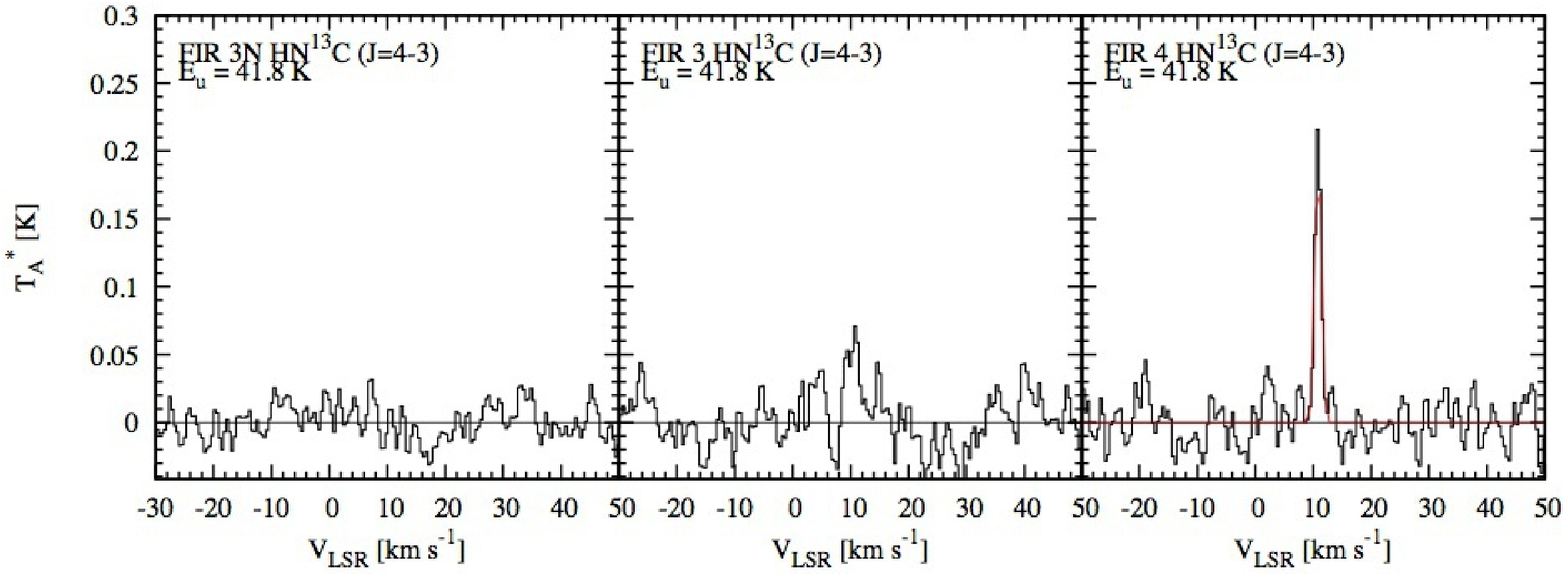}
\end{center}
\caption{HN$^{13}$C spectra}
\label{hn13c}
\end{figure}

\begin{figure}
\begin{center}
\includegraphics[angle=0,scale=.5]{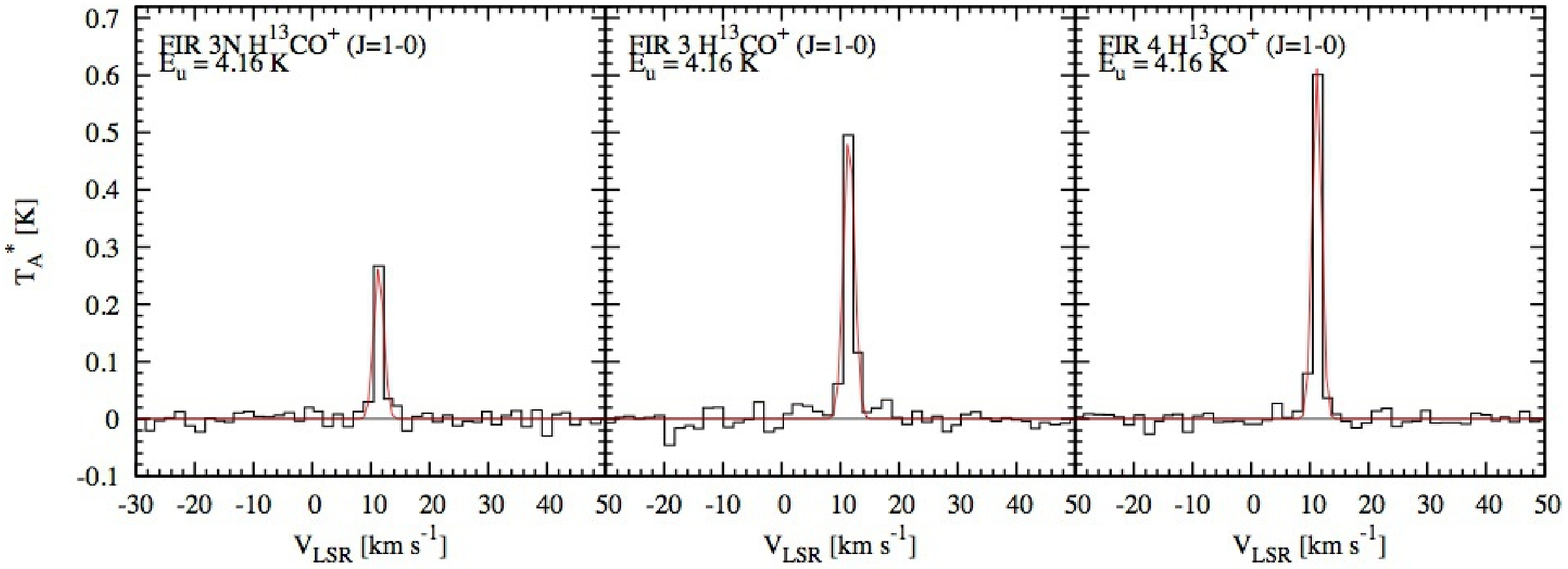}
\includegraphics[angle=0,scale=.5]{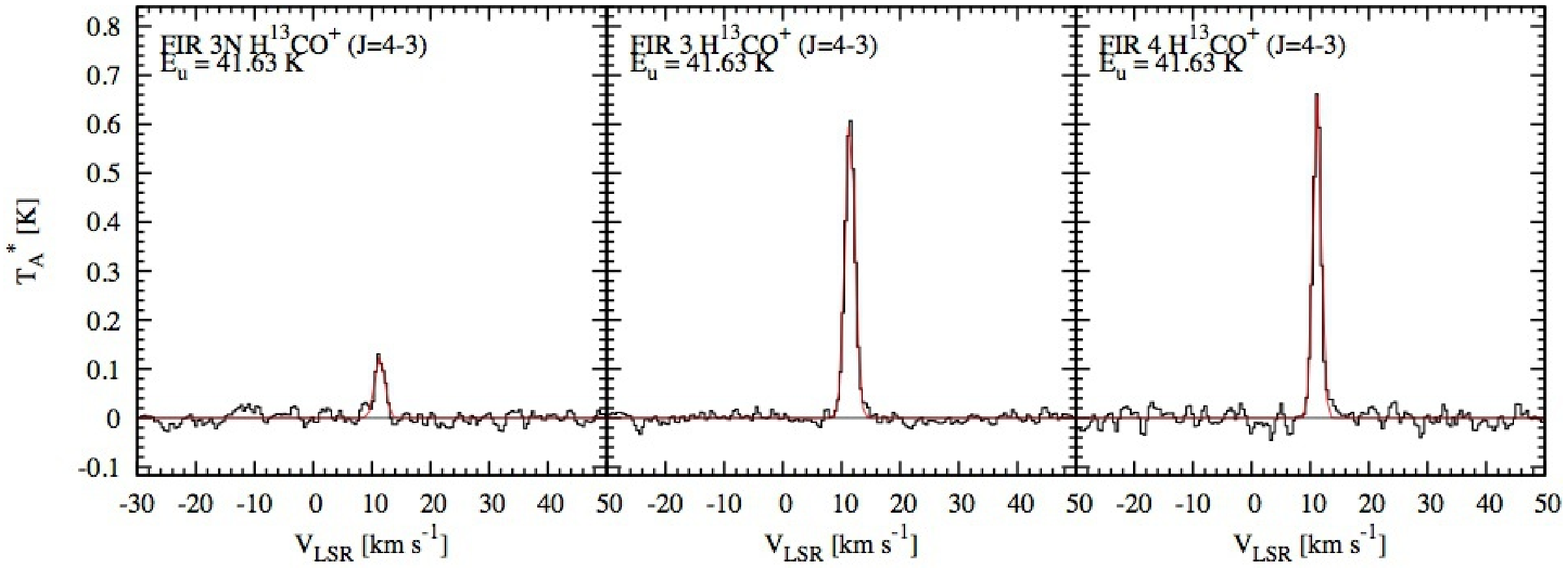}
\end{center}
\caption{H$^{13}$CO$^{+}$ spectra}
\label{h13cop}
\end{figure}

\begin{figure}
\begin{center}
\includegraphics[angle=0,scale=.5]{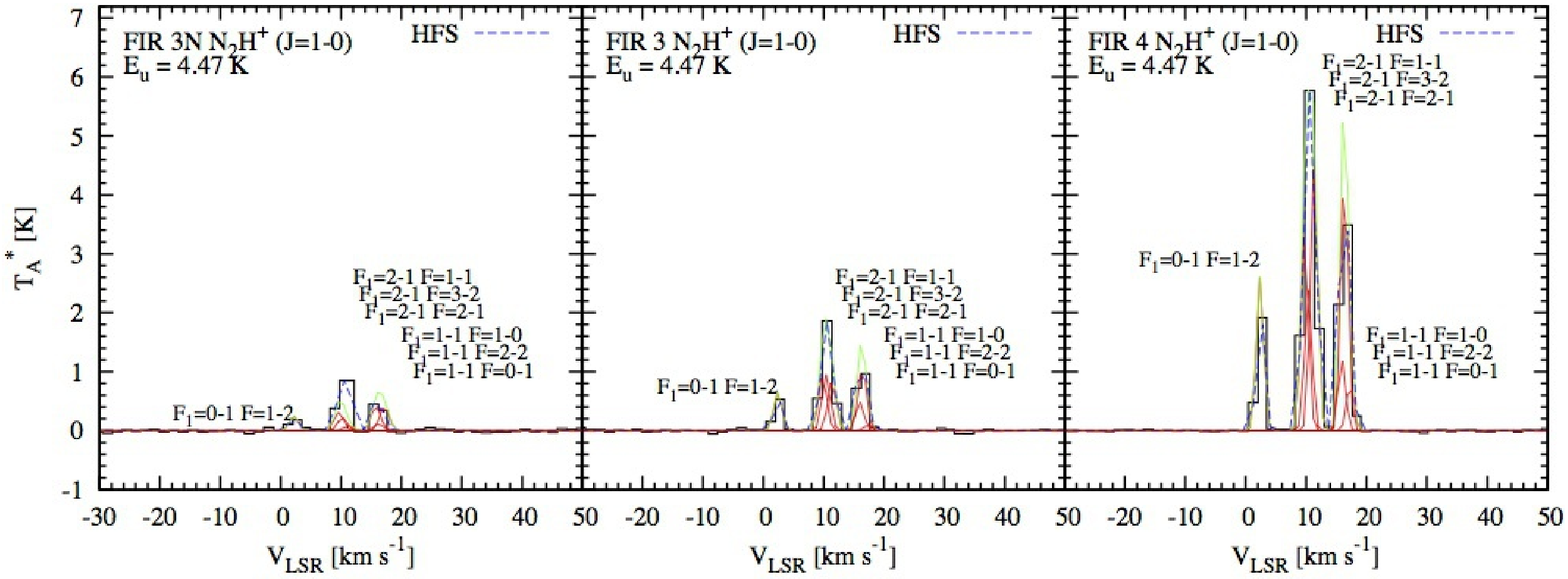}
\end{center}
\caption{N$_2$H$^+$ spectra. The HFS fitting result is shown by the blue dashed line.}
\label{n2hp}
\end{figure}

\begin{figure}
\begin{center}
\includegraphics[angle=0,scale=.52]{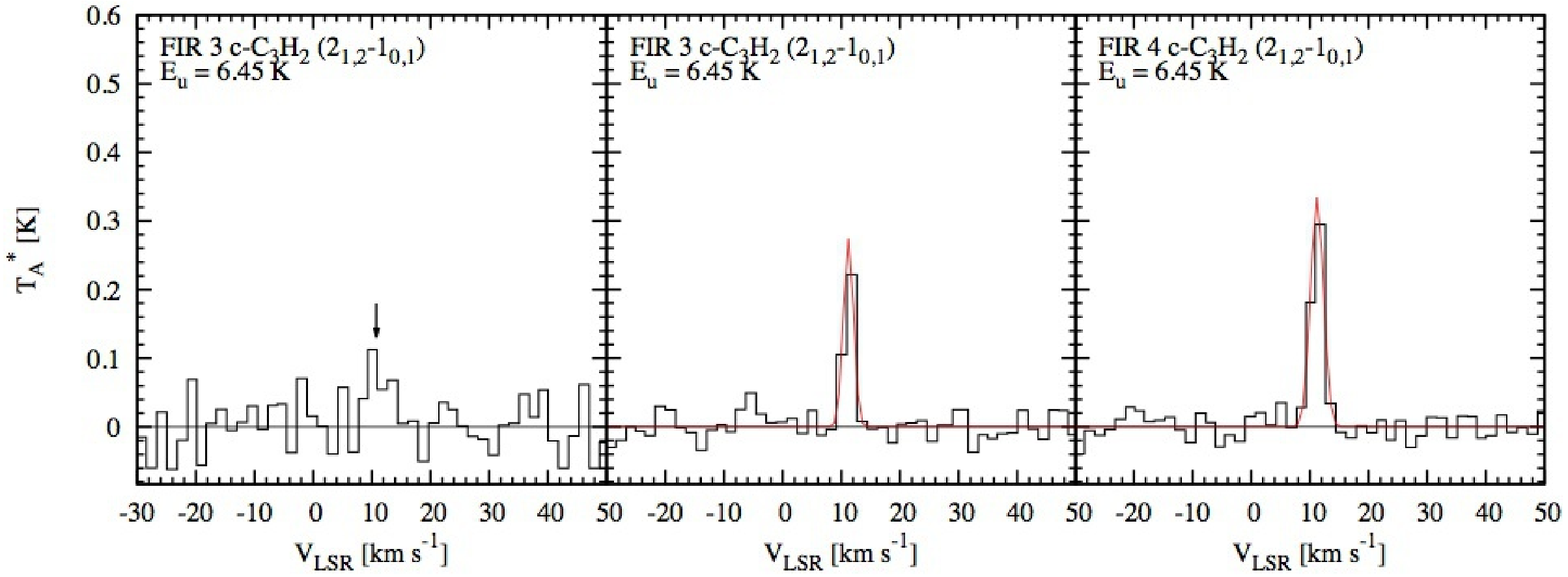}
\end{center}
\caption{c-C$_3$H$_2$ spectra. The vertical arrow in the left panel shows the detection, but the peak is not fitted well by the single Gaussian component due to insufficient velocity resolution.}
\label{c3h2}
\end{figure}

\begin{figure}
\begin{center}
\includegraphics[angle=0,scale=.5]{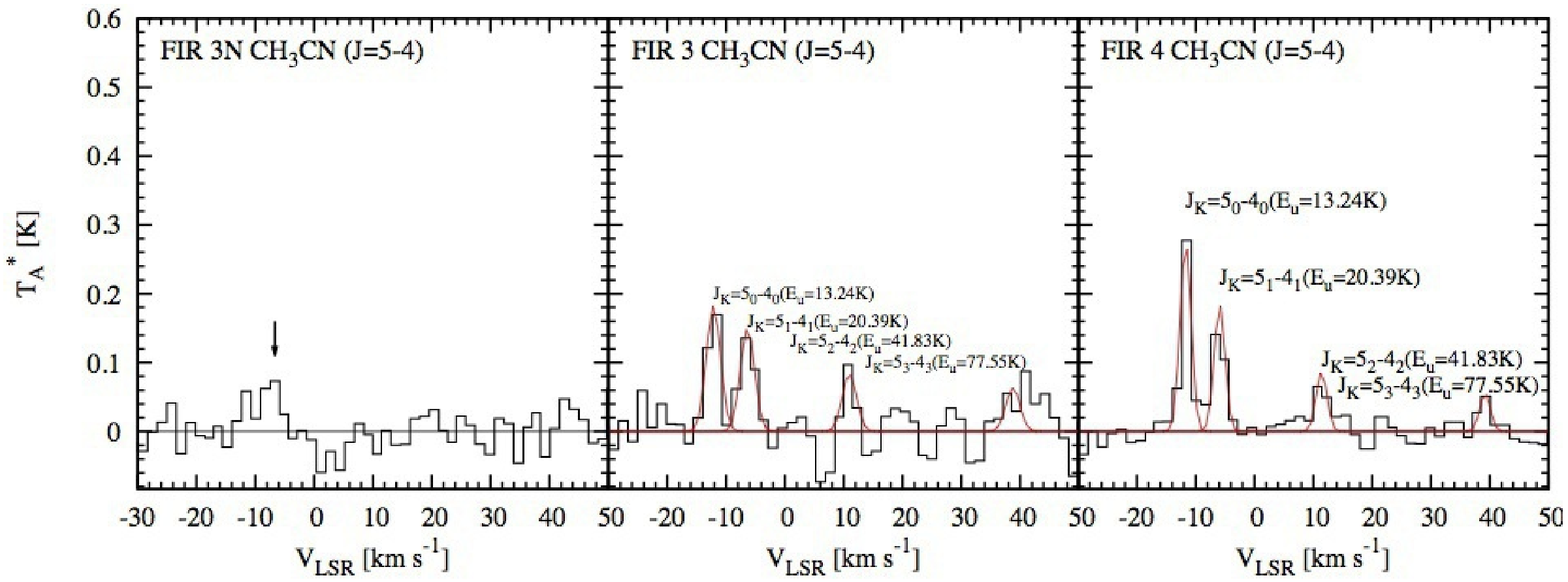}
\end{center}
\caption{CH$_3$CN spectra. The vertical arrow in the left panel shows the detection, but the peak is not fitted well by the single Gaussian component due to insufficient velocity resolution.}
\label{ch3cn}
\end{figure}

\begin{figure}
\begin{center}
\includegraphics[angle=0,scale=.5]{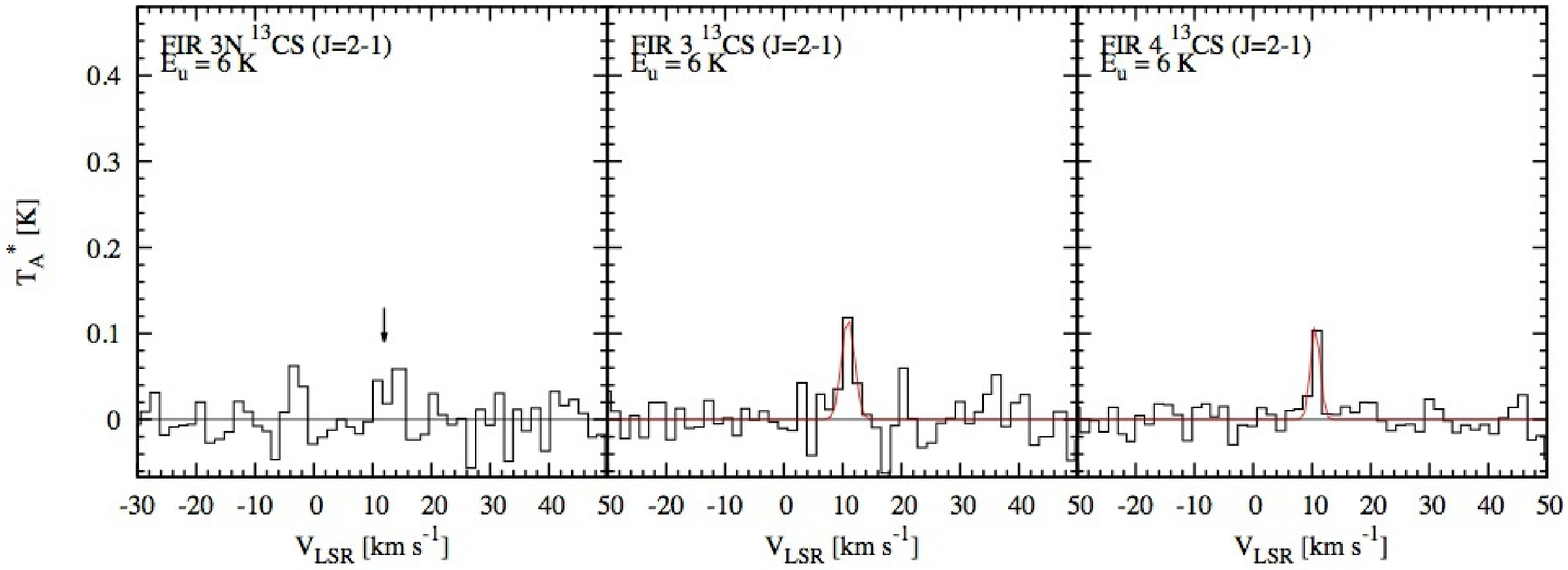}
\end{center}
\caption{$^{13}$CS spectra. The vertical arrow in the right panel shows the emission-like feature.}
\label{13cs}
\end{figure}

\begin{figure}
\begin{center}
\includegraphics[angle=0,scale=.5]{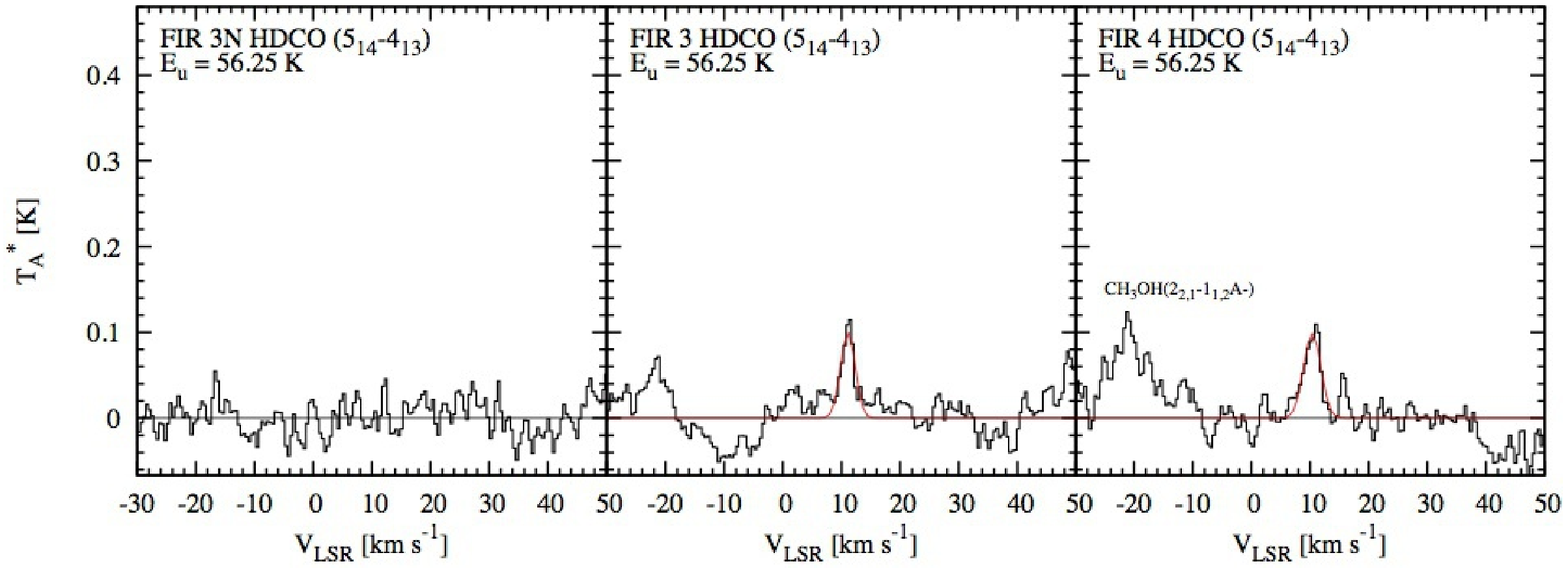}
\end{center}
\caption{HDCO spectra}
\label{hdco}
\end{figure}

\begin{figure}
\begin{center}
\includegraphics[angle=0,scale=.5]{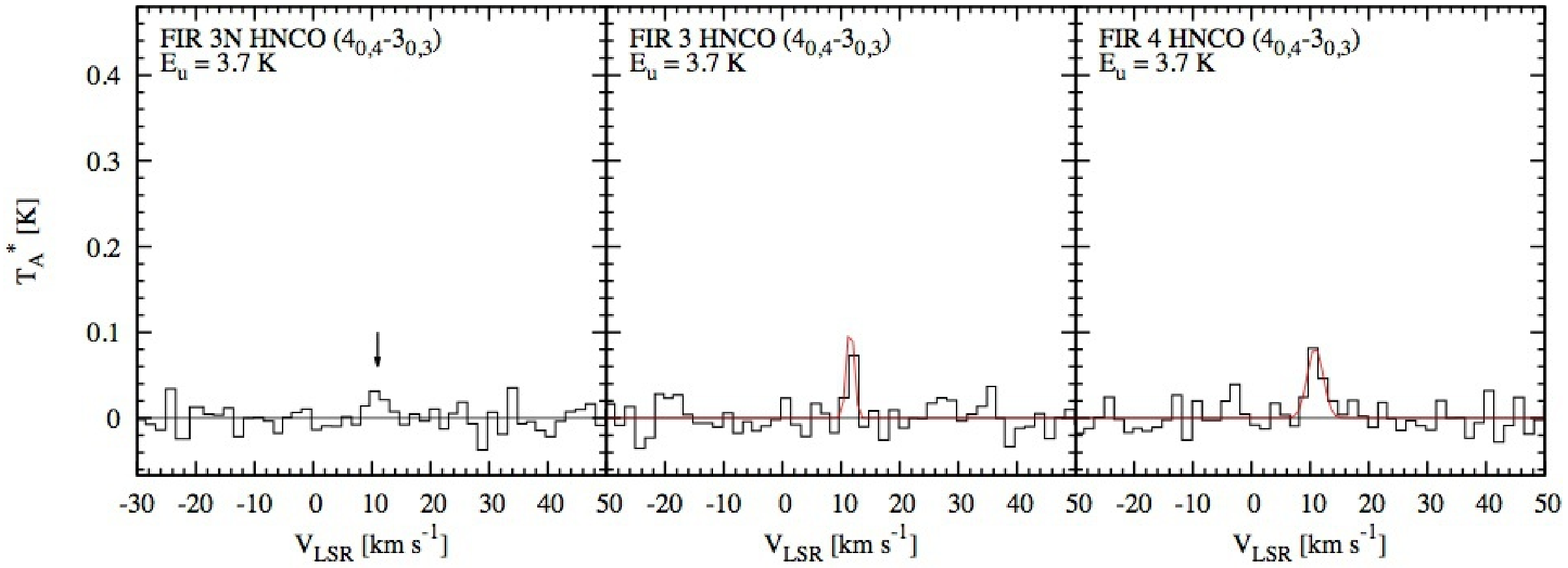}
\end{center}
\caption{HNCO spectra. The vertical arrow in the right panel shows the emission-like feature.}
\label{hnco}
\end{figure}

\begin{figure}
\begin{center}
\includegraphics[angle=0,scale=.5]{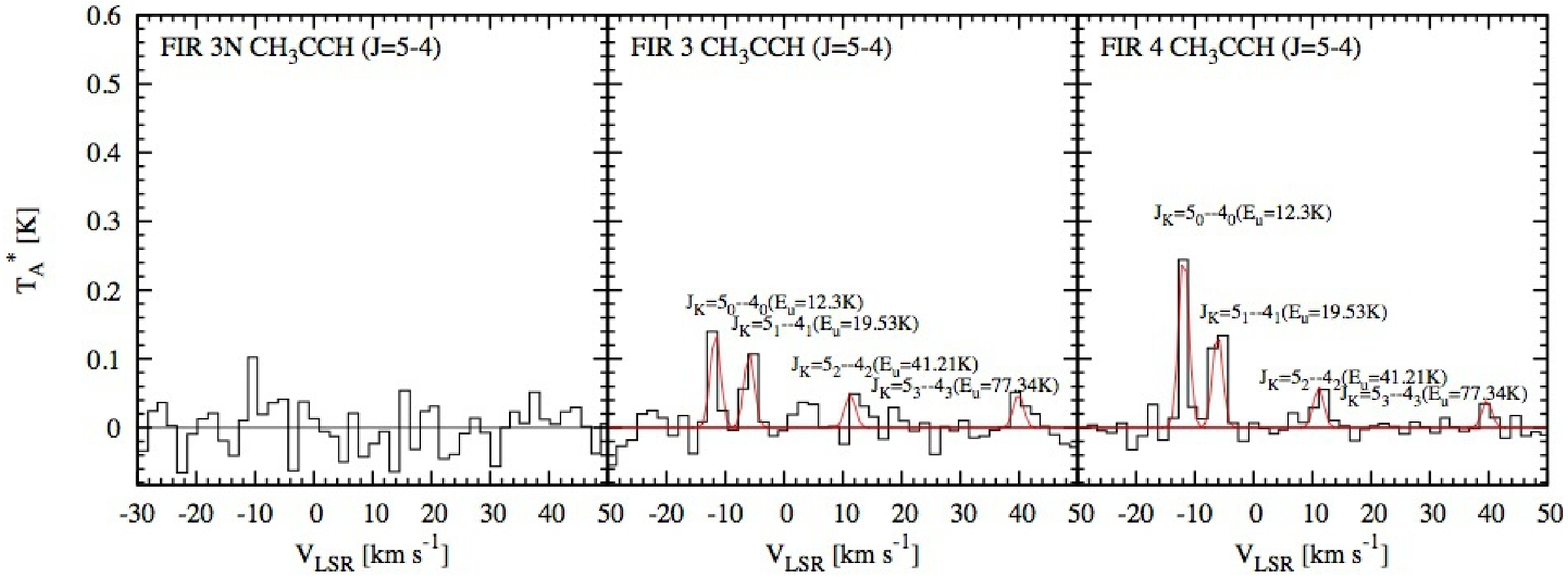}
\includegraphics[angle=0,scale=.51]{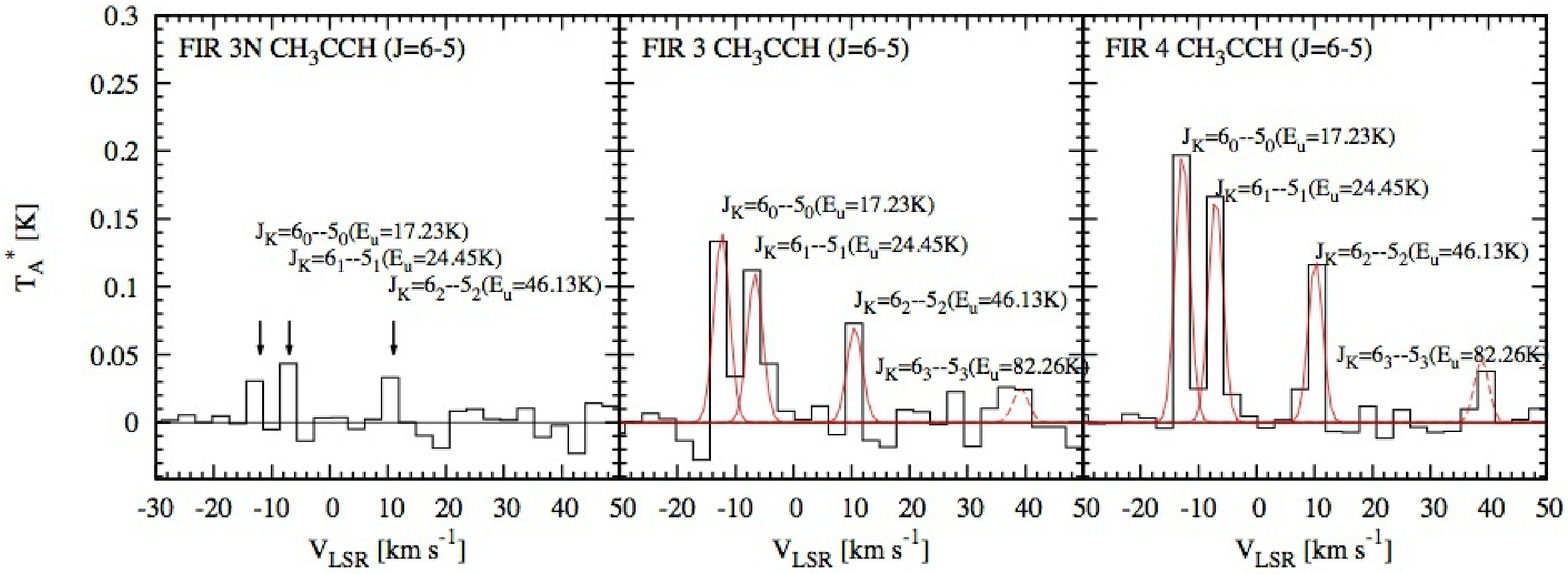}
\end{center}
\caption{CH$_3$CCH spectra. The vertical arrow in the right panel shows the emission-like feature.}
\label{ch3cch}
\end{figure}

\begin{figure}
\begin{center}
\includegraphics[angle=0,scale=.5]{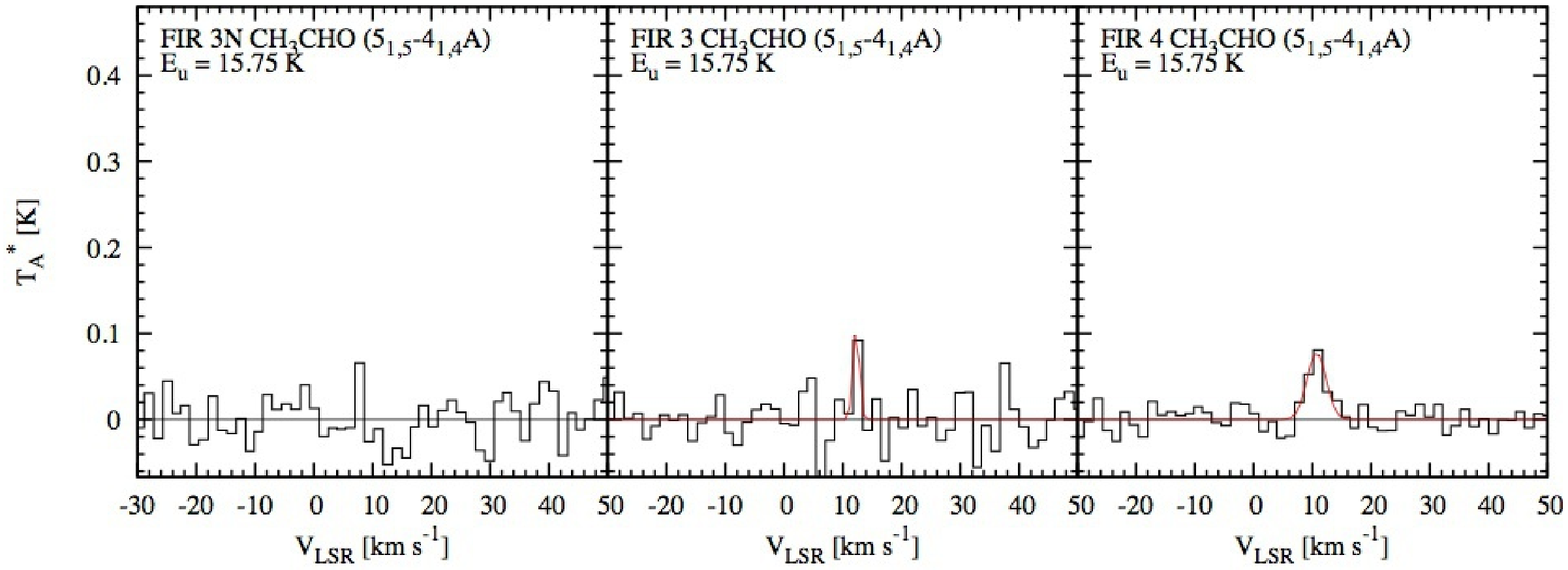}
\includegraphics[angle=0,scale=.5]{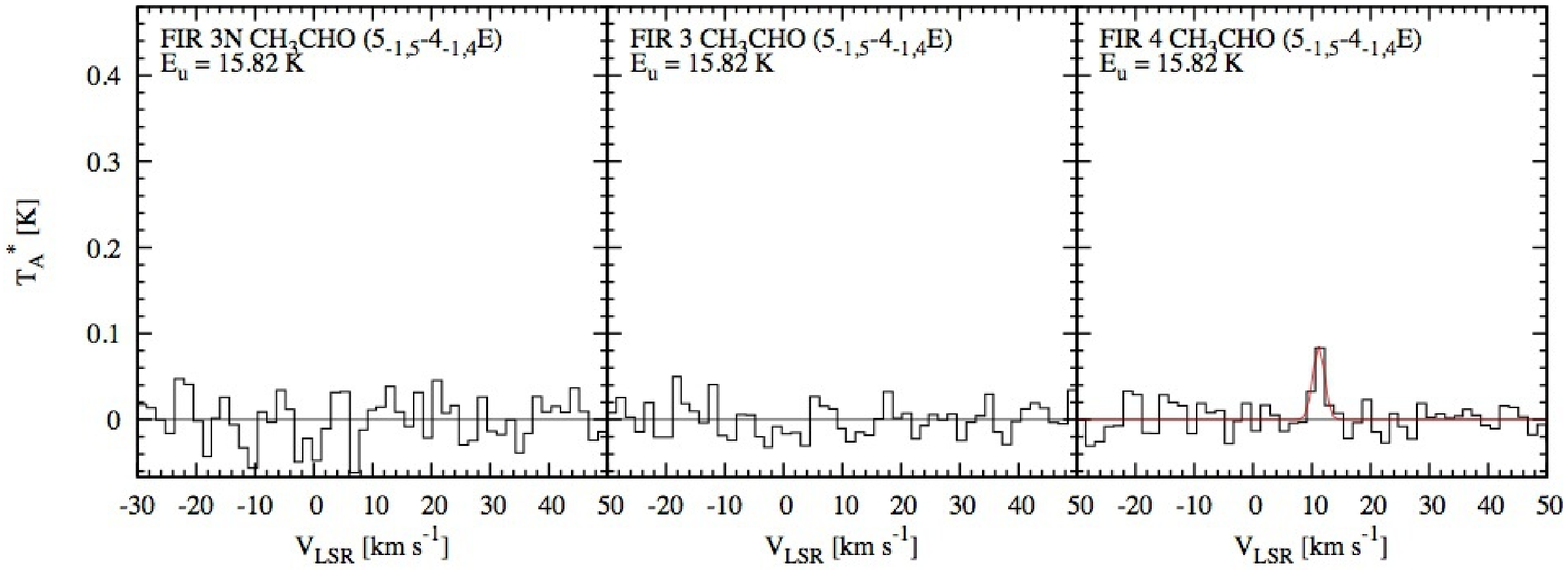}
\includegraphics[angle=0,scale=.51]{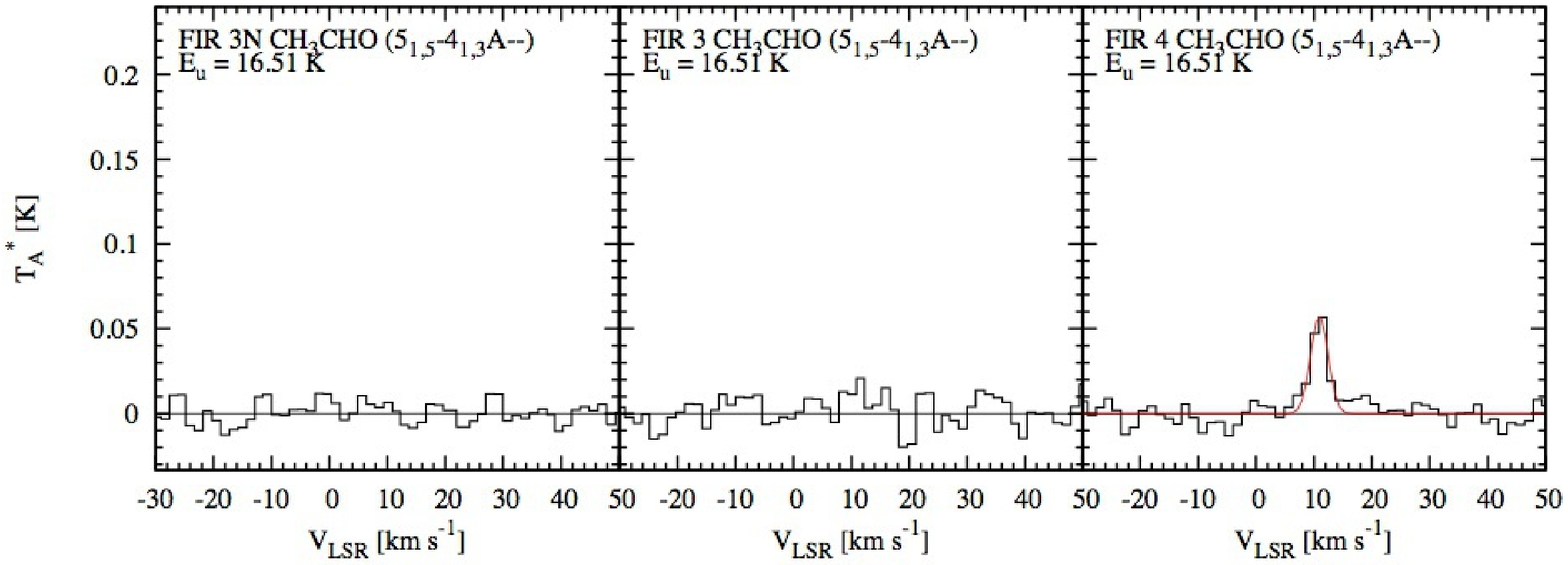}
\includegraphics[angle=0,scale=.5]{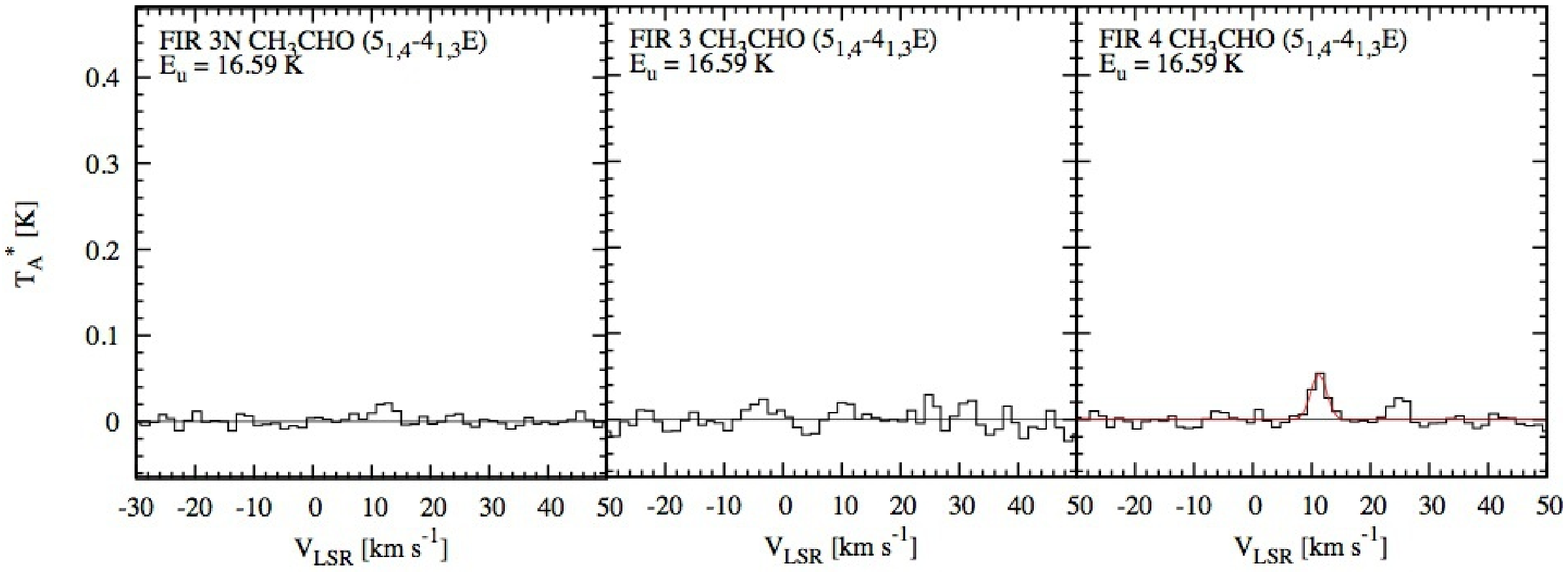}
\end{center}
\caption{CH$_3$CHO spectra}
\label{ch3cho}
\end{figure}

\clearpage

\begin{figure}
\begin{center}
\includegraphics[angle=0,scale=.5]{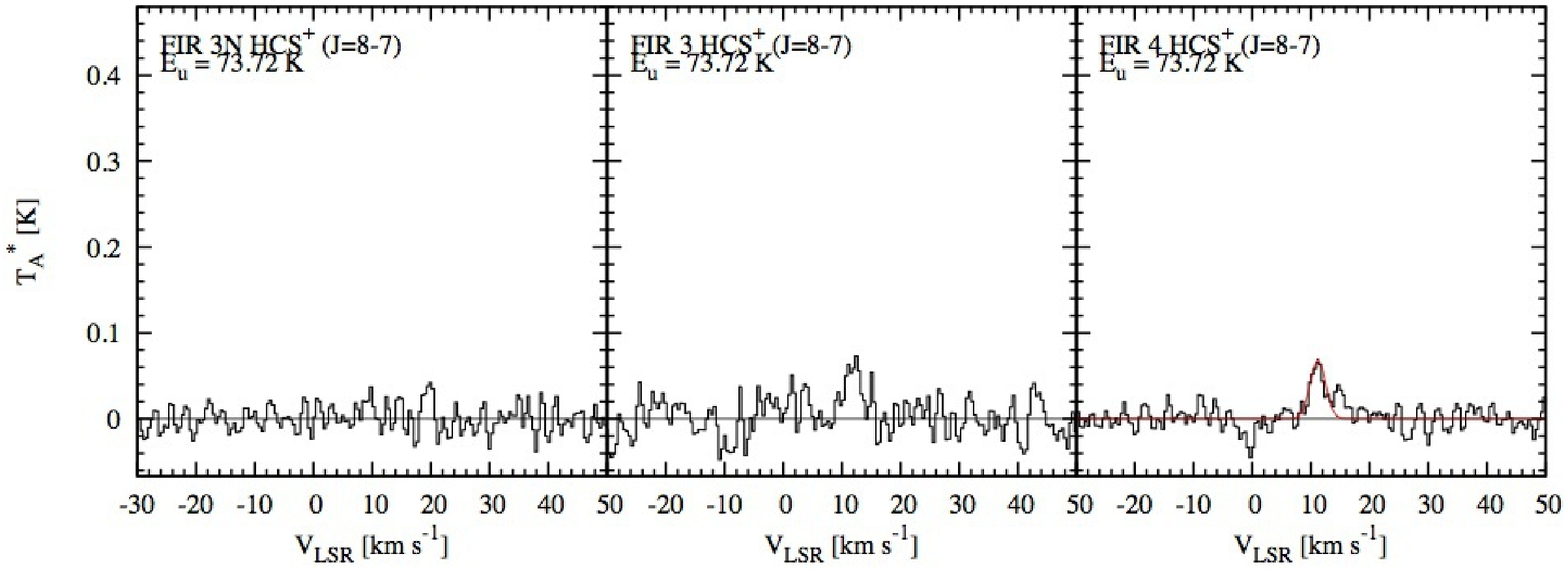}
\end{center}
\caption{HCS$^+$ spectra}
\label{hcsp}
\end{figure}

\begin{figure}
\begin{center}
\includegraphics[angle=0,scale=.5]{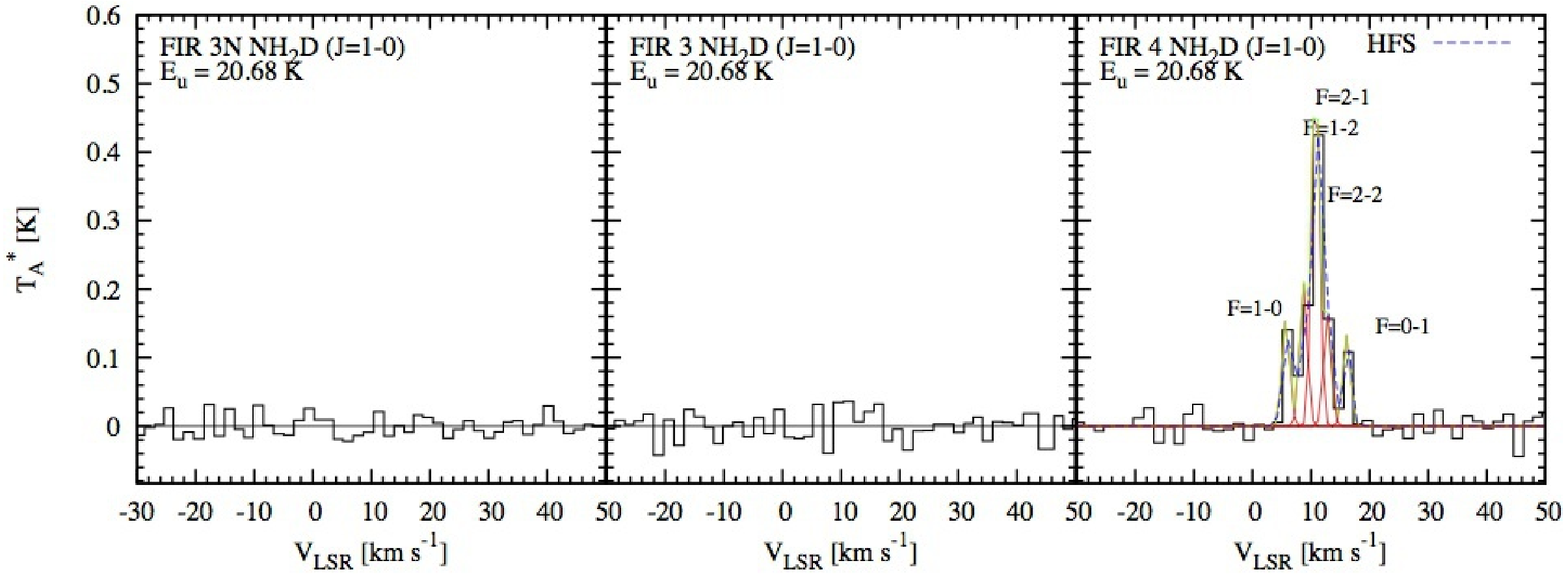}
\end{center}
\caption{NH$_2$D spectra. The HFS fitting result is shown by the blue dashed line.}
\label{nh2d}
\end{figure}

\begin{figure}
\begin{center}
\includegraphics[angle=0,scale=.51]{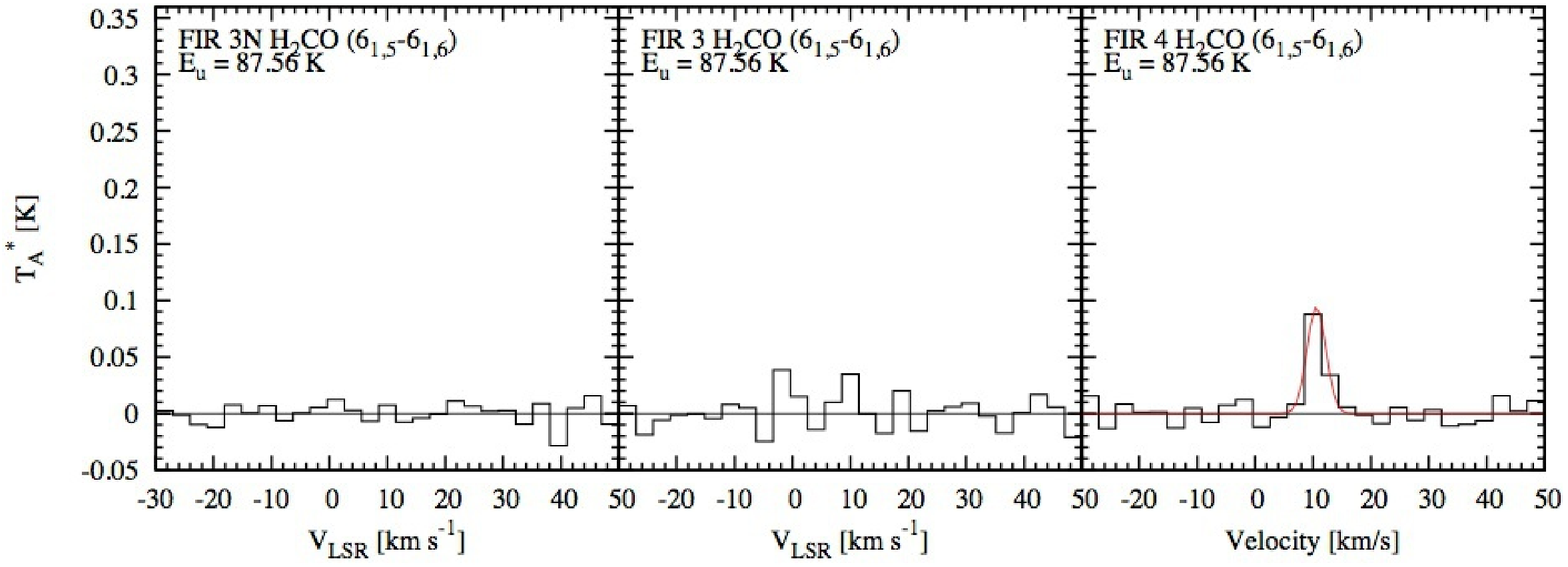}
\end{center}
\caption{H$_2$CO spectra}
\label{h2co}
\end{figure}

\begin{figure}
\begin{center}
\includegraphics[angle=0,scale=.5]{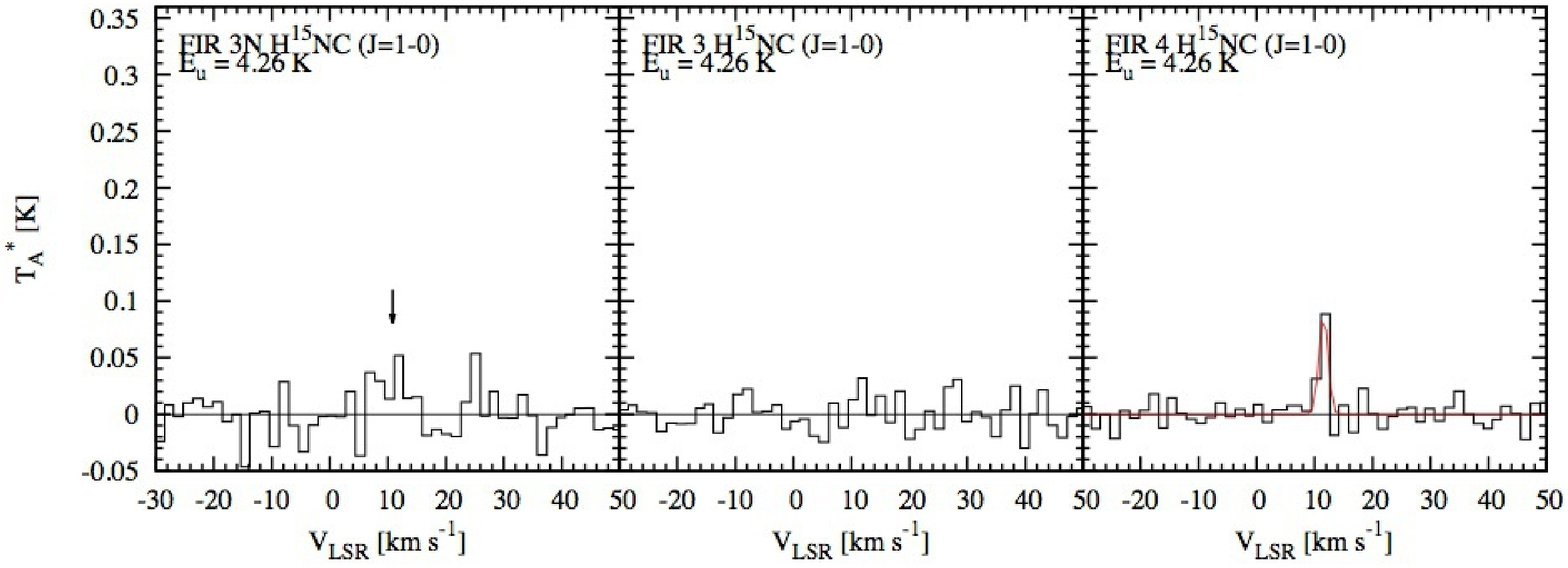}
\end{center}
\caption{H$^{15}$NC spectra. The vertical arrow in the left panel shows the detection, but the peak is not fitted well by the single Gaussian component due to insufficient velocity resolution.}
\label{h15nc}
\end{figure}

\begin{figure}
\begin{center}
\includegraphics[angle=0,scale=.5]{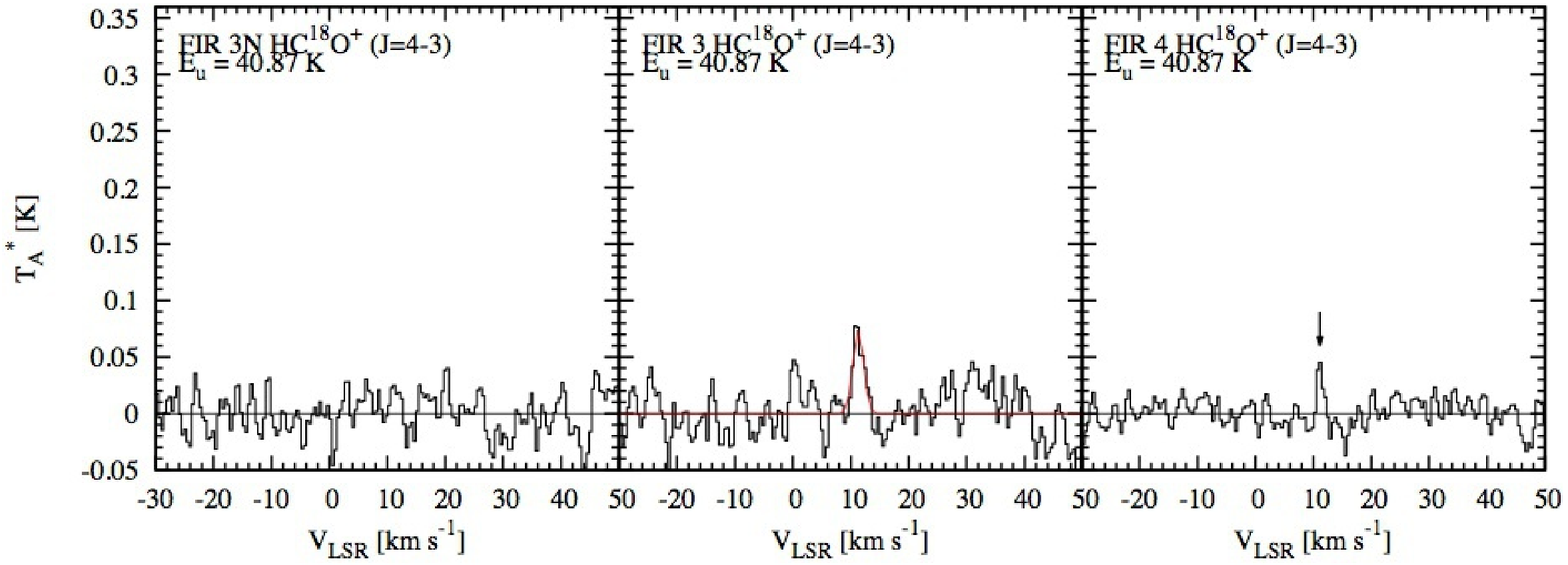}
\end{center}
\caption{HC$^{18}$O$^{+}$ spectra. The vertical arrow in the right panel shows the emission-like feature.}
\label{hc18op}
\end{figure}

\clearpage

\begin{figure}
\begin{center}
\includegraphics[angle=0,scale=.52]{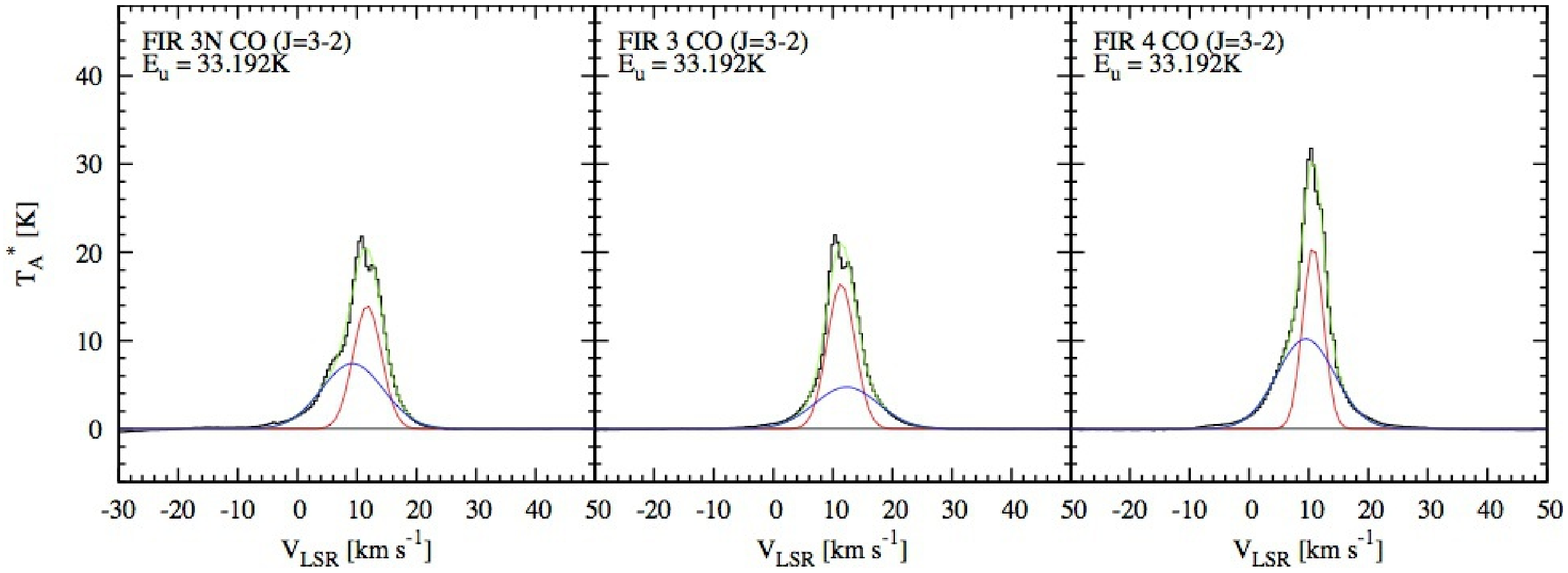}
\end{center}
\caption{CO spectra. The best-fit narrow and wide components are shown by the red and blue lines, respectively, and the green line shows the sum of the two components.}
\label{co}
\end{figure}

\begin{figure}
\begin{center}
\includegraphics[angle=0,scale=.5]{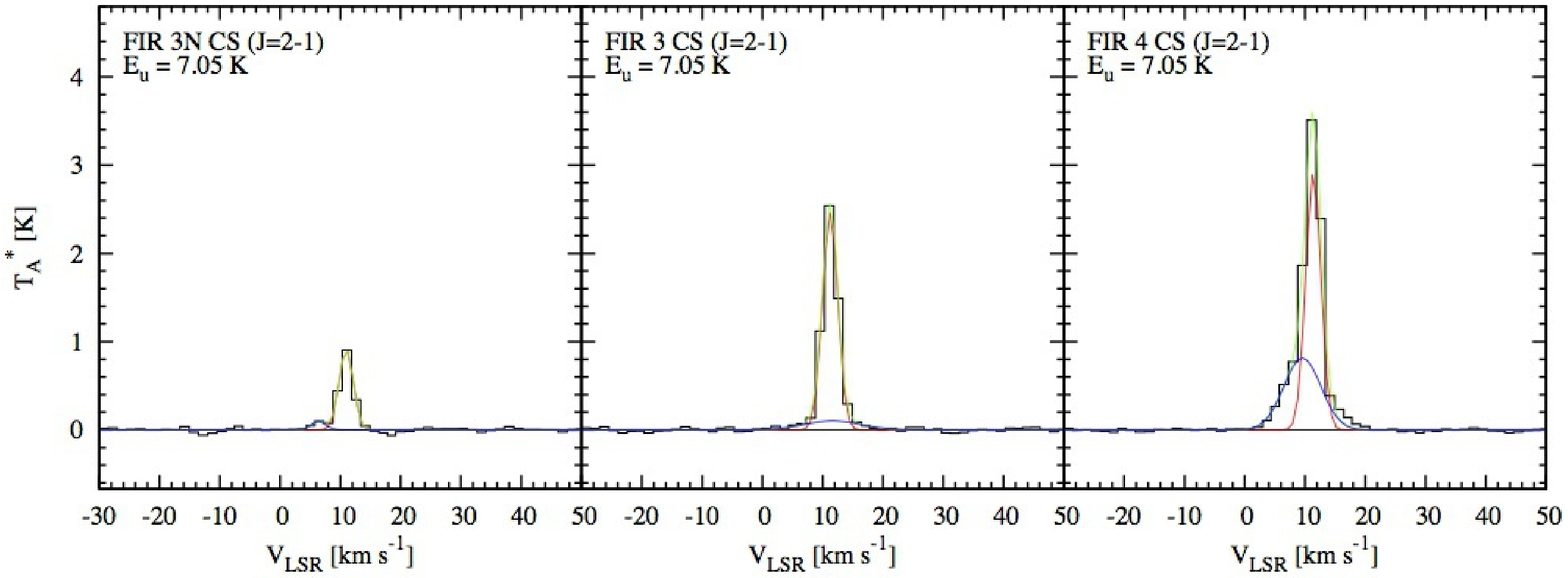}
\includegraphics[angle=0,scale=.53]{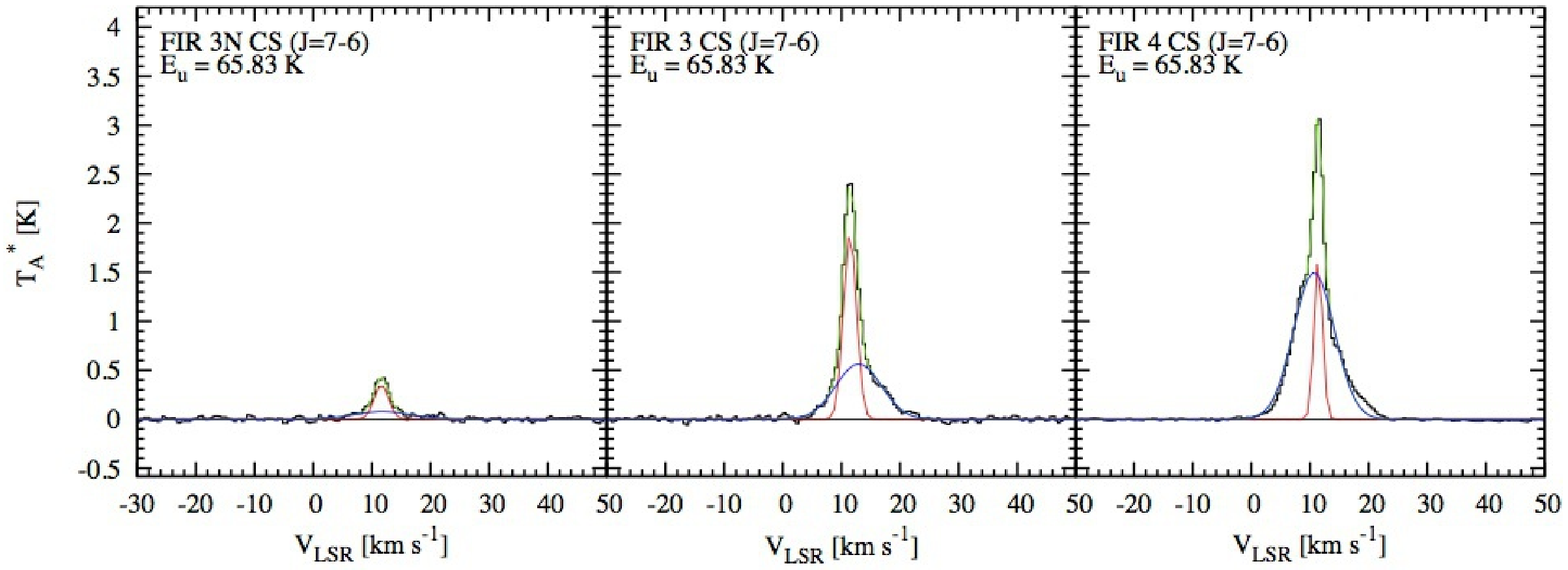}
\end{center}
\caption{CS spectra}
\label{cs}
\end{figure}

\begin{figure}
\begin{center}
\includegraphics[angle=0,scale=.5]{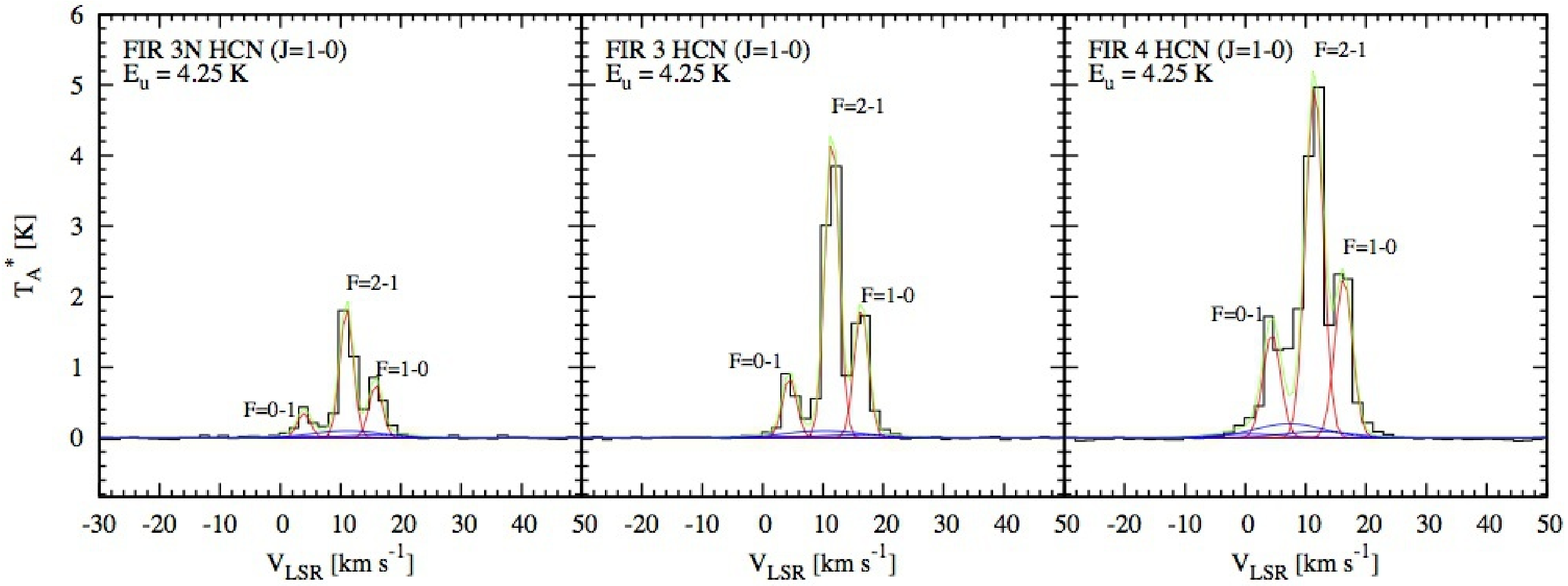}
\includegraphics[angle=0,scale=.5]{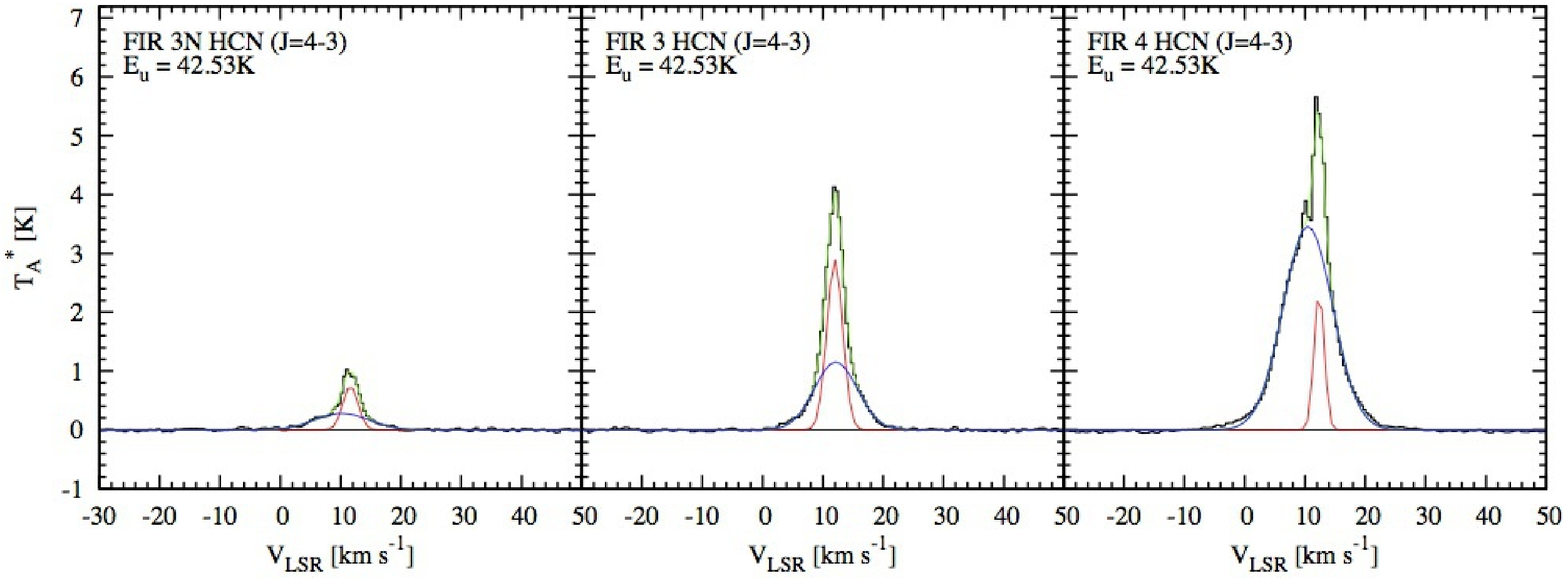}
\end{center}
\caption{HCN spectra}
\label{hcn}
\end{figure}

\begin{figure}
\begin{center}
\includegraphics[angle=0,scale=.5]{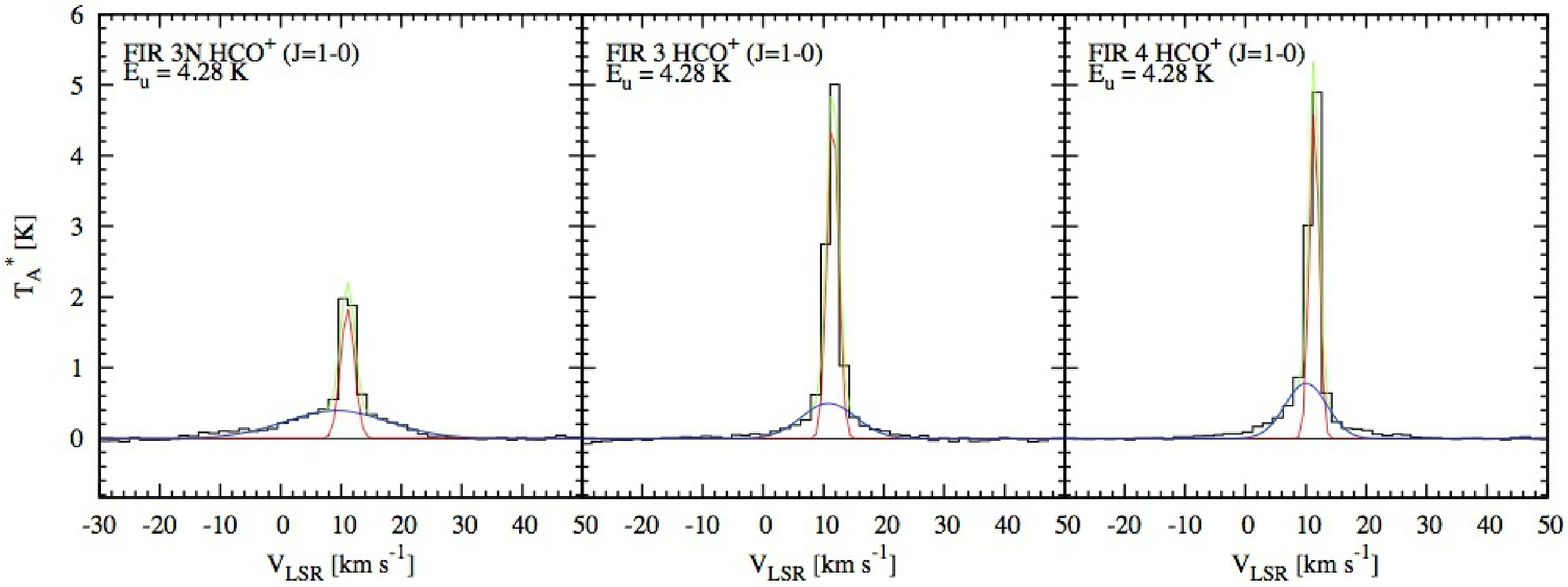}
\end{center}
\caption{HCO$^{+}$ spectra}
\label{hcop}
\end{figure}

\clearpage

\begin{figure}
\begin{center}
\includegraphics[angle=0,scale=.5]{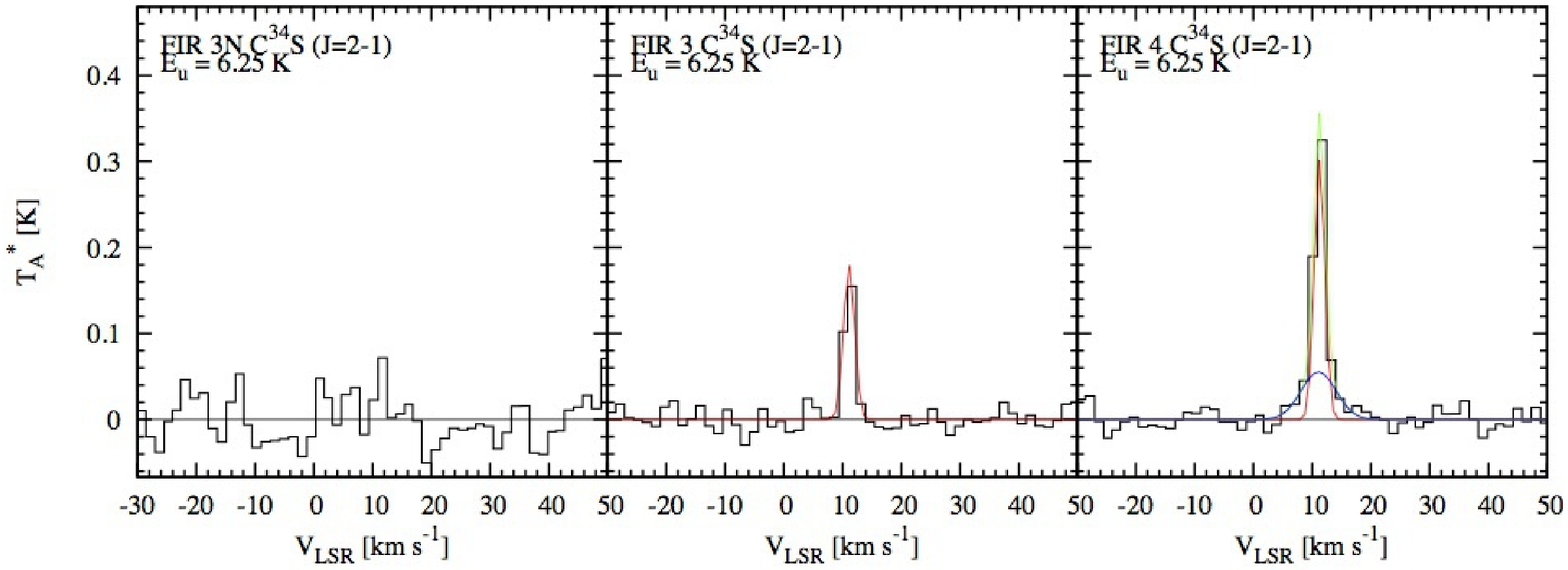}
\includegraphics[angle=0,scale=.5]{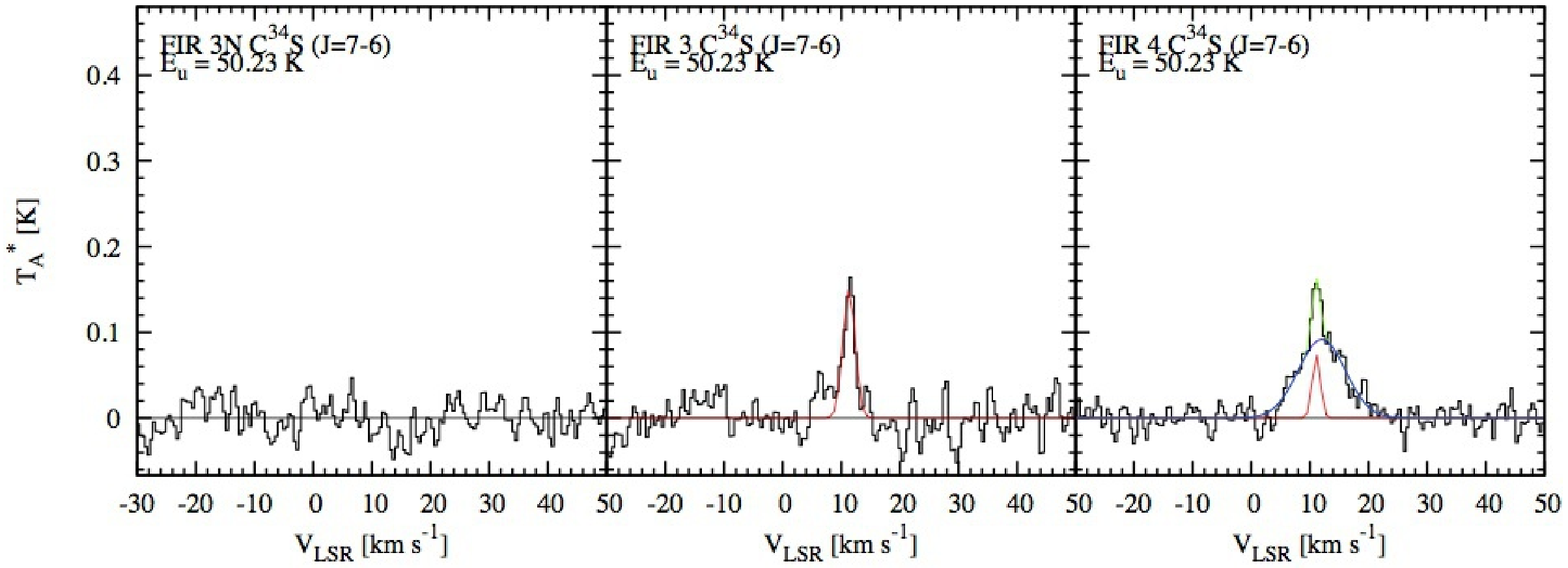}
\end{center}
\caption{C$^{34}$S spectra}
\label{c34s}
\end{figure}

\begin{figure}
\begin{center}
\includegraphics[angle=0,scale=.5]{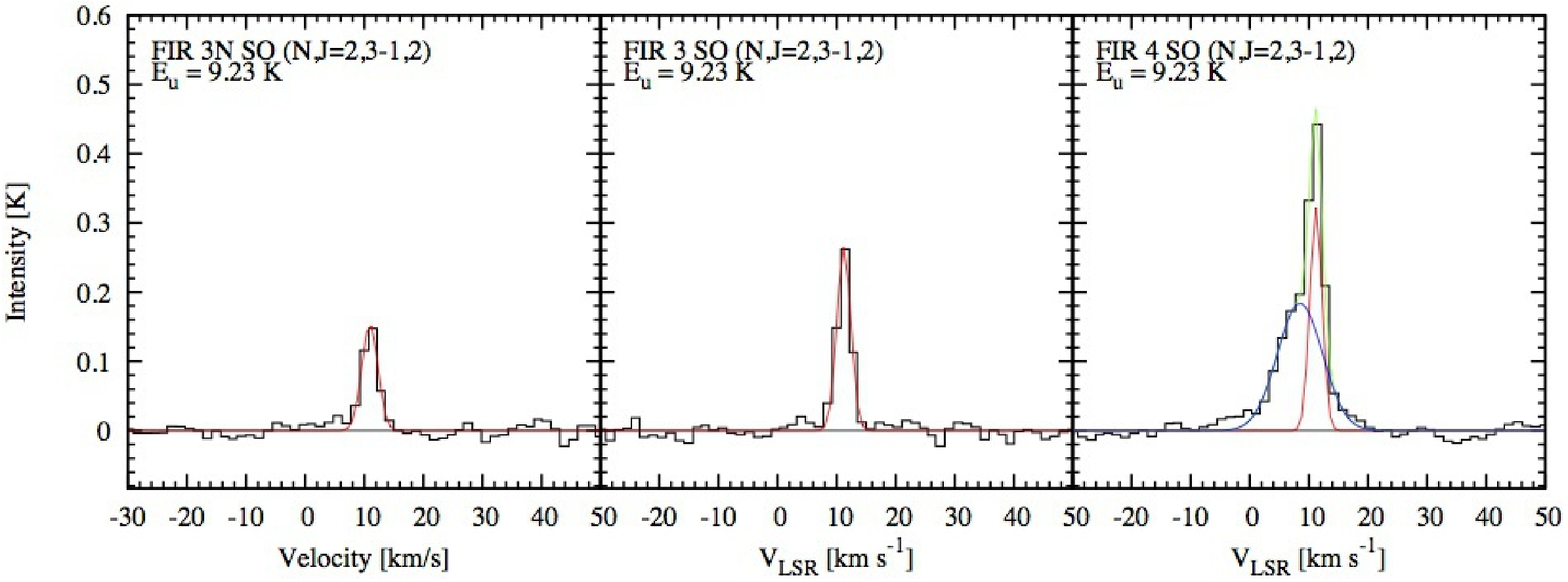}
\includegraphics[angle=0,scale=.5]{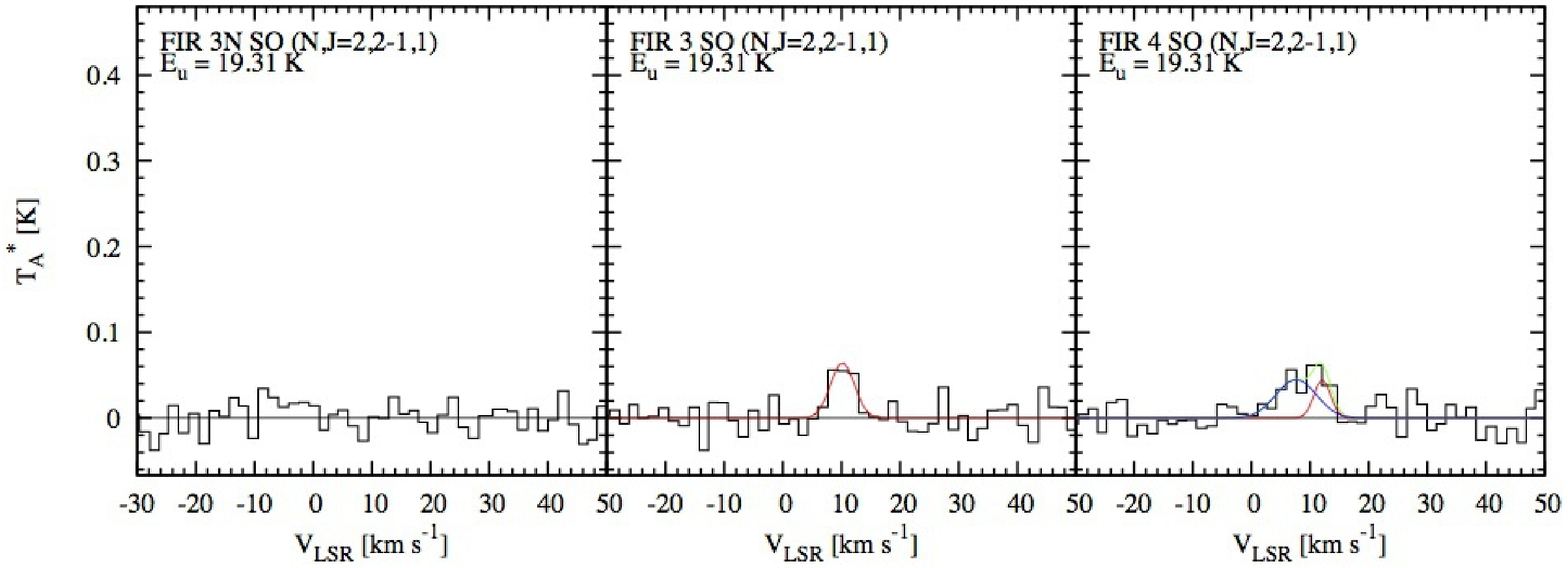}
\includegraphics[angle=0,scale=.5]{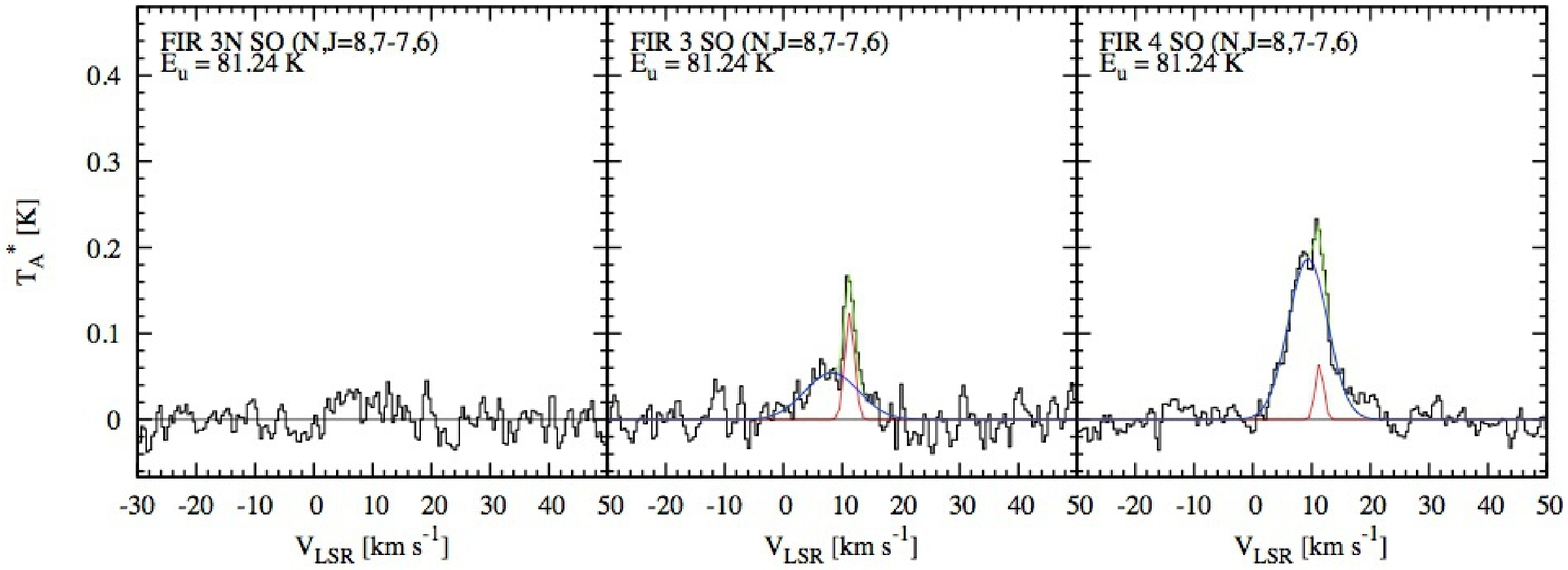}
\includegraphics[angle=0,scale=.51]{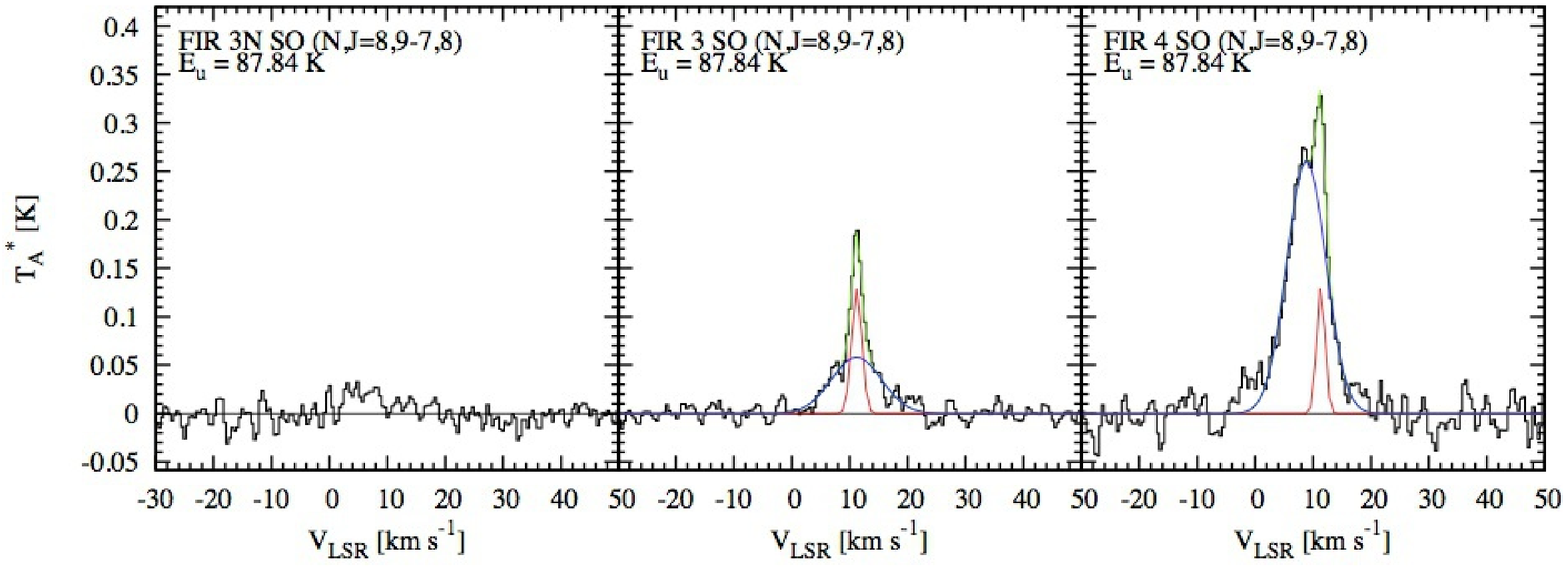}
\end{center}
\caption{SO spectra}
\label{so}
\end{figure}

\begin{figure}
\begin{center}
\includegraphics[angle=0,scale=.5]{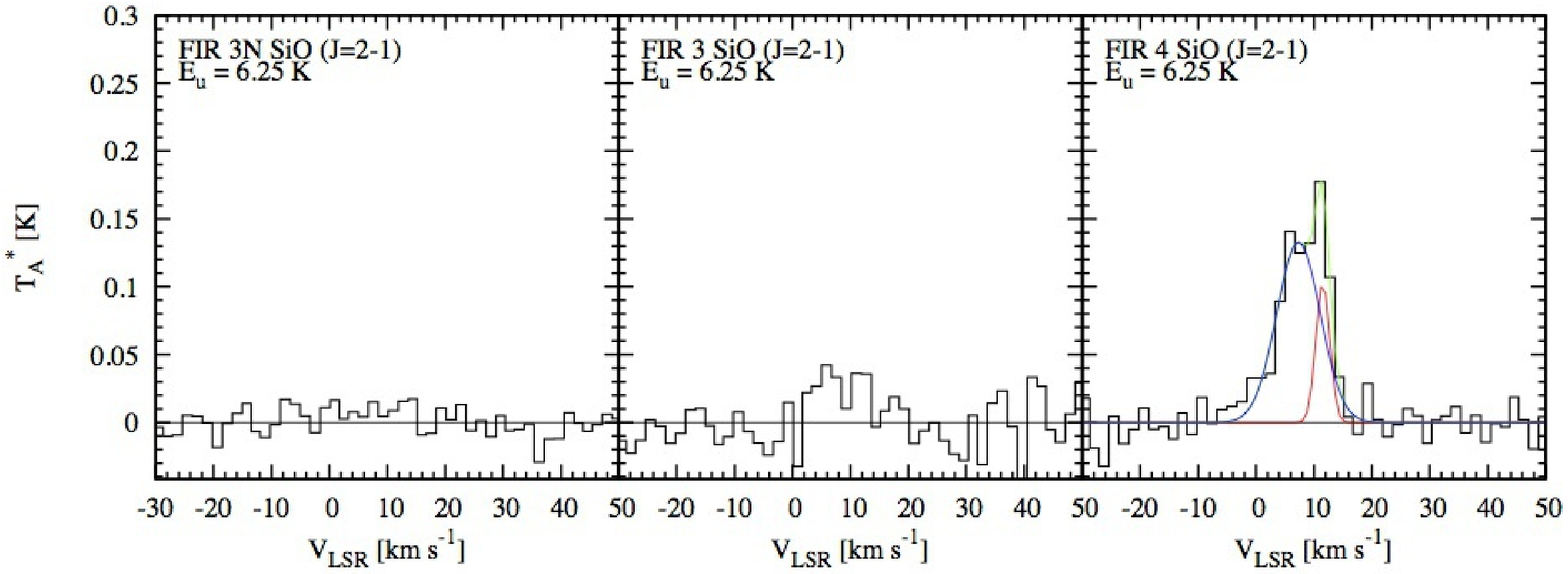}
\includegraphics[angle=0,scale=.5]{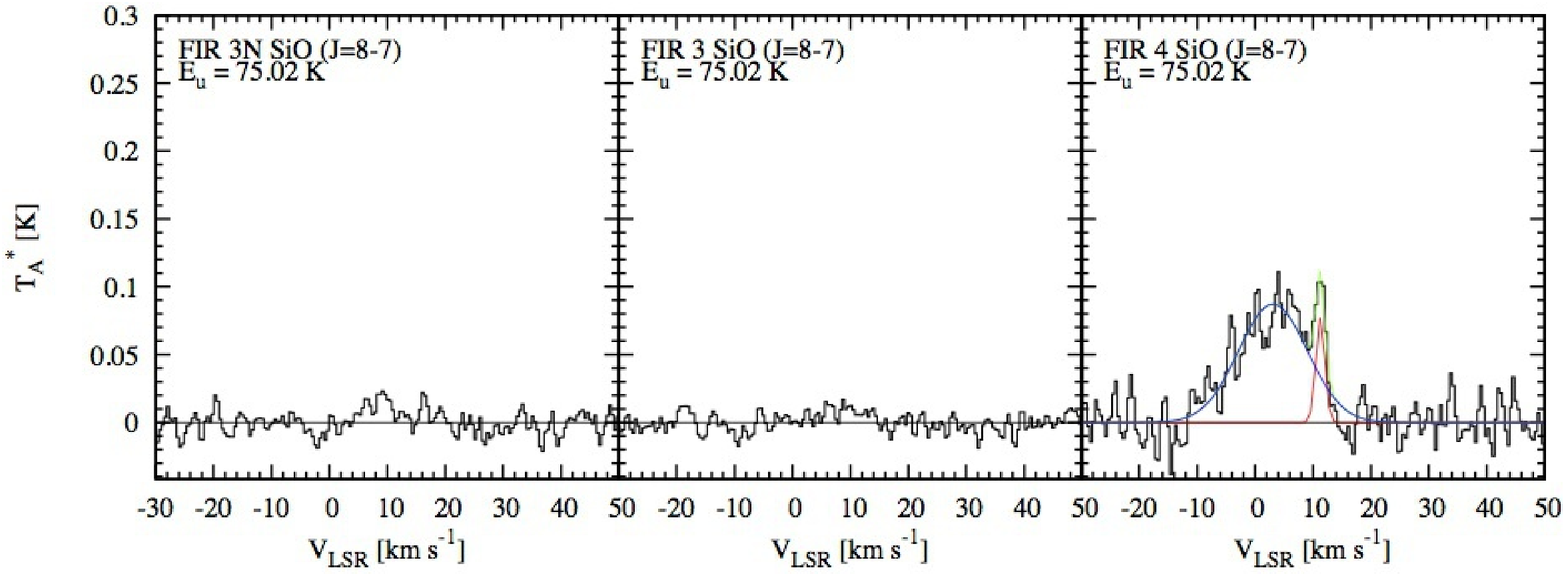}
\end{center}
\caption{SiO spectra}
\label{sio}
\end{figure}

\begin{figure}
\begin{center}
\includegraphics[angle=0,scale=.5]{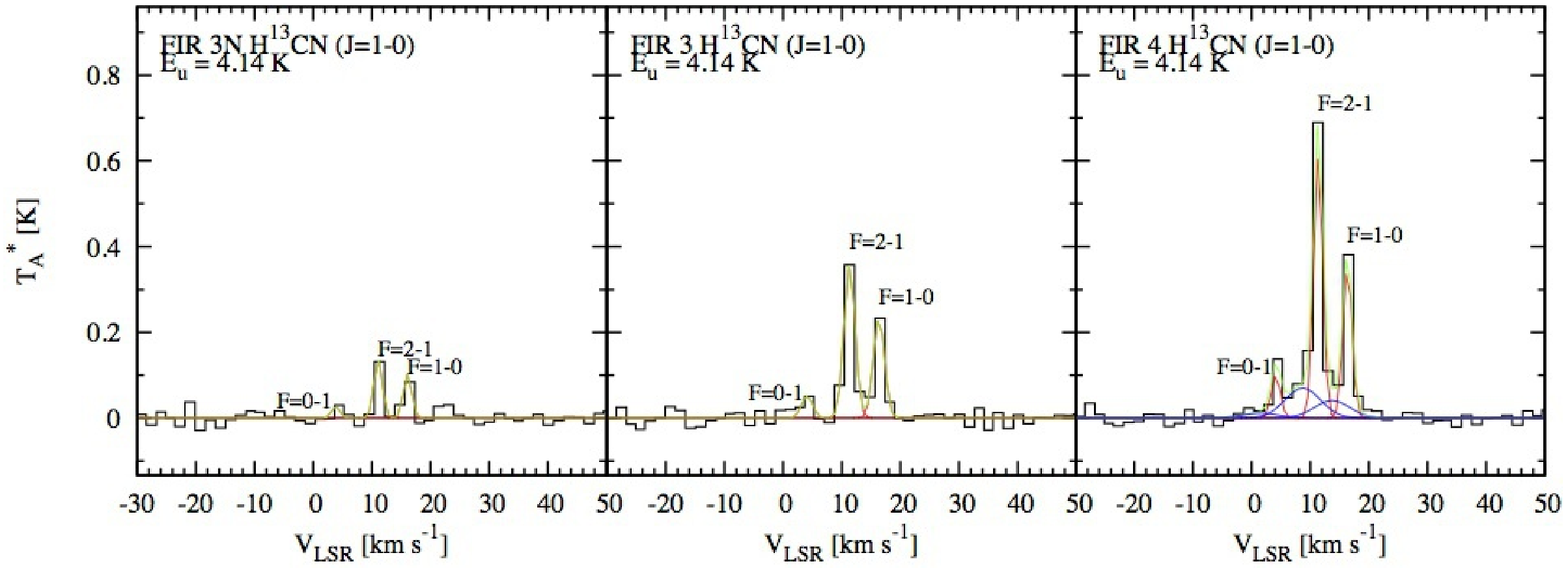}
\includegraphics[angle=0,scale=.5]{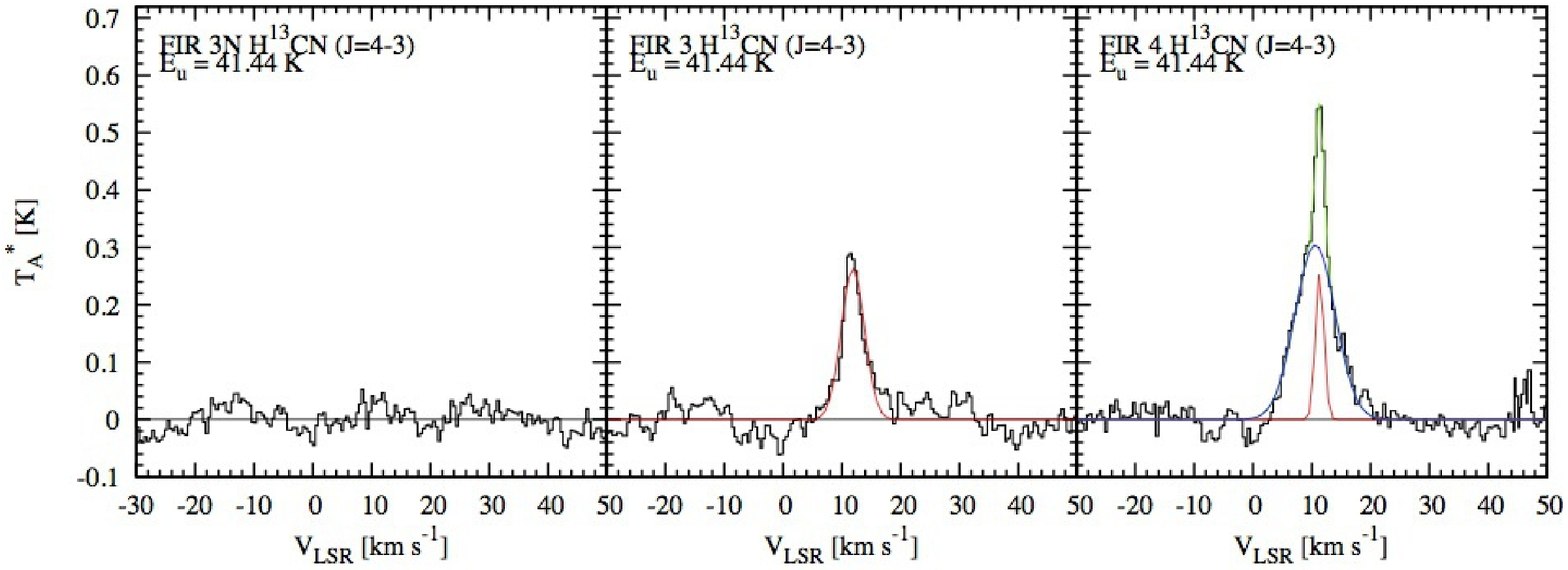}
\end{center}
\caption{H$^{13}$CN spectra}
\label{h13cn}
\end{figure}

\begin{figure}
\begin{center}
\includegraphics[angle=0,scale=.5]{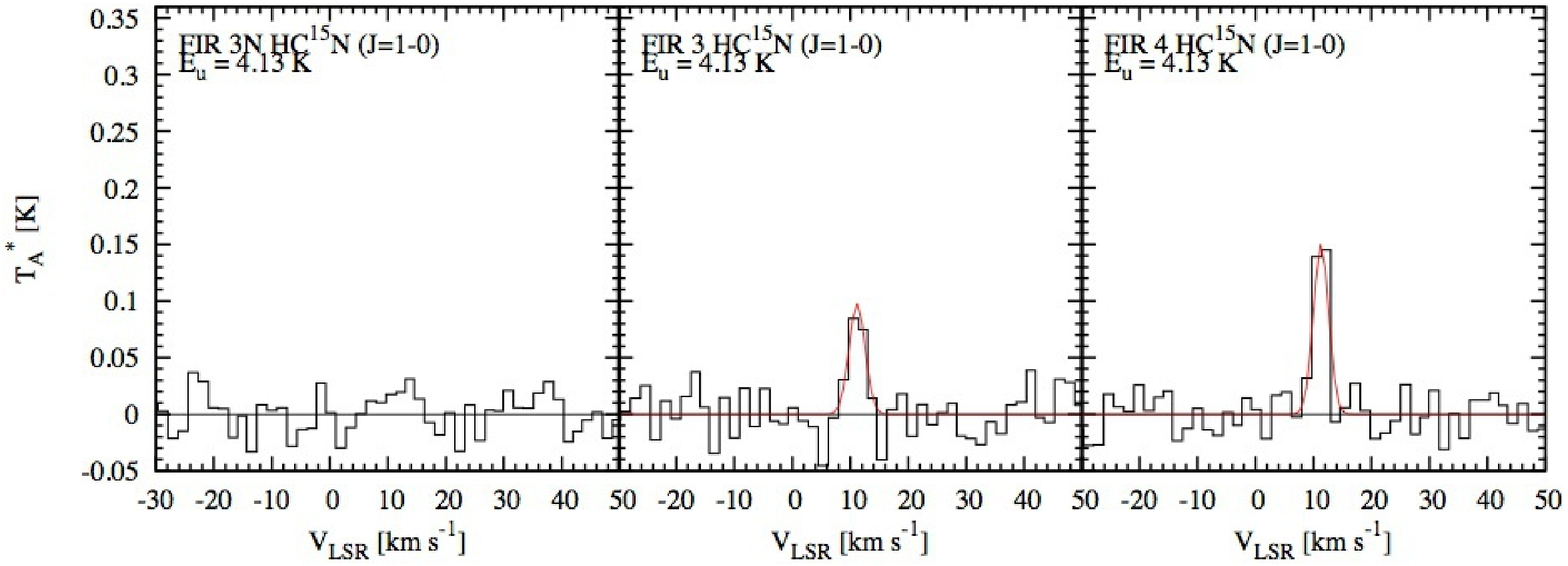}
\includegraphics[angle=0,scale=.5]{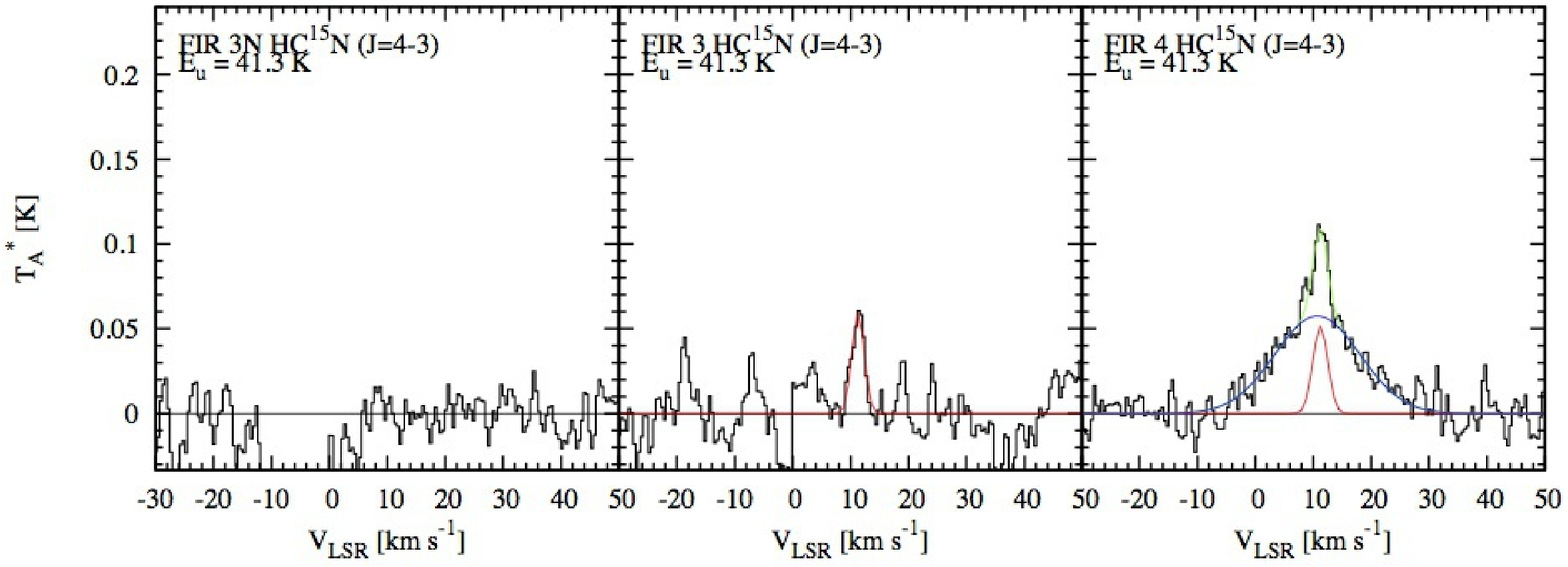}
\end{center}
\caption{HC$^{15}$N spectra}
\label{hc15n}
\end{figure}

\begin{figure}
\begin{center}
\includegraphics[angle=0,scale=.5]{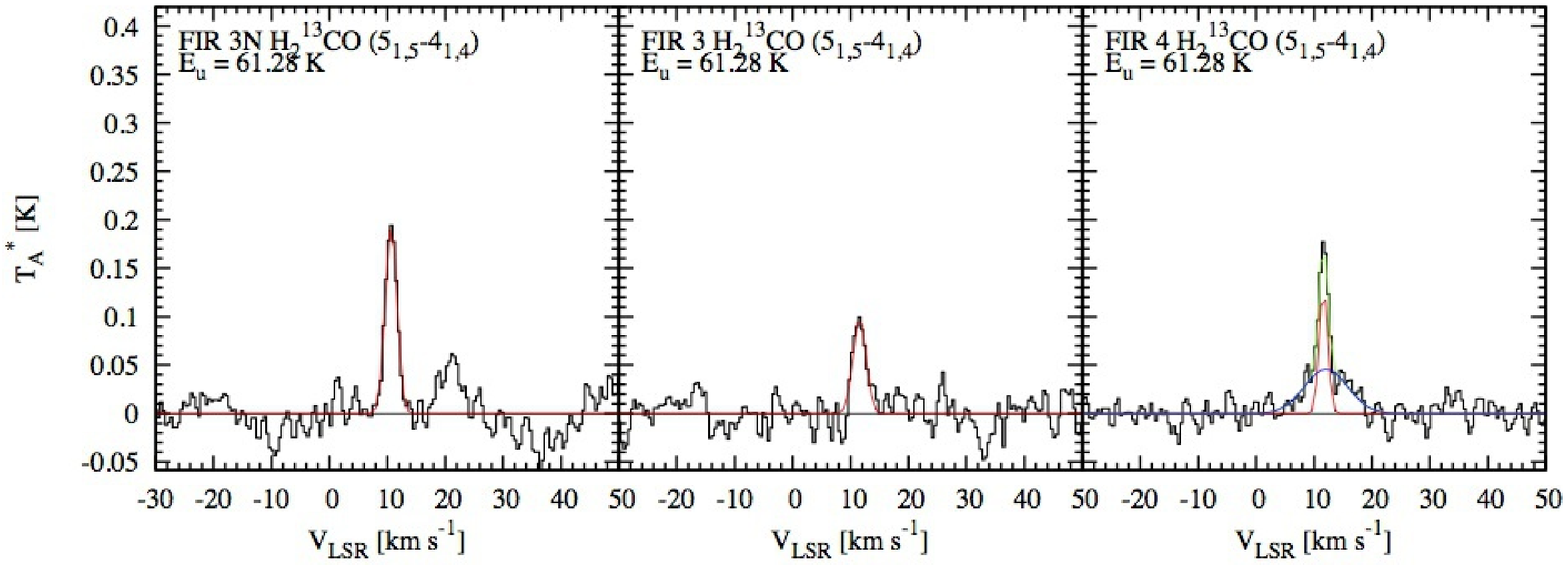}
\end{center}
\caption{H$_2^{13}$CO spectra}
\label{h213co}
\end{figure}

\begin{figure}
\begin{center}
\includegraphics[angle=0,scale=.5]{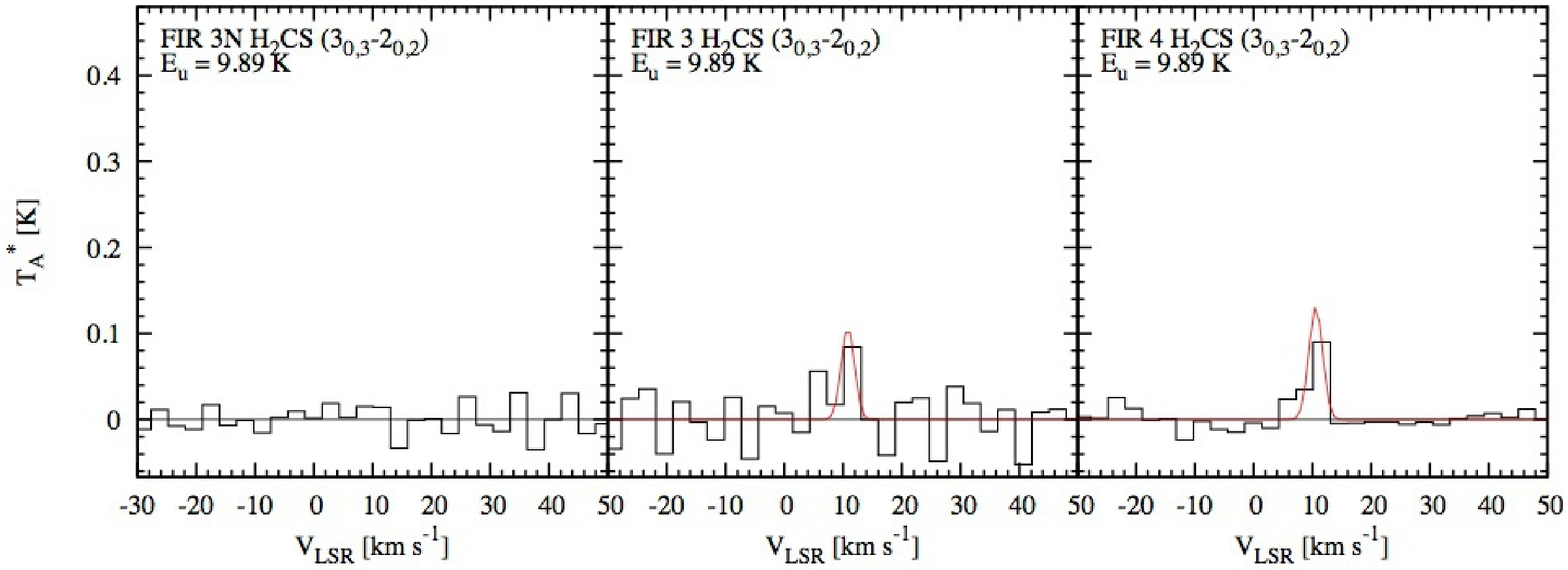}
\includegraphics[angle=0,scale=.5]{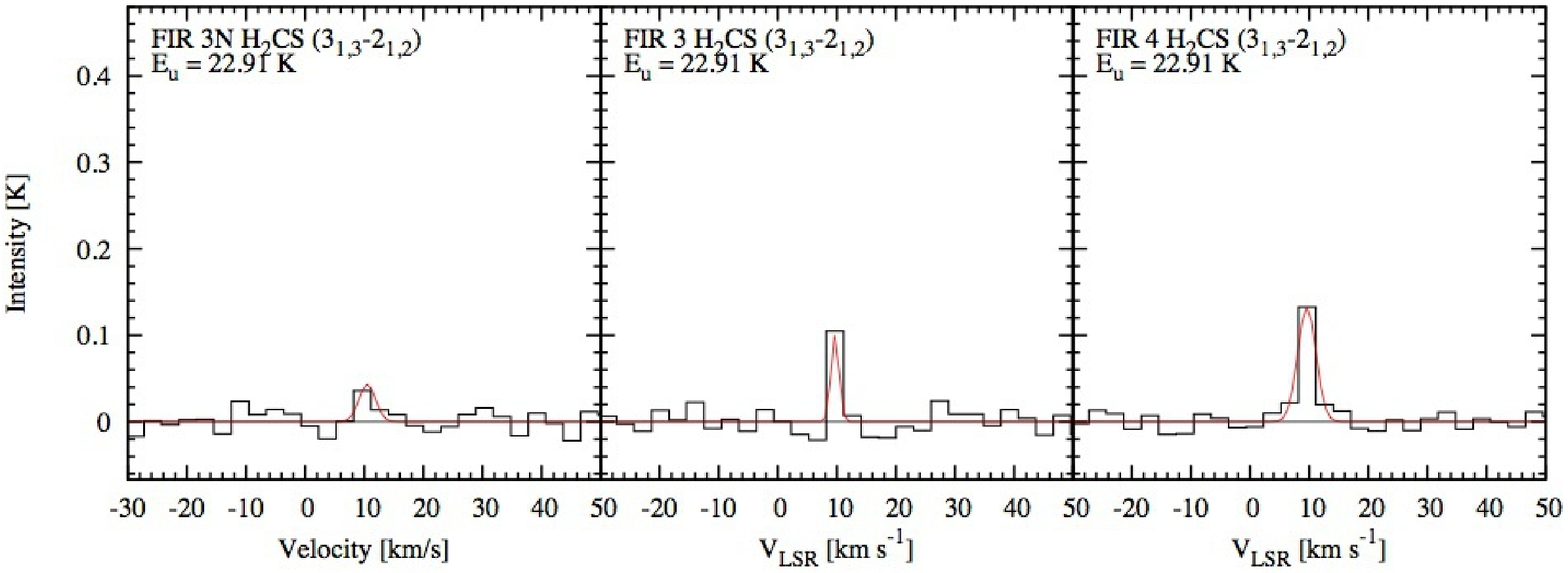}
\includegraphics[angle=0,scale=.51]{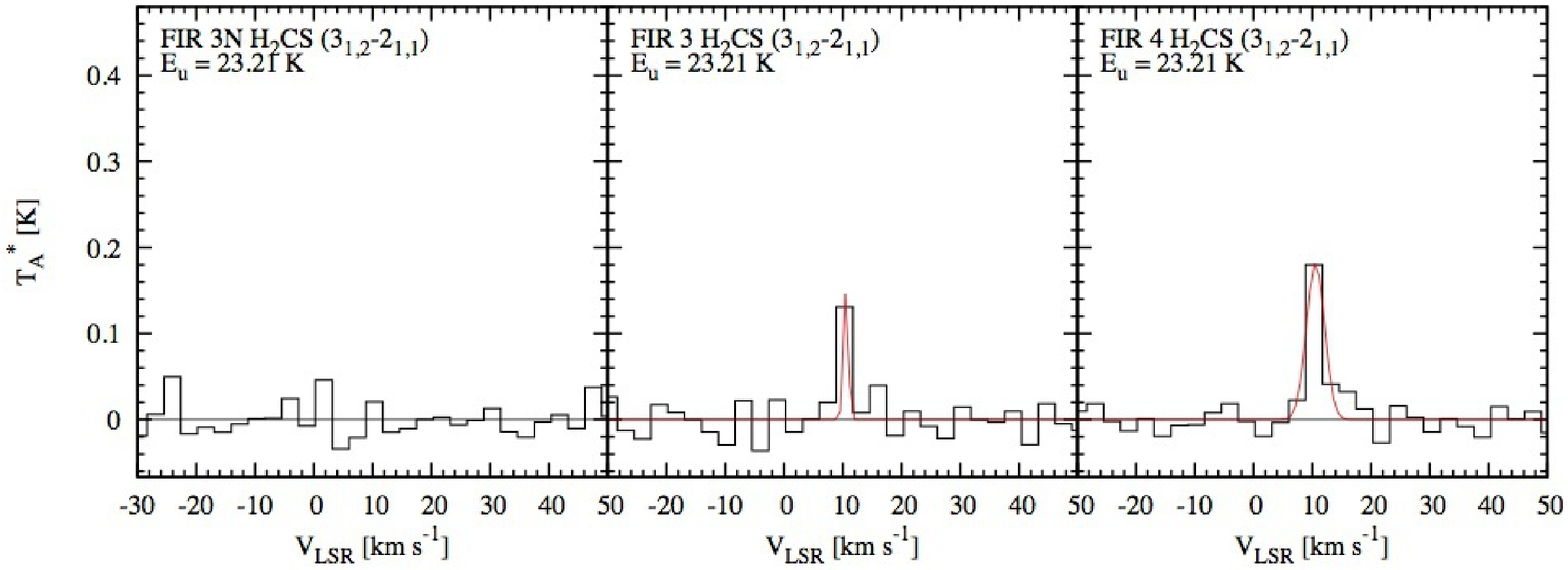}
\end{center}
\caption{H$_2$CS spectra}
\label{h2cs}
\end{figure}

\addtocounter{figure}{-1}
\begin{figure}
\begin{center}
\includegraphics[angle=0,scale=.5]{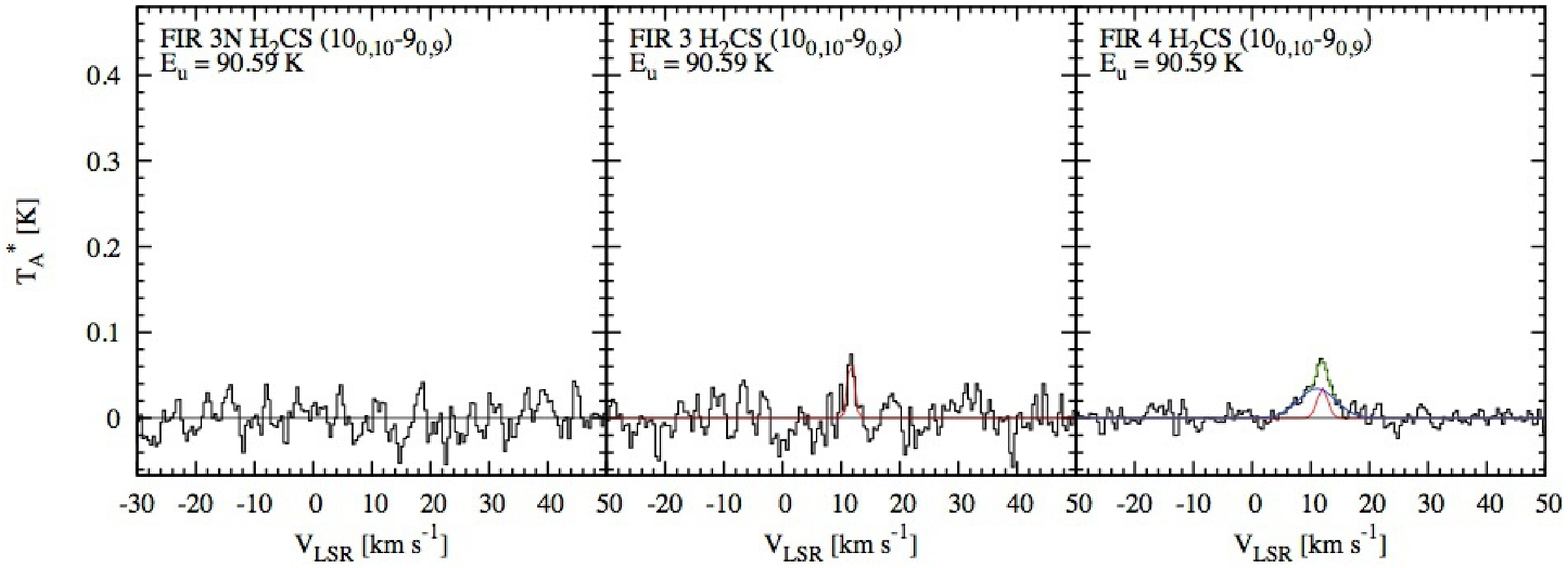}
\includegraphics[angle=0,scale=.5]{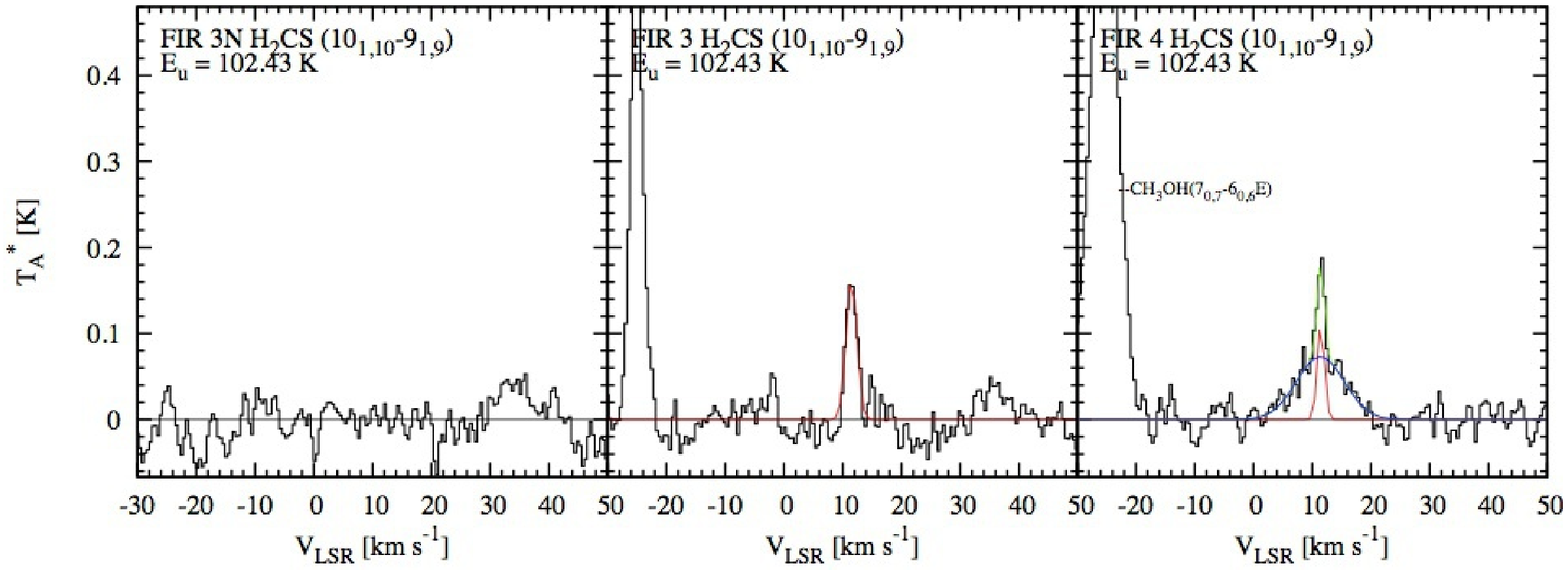}
\includegraphics[angle=0,scale=.5]{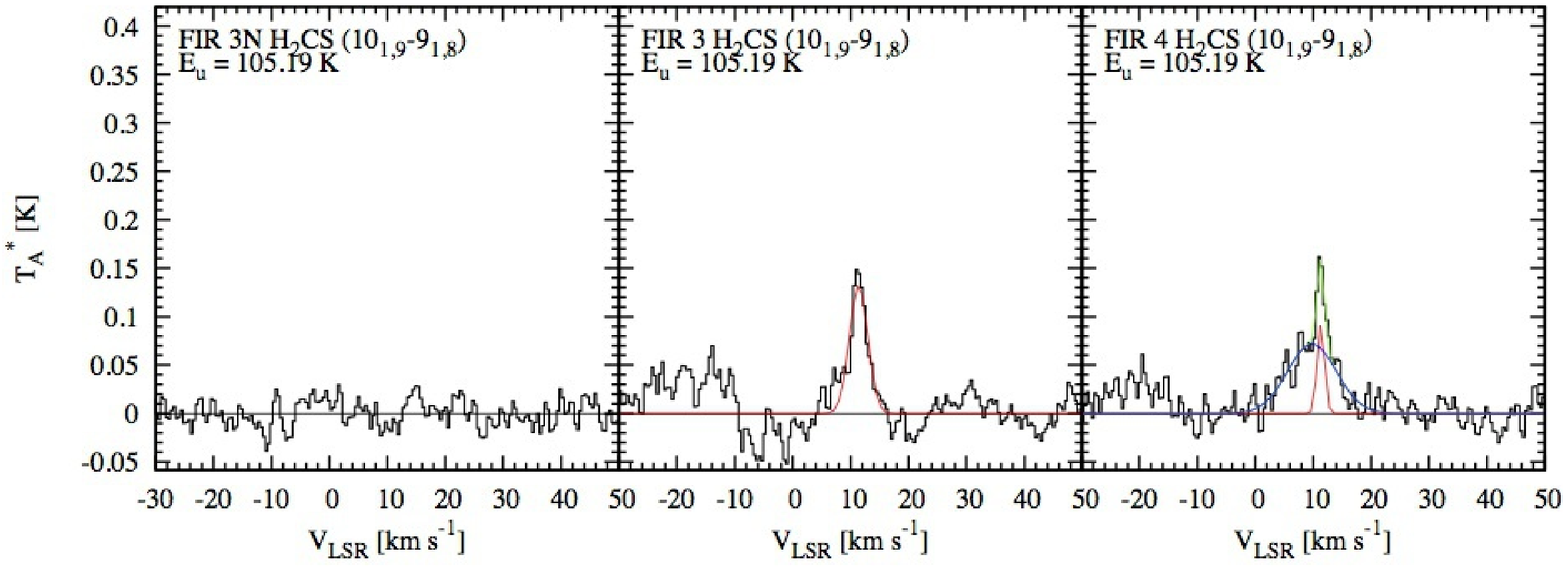}
\end{center}
\caption{\it Continued.}
\end{figure}

\begin{figure}
\begin{center}
\includegraphics[angle=0,scale=.5]{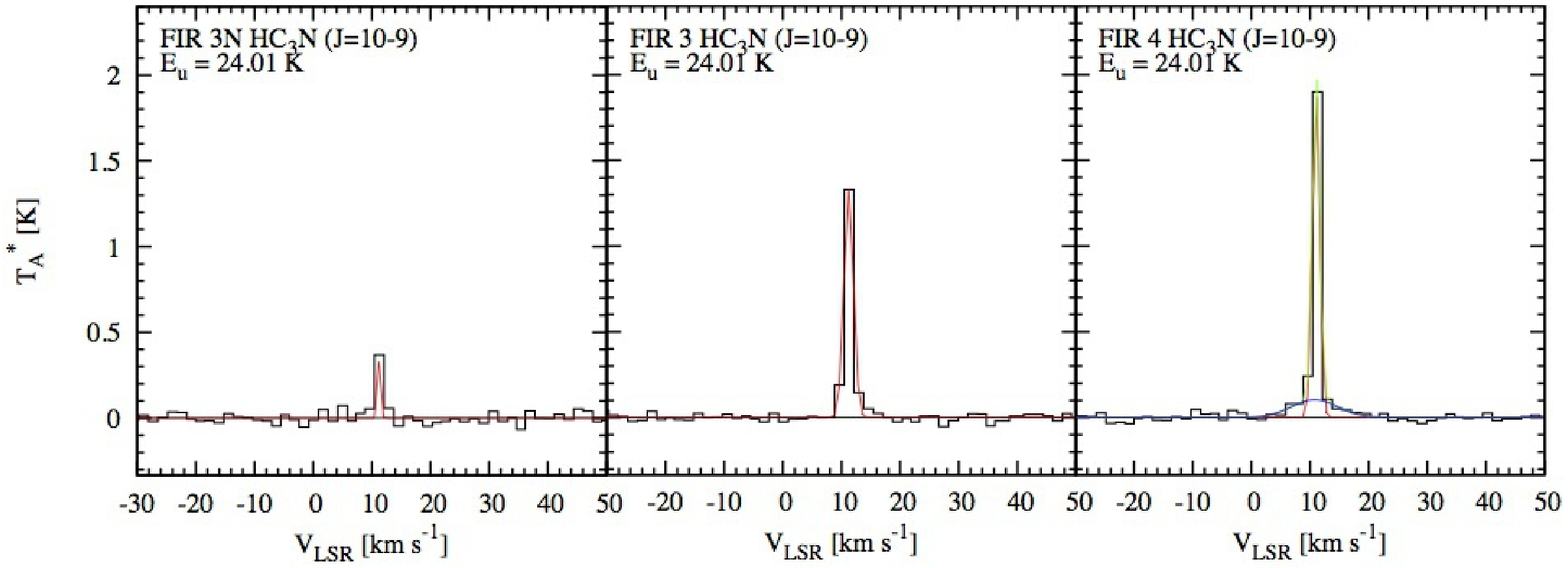}
\includegraphics[angle=0,scale=.5]{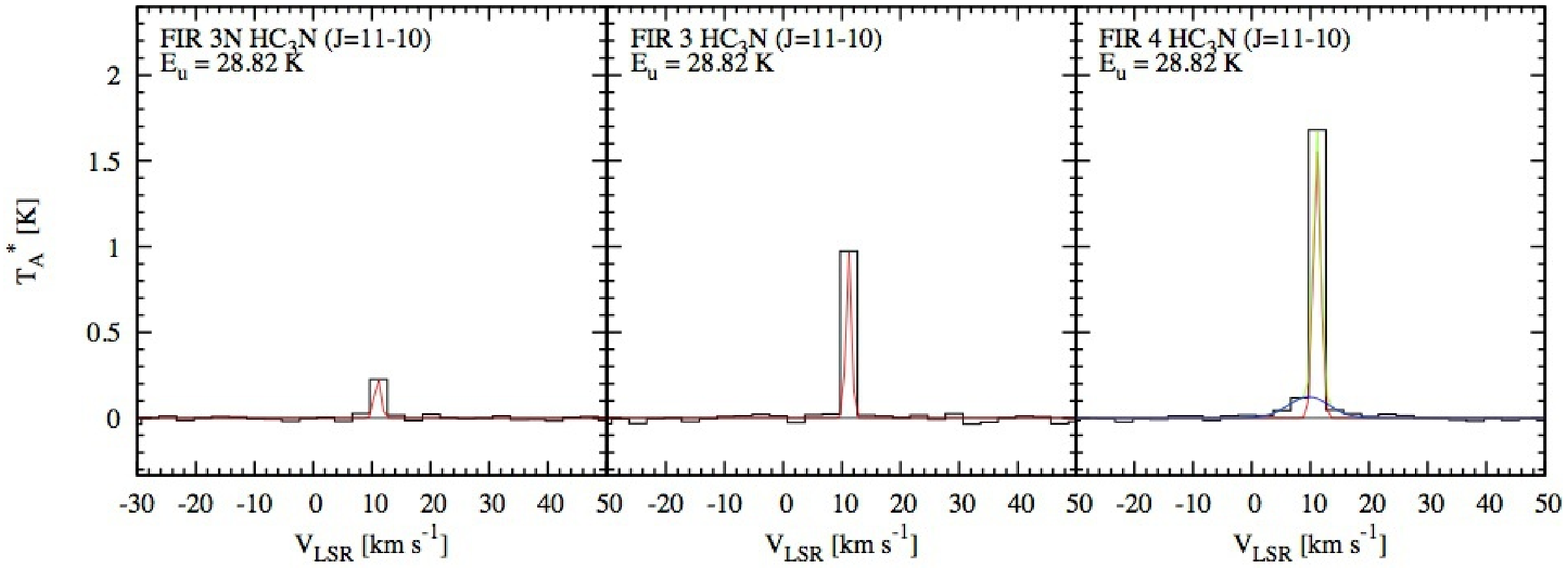}
\end{center}
\caption{HC$_3$N spectra}
\label{hc3n}
\end{figure}

\begin{figure}
\begin{center}
\includegraphics[angle=0,scale=.5]{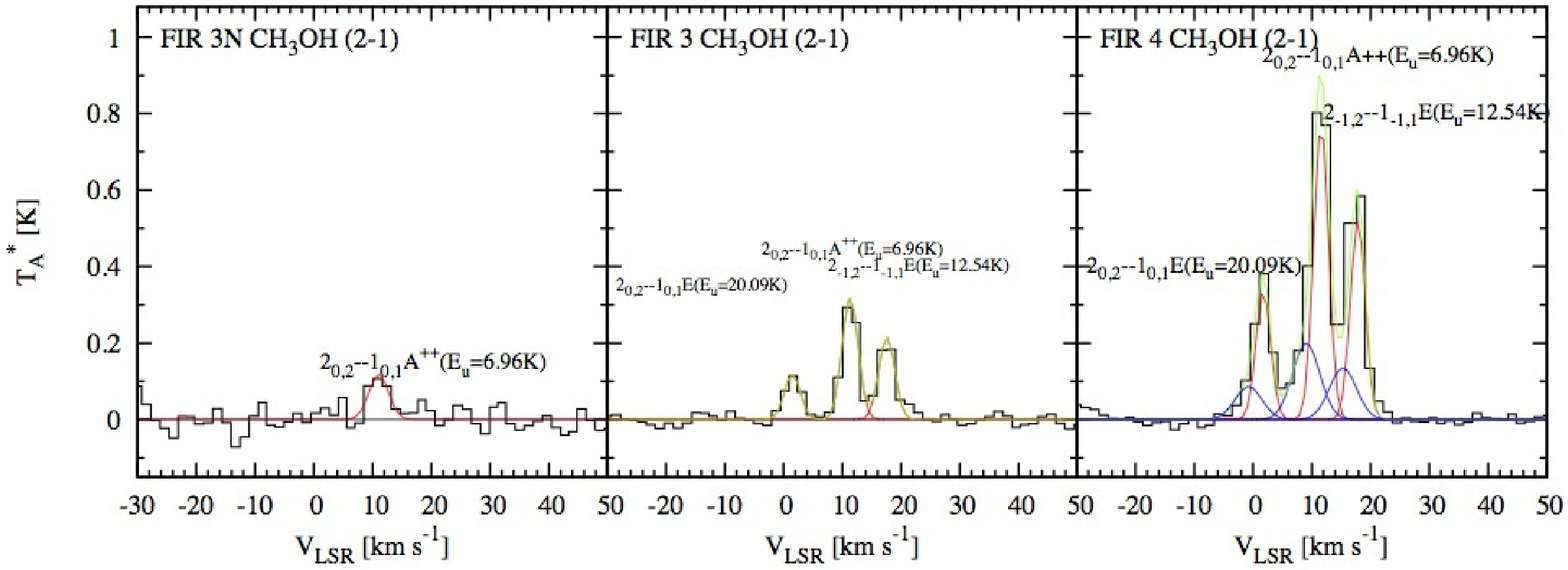}
\includegraphics[angle=0,scale=.5]{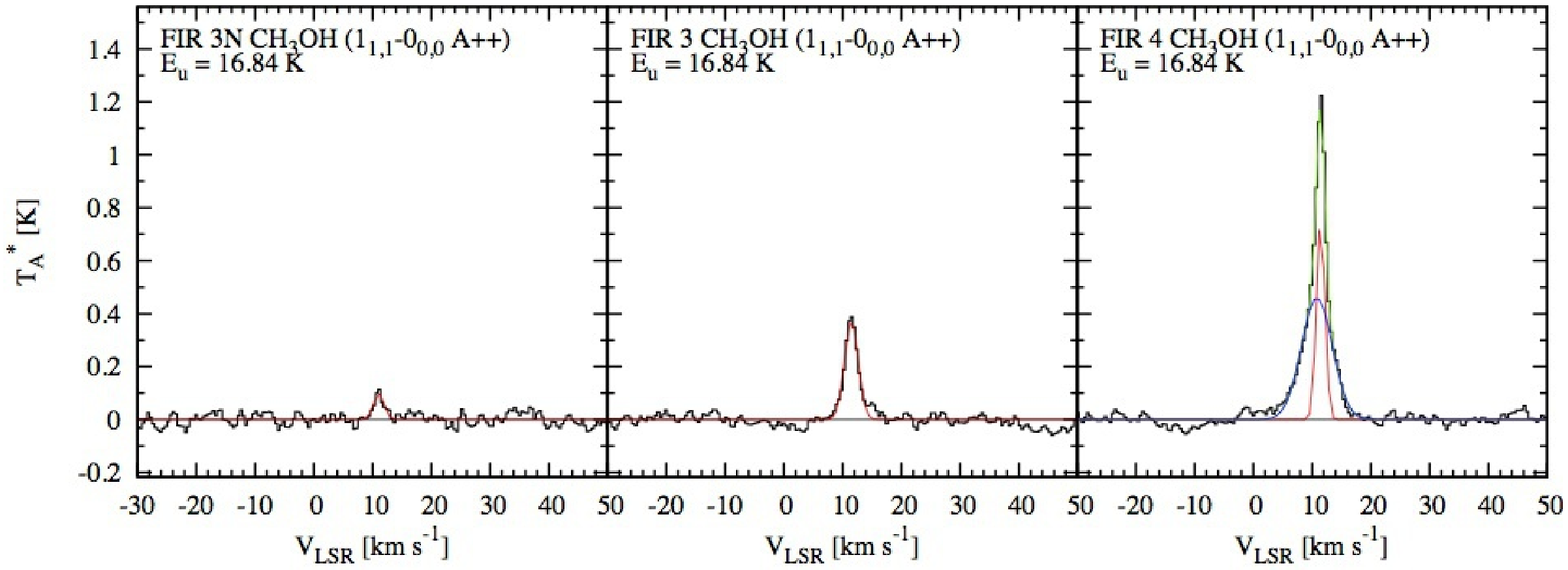}
\includegraphics[angle=0,scale=.51]{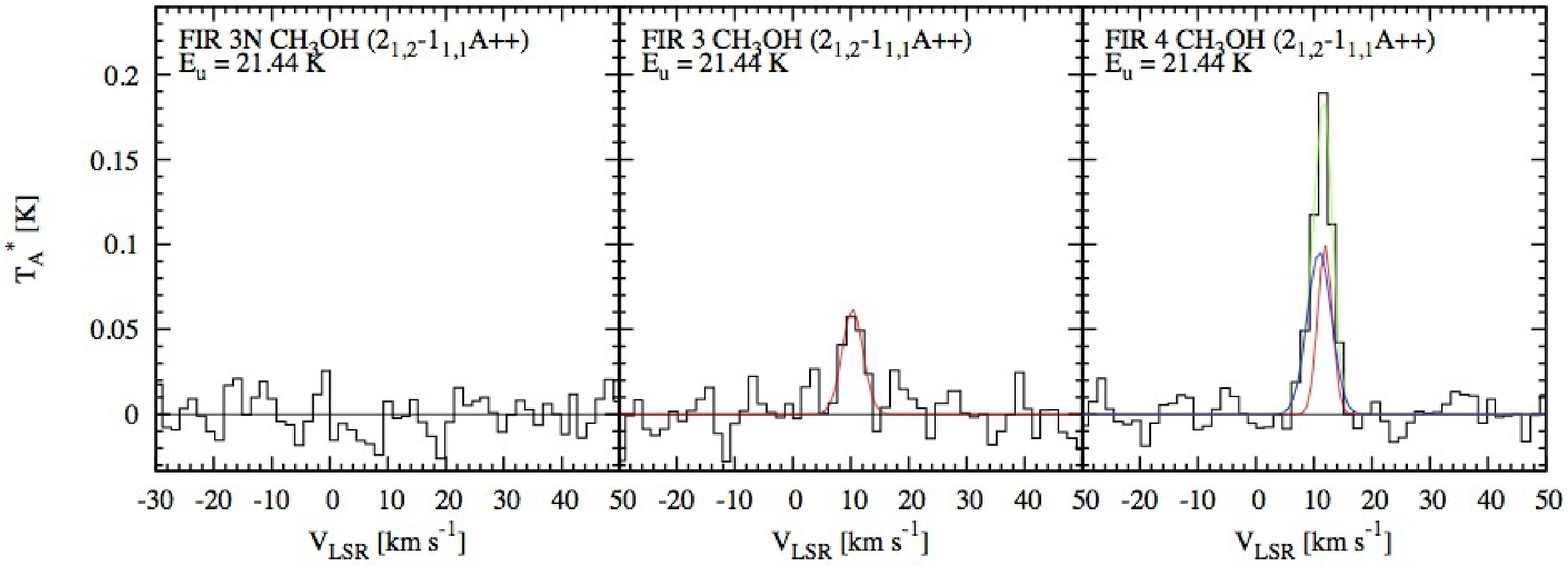}
\includegraphics[angle=0,scale=.5]{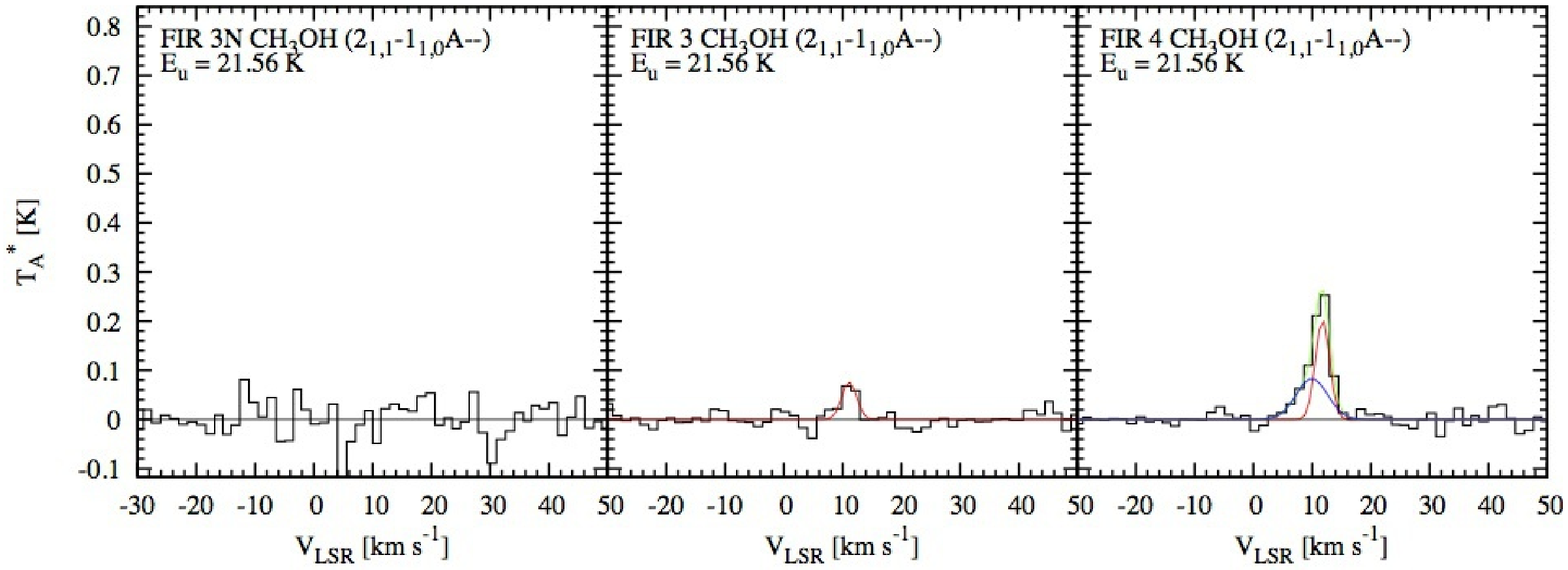}
\end{center}
\caption{CH$_3$OH spectra}
\label{ch3oh}
\end{figure}

\addtocounter{figure}{-1}
\begin{figure}
\begin{center}
\includegraphics[angle=0,scale=.5]{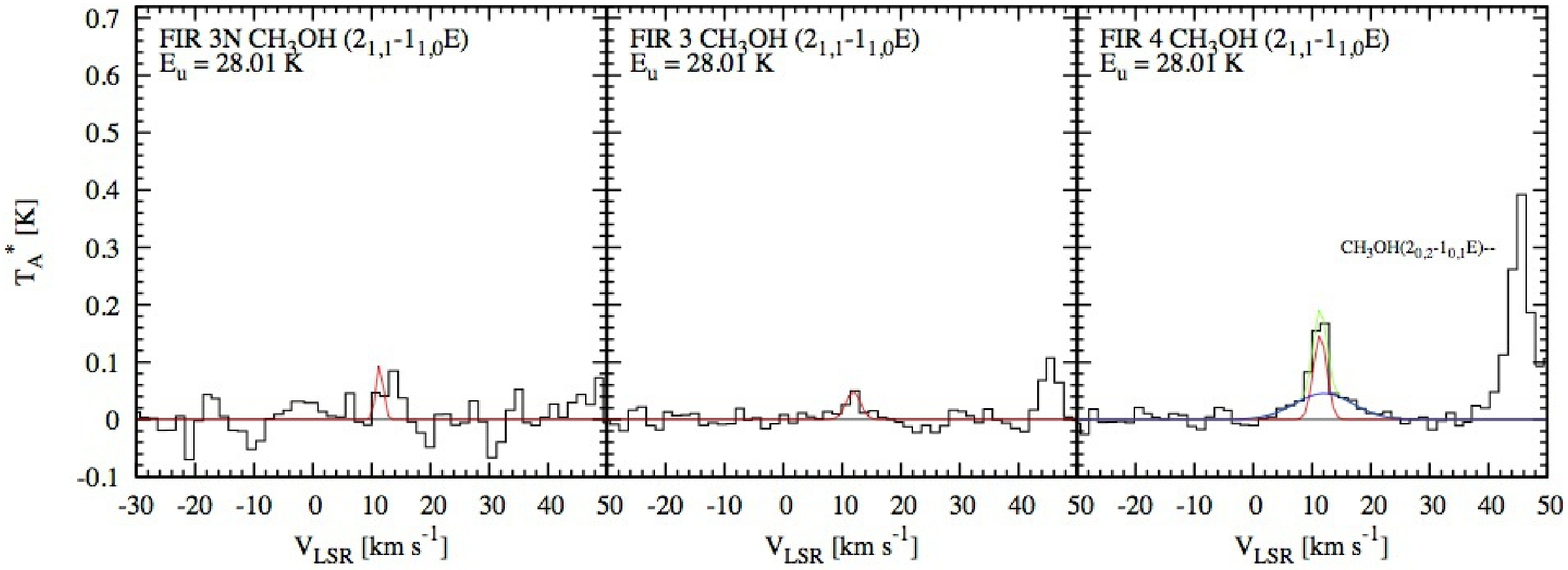}
\includegraphics[angle=0,scale=.5]{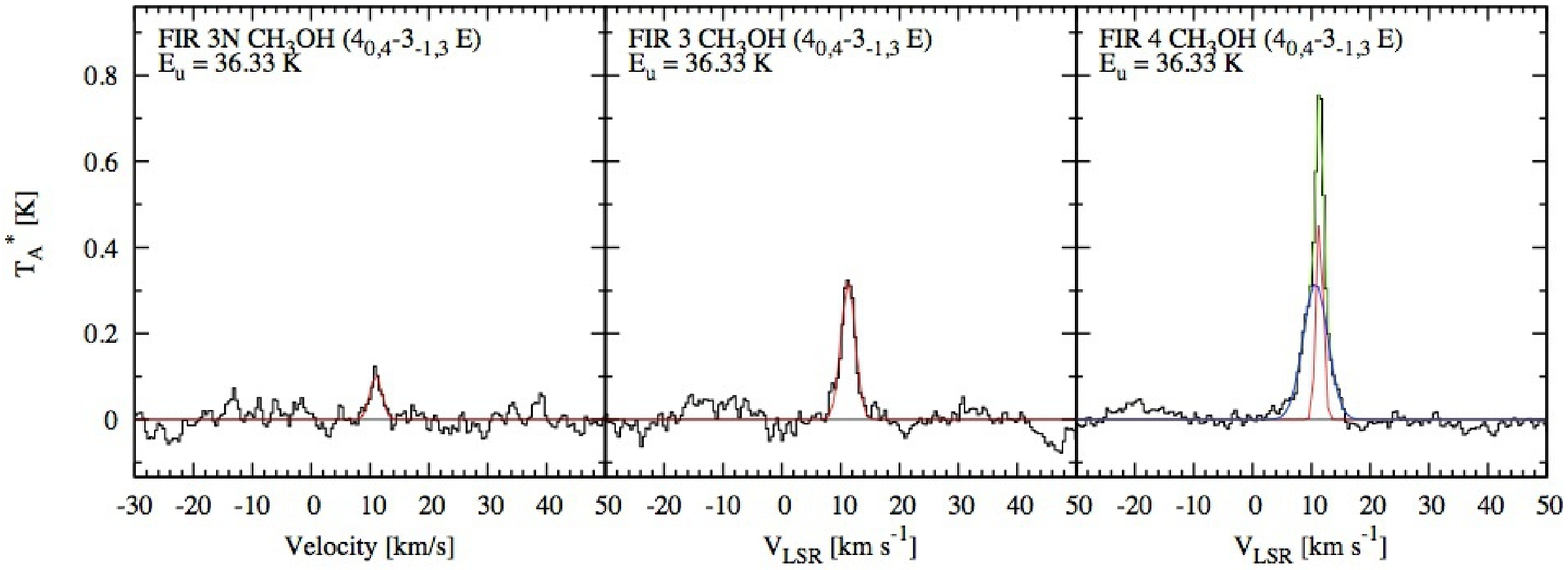}
\includegraphics[angle=0,scale=.49]{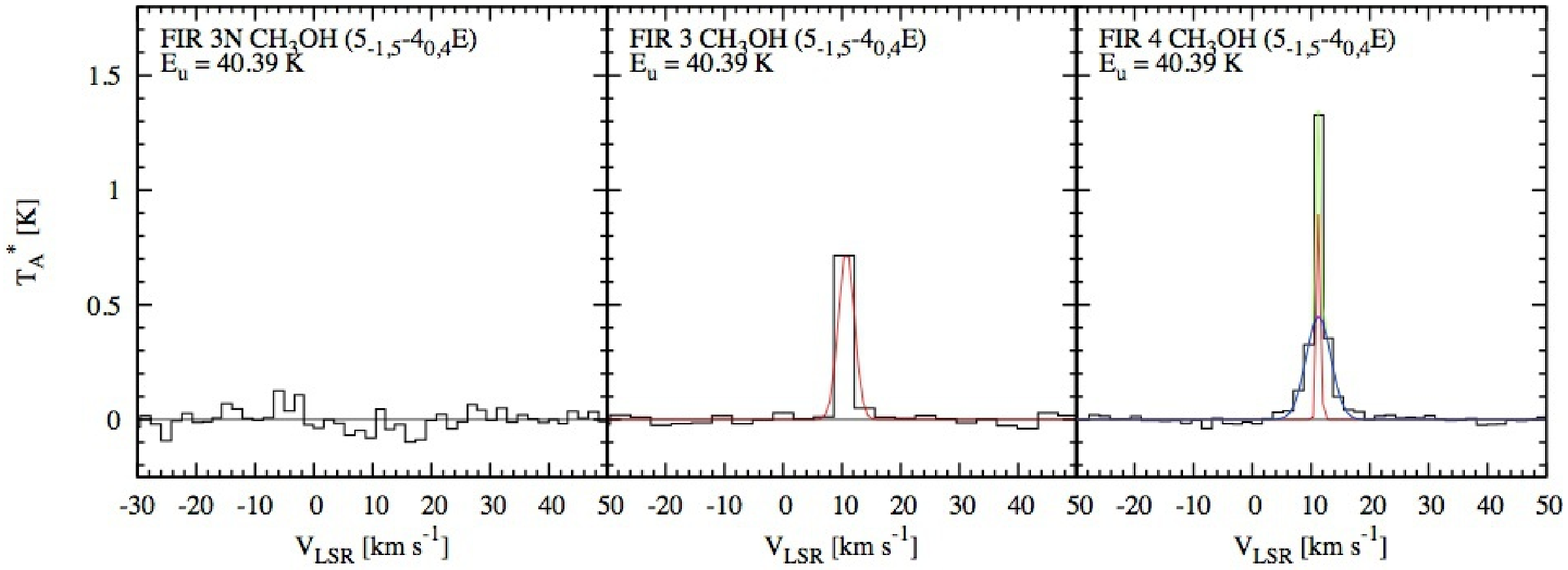}
\includegraphics[angle=0,scale=.5]{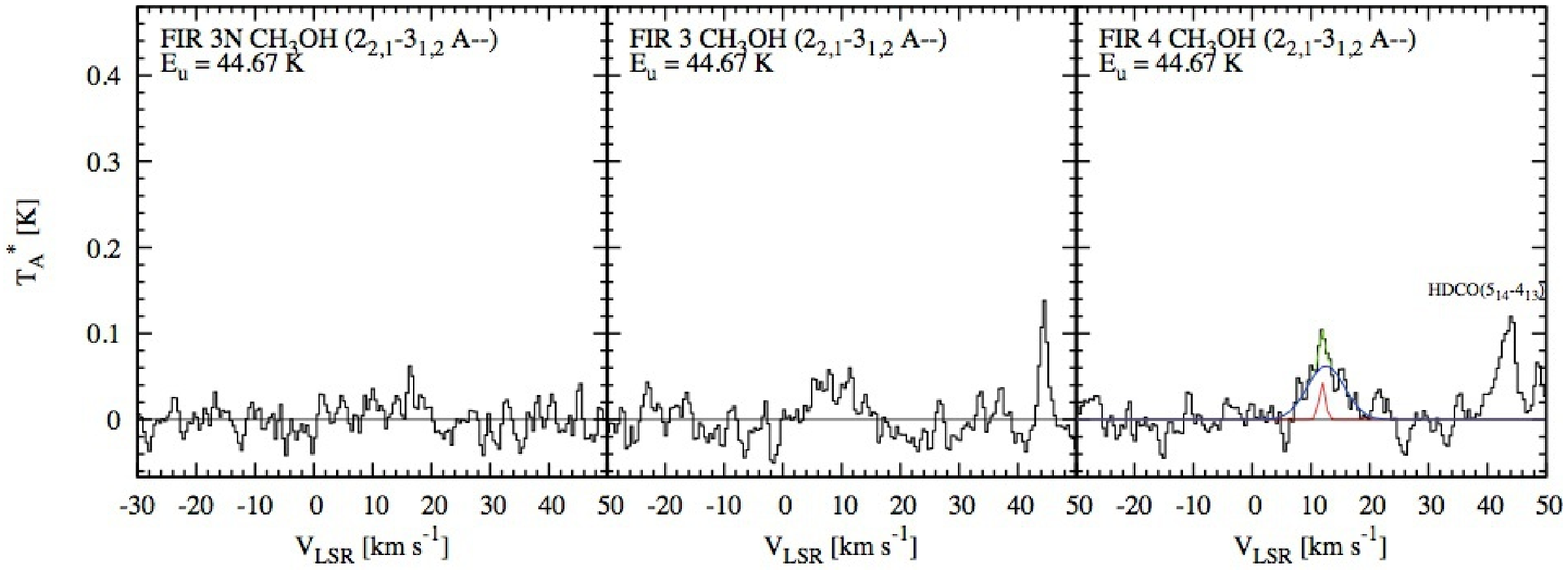}
\end{center}
\caption{\it Continued.}
\end{figure}

\addtocounter{figure}{-1}
\begin{figure}
\begin{center}
\includegraphics[angle=0,scale=.5]{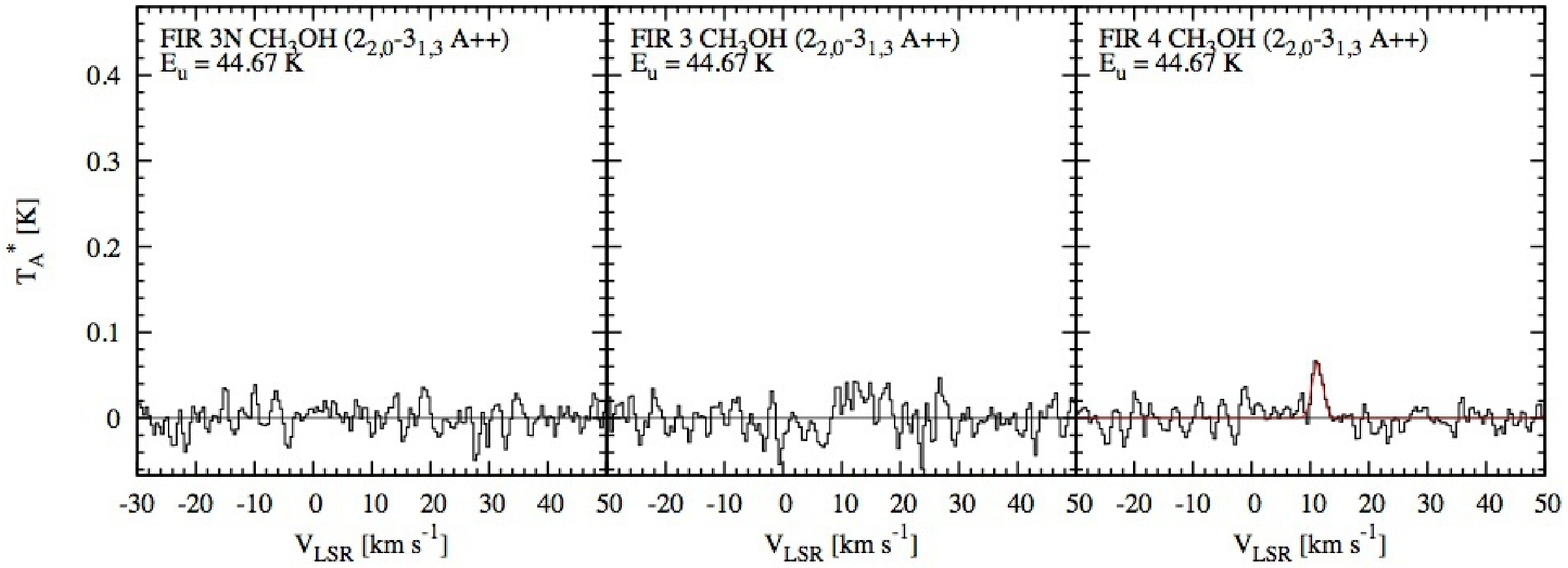}
\includegraphics[angle=0,scale=.5]{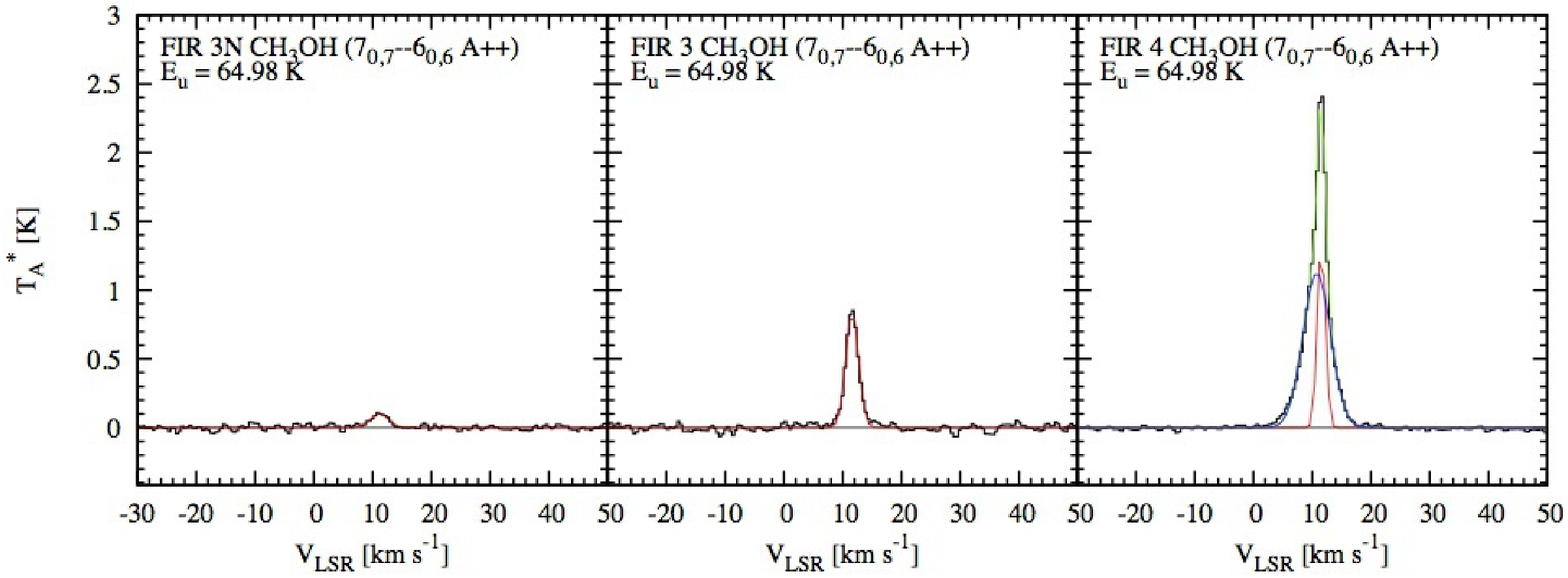}
\includegraphics[angle=0,scale=.5]{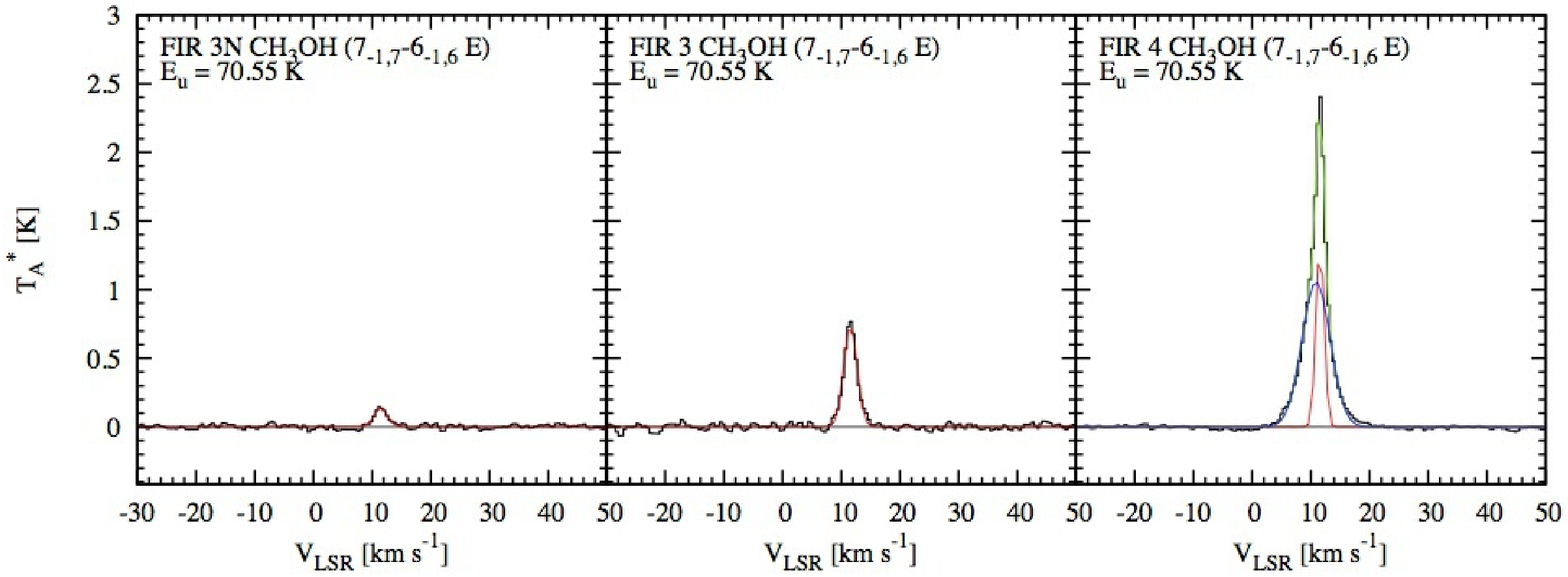}
\includegraphics[angle=0,scale=.5]{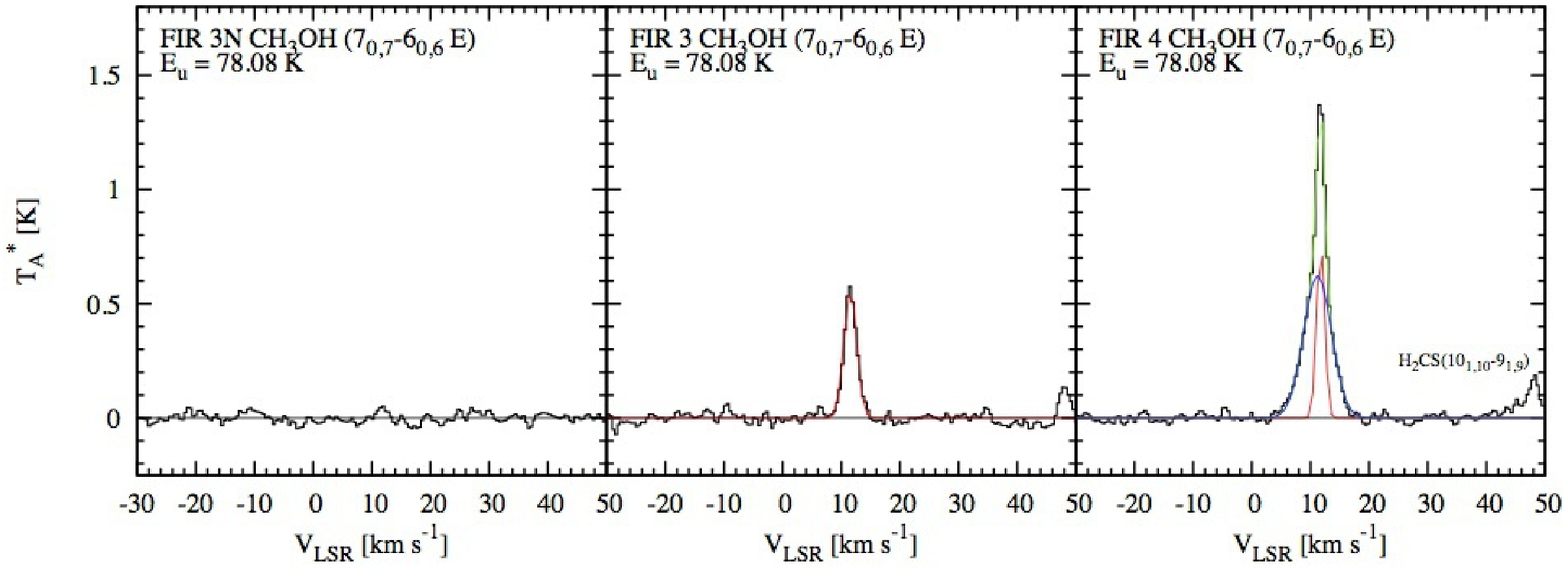}
\end{center}
\caption{\it Continued.}
\end{figure}

\addtocounter{figure}{-1}
\begin{figure}
\begin{center}
\includegraphics[angle=0,scale=.5]{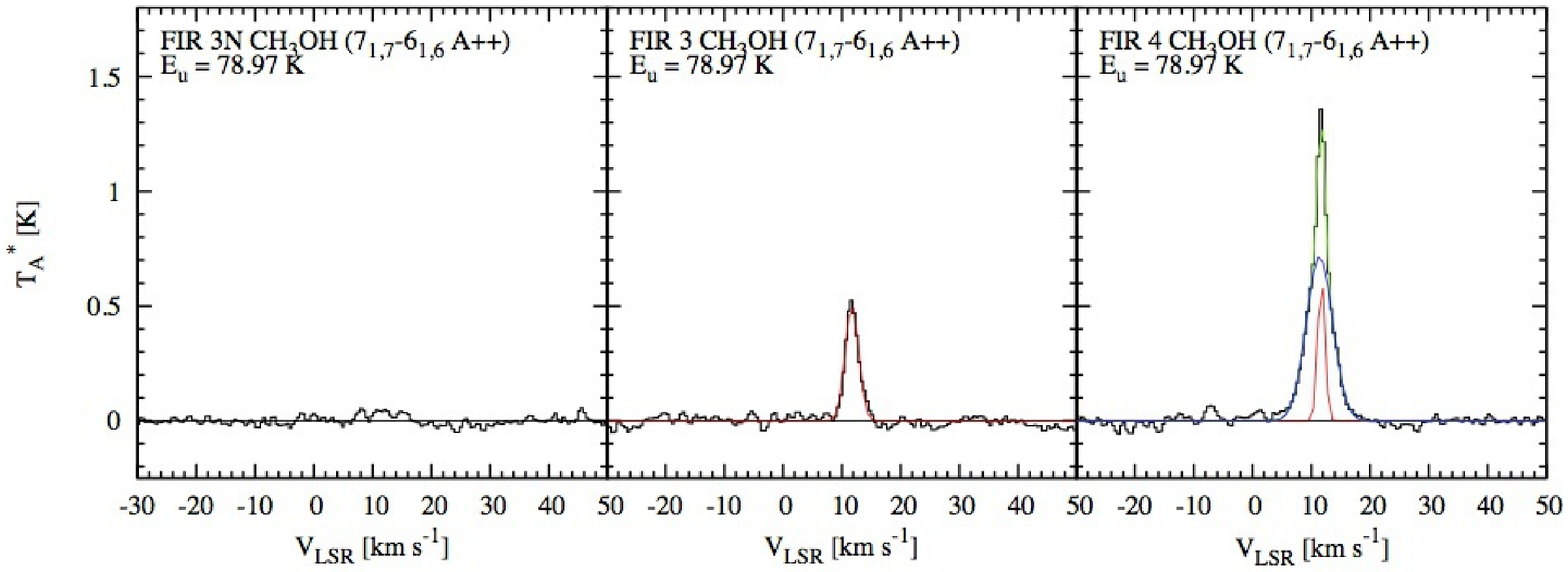}
\includegraphics[angle=0,scale=.5]{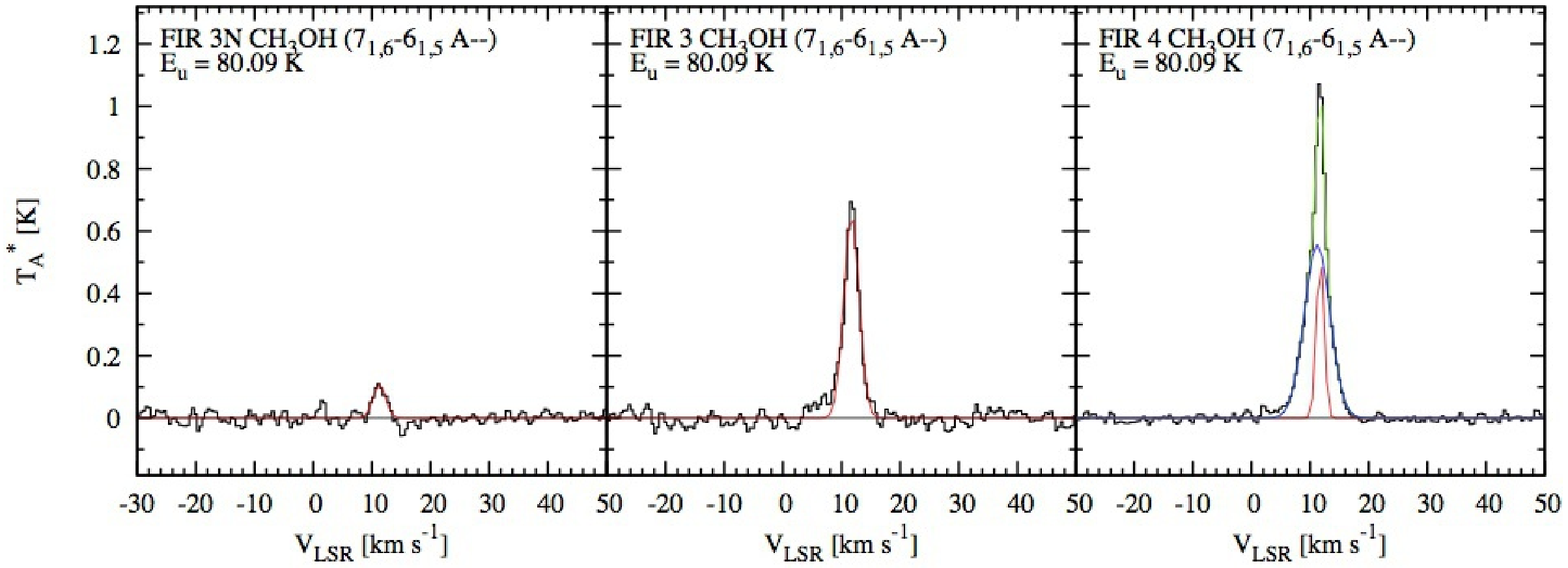}
\includegraphics[angle=0,scale=.5]{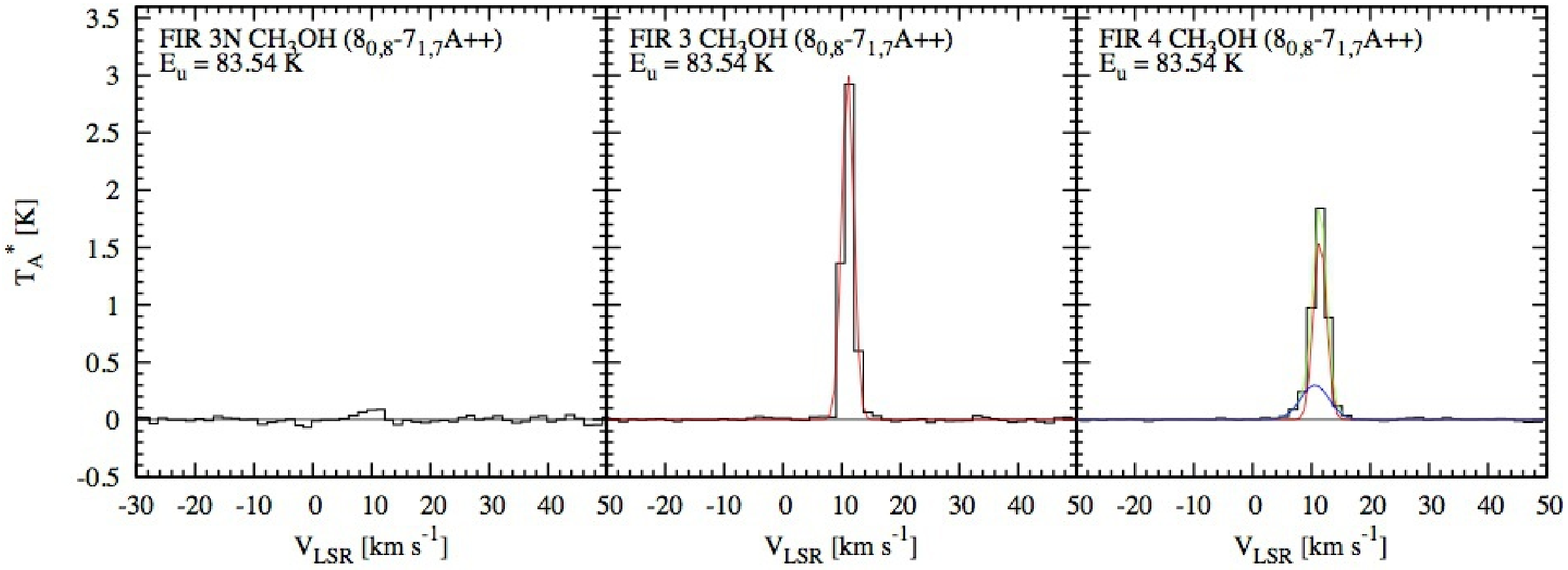}
\includegraphics[angle=0,scale=.5]{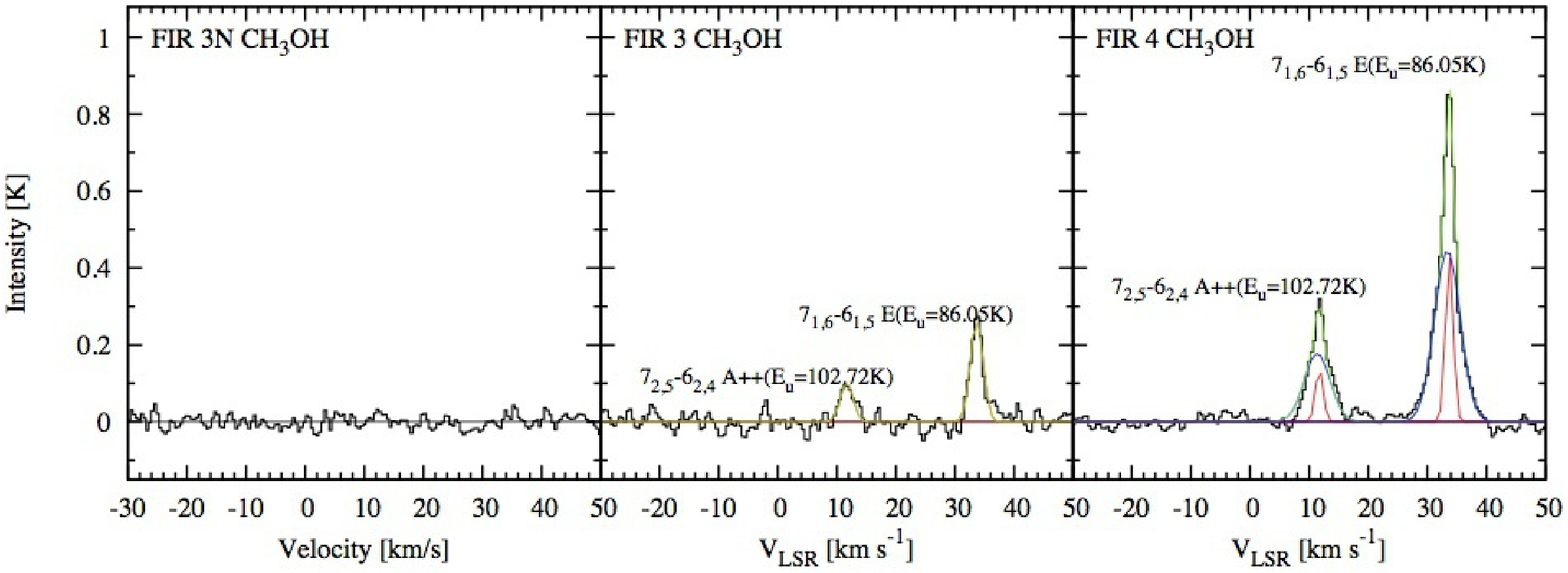}
\end{center}
\caption{\it Continued.}
\end{figure}

\addtocounter{figure}{-1}
\begin{figure}
\begin{center}
\includegraphics[angle=0,scale=.5]{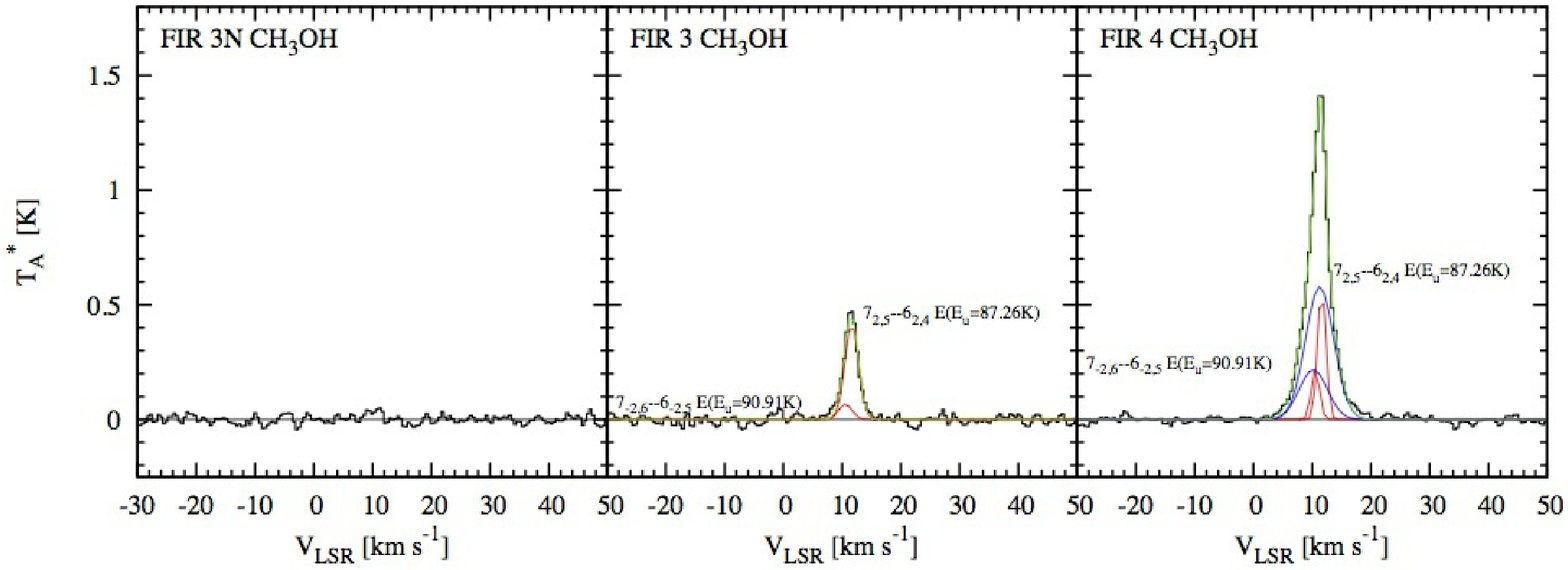} 
\includegraphics[angle=0,scale=.5]{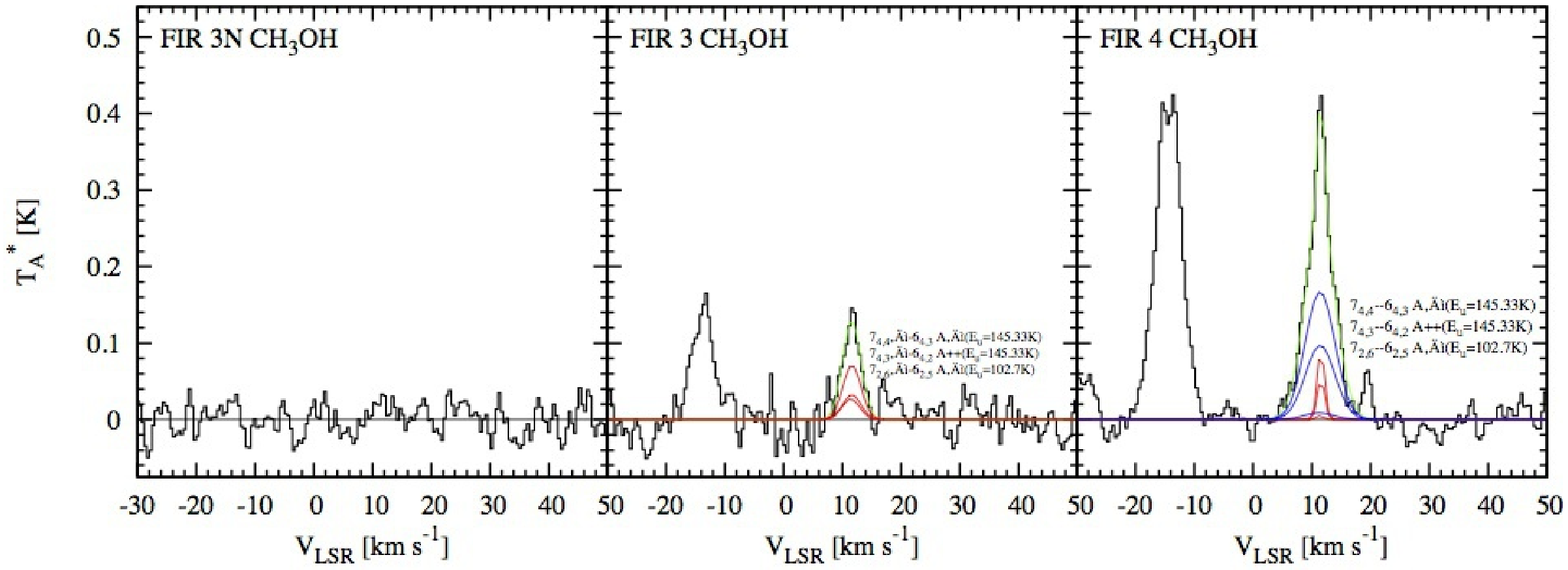}
\includegraphics[angle=0,scale=.5]{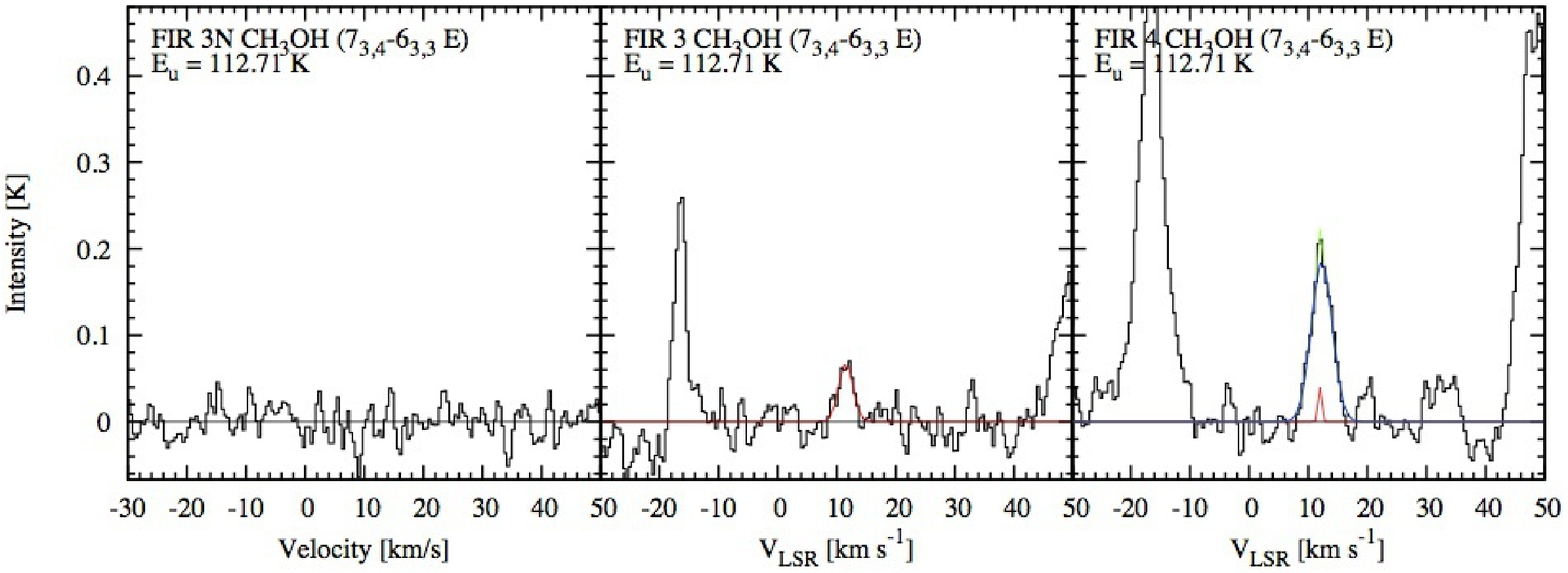}
\includegraphics[angle=0,scale=.5]{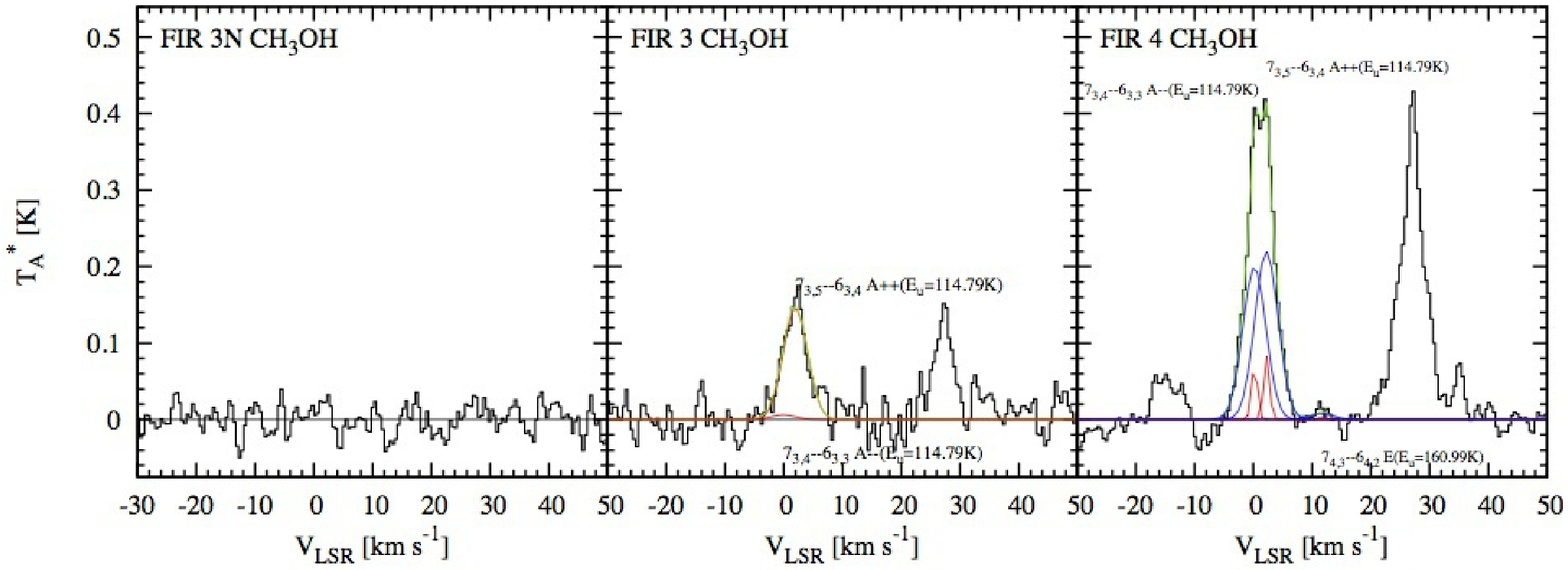}
\end{center}
\caption{\it Continued.}
\end{figure}

\addtocounter{figure}{-1}
\begin{figure}
\begin{center}
\includegraphics[angle=0,scale=.5]{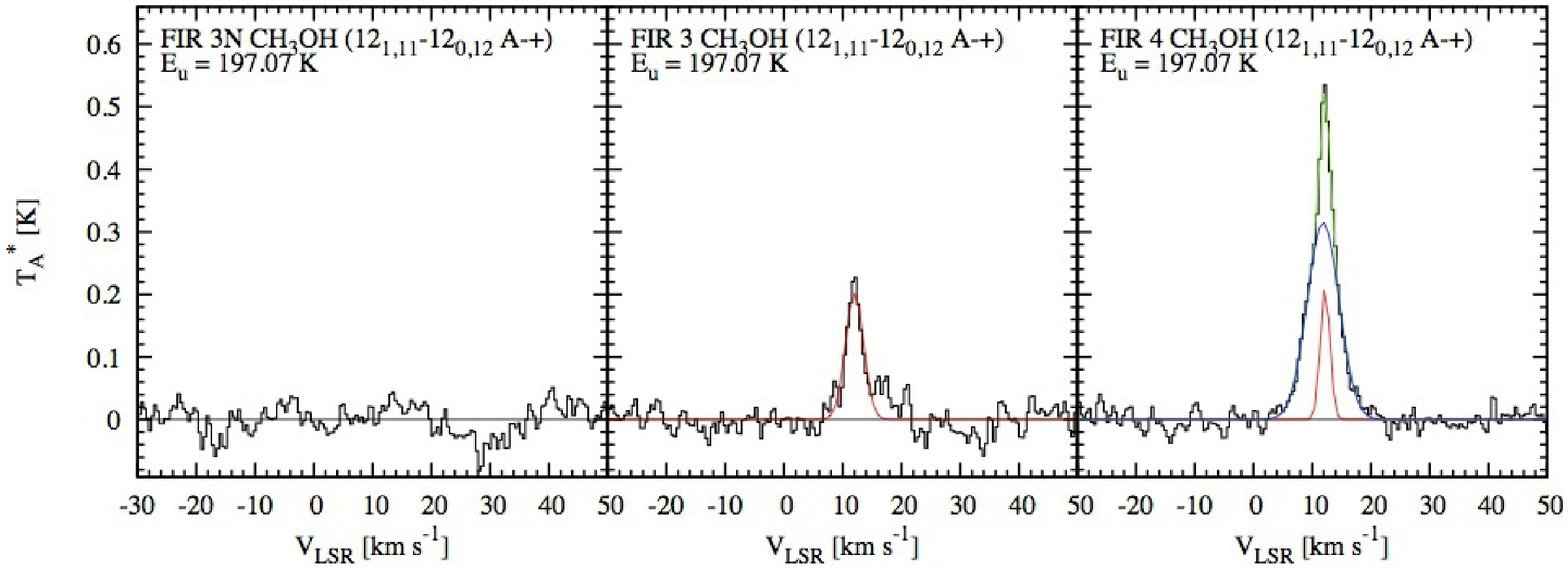}
\includegraphics[angle=0,scale=.5]{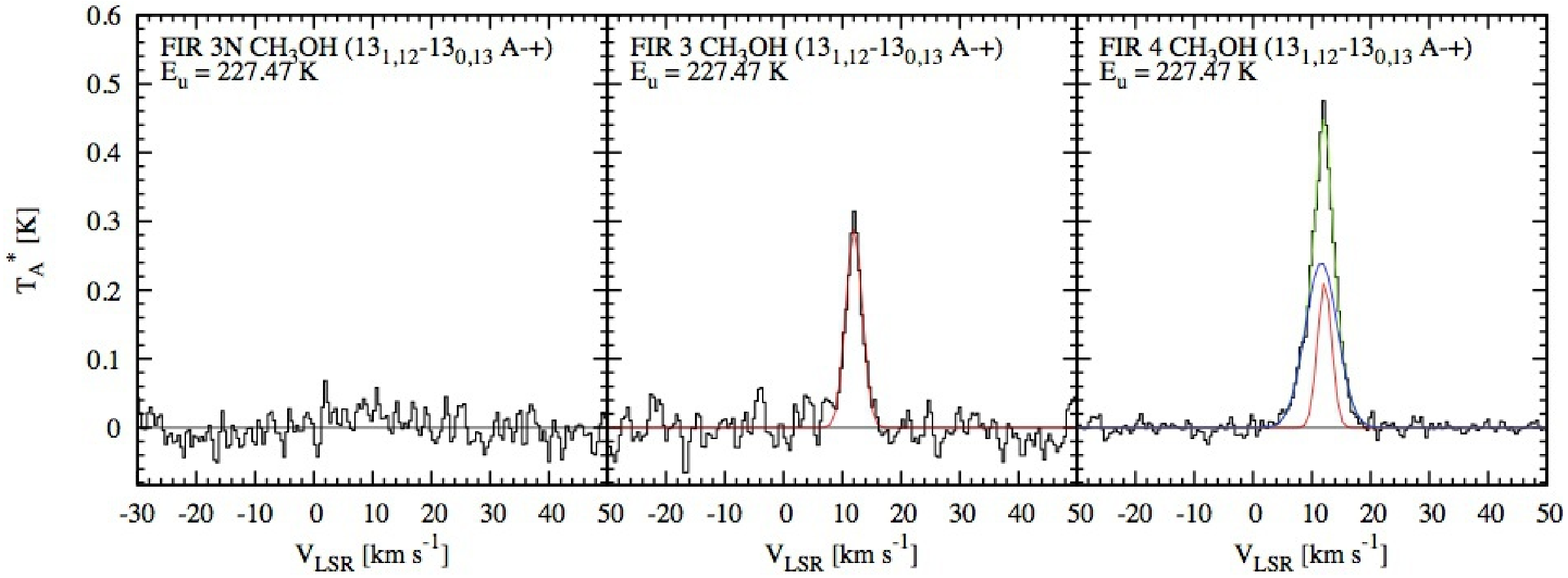}
\includegraphics[angle=0,scale=.5]{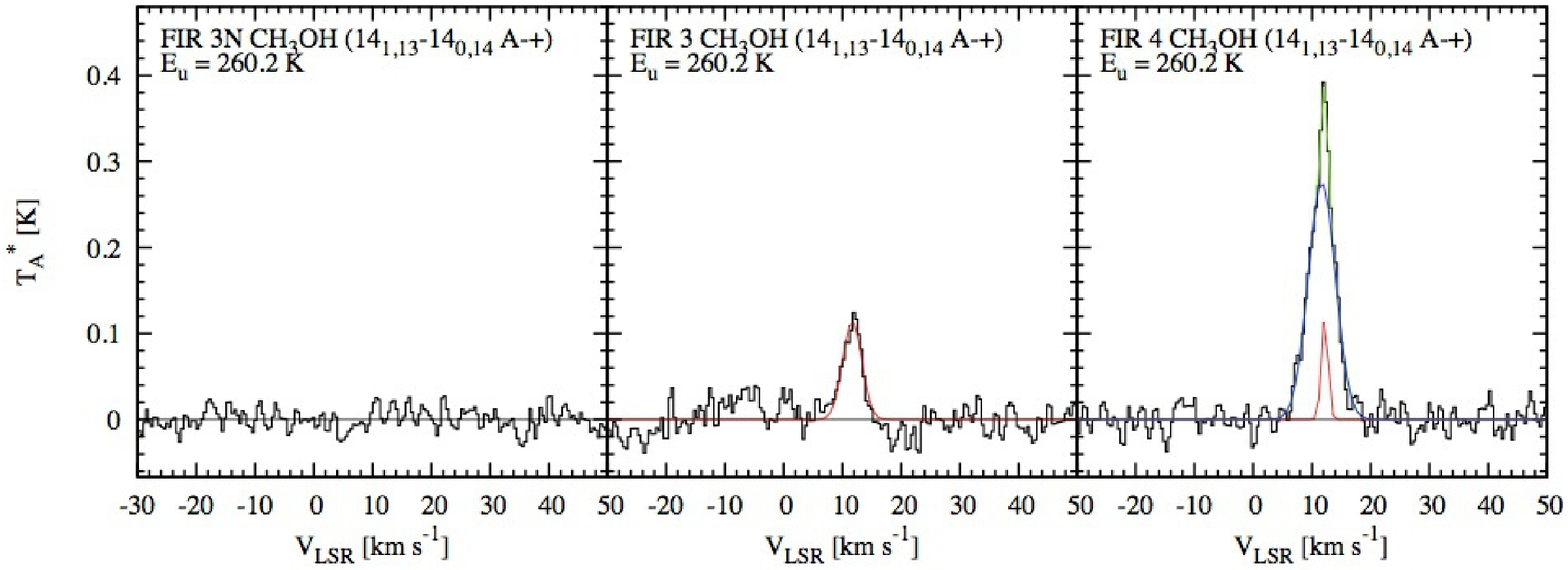}
\end{center}
\caption{\it Continued.}
\end{figure}


\clearpage

\begin{figure}
\begin{center}
\includegraphics[angle=0,scale=.35]{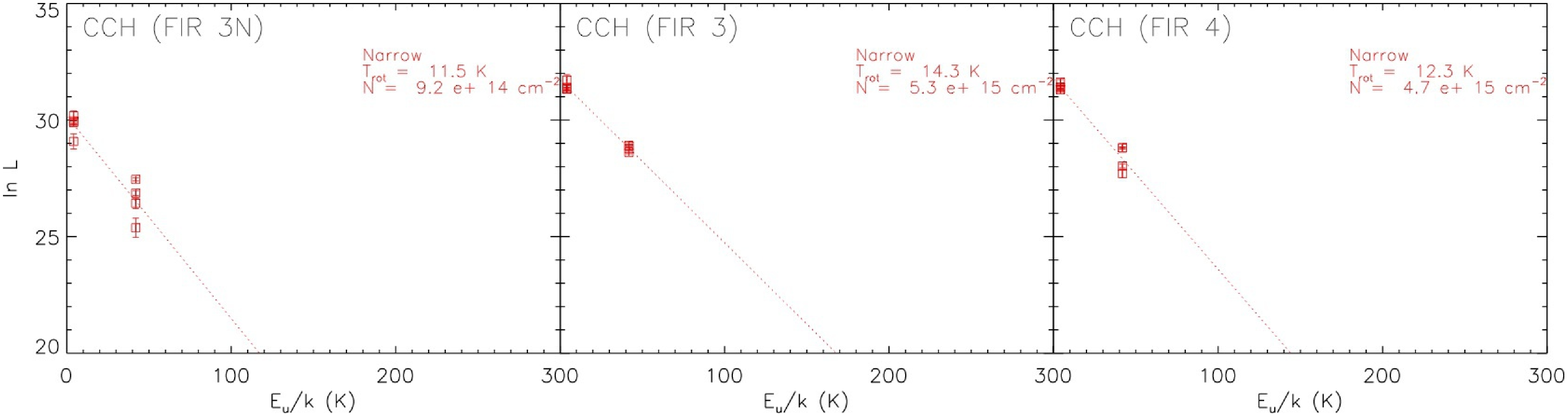}
\end{center}
\caption{C$_2$H rotation diagrams. The left, center, and right panels are for FIR 3N, FIR 3, and FIR 4, respectively. The error bars correspond to the 1$\sigma$ uncertainties.}
\label{cch_rotation}
\end{figure}

\begin{figure}
\begin{center}
\includegraphics[angle=0,scale=.35]{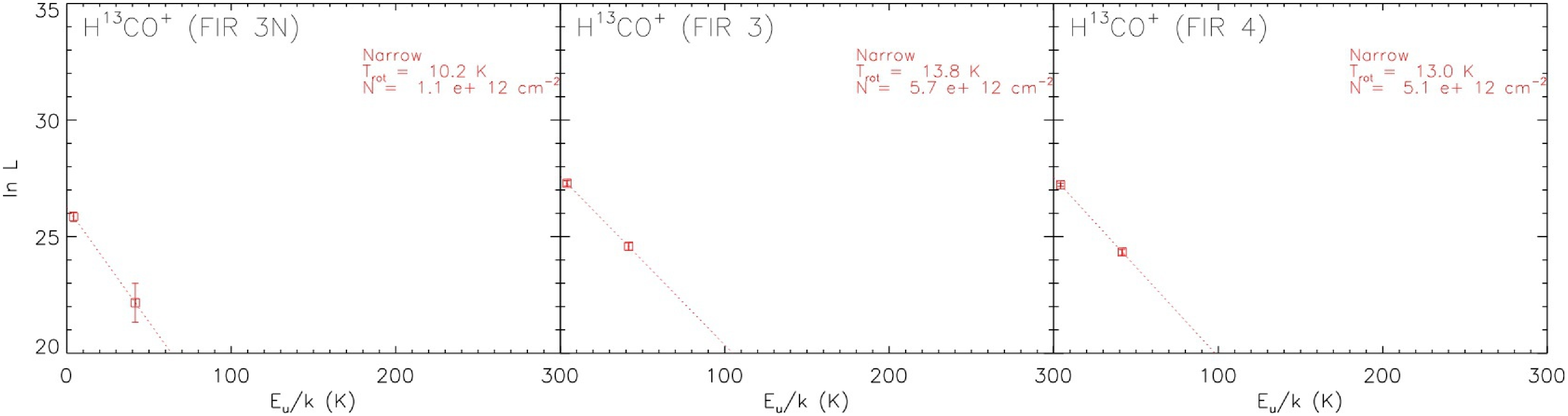}
\end{center}
\caption{H$^{13}$CO$^+$ Rotation Diagrams. }
\label{h13cop_rotation}
\end{figure}

\begin{figure}
\begin{center}
\includegraphics[angle=0,scale=.35]{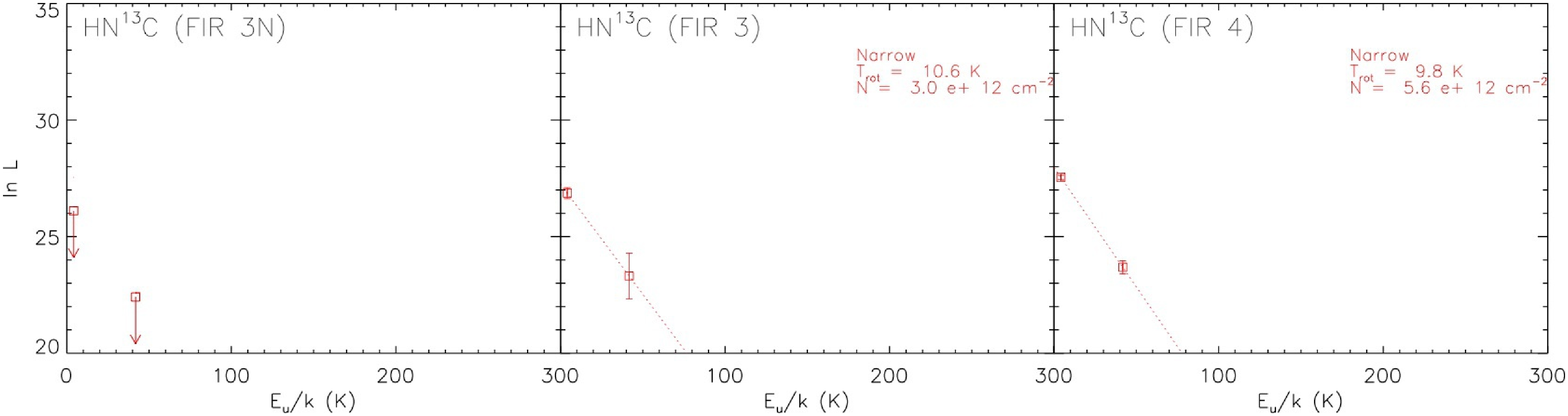}
\end{center}
\caption{HN$^{13}$C Rotation Diagrams. The downward arrow in the left panel means the 3$\sigma$ upper limit.}
\label{hn13c_rotation}
\end{figure}

\begin{figure}
\begin{center}
\includegraphics[angle=0,scale=.35]{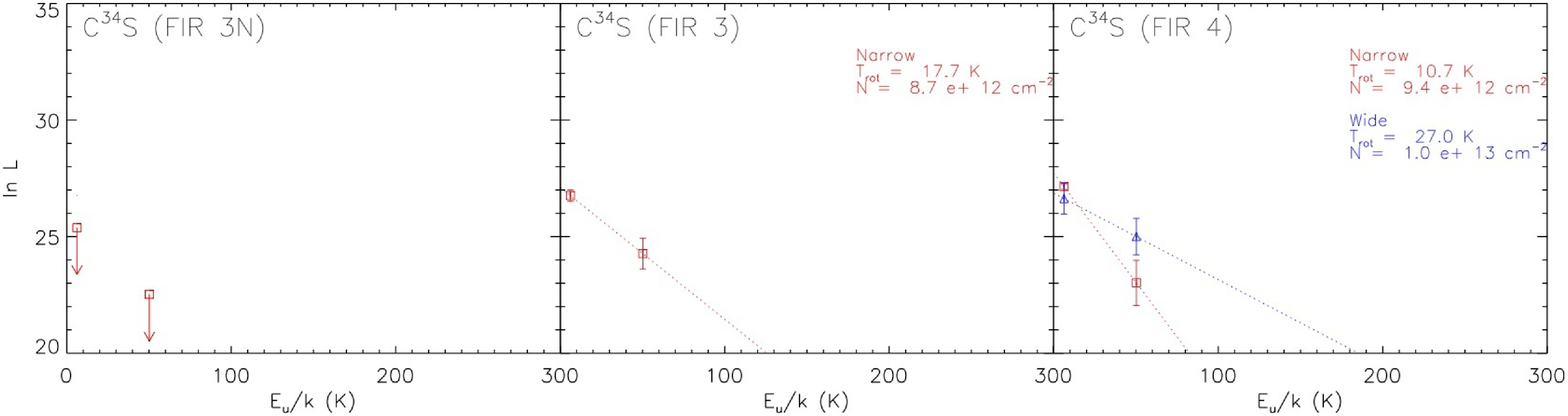}
\end{center}
\caption{C$^{34}$S Rotation Diagrams. }
\label{c34s_rotation}
\end{figure}

\begin{figure}
\begin{center}
\includegraphics[angle=0,scale=.35]{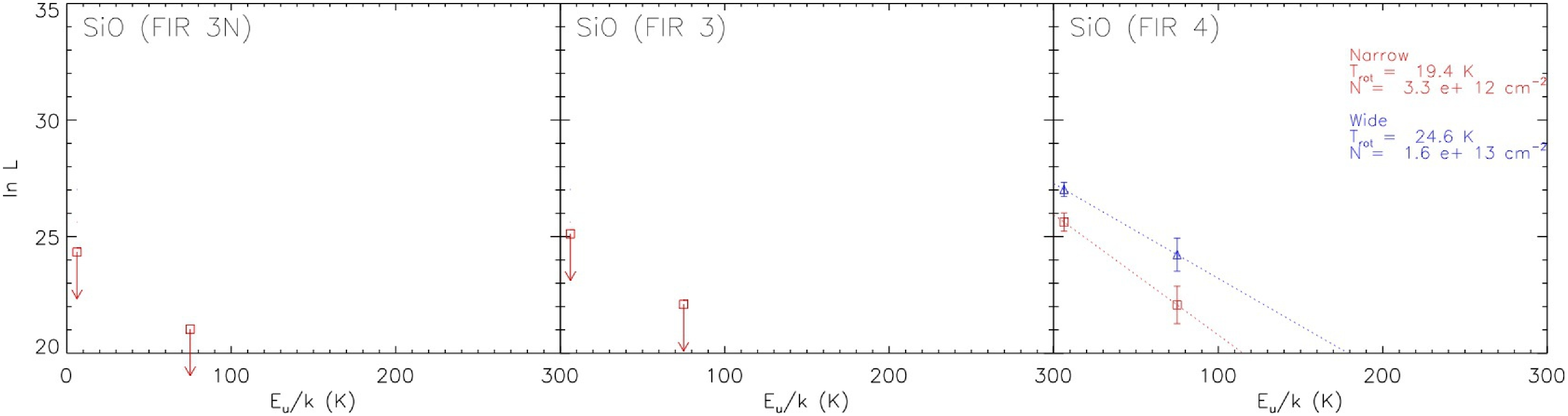}
\end{center}
\caption{SiO Rotation Diagrams. 
}
\label{sio_rotation}
\end{figure}

\begin{figure}
\begin{center}
\includegraphics[angle=0,scale=.35]{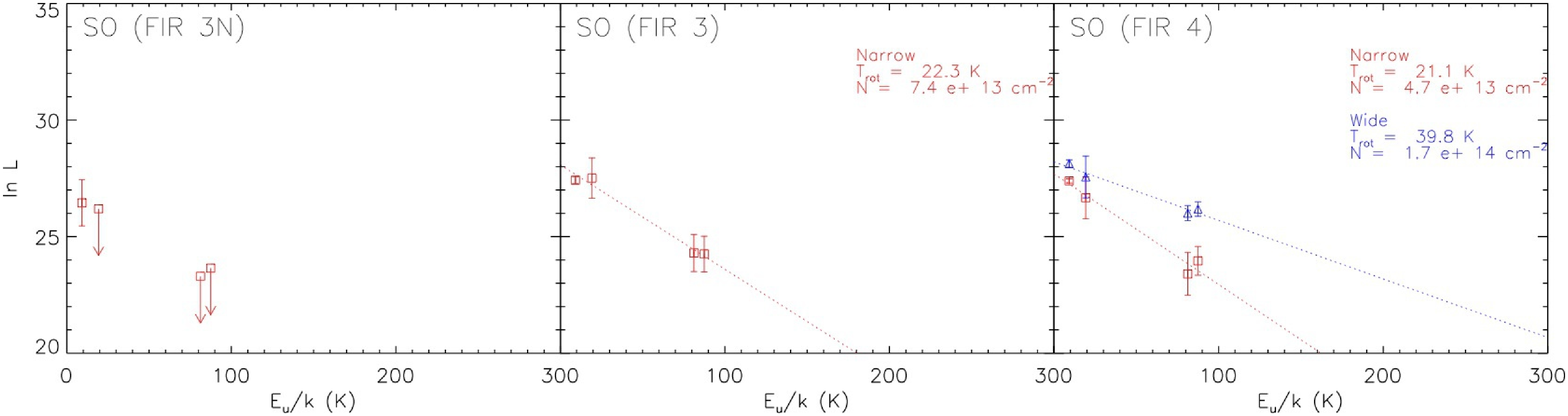}
\end{center}
\caption{SO Rotation Diagrams.}
\label{so_rotation}
\end{figure}

\begin{figure}
\begin{center}
\includegraphics[angle=0,scale=.35]{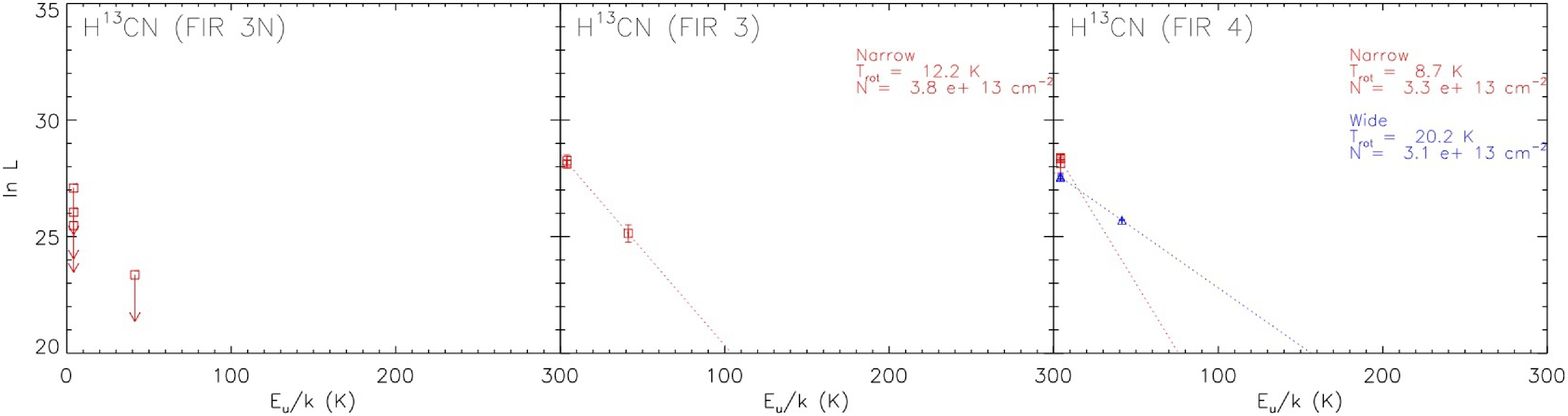}
\end{center}
\caption{H$^{13}$CN Rotation Diagrams. }
\label{h13cn_rotation}
\end{figure}

\begin{figure}
\begin{center}
\includegraphics[angle=0,scale=.35]{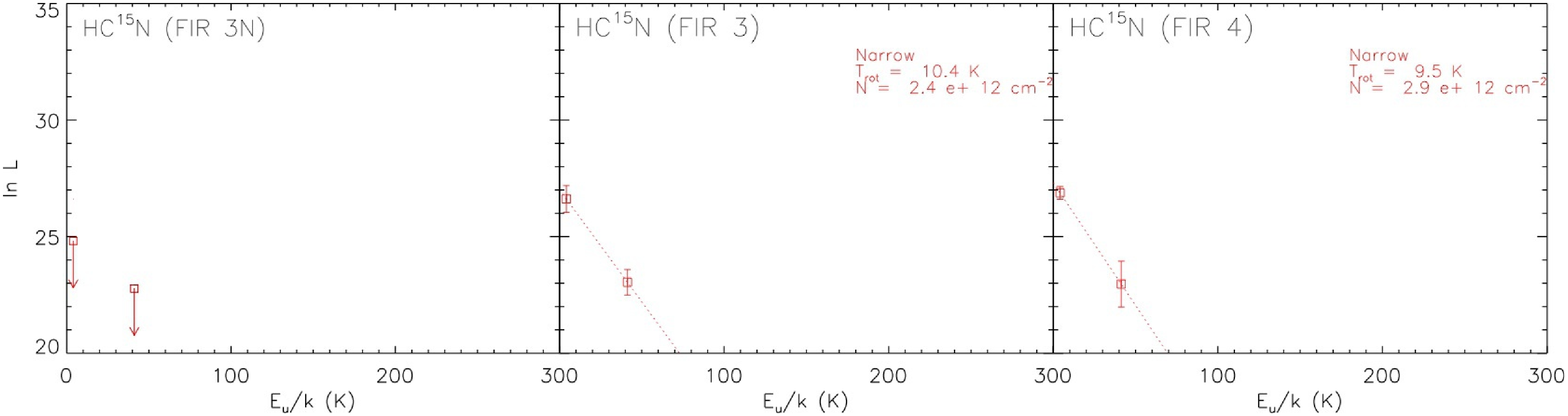}
\end{center}
\caption{HC$^{15}$N Rotation Diagrams. }
\label{hc15n_rotation}
\end{figure}

\begin{figure}
\begin{center}
\includegraphics[angle=0,scale=.35]{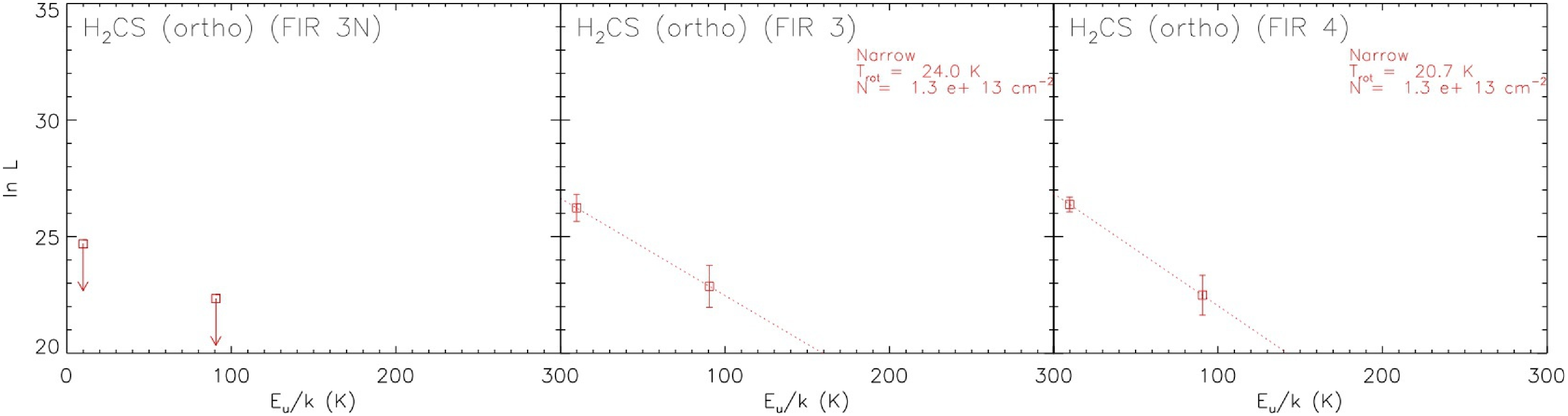}
\includegraphics[angle=0,scale=.35]{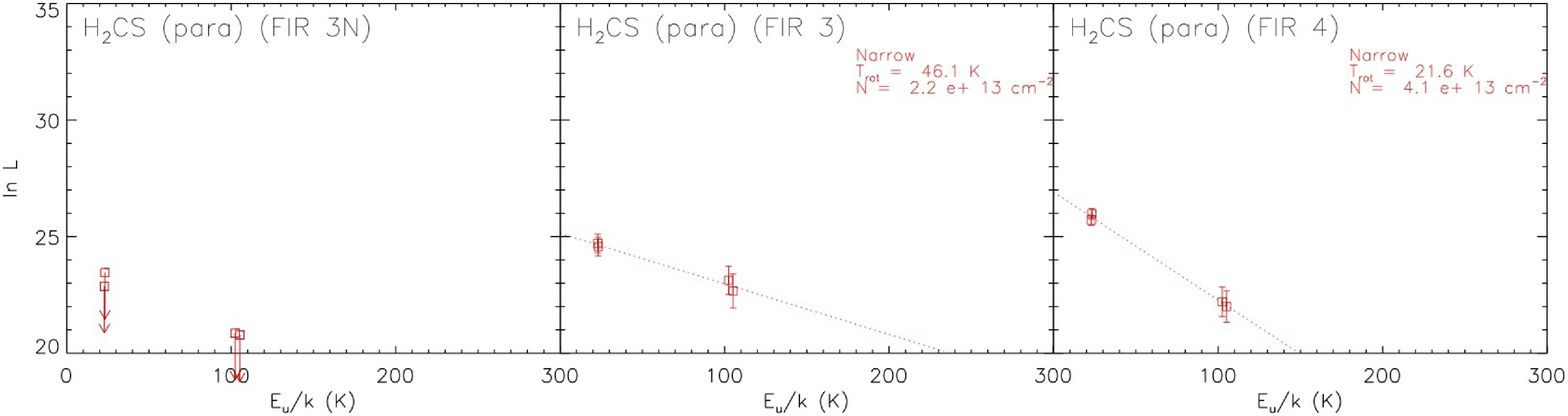}
\end{center}
\caption{H$_2$CS Rotation Diagrams. }
\label{h2cs_rotation}
\end{figure}

\begin{figure}
\begin{center}
\includegraphics[angle=0,scale=.35]{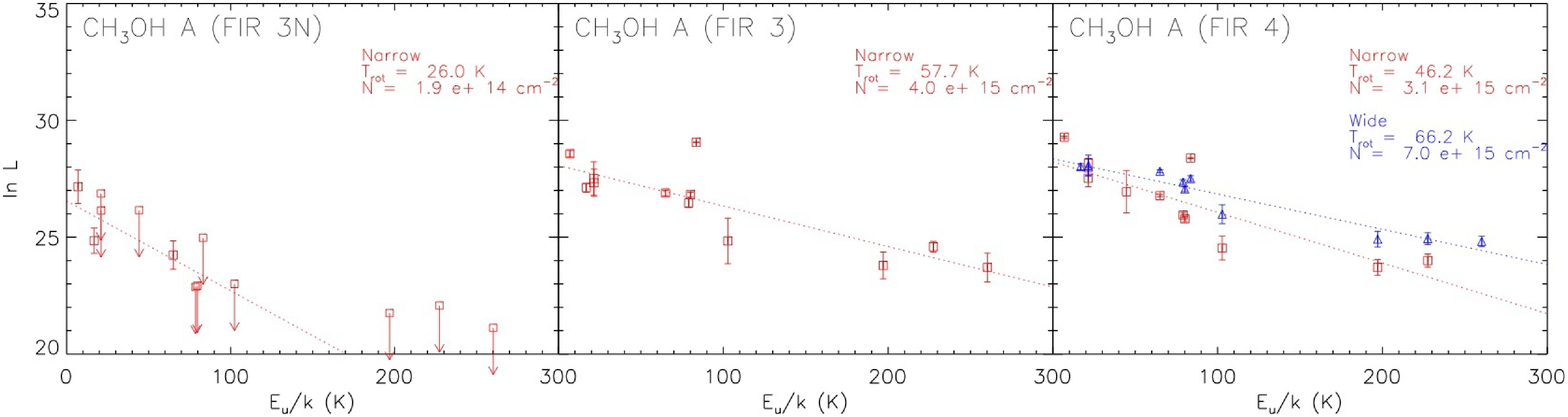}
\includegraphics[angle=0,scale=.35]{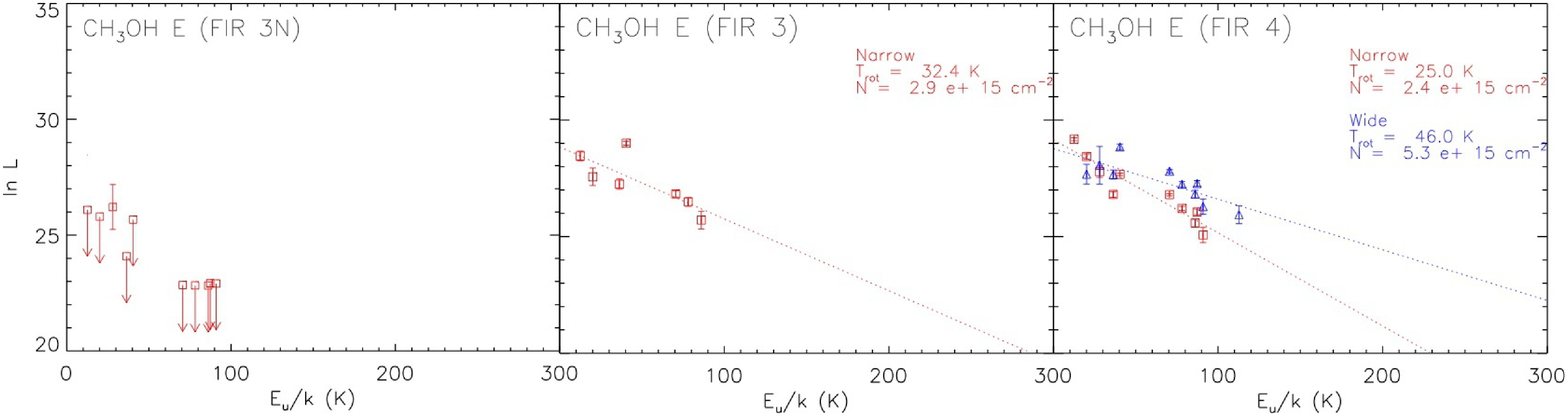}
\end{center}
\caption{CH$_3$OH Rotation Diagrams. }
\label{ch3oh_rotation}
\end{figure}

\begin{figure}
\begin{center}
\includegraphics[angle=0,scale=.35]{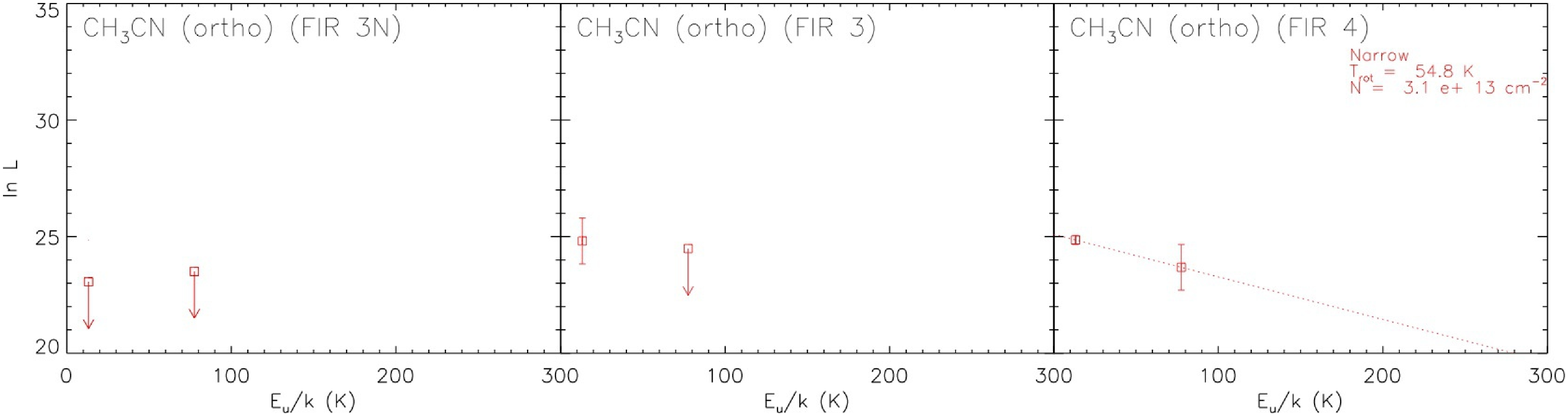}
\includegraphics[angle=0,scale=.35]{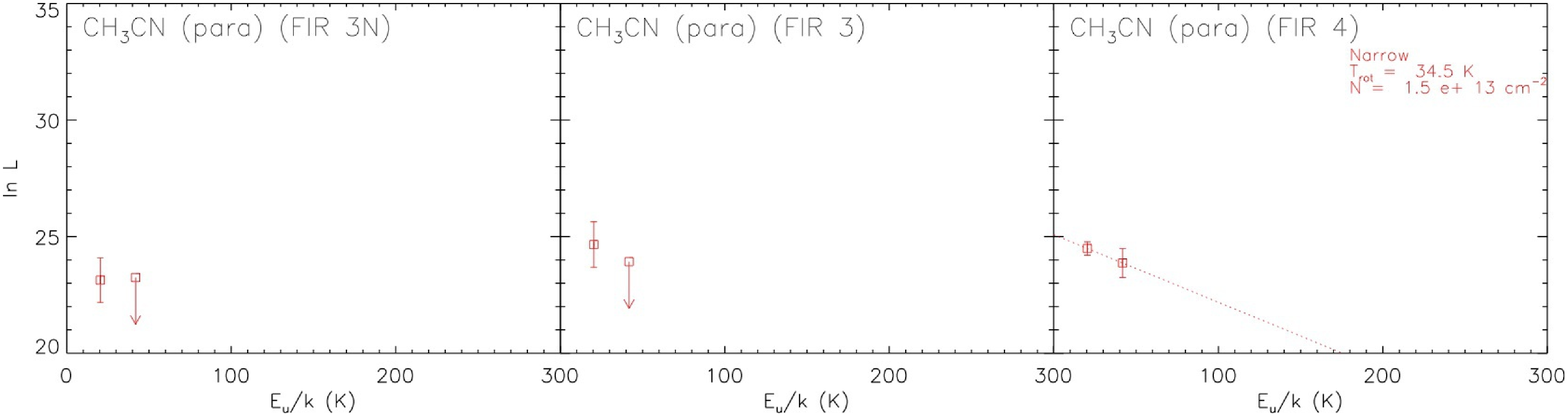}
\end{center}
\caption{CH$_3$CN Rotation Diagrams. }
\label{ch3cn_rotation}
\end{figure}

\begin{figure}
\begin{center}
\includegraphics[angle=0,scale=.35]{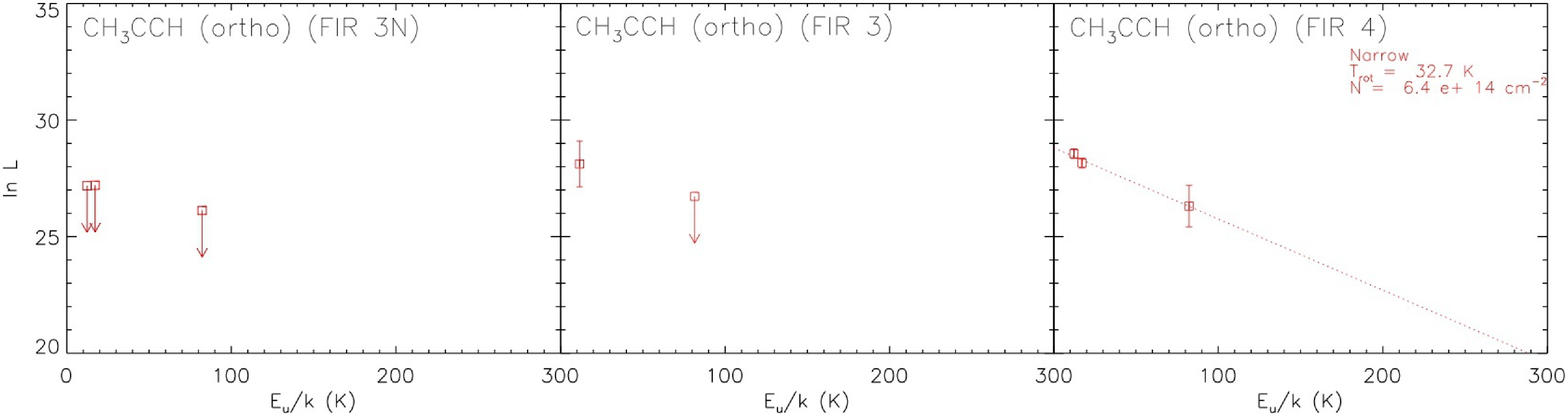}
\includegraphics[angle=0,scale=.35]{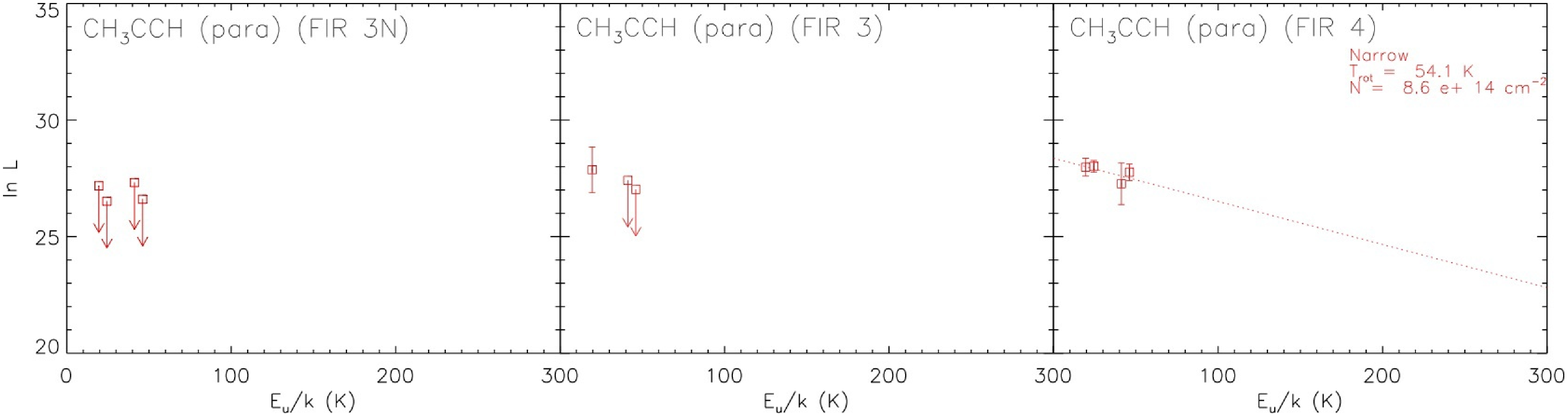}
\end{center}
\caption{CH$_3$CCH Rotation Diagrams. }
\label{ch3cch_rotation}
\end{figure}

\clearpage
\input{Table12.tex}
\input{Table13.tex}
\input{Table14.tex}

\end{document}

%% file: Table1.tex
\begin{deluxetable}{lcccccc}
\tablecolumns{3}
\tabletypesize{\scriptsize}
\tablecaption{Observed Positions and Flux Densities at FIR 3N, 3, and 4}
\tablewidth{0pt}
\tablehead{
\colhead{}
& \colhead{R.A.}
& \colhead{Dec. }
& \colhead{Physical environment}
& \colhead{$F_{\rm 1.1 mm}$$^{\dagger}$}
& \colhead{$N_{\rm H2}$$^{*}$}
& \colhead{$N_{\rm H2}$$^{**}$}
}
\startdata
             &        (J2000)                      &      (J2000)         &  & [Jy beam$^{-1}$] & [$\times$ 10$^{22}$ cm$^{-2}$] &  [$\times$ 10$^{22}$ cm$^{-2}$]  \\
\hline
\hline
FIR 3N & 5:35:28.7 & -5:09:15.6 & Northern lobe of the FIR 3 outflow   &  1.25 & 12.2 ($T_{\rm d}$=10.2 K) & 2.1 ($T_{\rm d}$=54 K) \\
FIR 3    & 5:35:27.6 & -5:09:34.0 & Outflow driving source   &  3.79 & 26.9 ($T_{\rm d}$=13.8 K) & 6.6 ($T_{\rm d}$=54 K) \\
FIR 4    & 5:35:26.8 & -5:09:57.4 & Outflow shock   &  5.95 & 45.0 ($T_{\rm d}$=13.0 K) & 10.0 ($T_{\rm d}$=56 K) \\
\enddata
\label{obstable}
\tablenotetext{\dagger}{1.1 mm dust continuum flux density estimated from the AzTEC 1.1 mm continuum data.}
\tablenotetext{*}{H$_2$ column density estimated on the assumption of $T_{\rm d}$=$T_{\rm rot}$ (H$^{13}$CO$^+$).}
\tablenotetext{**}{H$_2$ column density estimated on the assumption of $T_{\rm d}$=$T_{\rm CO\ peak}$.}
\end{deluxetable}

%% file: Table2.tex
\begin{deluxetable}{lcc}
\tablecolumns{3}
\tabletypesize{\scriptsize}
\tablecaption{Observational Parameters}
\tablewidth{0pt}
\tablehead{
\colhead{Telescope/Receiver}
& \colhead{NRO 45 m/T100}
& \colhead{ASTE/CATS345}
}
\startdata
$T_{\rm sys}$  & 150--200 K & 200--500 K \\
Frequency range         & 82--106 GHz & 335--355 GHz\\
Frequency resolution  & 488.2 kHz & 500.0 kHz \\
Velocity resolution (at 100/345 GHz) &  1.5 km s$^{-1}$  & 0.5 km s$^{-1}$ \\
Beam size (at 100/345 GHz) & 15.1$\arcsec$  & 19.7$\arcsec$  \\
\enddata
\label{obs}
\end{deluxetable}

%% file: Table3.tex
\begin{deluxetable}{lccc}
\tablecolumns{4}
\tabletypesize{\scriptsize}
\tablecaption{Rms Noise Levels in $T_{\rm A}^*$ at FIR 3N, 3, and 4}
\tablewidth{0pt}
\tablehead{
\colhead{Freq.}
& \colhead{FIR 3N}
& \colhead{FIR 3}
& \colhead{FIR 4}
}
\startdata
82--86  GHz        &  40.8 mK & 20.8 mK  &  17.8 mK \\
86--90  GHz        &  17.6 mK & 18.8 mK  &  13.5 mK   \\
90--94  GHz         & 26.8 mK  &  26.0 mK &  17.9 mK \\
94--98  GHz         & 27.9 mK  & 14.8 mK  &   12.1 mK\\
98--102  GHz    & 10.8 mK  & 13.3 mK  &  9.9 mK \\
102--106 GHz     & 20.8 mK  & 20.7 mK  &  13.9 mK \\
\hline
Mean (82--106 GHz)      & 24.1$\pm$10.3 mK  & 19.1$\pm$4.6 mK  &  14.2$\pm$3.2 mK \\
\hline
\hline
335--339  GHz    & 20.9 mK  &  33.2 mK & 23.9 mK  \\
339--343 GHz     & 21.1 mK  &  22.1 mK &  20.2 mK \\
343--347 GHz    &  34.9 mK & 32.9 mK  &   27.0 mK \\
347--351 GHz    &  18.6 mK & 23.3 mK  &   20.6 mK \\
351--355 GHz      & 37.0 mK  &  37.9 mK  &   42.6 mK  \\
\hline
Mean (335--355 GHz)      & 26.5$\pm$8.7 mK  & 29.9$\pm$6.9 mK  &  26.9$\pm$9.2 mK \\
\hline
\enddata
\label{rms_list}
\end{deluxetable}

%% file: Table4.tex
\begin{deluxetable}{lcccccc}
\tablecolumns{5}
\tabletypesize{\scriptsize}
\tablecaption{Number of the Detected Molecular Lines at FIR 3N, 3, and 4}
\tablewidth{0pt}
\tablehead{
\colhead{}
& \colhead{Num. of molecular lines}
& \colhead{Num. of the most abundant isotopic species}
& \colhead{Num. of the rare isotopic species}
}
\startdata
FIR 3N & 53   & 14 & 5 \\
FIR 3    & 100 & 17 & 10 \\
FIR 4    & 119 & 20 & 12 \\
\enddata
\label{number_detected_list}
\end{deluxetable}

%% file: Table5.tex
\begin{deluxetable}{lccccccc}
\tablecolumns{8}
\tabletypesize{\scriptsize}
\tablecaption{Detected Molecules toward FIR 3N, 3, and 4, and Relationship between the Physical Environment and the Chemical Composition}
\tablewidth{0pt}
\tablehead{
\colhead{Species}
& \multicolumn{2}{c}{FIR 3N}
& \multicolumn{2}{c}{FIR 3}
& \multicolumn{2}{c}{FIR 4}
& \colhead{Physical environment} \\
\colhead{}
& \colhead{Narrow}
& \colhead{Wide}
& \colhead{Narrow}
& \colhead{Wide}
& \colhead{Narrow}
& \colhead{Wide}
& \colhead{} 
}
\startdata
CN & $\circ$ &  \nodata &  $\circ$ &  \nodata &  $\circ$ &  \nodata & ambient dense gas \\
C$^{17}$O & $\circ$ &  \nodata &  $\circ$ &  \nodata &  $\circ$ &  \nodata & ambient dense gas \\
C$_2$H & $\circ$ &  \nodata &  $\circ$ &  \nodata &  $\circ$ &  \nodata & ambient dense gas \\
HNC & $\circ$ &  \nodata &  $\circ$ &  \nodata &  $\circ$ &  \nodata & ambient dense gas \\
HN$^{13}$C & $\circ$ &  \nodata &  $\circ$ &  \nodata &  $\circ$ &  \nodata & ambient dense gas \\
H$^{13}$CO$^+$ & $\circ$ &  \nodata &  $\circ$ &  \nodata &  $\circ$ &  \nodata & ambient dense gas \\
N$_2$H$^+$ & $\circ$ &  \nodata &  $\circ$ &  \nodata &  $\circ$ &  \nodata & ambient dense gas \\
c-C$_3$H$_2$ &  $\circ$ &  \nodata &  $\circ$ &  \nodata &  $\circ$ &  \nodata &  ambient dense gas \\
CH$_3$CN & $\circ$ &  \nodata &  $\circ$ &  \nodata &  $\circ$ &  \nodata & ambient dense gas \\
\hline
$^{13}$CS & \nodata &  \nodata &  $\circ$ &  \nodata &  $\circ$ &  \nodata & (possible) ambient dense gas \\
H$^{15}$NC & $\circ$ &  \nodata &   \nodata &  \nodata &  $\circ$ &  \nodata &  (possible) ambient dense gas \\
HC$^{18}$O$^+$ & \nodata &  \nodata &  $\circ$ &  \nodata &  \nodata &  \nodata & (possible) ambient dense gas \\
NH$_2$D & \nodata &  \nodata &  \nodata &  \nodata &  $\circ$ &  \nodata & (possible) ambient dense gas \\
CH$_3$CCH & \nodata  &  \nodata &  $\circ$ &  \nodata &  $\circ$ &  \nodata & (possible) ambient dense gas \\
\hline
CO & $\circ$ &  $\circ$ &  $\circ$ &  $\circ$ &  $\circ$ &  $\circ$ & molecular outflow \\
CS & $\circ$ &  $\circ$ &  $\circ$ &  $\circ$ &  $\circ$ &  $\circ$ & molecular outflow \\
HCN & $\circ$ &  $\circ$ &  $\circ$ &   $\circ$ &  $\circ$ &  $\circ$ & molecular outflow \\
HCO$^+$ & $\circ$ &  $\circ$ &  $\circ$ &   $\circ$ &  $\circ$ &  $\circ$ & molecular outflow \\
\hline
C$^{34}$S & \nodata &  \nodata &  $\circ$ &  \nodata &  $\circ$ &  $\circ$ & outflow shock \\
SO & $\circ$ &  \nodata &  $\circ$ &  $\circ$ &  $\circ$ &  $\circ$ & outflow shock \\
SiO & \nodata &  \nodata &  \nodata &  \nodata &  $\circ$ &  $\circ$ & outflow shock \\
H$^{13}$CN & $\circ$ &  \nodata &  $\circ$ &  \nodata &  $\circ$ &  $\circ$ & outflow shock \\
HC$^{15}$N & \nodata &  \nodata &  $\circ$ &  \nodata &  $\circ$ & $\circ$ & outflow shock \\
H$_2^{13}$CO & $\circ$ &  \nodata &  $\circ$ &  \nodata &  $\circ$ & $\circ$ & outflow shock \\
H$_2$CS & $\circ$ &  \nodata &  $\bullet$ &  \nodata &  $\circ$ &  $\circ$ & outflow shock \\
HC$_3$N & $\circ$ &  \nodata &  $\circ$ &  \nodata &  $\circ$ &  $\circ$ & outflow shock \\
CH$_3$OH & $\circ$ &  \nodata &  $\bullet$ &  \nodata &  $\circ$ &  $\circ$ & outflow shock \\
\hline
HCS$^+$ & \nodata &  \nodata &  \nodata &  \nodata &   $\bullet$ &  \nodata & (possible) outflow shock \\
H$_2$CO & \nodata &  \nodata &  \nodata &  \nodata &  $\bullet$ &  \nodata & (possible) outflow shock \\
HDCO & \nodata &  \nodata &  $\bullet$ &  \nodata &  $\bullet$ &  \nodata & (possible) outflow shock or (possible) molecular outflow {$^a$}\\
HNCO & \nodata &  \nodata &  $\circ$ &  \nodata &   $\bullet$ &  \nodata & (possible) outflow shock \\
CH$_3$CHO & \nodata &  \nodata &  $\circ$ &  \nodata &  $\bullet$ &  \nodata & (possible) outflow shock \\
\enddata
\label{detected_list}
\label{relation_between}
\tablenotetext{a}{See Appendix \ref{appendix_hdco}.} 
\tablenotetext{\dagger}{Open circle shows that the component is detected.} 
\tablenotetext{\ddagger}{Filled circle shows that the wide component could not be discerned, but the velocity width of the narrow component is significantly larger than those of the ambient dense gas tracers.}
\end{deluxetable}

%% file: Table6.tex
\begin{deluxetable}{lcccccccc}
\tablecolumns{5}
\tabletypesize{\scriptsize}
\tablecaption{Optical Depth}
\tablewidth{0pt}
\tablehead{
\colhead{}
& \colhead{}
& \colhead{}
& \multicolumn{3}{c}{$Narrow \  comp.$}
& \multicolumn{3}{c}{$Wide \  comp.$} \\
\colhead{Molecule}
& \colhead{Rarer isotopic species}
& \colhead{Transition}
& \colhead{FIR 3N}
& \colhead{FIR 3}
& \colhead{FIR 4}
& \colhead{FIR 3N}
& \colhead{FIR 3}
& \colhead{FIR 4}
}
\startdata
HCO$^{+}$ & H$^{13}$CO$^{+}$ & $J$=1--0           &  9.7            &  7.1            &  8.6        & \multicolumn{3}{c}{\nodata}\\
CS                 & C$^{34}$S   & $J$=2--1                       &   \nodata   &   0.4        &  2.1         &   \nodata   &   \nodata  &  0.9        \\ 
CS                 & C$^{34}$S   & $J$=7--6                       &   \nodata   &  1.0           &  $<<$ 1 &   \nodata   &  \nodata     &  0.7      \\ 
HCN             & H$^{13}$CN & $J$=1--0,$F$=1--1     &  9.1             &  8.5            &  10.7         &  \nodata    & \nodata    &   37.7      \\ 
HCN             & H$^{13}$CN & $J$=1--0,$F$=2--1     &  4.8             &  5.4            &  8.1         &  \nodata    &  \nodata   &   26.7         \\ 
HCN             & H$^{13}$CN &  $J$=1--0,$F$=0--1    &  4.9             &  3.9            &  4.3         &  \nodata    &  \nodata   &   12.8        \\ 
HCN             & H$^{13}$CN  & $J$=4--3                      &  \nodata   &  5.9            &  7.1          &  \nodata   & \nodata     &   5.7         \\ 
HNC             &  HN$^{13}$C & $J$=1--0                      &  3.1           & 4.5              &  6.2        & \multicolumn{3}{c}{\nodata}\\
\enddata
\label{optical}
\end{deluxetable}

%% file: Table7.tex
\begin{deluxetable}{lcccccccc}
\tablecolumns{9}
\rotate
\tabletypesize{\tiny}
\tablecaption{Rotational Temperatures, Column Densities, and Fractional Abundances of the Detected Molecules at FIR 3N, 3, and 4$^*$}
\tablewidth{0pt}
\tablehead{
 \colhead{Species}
&  \colhead{Position}
& \multicolumn{2}{c}{$Narrow \  comp.$}
& \multicolumn{2}{c}{$Wide \  comp.$}
& \multicolumn{3}{c}{}  \\
\cline{3-9} \\
\colhead{}
& \colhead{}
& \colhead{$T_{\rm rot}$ [K]$^{\dagger}$}
& \colhead{$N_{\rm mol}$ [cm$^{-2}$]$^{\ddagger}$}
& \colhead{$T_{\rm rot}$ [K]$^{\dagger}$}
& \colhead{$N_{\rm mol}$ [cm$^{-2}$]$^{\dagger}$}
& \colhead{$N_{\rm mol}$ total}
& \colhead{$X_{\rm mol}$$^{\ddagger}$ ($T_{\rm d}$ = $T_{\rm rot}$ (H$^{13}$CO$^{+}$))}
& \colhead{$X_{\rm mol}$$^{\ddagger}$ ($T_{\rm d}$ = $T_{\rm CO\ peak}$)}
}
\startdata
                                & FIR 3N  &  \nodata             & $<$ 1.5$\times$10$^{12}$  ($T$ = 10 K) & \multicolumn{4}{c}{\nodata} & \\
C$^{34}$S               & FIR 3    &  17.7$\pm$5.0   & (8.7$\pm$2.6)$\times$10$^{12}$             & \multicolumn{2}{c}{\nodata}     &  & (0.3$\pm$0.1)$\times$10$^{-10}$ &  (1.3$\pm$0.4)$\times$10$^{-10}$   \\
                                & FIR 4    &  10.7$\pm$2.6   & (9.4$\pm$1.8)$\times$10$^{12}$             &  27.0$\pm$16.9      & (1.1$\pm$0.8)$\times$10$^{13}$ &   (1.1$\pm$0.2)$\times$10$^{13}$ &  (2.4$\pm$0.4)$\times$10$^{-11}$ &  (1.1$\pm$0.2)$\times$10$^{-10}$ \\
\hline
                                & FIR 3N &  \nodata             & $<$ 9.4$\times$10$^{11}$ ($T$ = 20 K)  & \multicolumn{4}{c}{\nodata} & \\
SiO                          & FIR 3    &  \nodata             & $<$ 2.1$\times$10$^{12}$ ($T$ = 20 K)  & \multicolumn{4}{c}{\nodata} & \\
                                & FIR 4    & 19.4$\pm$4.9    & (3.3$\pm$1.4)$\times$10$^{12}$             &  24.6$\pm$6.8        & (1.6$\pm$0.5)$\times$10$^{13}$ & (1.9$\pm$0.5)$\times$10$^{13}$ & (4.2$\pm$1.1)$\times$10$^{-11}$ &  (1.9$\pm$0.5)$\times$10$^{-10}$ \\
\hline
                                & FIR 3N &  \nodata              & (2.0$\pm$0.4)$\times$10$^{13}$ ($T$ = 20 K) & \multicolumn{4}{c}{\nodata} &  \\
SO                           & FIR 3   &  22.3$\pm$2.1   & (7.4$\pm$1.9)$\times$10$^{13}$                      & \multicolumn{2}{c}{\nodata} &  & (0.3$\pm$0.1)$\times$10$^{-9}$  &  (1.1$\pm$0.3)$\times$10$^{-9}$  \\
                                & FIR 4   &  21.1$\pm$2.8   & (4.7$\pm$1.8)$\times$10$^{13}$ &   39.8 $\pm$5.2 & (1.7$\pm$0.3)$\times$10$^{14}$ & (2.2$\pm$0.3)$\times$10$^{14}$ & (4.9$\pm$0.7)$\times$10$^{-10}$ &  (2.2$\pm$0.3)$\times$10$^{-9}$  \\
\hline
                                & FIR 3N &  11.5 $\pm$1.5 & (9.2$\pm$3.0) $\times$10$^{14}$  & \multicolumn{2}{c}{\nodata} & & (0.8$\pm$0.3)$\times$10$^{-8}$ &  (4.4$\pm$0.4)$\times$10$^{-8}$  \\
 C$_2$H                  & FIR 3   &  14.3 $\pm$0.6 & (5.3$\pm$0.4) $\times$10$^{15}$  & \multicolumn{2}{c}{\nodata} &  & (2.0$\pm$0.2)$\times$10$^{-8}$ &  (8.0$\pm$0.6)$\times$10$^{-8}$ \\
                                & FIR 4   &  12.3 $\pm$1.0 & (4.7$\pm$0.9) $\times$10$^{15}$  & \multicolumn{2}{c}{\nodata} & & (1.0$\pm$0.2)$\times$10$^{-8}$ & (4.7$\pm$0.9)$\times$10$^{-8}$  \\
\hline
                                &  FIR 3N &  \nodata                 &  $<$ 4.3 $\times$ 10$^{12}$ ($T$ = 10 K) & \multicolumn{4}{c}{\nodata} & \\
H$^{13}$CN            &  FIR 3    &  12.2$\pm$0.5       & (3.8$\pm$0.3)$\times$10$^{13}$          & \multicolumn{2}{c}{\nodata}  & & (1.4$\pm$0.1)$\times$10$^{-10}$ &  (5.8$\pm$0.5)$\times$10$^{-10}$ \\
                                &  FIR 4    &    8.7$\pm$0.3       & (3.3$\pm$0.3)$\times$10$^{13}$          &   20.2 $\pm$  0.4 & (3.1$\pm$0.1)$\times$10$^{13}$ & (6.4$\pm$0.3)$\times$10$^{13}$ & (1.4$\pm$0.1)$\times$10$^{-10}$ & (6.4$\pm$0.3)$\times$10$^{-10}$  \\
 \hline
                                & FIR 3N &  \nodata                  &  $<$ 3.8$\times$10$^{11}$ ($T$ = 10 K)  & \multicolumn{4}{c}{\nodata} & \\
HC$^{15}$N            & FIR 3    &  10.4$\pm$2.3        & (2.4$\pm$1.5)$\times$10$^{12}$               & \multicolumn{2}{c}{\nodata} & & (0.9$\pm$0.6)$\times$10$^{-11}$  &  (3.6$\pm$2.3)$\times$10$^{-11}$  \\                         
                                & FIR 4    &    9.5$\pm$2.5        & (2.9$\pm$0.9)$\times$10$^{12}$               & \multicolumn{2}{c}{\nodata} & & (0.6$\pm$0.2)$\times$10$^{-11}$ & (2.9$\pm$0.9)$\times$10$^{-11}$ \\
\hline
                                 & FIR 3N &  10.2$\pm$2.4       & (1.1$\pm$0.3)$\times$10$^{12}$               & \multicolumn{2}{c}{\nodata}  &   & (8.8$\pm$2.1)$\times$10$^{-12}$ & (5.2$\pm$1.4)$\times$10$^{-11}$  \\
H$^{13}$CO$^+$     & FIR 3   &  13.8$\pm$1.0        & (5.7$\pm$0.7)$\times$10$^{12}$               & \multicolumn{2}{c}{\nodata}  &   & (2.1$\pm$0.3)$\times$10$^{-11}$ &  (8.6$\pm$1.1)$\times$10$^{-11}$   \\
                                 & FIR 4   &  13.0$\pm$0.6        & (5.1$\pm$0.4)$\times$10$^{12}$               & \multicolumn{2}{c}{\nodata}  &   & (1.1$\pm$0.1)$\times$10$^{-11}$ & (5.1$\pm$0.4)$\times$10$^{-11}$  \\
\hline
                                 & FIR 3N  &  \nodata                &  $<$ 6.7$\times$10$^{11}$ ($T$ = 10 K)    & \multicolumn{4}{c}{\nodata} & \\
HN$^{13}$C             & FIR 3    &  10.6$\pm$3.0       & (3.0$\pm$0.9)$\times$10$^{12}$               & \multicolumn{2}{c}{\nodata}  &   & (1.1$\pm$0.3)$\times$10$^{-11}$ &  (4.5$\pm$1.4)$\times$10$^{-11}$  \\
                                 & FIR 4    &    9.8$\pm$0.8       & (5.6$\pm$0.6)$\times$10$^{12}$               & \multicolumn{2}{c}{\nodata}  &   & (1.2$\pm$0.1)$\times$10$^{-11}$  & (5.6$\pm$0.6)$\times$10$^{-11}$\\
\hline
                                 & FIR 3N  &  \nodata                & $<$ 2.3$\times$10$^{12}$ ($T$ = 20 K)     & \multicolumn{4}{c}{\nodata} &  \\
H$_2$CS (ortho)      & FIR 3    &  24.0$\pm$7.6      & (1.3$\pm$0.8)$\times$10$^{13}$                & \multicolumn{2}{c}{\nodata}  &    & (0.5$\pm$0.3)$\times$10$^{-10}$ &  (2.0$\pm$1.2)$\times$10$^{-10}$  \\
                                 & FIR 4    &  20.7$\pm$4.9      & (1.3$\pm$0.5)$\times$10$^{13}$                & \multicolumn{2}{c}{\nodata}  &    & (0.3$\pm$0.1)$\times$10$^{-10}$  & (1.3$\pm$0.5)$\times$10$^{-10}$\\
\hline
                                 & FIR 3N &  \nodata                 &  $<$ 3.3$\times$10$^{12}$ ($T$ = 40 K)    & \multicolumn{4}{c}{\nodata} & \\
H$_2$CS (para)       & FIR 3   &  46.1$\pm$5.5       & (2.2$\pm$0.4)$\times$10$^{13}$                & \multicolumn{2}{c}{\nodata}  &     & (0.8$\pm$0.2)$\times$10$^{-10}$ &  (3.3$\pm$0.6)$\times$10$^{-10}$ \\
                                 & FIR 4   &  21.6$\pm$0.8       & (4.1$\pm$0.6)$\times$10$^{13}$                & \multicolumn{2}{c}{\nodata}  &     & (0.9$\pm$0.1)$\times$10$^{-10}$  & (4.1$\pm$0.6)$\times$10$^{-10}$ \\
\hline
                                             & FIR 3N  &  \nodata                 &   $<$ 3.5$\times$10$^{12}$ ($T$ = 40 K)               & \multicolumn{4}{c}{\nodata} & \\
CH$_3$CN (ortho)$^{**}$    & FIR 3     &  \nodata                 &  (2.0$\pm$0.9)$\times$10$^{13}$ ($T$ = 40 K)     & \multicolumn{4}{c}{\nodata} & \\ 
                                             & FIR 4     &  54.8$\pm$46.5     &  (3.1$\pm$1.0)$\times$10$^{13}$                          & \multicolumn{2}{c}{\nodata} &     &  (0.7$\pm$0.2)$\times$10$^{-10}$   & (3.1$\pm$1.0)$\times$10$^{-10}$\\
\hline
                                                                  &  FIR 3N   &  \nodata                  &  (4.5$\pm$4.4)$\times$10$^{12}$ ($T$ = 40 K)     & \multicolumn{4}{c}{\nodata} & \\
CH$_3$CN (para)$^{\dagger\dagger}$    &  FIR 3      &  \nodata                  &  (1.1$\pm$0.5)$\times$10$^{13}$ ($T$ = 40 K)     & \multicolumn{4}{c}{\nodata} & \\
                                                                 &  FIR 4      &  34.5$\pm$ 37.9      &  (1.5$\pm$1.3)$\times$10$^{13}$                          & \multicolumn{2}{c}{\nodata}  &    & (3.4$\pm$2.8)$\times$10$^{-11}$  & (1.5$\pm$1.3)$\times$10$^{-10}$ \\
\hline
                               & FIR 3N    &  26.0$\pm$21.3     & (1.9$\pm$2.3)$\times$10$^{14}$       & \multicolumn{2}{c}{\nodata}   & & (1.6$\pm$1.9)$\times$10$^{-9}$  &   (9.0$\pm$0.1)$\times$10$^{-9}$ \\
CH$_3$OH   A       & FIR 3       &  57.7$\pm$11.7     & (4.0$\pm$1.8)$\times$10$^{15}$       & \multicolumn{2}{c}{\nodata}   & & (0.2$\pm$ 0.1)$\times$10$^{-7}$ &   (6.1$\pm$2.3)$\times$10$^{-8}$  \\
                               & FIR 4      &  46.2$\pm$6.9       & (3.1$\pm$1.2)$\times$10$^{15}$       & 66.2$\pm$6.6                         & (7.0$\pm$1.4)$\times$10$^{15}$ & (1.0$\pm$0.2)$\times$10$^{16}$ & (2.2$\pm$0.4)$\times$10$^{-8}$ &  (1.0$\pm$0.2)$\times$10$^{-7}$\\
\hline
                               & FIR 3N    &  \nodata               & (3.4$\pm$2.9)$\times$10$^{14}$ ($T$ = 20 K)     & \multicolumn{4}{c}{\nodata} & \\
CH$_3$OH   E       & FIR 3      &  32.4$\pm$11.2    & (2.9$\pm$1.7)$\times$10$^{15}$                          & \multicolumn{2}{c}{\nodata} & & (1.1$\pm$0.6)$\times$10$^{-8}$ &  (4.4$\pm$2.6)$\times$10$^{-8}$  \\
                               & FIR 4      &  25.0$\pm$3.2      & (2.4$\pm$0.8)$\times$10$^{15}$                          & 46.0$\pm$12.3          & (5.3$\pm$2.2)$\times$10$^{15}$ & (7.7$\pm$2.3)$\times$10$^{15}$ & (1.7$\pm$0.5)$\times$10$^{-8}$  & (7.7$\pm$2.3)$\times$10$^{-8}$ \\                   
\hline
                                     & FIR 3N    &  \nodata              &  $<$ 2.3$\times$10$^{14}$ ($T$ = 40 K)               & \multicolumn{4}{c}{\nodata} & \\
CH$_3$CCH (ortho)    & FIR 3       &  \nodata              & (5.6$\pm$2.5)$\times$10$^{14}$  ($T$ = 40 K)    & \multicolumn{4}{c}{\nodata}  & \\
                                    & FIR 4       &  32.7$\pm$3.5    & (6.4 $\pm$1.0)$\times$10$^{14}$                          & \multicolumn{2}{c}{\nodata} & & (1.4$\pm$0.2)$\times$10$^{-9}$  & (6.4$\pm$1.0)$\times$10$^{-9}$ \\
\hline
                                                                    &  FIR 3N   &  \nodata                 &  $<$2.9$\times$10$^{14}$ ($T$ = 40 K)                  & \multicolumn{4}{c}{\nodata} & \\
CH$_3$CCH (para)$^{\dagger\dagger}$    & FIR 3      &  \nodata                 &  (5.4$\pm$3.1)$\times$10$^{14}$ ($T$ = 40 K)       & \multicolumn{4}{c}{\nodata} & \\
                                                                    & FIR 4      &  54.1$\pm$40.0     &  (8.6$\pm$4.1)$\times$10$^{14}$                            & \multicolumn{2}{c}{\nodata} & & (1.9$\pm$0.9)$\times$10$^{-9}$  & (8.6$\pm$4.1)$\times$10$^{-9}$\\
\hline
\enddata
\label{rotation_results}
\tablenotetext{*}{To estimated $T_{\rm rot}$, $N_{\rm mol}$, and $X_{\rm mol}$, we assumed the source size of 17$\arcsec$ for FIR 3 and 19$\arcsec$ for FIR 4 and $f$=1 for FIR 3N. }
\tablenotetext{\dagger}{To estimate the rotation temperature, $T_{\rm rot}$ and column density, $N_{\rm mol}$,  we assumed that the source size is equal to the minimum beam size in the detected transitions of each molecule.}
\tablenotetext{\ddagger}{To estimate the abundance relative to H$_2$, $X_{\rm mol}$ (= $N_{\rm mol}$/$N_{\rm H_2}$), we adopted $N_{\rm H_2}$ derived from the AzTEC 1.1-mm dust continuum data.} 
\tablenotetext{\dagger \dagger}{The fitting accuracy in the rotation diagram is poor due to the narrow $E_{\rm u}$ range ($\Delta E_{\rm u}$ $<$ 30 K (see Figs. \ref{ch3cn_rotation} and \ref{ch3cch_rotation}).}
\tablenotetext{\ddagger \ddagger}{ Large uncertainties in the Gaussian fitting due to the accuracy on the Gaussian fitting (see Fig. \ref{ch3cn}).}
\tablenotetext{**}{To estimate the values for CH$_3$CN at FIR 3, we used line parameters for CH$_3$CN ($J_K$=1$_0$--2$_0$)}
\end{deluxetable}

%% file: Table8.tex
\begin{deluxetable}{lcccccccc}
\tablecolumns{9}
\rotate
\tabletypesize{\tiny}
\tablecaption{Rotational Temperatures, Column Densities, and Fractional Abundances of the Detected Molecules at FIR 3N, 3, and 4, which are estimated on the assumption of $f$=1}
\tablewidth{0pt}
\tablehead{
 \colhead{Species}
&  \colhead{Position}
& \multicolumn{2}{c}{$Narrow \  comp.$}
& \multicolumn{2}{c}{$Wide \  comp.$} \\
\cline{3-9} \\
\colhead{}
& \colhead{}
& \colhead{$T_{\rm rot}$ [K]$^{\dagger}$}
& \colhead{$N_{\rm mol}$ [cm$^{-2}$]$^{\ddagger}$}
& \colhead{$T_{\rm rot}$ [K]$^{\dagger}$}
& \colhead{$N_{\rm mol}$ [cm$^{-2}$]$^{\dagger}$}
& \colhead{$N_{\rm mol}$ total}
& \colhead{$X_{\rm mol}$$^{\ddagger}$ ($T_{\rm d}$ = $T_{\rm rot}$ (H$^{13}$CO$^{+}$))}
& \colhead{$X_{\rm mol}$$^{\ddagger}$ ($T_{\rm d}$ = $T_{\rm CO\ peak}$)}
}
\startdata
                                & FIR 3N &  \nodata           & $<$1.5$\times$10$^{12}$  ($T$ = 10 K) & \multicolumn{4}{c}{\nodata} &\\
C$^{34}$S               & FIR 3   &  16.0$\pm$4.1 & (4.4$\pm$1.3)$\times$10$^{12}$  & \multicolumn{2}{c}{\nodata} & & (1.6$\pm$0.5)$\times$10$^{-11}$& (6.7$\pm$2.0)$\times$10$^{-11}$ \\
                                & FIR 4   &  10.1$\pm$2.3 & (5.5$\pm$1.1)$\times$10$^{12}$  &  23.6$\pm$12.9 & (5.6$\pm$4.3)$\times$10$^{12}$ &  (1.1$\pm$0.4)$\times$10$^{13}$   &  (2.4$\pm$0.9)$\times$10$^{-11}$& (1.1$\pm$0.4)$\times$10$^{-10}$ \\
\hline
                                & FIR 3N &  \nodata             & $<$9.4$\times$10$^{11}$ ($T$ = 20 K) & \multicolumn{4}{c}{\nodata}& \\
SiO                          & FIR 3   &  \nodata             & $<$1.0$\times$10$^{12}$ ($T$ = 20 K) & \multicolumn{4}{c}{\nodata} & \\
                                & FIR 4   &  18.8$\pm$ 4.6  & (1.8$\pm$0.8)$\times$10$^{12}$    & 23.6 $\pm$6.3 & (8.5$\pm$2.9)$\times$10$^{12}$ & (1.0$\pm$0.3)$\times$10$^{13}$ & (2.2$\pm$0.7)$\times$10$^{-11}$& (1.0$\pm$0.3)$\times$10$^{-10}$ \\
\hline
                                & FIR 3N &  \nodata           &  (2.0$\pm$0.4)$\times$10$^{13}$ ($T$ = 20 K) & \multicolumn{4}{c}{\nodata}& \\
SO                          & FIR 3    &  20.9$\pm$1.6 &  (3.7$\pm$0.8)$\times$10$^{13}$                      & \multicolumn{2}{c}{\nodata} & & (1.4$\pm$0.3)$\times$10$^{-10}$ & (5.6$\pm$1.2)$\times$10$^{-10}$ \\
                               & FIR 4    &  20.0$\pm$2.7 &  (2.6$\pm$1.1)$\times$10$^{13}$                      &  36.1$\pm$5.1 & (9.1$\pm$2.2)$\times$10$^{13}$ &   (1.2$\pm$0.2)$\times$10$^{14}$ &(2.7$\pm$0.4)$\times$10$^{-11}$& (1.2$\pm$0.2)$\times$10$^{-9}$ \\
\hline
                                  & FIR 3N &  11.5$\pm$1.5   & (9.2$\pm$3.0)$\times$10$^{14}$   & \multicolumn{2}{c}{\nodata} & & (7.5$\pm$2.5)$\times$10$^{-9}$& (4.4$\pm$ 1.4)$\times$10$^{-8}$  \\
 C$_2$H                    & FIR 3   &  13.7$\pm$0.5   & (2.5$\pm$0.2)$\times$10$^{15}$   & \multicolumn{2}{c}{\nodata} & & (9.4$\pm$0.7)$\times$10$^{-9}$& (3.8$\pm$0.3)$\times$10$^{-8}$\\
                                  & FIR 4   &  11.8$\pm$0.9   & (2.5$\pm$0.5)$\times$10$^{15}$   & \multicolumn{2}{c}{\nodata} & & (5.6$\pm$1.1)$\times$10$^{-9}$ & (2.5$\pm$0.5)$\times$10$^{-8}$\\
\hline
                                & FIR 3N  &  \nodata              & $<$4.3$\times$10$^{12}$ ($T$ = 10 K) & \multicolumn{4}{c}{\nodata} &\\
H$^{13}$CN            & FIR 3     &  11.7$\pm$0.5    & (1.8$\pm$0.2)$\times$10$^{13}$           & \multicolumn{2}{c}{\nodata} & & (6.7$\pm$0.6)$\times$10$^{-11}$ & (2.7$\pm$0.3)$\times$10$^{-10}$ \\
                                & FIR 4     &    8.4$\pm$0.3    & (1.8$\pm$0.1)$\times$10$^{13}$           & 19.0$\pm$0.3 & (15.9$\pm$0.3) $\times$ 10$^{12}$ & (3.4$\pm$0.1)$\times$10$^{13}$ & (7.6$\pm$0.2)$\times$10$^{-11}$ & (3.4$\pm$0.1)$\times$10$^{-10}$ \\
 \hline
                                & FIR 3N  &  \nodata             & $<$3.8$\times$10$^{11}$ ($T$ = 10 K) & \multicolumn{4}{c}{\nodata}& \\
HC$^{15}$N            & FIR 3     &  10.0$\pm$2.2  & (1.1$\pm$0.7)$\times$10$^{12}$ & \multicolumn{2}{c}{\nodata} & & (0.4$\pm$0.3)$\times$10$^{-11}$ & (1.7$\pm$1.1)$\times$10$^{-11}$ \\                         
                                & FIR 4    &    9.2$\pm$2.3   & (1.5$\pm$0.5)$\times$10$^{12}$ & \multicolumn{2}{c}{\nodata} & & (0.3$\pm$0.1)$\times$10$^{-11}$ & (1.5$\pm$0.5)$\times$10$^{-11}$\\
  \hline
                                 & FIR 3N &  10.2$\pm$2.4  & (1.1$\pm$0.3)$\times$10$^{12}$    & \multicolumn{2}{c}{\nodata} & & (0.9$\pm$ 0.2)$\times$10$^{-11}$ &(5.2$\pm$ 1.4)$\times$10$^{-11}$ \\
H$^{13}$CO$^+$     & FIR 3    &  13.2$\pm$0.9  & (2.7$\pm$0.3)$\times$10$^{12}$    & \multicolumn{2}{c}{\nodata} & & (1.0$\pm$ 0.1)$\times$10$^{-11}$ & (4.1$\pm$0.5)$\times$10$^{-11}$  \\
                                 & FIR 4    &  12.5$\pm$0.6  & (2.7$\pm$0.2)$\times$10$^{12}$    & \multicolumn{2}{c}{\nodata} & & (6.0$\pm$ 0.4)$\times$10$^{-12}$& (2.7$\pm$0.2)$\times$10$^{-11}$ \\
 \hline
                                & FIR 3N &  \nodata             &  $<$6.7$\times$10$^{11}$ ($T$ = 10 K) & \multicolumn{4}{c}{\nodata}& \\
HN$^{13}$C            & FIR 3   &  10.2$\pm$2.8   & (1.5$\pm$0.4)$\times$10$^{12}$   & \multicolumn{2}{c}{\nodata} & & (0.5$\pm$0.2)$\times$10$^{-11}$ & (2.3$\pm$0.6)$\times$10$^{-11}$ \\
                                & FIR 4   &    9.5$\pm$0.7   & (3.0$\pm$0.3)$\times$10$^{12}$   & \multicolumn{2}{c}{\nodata} & & (0.7$\pm$0.1)$\times$10$^{-11}$& (3.0$\pm$0.3)$\times$10$^{-11}$\\
\hline
                                &  FIR 3N &  \nodata           &  $<$2.3$\times$10$^{12}$ ($T$ = 20 K) & \multicolumn{4}{c}{\nodata}& \\
H$_2$CS (ortho)     & FIR 3    &  22.0$\pm$6.4  & (6.7$\pm$4.4)$\times$10$^{12}$           & \multicolumn{2}{c}{\nodata} & & (2.5$\pm$1.6)$\times$10$^{-11}$& (1.0$\pm$0.7)$\times$10$^{-10}$ \\
                                & FIR 4    &  19.4$\pm$4.3  & (7.6$\pm$2.8)$\times$10$^{12}$           & \multicolumn{2}{c}{\nodata} & & (1.7$\pm$0.6)$\times$10$^{-11}$& (7.6$\pm$2.8)$\times$10$^{-11}$\\ 
\hline
                                &  FIR 3N &  \nodata            &  $<$3.3$\times$10$^{12}$ ($T$ = 40 K) & \multicolumn{4}{c}{\nodata} &\\
H$_2$CS (para)      & FIR 3    &  39.4$\pm$3.5  & (1.1$\pm$0.2)$\times$10$^{13}$            & \multicolumn{2}{c}{\nodata} & & (4.1$\pm$0.7)$\times$10$^{-11}$& (1.7$\pm$0.3)$\times$10$^{-10}$\\
                                & FIR 4    &  20.2$\pm$0.8  & (2.4$\pm$0.3)$\times$10$^{13}$            & \multicolumn{2}{c}{\nodata} & & (5.4$\pm$0.8)$\times$10$^{-11}$& (2.4$\pm$0.3)$\times$10$^{-10}$ \\
\hline
                                 & FIR 3N  &  \nodata            &  $<$3.5 $\times$10$^{12}$ ($T$ = 40 K)           & \multicolumn{4}{c}{\nodata}& \\
CH$_3$CN (ortho)   & FIR 3    &  \nodata             &  (1.1$\pm$0.5)$\times$10$^{13}$ ($T$ = 40 K) & \multicolumn{4}{c}{\nodata}&  \\ 
                                 & FIR 4    &  54.8$\pm$46.5 &  (1.8$\pm$0.6)$\times$10$^{13}$                      & \multicolumn{2}{c}{\nodata} &  & (0.4$\pm$0.1)$\times$10$^{-10}$& (1.8$\pm$0.6)$\times$10$^{-10}$ \\
\hline
                                                                & FIR 3N &  \nodata & (4.5$\pm$4.1)$\times$10$^{12}$ ($T$ = 40 K) & \multicolumn{4}{c}{\nodata}& \\
CH$_3$CN (para)$^{\dagger\dagger}$  & FIR 3    &  \nodata & (1.1$\pm$0.6)$\times$10$^{13}$ ($T$ = 40 K) & \multicolumn{4}{c}{\nodata}& \\
                                                                & FIR 4   &  34.5$\pm$37.9 & (8.8$\pm$7.2)$\times$10$^{12}$   & \multicolumn{2}{c}{\nodata} & & (2.0$\pm$1.6)$\times$10$^{-11}$& (8.8$\pm$7.2)$\times$10$^{-11}$ \\
\hline
                              & FIR 3N  &  26.0$\pm$21.3 & (1.9$\pm$2.3)$\times$10$^{14}$  & \multicolumn{2}{c}{\nodata} & & (1.6$\pm$1.9)$\times$10$^{-9}$& (9.0$\pm$11.0)$\times$10$^{-9}$ \\
CH$_3$OH   A      & FIR 3     &  55.2$\pm$11.4 & (1.8$\pm$0.9)$\times$10$^{15}$                          & \multicolumn{2}{c}{\nodata} & & (0.7$\pm$0.3)$\times$10$^{-8}$& (2.7$\pm$1.4)$\times$10$^{-8}$ \\
                              & FIR 4    &  44.9$\pm$7.0   & (1.6$\pm$0.7)$\times$10$^{15}$                          &  63.9$\pm$6.7 & (3.5$\pm$0.8)$\times$10$^{15}$ & (5.1$\pm$1.1)$\times$10$^{15}$ & (1.1$\pm$0.2)$\times$10$^{-8}$& (5.1$\pm$1.1)$\times$10$^{-8}$\\
\hline
                              & FIR 3N &  \nodata           & (3.4$\pm$2.9)$\times$10$^{14}$ ($T$ = 20 K) & \multicolumn{4}{c}{\nodata}& \\
CH$_3$OH   E      & FIR 3   &  28.9$\pm$8.9 & (1.3$\pm$0.8)$\times$10$^{15}$  & \multicolumn{2}{c}{\nodata}  & & (0.5$\pm$0.3)$\times$10$^{-8}$& (2.0$\pm$1.2)$\times$10$^{-8}$ \\
                              & FIR 4   &  23.2$\pm$2.9 & (1.3$\pm$0.4)$\times$10$^{15}$  &  41.1$\pm$9.5 & (2.5$\pm$1.0)$\times$10$^{15}$ & (3.8$\pm$1.1)$\times$10$^{15}$ & (8.4$\pm$2.4)$\times$10$^{-9}$& (3.8$\pm$1.1)$\times$10$^{-8}$ \\                   
\hline
                                     &  FIR 3N &  \nodata          &  $<$2.3$\times$10$^{14}$ ($T$ = 40 K)             & \multicolumn{4}{c}{\nodata} &\\
CH$_3$CCH (ortho)    &  FIR 3    &  \nodata          & (2.7$\pm$1.2)$\times$10$^{14}$  ($T$ = 40 K) & \multicolumn{4}{c}{\nodata} &\\
                                    &  FIR 4    &  34.1$\pm$1.6 & (3.9$\pm$0.3)$\times$10$^{14}$                       & \multicolumn{2}{c}{\nodata} &  & (0.9$\pm$0.1)$\times$10$^{-9}$& (3.9$\pm$3.0)$\times$10$^{-9}$ \\
\hline
                                                                    & FIR 3N &  \nodata                                       & $<$2.9$\times$10$^{14}$ ($T$ = 40 K)           & \multicolumn{4}{c}{\nodata}& \\
CH$_3$CCH (para)$^{\dagger\dagger}$    & FIR 3   &  \nodata                                       & (2.6$\pm$1.5) $\times$10$^{14}$ ($T$ = 40 K) & \multicolumn{4}{c}{\nodata}& \\
                                                                    & FIR 4   &  58.9$\pm$62.0  & (5.4$\pm$3.3)$\times$10$^{14}$                       & \multicolumn{2}{c}{\nodata} & & (1.2$\pm$0.7)$\times$10$^{-9}$ & (5.4$\pm$3.3)$\times$10$^{-9}$\\
                                       
                                        \hline
\enddata
\label{rotation_results}
\tablenotetext{\dagger}{To estimate the rotation temperature, $T_{\rm rot}$ and column density, $N_{\rm mol}$,  we assumed $f$=1.}
\tablenotetext{\ddagger}{To estimate the abundance relative to H$_2$, $X_{\rm mol}$ (= $N_{\rm mol}$/$N_{\rm H_2}$), we adopted $N_{\rm H_2}$ derived from the AzTEC 1.1-mm dust continuum data.} 
\tablenotetext{\dagger \dagger}{The fitting accuracy in the rotation diagram is poor due to the narrow $E_{\rm u}$ range ($\Delta E_{\rm u}$ $<$ 30 K (see Figs. \ref{ch3cn_rotation} and \ref{ch3cch_rotation}).}
\tablenotetext{\ddagger \ddagger}{Large uncertainties in the Gaussian fitting due to the accuracy on the Gaussian fitting (see Fig. \ref{ch3cn}).}
\end{deluxetable}

%% file: Table9.tex
\begin{deluxetable}{lcccl}
\tablecolumns{5}
\tabletypesize{\scriptsize}
\tablecaption{Relationship between the physical environment and the chemical composition}
\tablewidth{0pt}
\tablehead{
\colhead{}
& \colhead{FIR 3N}
& \colhead{FIR 3}
& \colhead{FIR 4}
& \colhead{Molecules}
}
\startdata
\multirow{2}{*}{Ambient dense gas}   &      \multirow{2}{*}{Narrow}      &      \multirow{2}{*}{Narrow}       &   \multirow{2}{*}{Narrow}  & CN, C$^{17}$O, C$_2$H, HNC, HN$^{13}$C,  \\
                                                            & & & & H$^{13}$CO$^+$, N$_2$H$^+$, c-C$_3$H$_2$, CH$_3$CN \\
\hline
\multirow{4}{*}{{\bf Possible} ambient dense gas}   &       non-detection         &      Narrow             &   Narrow            & $^{13}$CS, CH$_3$CCH\\
                                                           &       non-detection         &      non-detection   &   Narrow           & NH$_2$D \\
                                                           &       Narrow                   &      non-detection   &   Narrow           & H$^{15}$NC  \\
                                                           &       non-detection         &      Narrow             &   non-detection & HC$^{18}$O$^+$ \\

\hline
Outflow                                    &      Narrow and Wide         &     Narrow and Wide     &   Narrow and Wide & CO, CS, HCN, HCO$^+$ \\
\hline
\multirow{2}{*}{Shock}  &        \multirow{2}{*}{Narrow}       &      \multirow{2}{*}{Narrow}       &   \multirow{2}{*}{Narrow and Wide} & C$^{34}$S, SO, SiO, H$^{13}$CN, HC$^{15}$N, \\
                                     &                                                    &                                                 &                                                         & H$_2^{13}$CO, H$_2$CS, HC$_3$N, CH$_3$OH \\

\hline
\multirow{2}{*}{Possible Shock}   &       Narrow                 &     Narrow      &  Narrow (large velocity width)  & HCS$^+$, H$_2$CO, HNCO, CH$_3$CHO   \\
                                                     &      non-detection        &     Narrow       &  Narrow (large velocity width) & HDCO    \\
\enddata
\label{phy_che_mol}
\end{deluxetable}

%% file: Table10.tex
\begin{deluxetable}{lccccccccccccc}
\tablecolumns{5}
\tabletypesize{\tiny}
\tablecaption{Column Densities of the Molecules Detected with Only One Transition at FIR 3N, 3, and 4$^*$}
\tablewidth{0pt}
\tablehead{
 \colhead{Species}
&  \colhead{Position}
&  \colhead{$T$ = 10 K}
&  \colhead{$T$ = 20 K}
&  \colhead{$T$ = 40 K}\\
\\
\colhead{}
& \colhead{}
& \colhead{$N_{\rm mol}$ \tablenotemark{a}}
& \colhead{$N_{\rm mol}$ \tablenotemark{a}}
& \colhead{$N_{\rm mol}$ \tablenotemark{a}}
\\
\colhead{}
& \colhead{}
& \colhead{[cm$^{-2}$]}
& \colhead{[cm$^{-2}$]}
& \colhead{[cm$^{-2}$]}
}
\startdata
C$^{34}$S                   & FIR 3N      & $<$1.5$\times$10$^{12}$                  & $<$2.3$\times$10$^{12}$                   & $<$4.0$\times$10$^{12}$ \\
\hline
SiO                              & FIR 3N      & $<$6.2$\times$10$^{11}$                  & $<$ 9.4$\times$10$^{11}$                  & $<$1.6$\times$10$^{12}$ \\
SiO                              & FIR 3         & $<$1.4$\times$10$^{12}$                  & $<$ 2.0$\times$10$^{12}$                 & $<$3.6$\times$10$^{12}$ \\
\hline
SO                               & FIR 3N      & (1.2$\pm$0.3)$\times$10$^{13}$       & (2.0$\pm$0.4)$\times$10$^{13}$        & (3.7$\pm$0.8)$\times$10$^{13}$ \\
\hline
H$^{13}$CN                & FIR 3N      & $<$4.3$\times$10$^{12}$                   & $<$7.6$\times$10$^{12}$                   & $<$1.4$\times$10$^{13}$ \\
 \hline
HC$^{15}$N                & FIR 3N      & $<$3.8$\times$10$^{11}$                   & $<$6.6$\times$10$^{11}$                   & $<$1.2$\times$10$^{12}$ \\
 \hline
HN$^{13}$C                & FIR 3N      & $<$6.7$\times$10$^{11}$                   & $<$1.2$\times$10$^{12}$                   & $<$2.2$\times$10$^{12}$ \\
 \hline
H$_2$CS (ortho)         & FIR 3N     & $<$1.6$\times$10$^{12}$                    & $<$2.3$\times$10$^{12}$                   & $<$4.9$\times$10$^{12}$ \\
\hline
H$_2$CS (para)          & FIR 3N     & $<$1.6$\times$10$^{12}$                    & $<$2.0$\times$10$^{12}$                   & $<$3.3$\times$10$^{12}$ \\
\hline
CH$_3$CN (ortho)       & FIR 3N     & $<$1.7$\times$10$^{12}$                   & $<$1.9$\times$10$^{12}$                    & $<$3.5$\times$10$^{12}$ \\
CH$_3$CN (ortho)       & FIR 3        & (5.1$\pm$2.2)$\times$10$^{12}$        & (5.7$\pm$2.4)$\times$10$^{12}$        & (1.1$\pm$0.5)$\times$10$^{13}$ \\
\hline
CH$_3$CN (para)        & FIR 3N     & (20.0$\pm$19.6)$\times$10$^{ 11}$    & (25.2$\pm$24.7)$\times$10$^{11}$    & (45.0$\pm$44.1)$\times$10$^{11}$ \\
CH$_3$CN (para)        & FIR 3       & (9.3$\pm$4.8)$\times$10$^{12}$         & (1.2$\pm$0.6)$\times$10$^{13}$        & (2.1$\pm$1.1)$\times$10$^{13}$ \\
\hline
CH$_3$OH E               & FIR 3N    & (3.7$\pm$3.2)$\times$10$^{14}$         & (3.4$\pm$2.9)$\times$10$^{14}$         & (6.3$\pm$5.4)$\times$10$^{14}$ \\
\hline
CH$_3$CCH (ortho)    & FIR 3N    & $<$1.0$\times$10$^{14}$                     & $<$1.2$\times$10$^{14}$                    & $<$ 2.3$\times$10$^{14}$ \\
CH$_3$CCH (ortho)    & FIR 3       & (2.5$\pm$1.1)$\times$10$^{14}$         & (2.9$\pm$1.3)$\times$10$^{14}$         & (5.6$\pm$2.5)$\times$10$^{14}$ \\
\hline
CH$_3$CCH (para)     & FIR 3N     & $<$ 1.2$\times$10$^{14}$                   & $<$ 1.6$\times$10$^{14}$                   & $<$ 2.9$\times$10$^{14}$ \\
CH$_3$CCH (para)     & FIR 3       & (2.3$\pm$1.3)$\times$10$^{14}$         & (3.0$\pm$1.7)$\times$10$^{14}$         & (5.4$\pm$3.1)$\times$10$^{14}$ \\
\hline
                                     & FIR 3N    & (6.2$\pm$0.4)$\times$10$^{15}$         & (2.7$\pm$0.2)$\times$10$^{15}$        & (2.5$\pm$0.2)$\times$10$^{15}$ \\
CN \tablenotemark{b}   &FIR 3       & (3.7$\pm$0.1)$\times$10$^{16}$         & (16.0$\pm$0.4)$\times$10$^{15}$      & (14.8$\pm$0.4)$\times$10$^{15}$ \\              
                                     & FIR 4       & (30.9$\pm$0.7)$\times$10$^{15}$       & (13.3$\pm$0.3)$\times$10$^{15}$     & (12.4$\pm$0.3)$\times$10$^{15}$ \\
\hline
                                     & FIR 3N    & (1.0$\pm$0.1)$\times$10$^{16}$         & (4.5$\pm$0.4)$\times$10$^{15}$      & (4.2$\pm$0.3)$\times$10$^{15}$ \\
C$^{17}$O                    & FIR 3      & (3.2$\pm$0.4)$\times$10$^{16}$         & (1.4$\pm$0.2)$\times$10$^{16}$      & (1.3$\pm$0.1)$\times$10$^{16}$ \\                 
                                     & FIR 4       & (3.7$\pm$0.2)$\times$10$^{16}$        & (1.6$\pm$0.1)$\times$10$^{16}$      & (1.5$\pm$0.1)$\times$10$^{16}$ \\
\hline
                                    & FIR 3N     & $<$ 2.7$\times$10$^{12}$                   & $<$4.2$\times$10$^{12}$                & $<$7.4$\times$10$^{12}$ \\
$^{13}$CS                   & FIR 3        & (1.1$\pm$0.7)$\times$10$^{13}$        & (1.6$\pm$1.1)$\times$10$^{13}$     & (2.9$\pm$1.9)$\times$10$^{13}$ \\                
                                    & FIR 4        & (6.6$\pm$3.2)$\times$10$^{12}$        & (1.0$\pm$0.5)$\times$10$^{13}$     & (1.8$\pm$0.9)$\times$10$^{13}$ \\
\hline
                                    & FIR 3N     & $<$9.8$\times$10$^{13}$                & $<$5.1$\times$10$^{12}$              & $<$1.6$\times$10$^{12}$ \\
HCS$^+$                     & FIR 3       & $<$2.2$\times$10$^{14}$                & $<$1.1$\times$10$^{13}$              & $<$3.7$\times$10$^{12}$ \\             
                                    & FIR 4       & (2.2$\pm$1.9)$\times$10$^{14}$     & (1.1$\pm$1.0)$\times$10$^{13}$   & (3.6$\pm$3.1)$\times$10$^{12}$ \\
\hline
                                                            &  FIR 3N   & (1.0$\pm$0.2)$\times$10$^{14}$    & (1.8$\pm$0.3)$\times$10$^{14}$     & (3.3$\pm$0.6)$\times$10$^{14}$ \\
N$_2$H$^+$ \tablenotemark{c}          &  FIR 3     & (3.7$\pm$0.3)$\times$10$^{14}$     & (6.3$\pm$0.5)$\times$10$^{14}$     & (1.2$\pm$0.1)$\times$10$^{15}$ \\      
                                                            &  FIR 4     & (12.2$\pm$0.2)$\times$10$^{14}$   & (21.2$\pm$0.3)$\times$10$^{14}$   & (39.4$\pm$0.5)$\times$10$^{14}$ \\
\hline
                                     & FIR 3N  & $<$5.9$\times$10$^{11}$                  & $<$1.6$\times$10$^{11}$                & $<$1.2$\times$10$^{11}$ \\
HC$^{18}$O$^+$         & FIR 3     & (1.6$\pm$1.5)$\times$10$^{12}$      & (4.6$\pm$4.1)$\times$10$^{11}$     & (3.4$\pm$3.0)$\times$10$^{11}$ \\                 
                                     & FIR 4     & $<$1.2$\times$10$^{12}$                 & $<$3.3$\times$10$^{11}$                 & $<$2.5$\times$10$^{11}$ \\
\hline
                                    & FIR 3N     & (4.0$\pm$3.8)$\times$10$^{11}$       & (7.0$\pm$6.7)$\times$10$^{11}$    & (13.1$\pm$12.5)$\times$10$^{11}$ \\
H$^{15}$NC                & FIR 3        & $<$8.2$\times$10$^{11}$                  & $<$1.4$\times$10$^{12}$               & $<$2.7$\times$10$^{12}$ \\
                                    & FIR 4        & (1.2$\pm$0.5)$\times$10$^{12}$      & (2.1$\pm$0.9)$\times$10$^{12}$    & (3.9$\pm$1.7)$\times$10$^{12}$ \\
\hline
                                   & FIR 3N      & $<$6.1$\times$10$^{15}$                 & $<$2.9$\times$10$^{14}$                & $<$9.7$\times$10$^{13}$ \\
H$_2$CO (para)        & FIR 3N      & $<$1.1$\times$10$^{16}$                 & $<$5.2$\times$10$^{14}$                & $<$1.7$\times$10$^{14}$ \\               
                                   & FIR 4        & (2.8$\pm$0.9)$\times$10$^{16}$      & (1.4$\pm$0.4)$\times$10$^{15}$     & (4.5$\pm$1.4)$\times$10$^{14}$ \\
\hline
                                         & FIR 3N & (2.7$\pm$1.4)$\times$10$^{13}$   & (4.8$\pm$2.6)$\times$10$^{12}$      & (3.0$\pm$1.6)$\times$10$^{12}$ \\
H$_2$$^{13}$CO (para)  & FIR 3    & (3.6$\pm$3.5)$\times$10$^{13}$   & (64.0$\pm$63.8)$\times$10$^{11}$  & (40.4$\pm$40.3)$\times$10$^{11}$ \\          
                                         & FIR 4    & (2.7$\pm$1.7)$\times$10$^{13}$   & (4.9$\pm$3.0)$\times$10$^{12}$      & (3.1$\pm$1.9)$\times$10$^{12}$ \\
%
\hline
                            & FIR 3N     & $<$5.5$\times$10$^{12}$              & $<$2.4$\times$10$^{12}$                 & $<$2.7$\times$10$^{12}$ \\
HDCO                 & FIR 3        & $<$1.8$\times$10$^{13}$              & $<$8.0$\times$10$^{12}$                 & $<$8.9$\times$10$^{12}$ \\          
                            & FIR 4       & (1.8$\pm$1.3)$\times$10$^{13}$   & (8.2$\pm$6.0)$\times$10$^{12}$      & (9.1$\pm$6.6)$\times$10$^{12}$ \\
\hline
                           & FIR 3N     & $<$4.3$\times$10$^{12}$               & $<$6.2$\times$10$^{12}$                & $<$1.3$\times$10$^{13}$ \\
HNCO                & FIR 3        & (8.8$\pm$4.4)$\times$10$^{12}$    & (1.3$\pm$0.6)$\times$10$^{13}$    & (2.7$\pm$1.3)$\times$10$^{13}$ \\
                           & FIR 4       & (1.2$\pm$0.6)$\times$10$^{13}$     & (1.8$\pm$0.9)$\times$10$^{13}$   & (3.8$\pm$1.9)$\times$10$^{13}$ \\
\hline
                                                   & FIR 3N   & $<$5.6$\times$10$^{13}$           & $<$5.5$\times$10$^{13}$              & $<$9.2$\times$10$^{13}$ \\
NH$_2$D \tablenotemark{d}      & FIR 3     & $<$5.9$\times$10$^{13}$            & $<$5.8$\times$10$^{13}$              & $<$9.7$\times$10$^{13}$ \\               
                                                   & FIR 4     & (4.6$\pm$0.5)$\times$10$^{14}$ & (4.5$\pm$0.5)$\times$10$^{14}$   & (7.4$\pm$0.7)$\times$10$^{14}$ \\
\hline
                           & FIR 3N   & (44.5$\pm$44.3)$\times$10$^{11}$       & (90.1$\pm$89.7)$\times$10$^{11}$   & (20.8$\pm$20.7)$\times$10$^{12}$ \\
c-C$_3$H$_2$   & FIR 3     & (1.6$\pm$0.4)$\times$10$^{13}$           & (3.3$\pm$0.7)$\times$10$^{13}$       & (7.5$\pm$1.7)$\times$10$^{13}$ \\                
                           & FIR 4     & (2.2$\pm$0.4)$\times$10$^{13}$           & (4.5$\pm$0.7)$\times$10$^{13}$       & (1.0$\pm$0.2)$\times$10$^{14}$ \\
\hline
                                                & FIR 3N   & (5.0$\pm$1.7)$\times$10$^{12}$     & (3.0$\pm$1.0)$\times$10$^{12}$    & (3.3$\pm$1.1)$\times$10$^{12}$ \\
HC$_3$N \tablenotemark{e}   & FIR 3      & (8.2$\pm$0.5)$\times$10$^{13}$     & (5.0$\pm$0.3)$\times$10$^{13}$    & (5.5$\pm$0.3)$\times$10$^{13}$ \\                   
                                                & FIR 4      & (8.2$\pm$0.2)$\times$10$^{13}$     & (5.0$\pm$0.1)$\times$10$^{13}$    & (5.5$\pm$0.2)$\times$10$^{13}$ \\
\hline
                                                     & FIR 3N   & $<$6.4$\times$10$^{13}$              & $<$8.2$\times$10$^{13}$                & $<$1.5$\times$10$^{14}$ \\
CH$_3$CHO \tablenotemark{f}    & FIR 3     & (5.6$\pm$3.9)$\times$10$^{13}$   & (7.1$\pm$5.0)$\times$10$^{13}$     & (1.3$\pm$0.9)$\times$10$^{14}$ \\                    
                                                     & FIR 4     & (1.1$\pm$0.7)$\times$10$^{14}$   & (1.4$\pm$0.9)$\times$10$^{14}$     & (2.6$\pm$1.8)$\times$10$^{14}$ \\
\enddata
\label{LTE_results}
\tablenotetext{*}{To estimated $T_{\rm rot}$, $N_{\rm mol}$, and $X_{\rm mol}$, we assumed the source size of 17$\arcsec$ for FIR 3 and 19$\arcsec$ for FIR 4 and $f$=1 for FIR 3N.} 
\tablenotetext{a}{To estimate column density, $N_{\rm mol}$,  we assumed that $T_{\rm ex}$ = 10, 20, and 40 K.}
\tablenotetext{b}{Values are estimated from the CN (3--2 7/2--5/2 $F$=7/2--5/2, 340.031567 GHz).  }
\tablenotetext{c}{Values are estimated from the N$_2$H$^+$ (1--0 $F_1$=2--1, $F_2$=2--1, 93.17348 GHz).  }
\tablenotetext{d}{Values are estimated from the NH$_2$D (1--0, 85.926263 GHz).}
\tablenotetext{e}{Values are estimated from the HC$_3$N (10--9, 90.978989 GHz).}
\tablenotetext{f}{Values are estimated from the CH$_3$CHO (5$_{1,5}$--4$_{1,4}$, 93.580914 GHz).}
\end{deluxetable}

%% file: Table11.tex
\begin{deluxetable}{lcccccccc}
\tablecolumns{9}
\rotate
\tabletypesize{\tiny}
\tablecaption{Rotational Temperatures, Column Densities, and Fractional Abundances Corrected for the Beam Dilution Effect at FIR 4}
\tablewidth{0pt}
\tablehead{
 \colhead{Species}
&  \colhead{Position}
& \multicolumn{2}{c}{$Narrow \  comp.$}
& \multicolumn{2}{c}{$Wide \  comp.$} \\
\cline{3-9} \\
\colhead{}
& \colhead{}
& \colhead{$T_{\rm rot}$ [K]$^{\dagger}$}
& \colhead{$N_{\rm mol}$ [cm$^{-2}$]$^{\dagger}$}
& \colhead{$T_{\rm rot}$ [K]$^{\dagger}$}
& \colhead{$N_{\rm mol}$ [cm$^{-2}$]$^{\dagger}$}
& \colhead{$N_{\rm mol}$ total}
& \colhead{$X_{\rm mol}$$^{\ddagger}$ ($T_{\rm d}$ = $T_{\rm rot}$ (H$^{13}$CO$^{+}$))}
& \colhead{$X_{\rm mol}$$^{\ddagger}$ ($T_{\rm d}$ = $T_{\rm CO\ peak}$)}
}
\startdata
C$^{34}$S             & FIR 4 &  11.4$\pm$ 2.9  & (1.6$\pm$0.3)$\times$10$^{14}$ & 32.0$\pm$23.8 & (2.0$\pm$1.5)$\times$10$^{14}$ & (3.6$\pm$1.5)$\times$10$^{14}$ & (8.0$\pm$3.3)$\times$10$^{-10}$ & (3.6$\pm$1.5)$\times$10$^{-9}$\\
SiO                        & FIR 4 &  20.0$\pm$ 5.2  & (6.4$\pm$2.8)$\times$10$^{13}$ & 25.6$\pm$7.4   & (3.1$\pm$1.1)$\times$10$^{14}$ & (3.7$\pm$1.1)$\times$10$^{14}$ & (8.2$\pm$2.4)$\times$10$^{-10}$  &  (3.7$\pm$1.1)$\times$10$^{-9}$ \\
SO                         & FIR 4 &  22.4$\pm$ 3.0  & (8.5$\pm$3.0)$\times$10$^{14}$ & 45.0$\pm$5.1   & (3.3$\pm$0.5)$\times$10$^{15}$ & (1.2$\pm$0.3)$\times$10$^{15}$ & (2.7$\pm$0.7)$\times$10$^{-9}$ &  (1.2$\pm$0.3)$\times$10$^{-8}$ \\
H$^{13}$CN          & FIR 4 &  8.9$\pm$ 0.3    & (6.3$\pm$0.5)$\times$10$^{14}$ & 21.5$\pm$0.4   & (6.2$\pm$0.1)$\times$10$^{14}$ & (12.5$\pm$0.5)$\times$10$^{14}$ & (2.8$\pm$0.1)$\times$10$^{-9}$ &  (12.5$\pm$0.5)$\times$10$^{-9}$ \\
HC$^{15}$N          & FIR 4 &  9.8$\pm$ 2.6    & (5.6$\pm$1.8)$\times$10$^{13}$ & \multicolumn{2}{c}{\nodata} && (1.2$\pm$0.4)$\times$10$^{-10}$ &  (5.6$\pm$1.8)$\times$10$^{-10}$ \\
H$_2$CS (ortho)   & FIR 4 &  22.6$\pm$ 5.8  & (2.2$\pm$0.8)$\times$10$^{14}$ & \multicolumn{2}{c}{\nodata} & & (4.9$\pm$1.8)$\times$10$^{-10}$ &  (2.2$\pm$0.8)$\times$10$^{-9}$ \\
H$_2$CS (para)    & FIR 4 & 23.6$\pm$ 0.9   & (6.7$\pm$0.8)$\times$10$^{14}$ & \multicolumn{2}{c}{\nodata} & & (1.5$\pm$0.2)$\times$10$^{-9}$ &  (6.7$\pm$0.8)$\times$10$^{-9}$ \\
CH$_3$OH A         & FIR 4 &  47.7$\pm$ 6.7  & (6.1$\pm$2.2)$\times$10$^{16}$ & 69.1$\pm$6.9   & (1.4$\pm$0.3)$\times$10$^{17}$ & (2.0$\pm$0.4)$\times$10$^{17}$ & (4.4$\pm$0.9)$\times$10$^{-7}$ &  (2.0$\pm$0.4)$\times$10$^{-6}$\\
CH$_3$OH E        & FIR 4 &  27.4$\pm$ 3.7  & (4.6$\pm$1.4)$\times$10$^{16}$ &  52.8$\pm$16.9 & (1.2$\pm$0.5)$\times$10$^{17}$ & (1.7$\pm$0.5)$\times$10$^{17}$ & (3.8$\pm$1.1)$\times$10$^{-7}$ & (1.7$\pm$0.5)$\times$10$^{-6}$\\
\hline
\enddata
\label{rotation_results_beam}
\tablenotetext{\dagger}{To correct for the beam dilution effect,  we assumed that the source size is 3$\arcsec$.}
\tablenotetext{\ddagger}{To estimate the abundance relative to H$_2$, $X_{\rm mol}$ (= $N_{\rm mol}$/$N_{\rm H_2}$), we adopted $N_{\rm H_2}$ derived from the AzTEC 1.1-mm dust continuum data.} 
\end{deluxetable}

%% file: TableA1.tex
\begin{deluxetable}{lcccccc}
\tablecolumns{7}
\tabletypesize{\scriptsize}
\tablecaption{Results of HFS fitting}
\tablewidth{0pt}
\tablehead{
\colhead{Molecule}
& \colhead{Target}
& \colhead{$T_{\rm ant}$ $\times$ $\tau$}
& \colhead{$V_{\rm sys}$ [km/s]}
& \colhead{$dV_{\rm FWHM}$ [km/s]}
& \colhead{$\tau_{\rm main}$}
& \colhead{$T_{\rm ex}$ [K]}
}
\startdata
N$_2$H$^+$ (1--0)            &       FIR 3N      &       2.4$\pm$0.4       &   11.10$\pm$0.04 & 1.72$\pm$0.16 & 0.82$\pm$0.5  & 16.4$\pm$15.3  \\
N$_2$H$^+$ (1--0)            &       FIR 3         &      6.1$\pm$0.5       &   11.31$\pm$0.01 & 1.57$\pm$0.07 & 1.4$\pm$0.4  & 14.3$\pm$2.8  \\
N$_2$H$^+$ (1--0)            &       FIR 4         &      23.0$\pm$0.3     &   11.37$\pm$0.01 & 1.49$\pm$0.01 & 2.4$\pm$0.1  & 26.4$\pm$0.5  \\
NH$_2$D (1--0)                 &       FIR 4         &      1.4$\pm$0.2       &  10.90$\pm$0.14 & 1.58$\pm$0.15 & 1.5$\pm$0.6 & 6.1$\pm$4.9 \\
\enddata
\label{HFS_result}
\end{deluxetable}

%% file: Table12.tex
\begin{deluxetable}{lccccccc}
\tablecolumns{6}
\tabletypesize{\tiny}
\tablecaption{Parameters of the Molecular Lines Detected in FIR 3N}
\tablewidth{0pt}
\tablehead{
\colhead{}
& \colhead{}
& \multicolumn{3}{c}{$Narrow \ Component$}
& \multicolumn{3}{c}{$Wide \ Component$}\\
\cline{3-8}\\
\colhead{Molecule with Transition}
& \colhead{Rest Freq.}
& \colhead{$V_{\rm sys}$}
& \colhead{$T_{\rm peak}$}
& \colhead{$dV_{\rm FWHM}$}
& \colhead{$V_{\rm sys}$}
& \colhead{$T_{\rm peak}$}
& \colhead{$dV_{\rm FWHM}$}\\
\colhead{}
& \colhead{[GHz]}
& \colhead{[km s$^{-1}$]}
& \colhead{[K]}
& \colhead{[km s$^{-1}$]}
& \colhead{[km s$^{-1}$]}
& \colhead{[K]}
& \colhead{[km s$^{-1}$]}
}
\startdata
c-C$_3$H$_2$ (2$_{1,2}$--1$_{0,1}$) &  85.338906 & \nodata  & 0.123$^d$  & \nodata & \multicolumn{3}{c}{\nodata} \\ 
H$^{13}$CN ($J$=1--0 $F$=1--1) & 86.3387367 & 11.1 & 0.103 & 1.69 & \multicolumn{3}{c}{\nodata} \\
H$^{13}$CN ($J$=1--0 $F$=2--1) & 86.3401764 & 11.1 & 0.137 & 1.69 & \multicolumn{3}{c}{\nodata} \\
H$^{13}$CN ($J$=1--0 $F$=0--1)   & 86.3422551 & 11.1 & 0.027 & 1.69 & \multicolumn{3}{c}{\nodata} \\
H$^{13}$CO$^+$ ($J$=1-0) & 86.754288 & 11.4 & 0.267 & 1.98 & \multicolumn{3}{c}{\nodata} \\
HN$^{13}$C ($J$=1--0) &  87.090859 & 11.2 & 0.080$^a$ & 2.01 & \multicolumn{3}{c}{\nodata} \\
C$_2$H (1--0 3/2--1/2 $F$=1--1) &  87.284156 & 11.1 & 0.161& 0.55 & \multicolumn{3}{c}{\nodata} \\
C$_2$H (1--0 3/2--1/2 $F$=2--1) &  87.316925 & 11.5 & 0.987 & 2.04 & \multicolumn{3}{c}{\nodata} \\
C$_2$H (1--0 3/2--1/2 $F$=1--0) &  87.328624 & 11.3 & 0.541 & 1.80 & \multicolumn{3}{c}{\nodata} \\
C$_2$H (1--0 1/2--1/2 $F$=1--1) &  87.402004 & 11.2 & 0.500 & 2.09 & \multicolumn{3}{c}{\nodata} \\
C$_2$H (1--0 1/2--1/2 $F$=0--1) &  87.407165 & 11.2 & 0.256 & 2.09 & \multicolumn{3}{c}{\nodata} \\
HCN ($J$=1--0 $F$=1-1) &  88.6304157 & 11.0 & 0.754 & 2.60 & 11.1 & 0.041$^{a,c}$  & 13.62  \\
HCN ($J$=1--0 $F$=2-1) & 88.6318473 & 11.0 & 1.833 & 2.60 & 11.1 & 0.100$^a$ & 13.62  \\
HCN ($J$=1--0 $F$=0-1) &  88.633936 & 11.0 & 0.350 & 2.60 & 11.1 & 0.019$^{a,c}$  & 13.62 \\
H$^{15}$NC ($J$=1-0) &  88.865692 & \nodata & 0.055$^d$ & \nodata & \multicolumn{3}{c}{\nodata} \\  
HCO$^+$ ($J$=1--0) &  89.188526 & 11.1 & 1.838 & 2.69 & 9.3 & 0.392 & 20.19  \\
HNC ($J$=1--0) &  90.663564 & 11.2 & 1.573 & 2.25 & \multicolumn{3}{c}{\nodata} \\
HC$_3$N ($J$=10--9) & 90.978989 & 11.3 & 0.381 & 0.54 & \multicolumn{3}{c}{\nodata} \\
CH$_3$CN (J$_{K}$=5$_1$--4$_1$) &  91.985313 & \nodata & 0.082$^d$ & \nodata & \multicolumn{3}{c}{\nodata} \\ 
N$_2$H$^+$ (1--0 $F_1$=1--1 $F$=0--1) &  93.171621 & 10.8 & 0.408 & 2.20 & \multicolumn{3}{c}{\nodata} \\
N$_2$H$^+$ (1--0 $F_1$=1--1 $F$=2--2) &  93.171917 & 10.8 & 0.110 & 2.20 & \multicolumn{3}{c}{\nodata} \\
N$_2$H$^+$ (1--0 $F_1$=1--1 $F$=1--0) &  93.172053 & 10.8 & 0.075 & 2.20 & \multicolumn{3}{c}{\nodata} \\
N$_2$H$^+$ (1--0 $F_1$=2--1 $F$=2--1) &  93.17348 & 10.8 & 0.416 & 2.20 & \multicolumn{3}{c}{\nodata} \\
N$_2$H$^+$ (1--0 $F_1$=2--1 $F$=3--2) &  93.173777 & 10.8 & 0.216 & 2.20 & \multicolumn{3}{c}{\nodata} \\
N$_2$H$^+$ (1--0 $F_1$=2--1 $F$=1--1) &  93.173967 & 10.8 & 0.312 & 2.20 & \multicolumn{3}{c}{\nodata} \\
N$_2$H$^+$ (1--0 $F_1$=0--1 $F$=1--2) &  93.176265 & 10.8 & 0.240 & 2.20 & \multicolumn{3}{c}{\nodata} \\
CH$_3$OH (2$_{0,2}$--1$_{0,1}$A++) &  96.741377 & 11.1 & 0.117 & 3.97 & \multicolumn{3}{c}{\nodata} \\
CH$_3$OH (2$_{1,1}$--1$_{1,0}$E)  & 96.755507 & 11.4 & 0.097 & 1.44 & \multicolumn{3}{c}{\nodata} \\
CS ($J$=2--1) &  97.980953 & 11.7 & 0.744 & 0.89 & 10.9 & 0.673 & 3.31 \\
SO ($N,J$=2,3-1,2) &  99.299905 & 10.9 & 0.154 & 3.20 & \multicolumn{3}{c}{\nodata} \\
HC$_3$N ($J$=11--10) &  100.076386 & 11.0 & 0.235 & 1.28 & \multicolumn{3}{c}{\nodata} \\
H$_2$CS (3$_{1,3}$--2$_{1,2}$)  &  101.477885 & 10.5 & 0.043 & 3.28 & \multicolumn{3}{c}{\nodata} \\
\hline
C$^{17}$O ($J$=3--2) &  337.060974 & 10.8 & 0.781 & 1.88 & \multicolumn{3}{c}{\nodata} \\
CH$_3$OH (7$_{-1,7}$--6$_{-1,6}$ E) &  338.344605 & 11.5 & 0.140 & 2.63 & \multicolumn{3}{c}{\nodata} \\
CH$_3$OH (7$_{0,7}$--6$_{0,6}$ A++) &  338.408718 & 11.2 & 0.103 & 3.52 & \multicolumn{3}{c}{\nodata} \\
CN (3--2 $J$=5/2--3/2 $F$=5/2--5/2) &  340.008097 & 11.3 & 0.023 & 2.48 & \multicolumn{3}{c}{\nodata} \\
CN (3--2 $J$=5/2--3/2 $F$=7/2--5/2) &  340.031567 & 11.3 & 0.374 & 2.48 & \multicolumn{3}{c}{\nodata} \\
CN (3--2 $J$=5/2--3/2 $F$=5/2--3/2) &  340.035525 & 11.3 & 0.286 & 2.48 & \multicolumn{3}{c}{\nodata} \\
CN (3--2 $J$=7/2--5/2 $F$=7/2--5/2) &  340.247625 & 11.1 & 1.020 & 2.48$^b$ & \multicolumn{3}{c}{\nodata} \\
CN (3--2 $J$=7/2--7/2 $F$=5/2--5/2) &  340.261818 & 11.1 & 0.039 & 2.48$^b$ & \multicolumn{3}{c}{\nodata} \\
CN (3--2 $J$=7/2--7/2 $F$=7/2--7/2) &  340.265025 & 11.1 & 0.089 & 2.48$^b$ & \multicolumn{3}{c}{\nodata} \\
CH$_3$OH (7$_{1,6}$--6$_{1,5}$ A--) &  341.415641 & 11.3 & 0.108 & 2.50 & \multicolumn{3}{c}{\nodata} \\
CS ($J$=7--6) &  342.882843 & 11.6 & 0.351 & 2.86 & 11.8 & 0.078 & 11.08  \\
H$_2^{13}$CO (5$_{1,5}$--4$_{1,4}$) &  343.325714 & 10.7 & 0.196 & 2.36 & \multicolumn{3}{c}{\nodata} \\
CO ($J$=3--2) &  345.79599 & 11.7 & 13.9 & 5.61 & 9.2 & 7.400 & 12.22 \\
H$^{13}$CO$^+$ ($J$=4--3) &  346.998352 & 11.4 & 0.126 & 2.00 & \multicolumn{3}{c}{\nodata} \\
C$_2$H (4--3 9/2--7/2 $F$=5--4) &  349.337741 & 11.8 & 0.135 & 2.22 & \multicolumn{3}{c}{\nodata} \\
C$_2$H (4--3 9/2--7/2 $F$=4--3) &  349.339067 & 11.8 & 0.859 & 2.22 & \multicolumn{3}{c}{\nodata} \\
C$_2$H (4--3 7/2--5/2 $F$=4--3) &  349.399342 & 11.1 & 0.516 & 2.02 & \multicolumn{3}{c}{\nodata} \\
C$_2$H (4--3 7/2--5/2 $F$=3--2)  &  349.400692 & 11.1 & 0.249 & 2.02 & \multicolumn{3}{c}{\nodata} \\
CH$_3$OH (4$_{0,4}$--3$_{-1,3}$ E) &  350.687651 & 11.0 & 0.103 & 2.37 & \multicolumn{3}{c}{\nodata} \\
CH$_3$OH (1$_{1,1}$--0$_{0,0}$ A++) &  350.90507 & 11.1 & 0.098 & 2.07 & \multicolumn{3}{c}{\nodata} \\
HCN ($J$=4--3) &  354.505493 & 11.6 & 0.742 & 2.96 & 10.2 & 0.280 & 10.49 \\
\enddata
\label{FIR3N_table}
\tablenotetext{a}{Peak intensity was fixed in the Gaussian fitting to obtain the best-fit solution.}
\tablenotetext{b}{Velocity width was assumed to be the same as that of CN (3--2 $J$=5/2--3/2 F=7/2--5/2, 340.031567 GHz)}
\tablenotetext{c}{In applying the narrow and wide components to the observed spectrum, the signal to noise ratio of each component became under the 3$\sigma$ noise level. }
\tablenotetext{d}{Although the signal to noise ratio of the peak temperature is over 3, the spectrum is not fitted well by the single Gaussian component due to lack of channel numbers. }
\end{deluxetable}

%% file: Table13.tex
\begin{deluxetable}{lccccccc}
\tablecolumns{6}
\tabletypesize{\tiny}
\tablecaption{Parameters of the Molecular Lines Detected in FIR 3}
\tablewidth{0pt}
\tablehead{
\colhead{}
& \colhead{}
& \multicolumn{3}{c}{$Narrow \ Component$}
& \multicolumn{3}{c}{$Wide \ Component$}\\
\cline{3-8}\\
\colhead{Molecule with Transition}
& \colhead{Rest Freq.}
& \colhead{$V_{\rm sys}$}
& \colhead{$T_{\rm peak}$}
& \colhead{$dV_{\rm FWHM}$}
& \colhead{$V_{\rm sys}$}
& \colhead{$T_{\rm peak}$}
& \colhead{$dV_{\rm FWHM}$}\\
\colhead{}
& \colhead{[GHz]}
& \colhead{[km s$^{-1}$]}
& \colhead{[K]}
& \colhead{[km s$^{-1}$]}
& \colhead{[km s$^{-1}$]}
& \colhead{[K]}
& \colhead{[km s$^{-1}$]}
}
\startdata
CH$_3$OH (5$_{-1,5}$--4$_{0,4}$E) &  84.521206 & 10.8 & 0.744 & 3.13 & \multicolumn{3}{c}{\nodata} \\
c-C$_3$H$_2$ (2$_{1,2}$--1$_{0,1}$) & 85.338906 & 11.2 & 0.274 & 2.03 & \multicolumn{3}{c}{\nodata} \\
CH$_3$CCH (J$_K$=5$_3$--4$_3$) &  85.4426 & 11.2 & 0.050$^{b,e}$  & 2.02 & \multicolumn{3}{c}{\nodata} \\
CH$_3$CCH (J$_K$=5$_2$--4$_2$) & 85.450765 & 11.2 & 0.050$^{b,e}$  & 2.02 & \multicolumn{3}{c}{\nodata} \\
CH$_3$CCH (J$_K$=5$_1$--4$_1$) & 85.455665 & 11.2 & 0.110$^b$ & 2.02 & \multicolumn{3}{c}{\nodata} \\
CH$_3$CCH (J$_K$=5$_0$--4$_0$) &  85.457299 & 11.2 & 0.140$^b$ & 2.02 & \multicolumn{3}{c}{\nodata} \\
HC$^{15}$N ($J$=1--0) & 86.054967 & 11.2 & 0.098 & 3.26 & \multicolumn{3}{c}{\nodata} \\
SO ($N,J$=2,2--1,1) &  86.093983 & 10.1 & 0.065 & 4.78 & \multicolumn{3}{c}{\nodata} \\
H$^{13}$CN ($J$=1--0 $F$=1--1) & 86.3387367 & 11.3 & 0.236 & 2.22 & \multicolumn{3}{c}{\nodata} \\
H$^{13}$CN ($J$=1--0 $F$=2--1) & 86.3401764 & 11.3 & 0.358 & 2.22 & \multicolumn{3}{c}{\nodata} \\
H$^{13}$CN ($J$=1--0 $F$=0--1) & 86.3422551 & 11.3 & 0.052$^e$ & 2.22 & \multicolumn{3}{c}{\nodata} \\
H$^{13}$CO$^+$ ($J$=1-0) & 86.754288 & 11.5 & 0.502 & 2.16 & \multicolumn{3}{c}{\nodata} \\
HN$^{13}$C ($J$=1--0) & 87.090859 & 11.4 & 0.240 & 1.43 & \multicolumn{3}{c}{\nodata} \\
C$_2$H (1--0 3/2--1/2 $F$=1--1) & 87.284156 & 11.3 & 0.247 & 2.45 & \multicolumn{3}{c}{\nodata} \\
C$_2$H (1--0 3/2--1/2 $F$=2--1) & 87.316925 & 11.5 & 1.924 & 2.17 & \multicolumn{3}{c}{\nodata} \\
C$_2$H (1--0 3/2--1/2 $F$=1--0) & 87.328624 & 11.4 & 1.031 & 1.95 & \multicolumn{3}{c}{\nodata} \\
C$_2$H (1--0 1/2--1/2 $F$=1--1) & 87.402004 & 11.3 & 0.943 & 2.17 & \multicolumn{3}{c}{\nodata} \\
C$_2$H (1--0 1/2--1/2 $F$=0--1) & 87.407165 & 11.3 & 0.410 & 2.17 & \multicolumn{3}{c}{\nodata} \\
HNCO (4$_{0,4}$-3$_{0,3}$) & 87.925238 & 11.6 & 0.114 & 1.48 & \multicolumn{3}{c}{\nodata} \\
HCN ($J$=1--0 $F$=1-1) & 88.6304157 & 11.5 & 1.847 & 2.77 & 10.2 & 0.043$^{b,c}$  & 15.32  \\
HCN ($J$=1--0 $F$=2-1) & 88.6318473 & 11.5 & 4.285 & 2.77 & 10.2 & 0.100$^b$ & 15.32 \\
HCN ($J$=1--0 $F$=0-1) & 88.633936 & 11.5 & 0.844 & 2.77 & 10.2 & 0.0197$^{b,c}$  & 15.32 \\
HCO$^+$ ($J$=1--0) &  89.188526 & 11.6 & 4.631 & 2.22 & 10.9 & 0.499 & 10.67 \\
HNC ($J$=1--0) &  90.663564 & 11.3 & 3.372 & 2.24 & \multicolumn{3}{c}{\nodata} \\
HC$_3$N ($J$=10--9) &  90.978989 & 11.2 & 1.331 & 1.90 & \multicolumn{3}{c}{\nodata} \\
CH$_3$CN (J$_{K}$=5$_3$--4$_3$) & 91.971465 & 11.0 & 0.063$^e$ & 2.85 & \multicolumn{3}{c}{\nodata} \\
CH$_3$CN (J$_{K}$=5$_2$--4$_2$) & 91.979994 & 11.0 & 0.084 & 2.85 & \multicolumn{3}{c}{\nodata} \\
CH$_3$CN (J$_{K}$=5$_1$--4$_1$) & 91.985313 & 11.0 & 0.150 & 2.85 & \multicolumn{3}{c}{\nodata} \\
CH$_3$CN (J$_{K}$=5$_0$--4$_0$) & 91.987086 & 11.0 & 0.182 & 2.85 & \multicolumn{3}{c}{\nodata} \\
$^{13}$CS ($J$=2--1) &  92.49427 & 10.9 & 0.118 & 2.59 & \multicolumn{3}{c}{\nodata} \\
N$_2$H$^+$ (1--0 $F_1$=1--1 $F$=0--1) &  93.171621 & 11.4 & 0.088 & 1.64 & \multicolumn{3}{c}{\nodata} \\
N$_2$H$^+$ (1--0 $F_1$=1--1 $F$=2--2) &  93.171917 & 11.4 & 1.060 & 1.64 & \multicolumn{3}{c}{\nodata} \\
N$_2$H$^+$ (1--0 $F_1$=1--1 $F$=1--0) &  93.172053 & 11.4 & 0.490 & 1.64 & \multicolumn{3}{c}{\nodata} \\
N$_2$H$^+$ (1--0 $F_1$=2--1 $F$=2--1) & 93.17348 & 11.4 & 0.818 & 1.64 & \multicolumn{3}{c}{\nodata} \\
N$_2$H$^+$ (1--0 $F_1$=2--1 $F$=3--2) &  93.173777 & 11.4 & 0.952 & 1.64 & \multicolumn{3}{c}{\nodata} \\
N$_2$H$^+$ (1--0 $F_1$=2--1 $F$=1--1) & 93.173967 & 11.4 & 0.931 & 1.64 & \multicolumn{3}{c}{\nodata} \\
N$_2$H$^+$ (1--0 $F_1$=0--1 $F$=1--2) &  93.176265 & 11.4 & 0.644 & 1.64 & \multicolumn{3}{c}{\nodata} \\
CH$_3$CHO (5$_{1,5}$-4$_{1,4}$A) &  93.580914 & 12.3 & 0.111 & 1.26 & \multicolumn{3}{c}{\nodata} \\
CH$_3$OH (8$_{0,8}$--7$_{1,7}$A++) &  95.169516 & 11.0 & 3.038 & 2.39 & \multicolumn{3}{c}{\nodata} \\
CH$_3$OH (2$_{1,2}$--1$_{1,1}$A++) &  95.91431 & 10.3 & 0.062 & 4.25 & \multicolumn{3}{c}{\nodata} \\
C$^{34}$S ($J$=2--1) & 96.41295 & 11.1 & 0.180 & 2.21 & \multicolumn{3}{c}{\nodata} \\
CH$_3$OH (2$_{-1,2}$--1$_{-1,1}$E) & 96.739363 & 11.3 & 0.215 & 3.21& \multicolumn{3}{c}{\nodata} \\
CH$_3$OH (2$_{0,2}$--1$_{0,1}$A++) &  96.741377 & 11.3 & 0.321 & 3.21 & \multicolumn{3}{c}{\nodata} \\
CH$_3$OH (2$_{0,2}$--1$_{0,1}$E) &  96.744549 & 11.3 & 0.118 & 3.21 & \multicolumn{3}{c}{\nodata} \\
CH$_3$OH (2$_{1,1}$--1$_{1,0}$A--) &  97.582808 & 11.1 & 0.076 & 2.99 & \multicolumn{3}{c}{\nodata} \\
CS ($J$=2--1) &  97.980953 & 11.2 & 2.460 & 3.01 & 11.6 & 0.101 & 10.99 \\
SO ($N,J$=2,3-1,2) &  99.299905 & 11.2 & 0.264 & 2.79 & \multicolumn{3}{c}{\nodata} \\
HC$_3$N ($J$=11--10) &  100.076386 & 11.1 & 0.973 & 1.08 & \multicolumn{3}{c}{\nodata} \\
H$_2$CS (3$_{1,3}$--2$_{1,2}$) &  101.477885 & 9.7 & 0.100$^b$ & 1.60 & \multicolumn{3}{c}{\nodata} \\
CH$_3$CCH (J$_K$=6$_3$--5$_3$) &  102.530346 & 10.6 & 0.025$^{b,e}$ & 3.14 & \multicolumn{3}{c}{\nodata} \\
CH$_3$CCH (J$_K$=6$_2$--5$_2$) & 102.540143 & 10.6 & 0.070$^b$ & 3.14 & \multicolumn{3}{c}{\nodata} \\
CH$_3$CCH (J$_K$=6$_1$--5$_1$) &  102.546023 & 10.6 & 0.110$^b$ & 3.14 & \multicolumn{3}{c}{\nodata} \\
CH$_3$CCH (J$_K$=6$_0$--5$_0$) &  102.547983 & 10.6 & 0.140$^b$ & 3.14 & \multicolumn{3}{c}{\nodata} \\
H$_2$CS (3$_{0,3}$--2$_{0,2}$) & 103.040548 & 10.8 & 0.108 & 2.62 & \multicolumn{3}{c}{\nodata} \\
H$_2$CS (3$_{1,2}$--2$_{1,1}$) &  104.617109 & 10.5 & 0.155 & 0.95 & \multicolumn{3}{c}{\nodata} \\
\hline
HDCO (5$_{14}$--4$_{13}$) &  335.096771 & 11.1 & 0.100 & 2.93 & \multicolumn{3}{c}{\nodata} \\
CH$_3$OH (7$_{1,7}$--6$_{1,6}$ A++) &  335.582022 & 11.7 & 0.505 & 2.69 & \multicolumn{3}{c}{\nodata} \\
CH$_3$OH (12$_{1,11}$--12$_{0,12}$ A-+)  &  336.865153 & 12.0 & 0.202 & 3.65 & \multicolumn{3}{c}{\nodata} \\
C$^{17}$O ($J$=3--2) & 337.060974 & 11.1 & 0.898 & 2.10 & \multicolumn{3}{c}{\nodata} \\
C$^{34}$S ($J$=7--6) & 337.396454 & 11.3 & 0.150 & 2.47 & \multicolumn{3}{c}{\nodata} \\
H$_2$CS (10$_{1,10}$--9$_{1,9}$) &  338.083191 & 11.5 & 0.164 & 2.17 & \multicolumn{3}{c}{\nodata} \\
CH$_3$OH (7$_{0,7}$--6$_{0,6}$ E) &  338.124498 & 11.5 & 0.560 & 2.57 & \multicolumn{3}{c}{\nodata} \\
CH$_3$OH (7$_{-1,7}$--6$_{-1,6}$ E) &  338.344605 & 11.6 & 0.746 & 2.66 & \multicolumn{3}{c}{\nodata} \\
CH$_3$OH (7$_{0,7}$--6$_{0,6}$ A++) &  338.408718 & 11.6 & 0.843 & 2.56 & \multicolumn{3}{c}{\nodata} \\
CH$_3$OH (7$_{4,4}$--6$_{4,3}$ A--)$^a$ &  338.512627 & 11.6 & 0.072 & 3.60 & \multicolumn{3}{c}{\nodata} \\
CH$_3$OH (7$_{4,3}$--6$_{4,2}$ A++)$^a$ &  338.512639 & 11.6 & 0.033 & 3.60 & \multicolumn{3}{c}{\nodata} \\
CH$_3$OH (7$_{3,5}$--6$_{3,4}$ A++)$^a$ & 338.540824 & 11.3 & 0.146 & 4.88 & \multicolumn{3}{c}{\nodata} \\
CH$_3$OH (7$_{3,4}$--6$_{3,3}$ E) &  338.583223 & 11.6 & 0.068 & 2.85 & \multicolumn{3}{c}{\nodata} \\
CH$_3$OH (7$_{1,6}$--6$_{1,5}$ E) &  338.614953 & 11.7 & 0.264 & 2.52 & \multicolumn{3}{c}{\nodata} \\
CH$_3$OH (7$_{2,5}$--6$_{2,4}$ A++) &  338.639807 & 11.7 & 0.102 & 2.52 & \multicolumn{3}{c}{\nodata} \\
CH$_3$OH (7$_{2,5}$--6$_{2,4}$ E)$^a$ &  338.721694 & 11.6 & 0.418 & 2.70 & \multicolumn{3}{c}{\nodata} \\
CH$_3$OH (7$_{-2,6}$--6$_{-2,5}$ E)$^a$ &  338.722914 & 11.6 & 0.067 & 2.70 & \multicolumn{3}{c}{\nodata} \\
CN (3--2 $J$=5/2--5/2 $F$=7/2--7/2) &  339.51669 & 11.5 & 0.131 & 1.96 & \multicolumn{3}{c}{\nodata} \\
CN (3--2 $J$=5/2--3/2 $F$=5/2--5/2) &  340.008097 & 11.6 & 0.149 & 2.53 & \multicolumn{3}{c}{\nodata} \\
CN (3--2 $J$=5/2--3/2 $F$=3/2--3/2) &  340.019605 & 11.6 & 0.108 & 2.53 & \multicolumn{3}{c}{\nodata} \\
CN (3--2 $J$=5/2--3/2 $F$=7/2--5/2) &  340.031567 & 11.6 & 1.064 & 2.53 & \multicolumn{3}{c}{\nodata} \\
CN (3--2 $J$=5/2--3/2 $F$=5/2--3/2) &  340.035525 & 11.6 & 0.834 & 2.53 & \multicolumn{3}{c}{\nodata} \\
CN (3--2 $J$=7/2--5/2 $F$=7/2--5/2) &  340.247625 & 11.3 & 2.488 & 2.53$^d$ & \multicolumn{3}{c}{\nodata} \\
CN (3--2 $J$=7/2--7/2 $F$=5/2--5/2) &  340.261818 & 11.3 & 0.165 & 2.53$^d$ & \multicolumn{3}{c}{\nodata} \\
CN (3--2 $J$=7/2--7/2 $F$=7/2--7/2) &  340.265025 & 11.3 & 0.205 & 2.53$^d$ & \multicolumn{3}{c}{\nodata} \\
HC$^{18}$O$^+$ ($J$=4--3) &  340.630707 & 11.3 & 0.074 & 2.16$^d$ & \multicolumn{3}{c}{\nodata} \\
SO ($N,J$=8,7--7,6) & 340.71435 & 11.3 & 0.124 & 1.90 & 8.3 & 0.054$^e$ & 10.81 \\
CH$_3$OH (7$_{1,6}$--6$_{1,5}$ A--) &  341.415641 & 11.7 & 0.656 & 3.10 & \multicolumn{3}{c}{\nodata} \\
CH$_3$OH (13$_{1,12}$--13$_{0,13}$ A-+) &  342.729781 & 12.0 & 0.288 & 3.33 & \multicolumn{3}{c}{\nodata} \\
CS ($J$=7--6) &  342.882843 & 11.5 & 1.899 & 2.59  & 12.9 & 0.563 & 9.09 \\
H$_2$CS (10$_{0,10}$--9$_{0,9}$) &  342.94646 & 11.7 & 0.074 & 1.29 & \multicolumn{3}{c}{\nodata} \\
H$_2^{13}$CO (5$_{1,5}$--4$_{1,4}$) &  343.325714 & 11.6 & 0.099 & 2.64 & \multicolumn{3}{c}{\nodata} \\
H$^{13}$CN ($J$=4--3) &  345.339752 & 11.9 & 0.264 & 4.64 & \multicolumn{3}{c}{\nodata} \\
CO ($J$=3--2) &  345.79599 & 11.4 & 16.419 & 5.50 & 12.3 & 4.727 & 12.56 \\
SO ($N,J$=8,9--7,8) &  346.528587 & 11.2 & 0.129 & 2.07 & 11.2 & 0.058$^e$ & 10.16 \\
H$^{13}$CO$^+$ ($J$=4--3) & 346.998352 & 11.4 & 0.615 & 1.97 & \multicolumn{3}{c}{\nodata} \\
H$_2$CS (10$_{1,9}$--9$_{1,8}$) &  348.534363 & 11.4 & 0.133 & 3.73 & \multicolumn{3}{c}{\nodata} \\
CH$_3$OH (14$_{1,13}$--14$_{0,14}$ A-+) &  349.106954 & 11.7 & 0.114 & 3.81 & \multicolumn{3}{c}{\nodata} \\
C$_2$H (4--3 9/2--7/2 $F$=5--4) &  349.337741 & 11.4 & 1.910 & 2.27 & \multicolumn{3}{c}{\nodata} \\
C$_2$H (4--3 9/2--7/2 $F$=4--3) &  349.339067 & 11.4 & 1.367 & 2.27 & \multicolumn{3}{c}{\nodata} \\
C$_2$H (4--3 7/2--5/2 $F$=4--3) &  349.399342 & 11.6 & 1.132 & 2.28 & \multicolumn{3}{c}{\nodata} \\
C$_2$H (4--3 7/2--5/2 $F$=3--2) &  349.400692 & 11.6 & 1.145 & 2.28 & \multicolumn{3}{c}{\nodata} \\
CH$_3$OH (4$_{0,4}$--3$_{-1,3}$ E) &  350.687651 & 11.3 & 0.320 & 2.86 & \multicolumn{3}{c}{\nodata} \\
CH$_3$OH (1$_{1,1}$--0$_{0,0}$ A++) &  350.90507 & 11.5 & 0.376 & 2.72 & \multicolumn{3}{c}{\nodata} \\
HCN ($J$=4--3) &  354.505493 & 12.0 & 2.890 & 3.08  & 12.1 & 1.151& 8.37 \\
\enddata
\label{FIR3_table}
\tablenotetext{a}{Multi transitions heavily overlap each other, and the line parameters have large uncertainties. Thus, we did not make the rotation diagram.}
\tablenotetext{b}{Peak intensity was fixed in the Gaussian fitting to obtain the best-fit solution.}
\tablenotetext{c}{In the fitting, we assumed that the intensity ratio of the narrow to wide components is the same among the hfs components.}
\tablenotetext{d}{Velocity width was assumed to be the same as that of CN (3--2 $J$=5/2--3/2 F=7/2--5/2, 340.031567 GHz)}
\tablenotetext{e}{In applying the narrow and wide components to the observed spectrum, the signal to noise ratio of each component became under the 3$\sigma$ noise level. Thus, we did not make the rotation diagram.}
\end{deluxetable}

%% file: Table14.tex
\begin{deluxetable}{lccccccc}
\tablecolumns{6}
\tabletypesize{\tiny}
\tablecaption{Parameters of the Molecular Lines Detected in FIR 4}
\tablewidth{0pt}
\tablehead{
\colhead{}
& \colhead{}
& \multicolumn{3}{c}{$Narrow \ Component$}
& \multicolumn{3}{c}{$Wide \ Component$}\\
\cline{3-8}\\
\colhead{Molecule with Transition}
& \colhead{Rest Freq.}
& \colhead{$V_{\rm sys}$}
& \colhead{$T_{\rm peak}$}
& \colhead{$dV_{\rm FWHM}$}
& \colhead{$V_{\rm sys}$}
& \colhead{$T_{\rm peak}$}
& \colhead{$dV_{\rm FWHM}$}\\
\colhead{}
& \colhead{[GHz]}
& \colhead{[km s$^{-1}$]}
& \colhead{[K]}
& \colhead{[km s$^{-1}$]}
& \colhead{[km s$^{-1}$]}
& \colhead{[K]}
& \colhead{[km s$^{-1}$]}
}
\startdata
CH$_3$OH (5$_{-1,5}$--4$_{0,4}$E) &  84.521206 & 11.3 & 0.950 & 0.72 & 11.3 & 0.451 & 4.93 \\
c-C$_3$H$_2$ (2$_{1,2}$--1$_{0,1}$) &  85.338906 & 11.2  & 0.333 & 2.59 & \multicolumn{3}{c}{\nodata} \\
CH$_3$CCH (J$_K$=5$_3$--4$_3$) &  85.4426 & 11.1 & 0.040$^b$$^,$$^e$  & 1.98 & \multicolumn{3}{c}{\nodata} \\
CH$_3$CCH (J$_K$=5$_2$--4$_2$) &  85.450765 & 11.1 & 0.060$^b$& 1.98 & \multicolumn{3}{c}{\nodata} \\
CH$_3$CCH (J$_K$=5$_1$--4$_1$) &  85.455665 & 11.1 & 0.140$^b$ & 1.98 & \multicolumn{3}{c}{\nodata} \\
CH$_3$CCH (J$_K$=5$_0$--4$_0$) &  85.457299 & 11.1 & 0.260$^b$ & 1.98 & \multicolumn{3}{c}{\nodata} \\
NH$_2$D (1(1,1)0+ -1(0,1)0- $F$=0--1) & 85.924747 & 10.8 & 0.133 & 1.54 & \multicolumn{3}{c}{\nodata} \\
NH$_2$D (1(1,1)0+ -1(0,1)0- $F$=2--1) & 85.925684 & 10.8 & 0.161 & 1.54 & \multicolumn{3}{c}{\nodata} \\
NH$_2$D (1(1,1)0+ -1(0,1)0- $F$=2--2) &  85.926263 & 10.8 & 0.534 & 1.54 & \multicolumn{3}{c}{\nodata} \\
NH$_2$D (1(1,1)0+ -1(0,1)0- $F$=1--2) &  85.926858 & 10.8 & 0.208 & 1.54 & \multicolumn{3}{c}{\nodata} \\
NH$_2$D (1(1,1)0+ -1(0,1)0- $F$=1--0) &  85.927721 & 10.8 & 0.157 & 1.54 & \multicolumn{3}{c}{\nodata} \\
HC$^{15}$N ($J$=1--0) &  86.054967 & 11.3 & 0.15$^b$ & 3.08 & \multicolumn{3}{c}{\nodata} \\
SO ($N,J$=2,2--1,1) &  86.093983 & 12.1 & 0.045$^b$ & 3.29 & 7.6 & 0.045$^b$ & 8.05 \\
H$^{13}$CN ($J$=1--0 $F$=1--1) &  86.3387367 & 11.3 & 0.358 & 1.83  & 8.7 & 0.041$^b$ & 7.12 \\
H$^{13}$CN ($J$=1--0 $F$=2--1) &  86.3401764 & 11.3 & 0.613 & 1.83 & 8.7 & 0.07$^b$ & 7.12 \\
H$^{13}$CN ($J$=1--0 $F$=0--1) &  86.3422551 & 11.3 & 0.100 & 1.83 & 8.7 & 0.011$^b$$^,$$^e$  & 7.12 \\
H$^{13}$CO$^+$ ($J$=1-0) &  86.754288 & 11.2 & 0.610 & 1.85 & \multicolumn{3}{c}{\nodata} \\
SiO ($J$=2--1) & 86.846995 & 11.5 & 0.104 & 2.78 & 7.3 & 0.133 & 8.84 \\
HN$^{13}$C ($J$=1--0) & 87.090859 & 11.4 & 0.463 & 1.62 & \multicolumn{3}{c}{\nodata} \\
C$_2$H (1--0 3/2--1/2 $F$=1--1) &  87.284156 & 11.2 & 0.331 & 1.85 & \multicolumn{3}{c}{\nodata} \\
C$_2$H (1--0 3/2--1/2 $F$=2--1) &  87.316925 & 11.4 & 2.415 & 1.84 & \multicolumn{3}{c}{\nodata} \\
C$_2$H (1--0 3/2--1/2 $F$=1--0) &  87.328624 & 11.3 & 1.310 & 1.67 & \multicolumn{3}{c}{\nodata} \\
C$_2$H (1--0 1/2--1/2 $F$=1--1) &  87.402004 & 11.2 & 1.420 & 1.68 & \multicolumn{3}{c}{\nodata} \\
C$_2$H (1--0 1/2--1/2 $F$=0--1) &  87.407165 & 11.2 & 0.600 & 1.68 & \multicolumn{3}{c}{\nodata} \\
HNCO (4$_{0,4}$-3$_{0,3}$) &  87.925238 & 10.9 & 0.083 & 3.15 & \multicolumn{3}{c}{\nodata} \\
HCN ($J$=1--0 $F$=1-1) &  88.6304157 & 11.5 & 2.256 & 3.43  & 7.3 & 0.090$^c$ & 13.64 \\
HCN ($J$=1--0 $F$=2-1) &  88.6318473 & 11.5 & 5.001 & 3.43 & 7.3 & 0.2$^c$ & 13.64 \\
HCN ($J$=1--0 $F$=0-1) & 88.633936 & 11.5 & 1.470 & 3.43 & 7.3 & 0.0588$^c$ & 13.64 \\
H$^{15}$NC ($J$=1-0) &  88.865692 & 11.5 & 0.090$^b$ & 1.82 & \multicolumn{3}{c}{\nodata} \\
HCO$^+$ ($J$=1--0) & 89.188526 & 11.4 & 4.696 & 1.85 & 10.0 & 0.786 & 8.12 \\
HNC ($J$=1--0) &  90.663564 & 11.3 & 4.891 & 2.01 & \multicolumn{3}{c}{\nodata} \\
HC$_3$N ($J$=10--9) &  90.978989 & 11.0 & 1.946 & 1.44 & 10.8 & 0.104 & 9.27 \\
CH$_3$CN (J$_{K}$=5$_3$--4$_3$ $F$=4--3) &  91.971465 & 11.4 & 0.055 & 2.16 & \multicolumn{3}{c}{\nodata} \\
CH$_3$CN (J$_{K}$=5$_2$--4$_2$ $F$=6--5) &  91.979994 & 11.4 & 0.087 & 2.16 & \multicolumn{3}{c}{\nodata} \\
CH$_3$CN (J$_{K}$=5$_1$--4$_1$) &  91.985313 & 11.4 & 0.185 & 2.16 & \multicolumn{3}{c}{\nodata} \\
CH$_3$CN (J$_{K}$=5$_0$--4$_0$) &  91.987086 & 11.4 & 0.278 & 2.16 & \multicolumn{3}{c}{\nodata} \\
$^{13}$CS ($J$=2--1) & 92.49427 & 10.6 & 0.111 & 1.89 & \multicolumn{3}{c}{\nodata} \\
N$_2$H$^+$ (1--0 $F_1$=1--1 $F$=0--1) &  93.171621 & 11.3 & 0.823 & 1.43 & \multicolumn{3}{c}{\nodata} \\
N$_2$H$^+$ (1--0 $F_1$=1--1 $F$=2--2) &  93.171917 & 11.3 & 4.477 & 1.43 & \multicolumn{3}{c}{\nodata} \\
N$_2$H$^+$ (1--0 $F_1$=1--1 $F$=1--0) &  93.172053 & 11.3 & 1.211 & 1.43 & \multicolumn{3}{c}{\nodata} \\
N$_2$H$^+$ (1--0 $F_1$=2--1 $F$=2--1) &  93.17348 & 11.3 & 4.572 & 1.43 & \multicolumn{3}{c}{\nodata} \\
N$_2$H$^+$ (1--0 $F_1$=2--1 $F$=3--2) &  93.173777 & 11.3 & 2.380 & 1.43 & \multicolumn{3}{c}{\nodata} \\
N$_2$H$^+$ (1--0 $F_1$=2--1 $F$=1--1) &  93.173967 & 11.3 & 3.423 & 1.43  & \multicolumn{3}{c}{\nodata} \\
N$_2$H$^+$ (1--0 $F_1$=0--1 $F$=1--2) &  93.176265 & 11.3 & 2.633 & 1.43 & \multicolumn{3}{c}{\nodata} \\
CH$_3$CHO (5$_{1,5}$-4$_{1,4}$A) &  93.580914 & 10.7 & 0.078 & 3.81 & \multicolumn{3}{c}{\nodata} \\
CH$_3$CHO (5$_{-1,5}$--4$_{-1,4}$E) &  93.595238 & 11.2 & 0.085 & 2.34 & \multicolumn{3}{c}{\nodata} \\
CH$_3$OH (8$_{0,8}$--7$_{1,7}$A++) &  95.169516 & 11.5 & 1.581 & 2.59  & 10.6 & 0.302 & 5.63 \\
CH$_3$OH (2$_{1,2}$--1$_{1,1}$A++) &  95.91431 & 11.9 & 0.100 & 2.93 & 11.0 & 0.096 & 4.89 \\
C$^{34}$S ($J$=2--1) &  96.41295 & 11.3 & 0.303 & 2.12 & 11.1 & 0.055 & 6.99 \\
CH$_3$OH (2$_{-1,2}$--1$_{-1,1}$E) &  96.739363 & 11.6 & 0.524 & 2.97 & 9.1 & 0.135$^c$ & 5.42 \\
CH$_3$OH (2$_{0,2}$--1$_{0,1}$A++) &  96.741377 & 11.6 & 0.776 & 2.97 & 9.1 & 0.2$^c$ & 5.42 \\
CH$_3$OH (2$_{0,2}$--1$_{0,1}$E) &  96.744549 & 11.6 & 0.334 & 2.97 & 9.1 & 0.086$^c$ & 5.42 \\
CH$_3$OH (2$_{1,1}$--1$_{1,0}$E) &  96.755507 & 11.4 & 0.148 & 2.67 & 12.0 & 0.045 & 11.64 \\
CH$_3$OH (2$_{1,1}$--1$_{1,0}$A--) &  97.582808 & 11.8 & 0.204 & 2.83 & 10.0 & 0.083 & 6.31 \\
CS ($J$=2--1) &  97.980953 & 11.4 & 2.919 & 2.97 & 9.6 & 0.813 & 7.23 \\
CH$_3$CHO (5$_{1,4}$--4$_{1,3}$E) &  98.863314 & 11.3 & 0.054 & 3.11 & \multicolumn{3}{c}{\nodata} \\
CH$_3$CHO (5$_{1,5}$--4$_{1,3}$A--) &  98.900951 & 10.9 & 0.058 & 3.52 & \multicolumn{3}{c}{\nodata} \\
SO ($N,J$=2,3-1,2) &  99.299905 & 11.2 & 0.322 & 2.42 & 8.5 & 0.184 & 8.93 \\
HC$_3$N ($J$=11--10) &  100.076386 & 11.1 & 1.568 & 1.52 & 9.7 & 0.124 & 8.20 \\
H$_2$CO (6$_{1,5}$-6$_{1,6}$) &  101.332993 & 10.6 & 0.094 & 3.84 & \multicolumn{3}{c}{\nodata} \\
H$_2$CS (3$_{1,3}$--2$_{1,2}$)   & 101.477885 & 9.6 & 0.132 & 3.62 & \multicolumn{3}{c}{\nodata}\\
CH$_3$CCH (J$_K$=6$_3$--5$_3$) &  102.530346 & 10.2 & 0.047 & 2.91 & \multicolumn{3}{c}{\nodata} \\
CH$_3$CCH (J$_K$=6$_2$--5$_2$) & 102.540143 & 10.2 & 0.119 & 2.91 & \multicolumn{3}{c}{\nodata} \\
CH$_3$CCH (J$_K$=6$_1$--5$_1$) &  102.546023 & 10.2 & 0.167 & 2.91 & \multicolumn{3}{c}{\nodata} \\
CH$_3$CCH (J$_K$=6$_0$--5$_0$) &  102.547983 & 10.2 & 0.199 & 2.91 & \multicolumn{3}{c}{\nodata} \\
H$_2$CS (3$_{0,3}$--2$_{0,2}$) &      103.040548 & 10.6 & 0.131 & 2.74 & \multicolumn{3}{c}{\nodata}   \\
H$_2$CS (3$_{1,2}$--2$_{1,1}$) & 104.617109 & 10.6 & 0.182 & 3.59 & \multicolumn{3}{c}{\nodata}  \\
\hline
HDCO (5$_{14}$--4$_{13}$) &  335.096771 & 10.4 & 0.099 & 3.45  & \multicolumn{3}{c}{\nodata} \\
CH$_3$OH (2$_{2,1}$--3$_{1,2}$ A--) &  335.133686 & 11.9 & 0.043$^e$  & 1.26 & 12.5 & 0.062$^e$  & 7.48 \\
CH$_3$OH (7$_{1,7}$--6$_{1,6}$ A++) &  335.582022 & 11.7 & 0.649 & 1.41  & 11.4 & 0.716 & 5.25 \\
CH$_3$OH (12$_{1,11}$--12$_{0,12}$ A-+)  &  336.865153 & 12.2 & 0.213 & 1.90 & 11.9 & 0.315 & 6.37 \\
C$^{17}$O ($J$=3--2) &  337.060974 & 11.0 & 1.412 & 1.78  & \multicolumn{3}{c}{\nodata} \\
C$^{34}$S ($J$=7--6) &  337.396454 & 11.1 & 0.074 & 1.62 & 12.0 & 0.092 & 9.45 \\
H$_2$CS (10$_{1,10}$--9$_{1,9}$) & 338.083191 & 11.5 & 0.114 & 1.41 & 11.4 & 0.0728 & 9.60 \\
CH$_3$OH (7$_{0,7}$--6$_{0,6}$ E) &  338.124498 & 11.7  & 0.780 & 1.58 & 11.2 & 0.624 & 5.59 \\
CH$_3$OH (7$_{-1,7}$--6$_{-1,6}$ E) &  338.344605 & 11.6 & 1.374 & 1.59 & 10.9 & 1.057 & 5.64 \\
CH$_3$OH (7$_{0,7}$--6$_{0,6}$ A++) &  338.408718 & 11.6 & 1.388 & 1.57 & 10.8 & 1.128 & 5.51 \\
CH$_3$OH (7$_{4,4}$--6$_{4,3}$ A-)$^a$ &  338.512627 & 11.6 & 0.055 & 1.42  & 11.4 & 0.097$^c$ & 5.93 \\
CH$_3$OH (7$_{4,3}$--6$_{4,2}$ A++)$^a$ & 338.512639 & 11.6 & 0.094 & 1.42 & 11.4 & 0.167$^c$ & 5.93 \\
CH$_3$OH (7$_{3,5}$--6$_{3,4}$ A++)$^a$ &  338.540824 & 11.6 & 0.083 & 1.17 & 11.6 & 0.220$^c$ & 4.66 \\
CH$_3$OH (7$_{3,4}$--6$_{3,3}$ A--)$^a$ &  338.543149 & 11.6 & 0.075 & 1.17 & 11.6 & 0.20$^c$ & 4.66 \\
CH$_3$OH (7$_{3,4}$--6$_{3,3}$ E) & 338.583223 & 11.9 & 0.042$^e$ & 0.59 & 12.3 & 0.186 & 4.15 \\
CH$_3$OH (7$_{1,6}$--6$_{1,5}$ E) & 338.614953 & 11.7 & 0.434 & 1.53 & 11.4 & 0.446 & 5.13 \\
CH$_3$OH (7$_{2,5}$--6$_{2,4}$ A++) &  338.639807 & 11.7 & 0.140 & 1.53 & 11.4 & 0.178 & 5.13 \\
CH$_3$OH (7$_{2,5}$--6$_{2,4}$ E) &  338.721694 & 11.6 & 0.586 & 1.65 & 11.3 & 0.579 & 5.66 \\
CH$_3$OH (7$_{-2,6}$--6$_{-2,5}$ E) &  338.722914 & 11.6  & 0.220 & 1.65 & 11.3 & 0.217 & 5.66 \\
CN (3--2 $J$=5/2--5/2 $F$=7/2--7/2) &  339.51669 & 11.3 & 0.214 & 1.83  & \multicolumn{3}{c}{\nodata} \\
CN (3--2 $J$=5/2--3/2 F=5/2--5/2) &  340.008097 & 11.4 & 0.219 & 2.07  & \multicolumn{3}{c}{\nodata} \\
CN (3--2 $J$=5/2--3/2 F=3/2--3/2) &  340.019605 & 11.4 & 0.206 & 2.07  & \multicolumn{3}{c}{\nodata} \\
CN (3--2 $J$=5/2--3/2 F=7/2--5/2) &  340.031567 & 11.4 & 1.323 & 2.07  & \multicolumn{3}{c}{\nodata} \\
CN (3--2 $J$=5/2--3/2 F=3/2--1/2)$^a$ &  340.035281 & 11.4 & 0.390 & 2.07  & \multicolumn{3}{c}{\nodata} \\
CN (3--2 $J$=5/2--3/2 F=5/2--3/2)$^a$ &  340.035525 & 11.4 & 0.812 & 2.07   & \multicolumn{3}{c}{\nodata} \\
CH$_3$OH (2$_{2,0}$--3$_{1,3}$ A++) &  340.141288 & 11.2 & 0.067 & 2.06  & \multicolumn{3}{c}{\nodata} \\
CN (3--2 $J$=7/2--5/2 F=7/2--5/2) & 340.247625 & 11.2 & 2.874 & 2.07$^d$  & \multicolumn{3}{c}{\nodata} \\
CN (3--2 $J$=7/2--7/2 F=5/2--5/2) & 340.261818 & 11.2 & 0.201 & 2.07$^d$  & \multicolumn{3}{c}{\nodata} \\
CN (3--2 $J$=7/2--7/2 F=7/2--7/2) &  340.265025 & 11.2 & 0.322 & 2.07$^d$  & \multicolumn{3}{c}{\nodata} \\
SO ($N,J$=8,7--7,6) &  340.71435 & 11.4 & 0.066 & 1.64 & 9.4 & 0.187 & 7.85 \\
HCS$^{+}$ ($J$=8--7) &  341.35022 & 11.2 & 0.070$^b$ & 3.01  & \multicolumn{3}{c}{\nodata} \\
CH$_3$OH (7$_{1,6}$--6$_{1,5}$ A--) & 341.415641 & 11.7 & 0.539 & 1.49 & 11.2 & 0.557 & 5.16 \\
CH$_3$OH (13$_{1,12}$--13$_{0,13}$ A-+) & 342.729781 & 12.2 & 0.212 & 2.82 & 11.7 & 0.240 & 6.40 \\
CS ($J$=7--6) &  342.882843 & 11.4 & 1.647 & 1.67 & 10.7 & 1.497 & 8.20 \\
H$_2$CS (10$_{0,10}$--9$_{0,9}$) & 342.94646 & 12.0 & 0.035$^e$ & 2.11 & 11.0 & 0.034$^e$ & 7.50 \\
H$_2^{13}$CO (5$_{1,5}$--4$_{1,4}$) &  343.325714 & 11.6 & 0.134 & 1.70 & 12.1 & 0.046$^e$ & 8.75 \\
HC$^{15}$N ($J$=4--3) &  344.200104 & 11.2 & 0.052 & 3.02 & 10.7  & 0.058 & 17.31 \\
H$^{13}$CN ($J$=4--3) & 345.339752 & 11.4 & 0.259 & 1.68 & 10.6 & 0.305 & 8.02 \\
CO ($J$=3--2) &  345.79599 & 10.8 & 20.649 & 4.13 & 9.5 & 10.186 & 11.45 \\
SO ($N,J$=8,9--7,8) &  346.528587 & 11.4 & 0.132 & 1.70  & 8.9 & 0.261 & 7.92 \\
H$^{13}$CO$^+$ ($J$=4--3) &  346.998352 & 11.2 & 0.660 & 1.63  & \multicolumn{3}{c}{\nodata} \\
SiO ($J$=8--7) &  347.330688 & 11.3 & 0.077 & 1.83 & 3.1 & 0.087 & 13.99 \\
HN$^{13}$C ($J$=4--3) &  348.340485 & 10.8 & 0.218 & 1.24  & \multicolumn{3}{c}{\nodata} \\
H$_2$CS (10$_{1,9}$--9$_{1,8}$) &  348.534363 & 11.3 & 0.092 & 1.51  & 9.8 & 0.0717 & 9.95 \\
CH$_3$OH (14$_{1,13}$--14$_{0,14}$ A-+) &  349.106954 & 12.2 & 0.120 & 1.32 & 11.7 & 0.276 & 5.42 \\
C$_2$H (4--3 9/2--7/2 $F$=5--4) &  349.337741 & 11.1 & 2.054 & 2.22   & \multicolumn{3}{c}{\nodata} \\
C$_2$H (4--3 9/2--7/2 $F$=4--3) &  349.339067 & 11.1 & 0.746 & 2.22  & \multicolumn{3}{c}{\nodata} \\
C$_2$H (4--3 7/2--5/2 $F$=4--3) & 349.399342 & 11.0 & 1.456 & 2.44  & \multicolumn{3}{c}{\nodata} \\
C$_2$H (4--3 7/2--5/2 $F$=3--2) &  349.400692 & 11.0 & 0.360 & 2.44  & \multicolumn{3}{c}{\nodata} \\
CH$_3$OH (4$_{0,4}$--3$_{-1,3}$ E) &  350.687651 & 11.4 & 0.487 & 1.36 & 10.7 & 0.316 & 4.93 \\
CH$_3$OH (1$_{1,1}$--0$_{0,0}$ A++) &  350.90507 & 11.5 & 0.769 & 1.65 & 10.8  & 0.460 & 6.16 \\
HCN ($J$=4--3) & 354.505493 & 12.4 & 2.385 & 1.96 & 10.5 & 3.459 & 9.69 \\
\enddata
\label{FIR4_table}
\tablenotetext{a}{Multi transitions heavily overlap each other, and the line parameters have large uncertainties. Thus, we did not make the rotation diagram.}
\tablenotetext{b}{Peak intensity was fixed in the Gaussian fitting to obtain the best-fit solution.}
\tablenotetext{c}{In the fitting, we assumed that the intensity ratio of the narrow to wide components is the same among the hfs components.}
\tablenotetext{d}{Velocity width was assumed to be the same as that of CN (3--2 $J$=5/2--3/2 F=7/2--5/2, 340.031567 GHz)}
\tablenotetext{e}{In applying the narrow and wide components to the observed spectrum, the signal to noise ratio of each component became under the 3$\sigma$ noise level. Thus, we did not make the rotation diagram.}
\end{deluxetable}